%% file: main.tex
\documentclass[floatfix,superscriptaddress,twocolumn,pra,nofootinbib]{revtex4-2}

\usepackage{amsmath,amssymb,amsthm,bm,mathrsfs,stmaryrd,physics}
\usepackage[dvipsnames,table]{xcolor}
\usepackage{graphicx,adjustbox,multirow,booktabs,array,tikz,pgfplots,comment,fvextra}
\usepackage[caption=false]{subfig}
\usepackage[inline]{enumitem}
\usepackage[yyyymmdd]{datetime}
\usepackage[ruled,vlined,linesnumbered]{algorithm2e}
\usepackage[most]{tcolorbox}
\usepackage[
  colorlinks=true,
  hypertexnames=false,
  linkcolor=RoyalBlue,
  citecolor=ForestGreen,
  urlcolor=BrickRed
]{hyperref}
\newsavebox{\widetablebox}
\newcommand{\widetable}[1]{\sbox{\widetablebox}{#1}%
  \ifdim0.8889\wd\widetablebox>\textwidth\resizebox{\textwidth}{!}{\usebox{\widetablebox}}%
  \else\scalebox{0.8889}{\usebox{\widetablebox}}\fi}
\input{includes/generated/appendix_numbers.tex}

\pgfplotsset{compat=1.18}
\usetikzlibrary{quantikz,fit,backgrounds,positioning,decorations.pathreplacing,calc,arrows.meta,shapes.geometric}
\setlist[enumerate]{label=(\arabic*)}

\hypersetup{breaklinks=true,colorlinks=true,allcolors=BlueViolet}
\tcbset{
  qrmblock/.style={breakable,enhanced,colback=black!2,boxrule=0pt,
    left=6pt,right=6pt,top=6pt,bottom=6pt,sharp corners},
  qrmblocksmall/.style={qrmblock,fontupper=\scriptsize}
}
\DefineVerbatimEnvironment{QRMList}{Verbatim}{
  breaklines,breakanywhere,fontsize=\scriptsize
}

\definecolor{greenBg}{RGB}{200,235,200}
\definecolor{burntOrangeBg}{RGB}{235,190,140}
\definecolor{purpleBg}{RGB}{220,200,240}
\definecolor{blueBg}{RGB}{200,220,255}
\definecolor{yellowBg}{RGB}{255,255,190}
\definecolor{oliveBg}{RGB}{215,230,180}
\definecolor{goldBg}{RGB}{255,235,170}
\definecolor{redBg}{RGB}{255,195,195}
\definecolor{darkblue}{RGB}{0,0,139}
\definecolor{darkred}{RGB}{150,0,0}
\newcolumntype{G}{>{\columncolor{green!15}}c}
\newcolumntype{B}{>{\columncolor{blue!12}}c}
\newcolumntype{L}[1]{>{\raggedright\arraybackslash}p{#1}}

\theoremstyle{definition}
\newtheorem{theorem}{Theorem}[section]
\newtheorem{corollary}[theorem]{Corollary}
\newtheorem{proposition}[theorem]{Proposition}
\newtheorem{definition}[theorem]{Definition}

\newtheorem{remark}[theorem]{Remark}
\newtheorem{lemma}[theorem]{Lemma}
\newtheorem{construction}[theorem]{Construction}

\usepackage[capitalise]{cleveref}
\crefname{theorem}{Theorem}{Theorems}
\crefname{corollary}{Corollary}{Corollaries}
\crefname{proposition}{Proposition}{Propositions}
\crefname{definition}{Definition}{Definitions}
\crefname{example}{Example}{Examples}
\crefname{notation}{Notation}{Notations}
\crefname{remark}{Remark}{Remarks}
\crefname{lemma}{Lemma}{Lemmas}
\crefname{construction}{Construction}{Constructions}
\crefname{appendix}{Appendix}{Appendices}
\crefname{algocf}{Algorithm}{Algorithms}

\newcommand{\code}[1]{\llbracket #1 \rrbracket}
\DeclareRobustCommand{\Mfac}[1]{\ensuremath{\overline{\mathcal{M}}_{\code{#1}}}}
\DeclareRobustCommand{\Tofs}{\ensuremath{\mathrm{Tof}_{\#}}}
\DeclareRobustCommand{\twoCCZ}{\ensuremath{2CCZ}}

\newcommand{\panelimage}[3][]{%
  \begin{tikzpicture}
    \node[inner sep=0pt,anchor=south west] (img) at (0,0)
      {\includegraphics[#1]{#2}};
    \node[overlay,anchor=north west,font=\bfseries\small]
      at ([xshift=-7pt,yshift=8pt]img.north west) {#3};
  \end{tikzpicture}%
}

\newcommand{\GTAR}{Global Technology Applied Research, JPMorganChase, New York, NY 10017, USA}
\newcommand{\VADP}{Department of Physics, Virginia Tech, Blacksburg, VA, USA 24061}

\begin{document}
\title{Tailored Magic States for Algorithmic Primitives}
\author{Matthew Steinberg}
\affiliation{\GTAR}
\author{Joseph Sullivan}
\affiliation{\GTAR}
\author{Chunjun Cao}
\affiliation{\VADP}
\author{Michael A. Perlin}
\affiliation{\GTAR}

\input{sections/00_abstract}

\maketitle
{\footnotesize
\tableofcontents
}

\input{sections/01_introduction}

\input{sections/02_background}

\input{sections/03_csstp_codes}

\input{sections/04_tailored_primitives}

\input{sections/05_discussion}

\input{sections/06_backmatter}

\renewcommand{\bibfont}{\footnotesize}
\bibliography{main.bib}

\onecolumngrid
\appendix
\crefalias{section}{appendix}
\crefalias{subsection}{appendix}
\crefalias{subsubsection}{appendix}

\input{sections/A_conjoining_pcm}

\input{sections/B_protocol_proofs}

\input{sections/C_csstp_examples}

\input{sections/D_magic_properties}

\input{sections/E_distance_growth}

\input{sections/F_random_walk}

\input{sections/G_resource_model}

\input{sections/H_extended_benchmarks}
\input{sections/I_fourth_level}

\end{document}

%% file: includes/generated/appendix_numbers.tex
\newcommand{\apxaccept}{0.280}
\newcommand{\apxattemptsperkept}{3.58}

\newcommand{\apxd}{21}
\newcommand{\apxdeltar}{21}

\newcommand{\apxgidneyqr}{30{,}000}

\newcommand{\apxkeptlen}{18}
\newcommand{\apxkeptmedian}{54}
\newcommand{\apxkeptmin}{18}
\newcommand{\apxkeptpfirst}{28.0}
\newcommand{\apxkeptpninety}{144}
\newcommand{\apxlaneq}{886}
\newcommand{\apxlanes}{32}
\newcommand{\apxlitsteps}{11}
\newcommand{\apxlittiles}{11.5}
\newcommand{\apxmaxreldiff}{0.1}
\newcommand{\apxmaxrelvol}{0.2}
\newcommand{\apxmeankept}{64.4}

\newcommand{\apxpLd}{1.0\times10^{-12}}
\newcommand{\apxpLtwentythree}{1.0\times10^{-13}}
\newcommand{\apxpT}{1.45\times10^{-6}}
\newcommand{\apxpphys}{1.0\times10^{-3}}
\newcommand{\apxpth}{1.0\times10^{-2}}
\newcommand{\apxqrperkept}{57{,}042}
\newcommand{\apxrate}{0.50}

\newcommand{\apxseightAfour}{56}
\newcommand{\apxseightAtwo}{28}
\newcommand{\apxseightEthree}{4.19\times10^{-9}}
\newcommand{\apxseightEtwo}{1.18\times10^{-10}}
\newcommand{\apxseightF}{10}
\newcommand{\apxseightTR}{4{,}193}
\newcommand{\apxseightactive}{16}
\newcommand{\apxseightbuffp}{14{,}112}

\newcommand{\apxseighteE}{5.89\times10^{-11}}
\newcommand{\apxseightendone}{192.4}

\newcommand{\apxseightfacTR}{2{,}940}
\newcommand{\apxseightfacfp}{11{,}466}
\newcommand{\apxseightfp}{53{,}930}
\newcommand{\apxseightk}{3}
\newcommand{\apxseightlaneR}{1{,}104}
\newcommand{\apxseightlaneV}{978{,}395}
\newcommand{\apxseightlanefp}{28{,}352}
\newcommand{\apxseightlaneseight}{384}
\newcommand{\apxseightlanesone}{1{,}156}
\newcommand{\apxseightlanessixtyfour}{359}
\newcommand{\apxseightlat}{360}
\newcommand{\apxseightlead}{5.89\times10^{-11}}
\newcommand{\apxseightmx}{1}
\newcommand{\apxseightn}{8}
\newcommand{\apxseightq}{0.999988}
\newcommand{\apxseightrawTR}{1{,}127}
\newcommand{\apxseightrej}{1.16\times10^{-5}}
\newcommand{\apxseightretry}{0.00}
\newcommand{\apxseightrunEthree}{1.47\times10^{-9}}

\newcommand{\apxseightrunTR}{1{,}470}

\newcommand{\apxseightstoreTR}{126}
\newcommand{\apxseighttau}{7}
\newcommand{\apxseighttaur}{147}
\newcommand{\apxseighttileV}{3{,}697{,}952}
\newcommand{\apxseighttn}{45.4}
\newcommand{\apxseighttnexact}{45.4}

\newcommand{\apxseightvol}{4.68}
\newcommand{\apxstep}{21}
\newcommand{\apxsupplyratio}{8}

\newcommand{\apxtfourteenactive}{28}

\newcommand{\apxtfourteenlaneseight}{416}
\newcommand{\apxtfourteenlanesone}{1{,}161}
\newcommand{\apxtfourteenlanessixtyfour}{341}

\newcommand{\apxtfourteenlatexact}{340.0}

\newcommand{\apxtfourteentau}{13}
\newcommand{\apxtfourteentaur}{273}

\newcommand{\apxtfourteentnexact}{46.0}

\newcommand{\apxtile}{882}

\newcommand{\apxttwelveEthree}{3.32\times10^{-9}}
\newcommand{\apxttwelveEtwo}{1.39\times10^{-10}}

\newcommand{\apxttwelveTR}{3{,}323}
\newcommand{\apxttwelveactive}{24}

\newcommand{\apxttwelvefacTR}{3{,}003}

\newcommand{\apxttwelvelaneR}{887}

\newcommand{\apxttwelvelatexact}{297.9}

\newcommand{\apxttwelverawTR}{215}

\newcommand{\apxttwelvestoreTR}{105}
\newcommand{\apxttwelvetau}{11}
\newcommand{\apxttwelvetaur}{231}

\newcommand{\apxttwelvetnexact}{45.9}
\newcommand{\apxttwelvevol}{3.72}
\newcommand{\apxvtwoperiod}{484}
\newcommand{\apxvtwoq}{121}

%% file: sections/00_abstract.tex
\begin{abstract}
Realizing quantum algorithms in fault-tolerant quantum computers requires the frequent use of algorithmic primitives such as adders, multipliers, and Boolean logic, which typically contain clusters of multi-qubit non-Clifford gates.
Decomposition of such clusters into $T$ and $CCZ$ gates can add scheduling and spacetime overhead.
Synthillation addresses these issues for diagonal circuits at the third level of the Clifford hierarchy by jointly synthesizing and distilling their non-Clifford content.
In this work, we develop a complementary, code-theoretic approach.
We introduce CSS-T-Product (CSS-TP) codes, whose transversal single-qubit phase gates implement multi-qubit logical diagonal gates.
We develop analytical rules to predict the resulting logical gates from tensor-network operations.
Using quantum LEGO, we generate code families with logical actions at the third and fourth levels of the Clifford hierarchy.
These resource states are classified into three categories: \emph{canonical} (such as $T$ or $CCZ$), tensor products of canonical, and \emph{irreducible} states that cannot be reduced to a product of non-Clifford gates via Clifford operations and logical basis transformations.
Conversion into Encode-$T$-Decode factories recovers known synthillation resources and gives procedures for preparing magic states at higher levels of the Clifford hierarchy; the corresponding magic-state factories are then analyzed with a surface-code-based model for various algorithmic primitives.
Our findings show that tailored magic factories can provide significant latency improvements at a similar or smaller occupied spacetime volume.
As an example, at code distance $21$ with one factory per resource family, tailored resources reduce the latency of a $GF(2^{8})$ multiplier and a shared-addend carry-lookahead adder by up to $48\%$ and $38\%$, respectively, and the smallest tailored factory also lowers the occupied spacetime volume.
CSS-TP codes therefore connect code construction to resource factories and the realization of specific algorithmic primitives.
\end{abstract}

%% file: sections/01_introduction.tex
\section{Introduction}

The production of high-fidelity non-Clifford resources is a major source of overhead in fault-tolerant quantum computation~\cite{campbell2017roads,ogorman2017quantum,cain2026shor}.
In architectures where Clifford operations are comparatively inexpensive, universal computation is commonly supplied by preparing and consuming a small set of canonical magic states, such as $\ket{T}$ and $\ket{CCZ}$~\cite{bravyi2005universal,bravyi2012magic,jones2013multilevel,gidney2019efficient,litinski2019magic}.
This abstraction has proved extremely effective: factories can be optimized independently of the target algorithm, while algorithms can be compiled into a standard gate set.
At the same time, this approach treats the non-Clifford content of an algorithm as a collection of elementary gates, even when those gates originate from a larger structured primitive.

Many quantum algorithms, including Shor's algorithm~\cite{shor1999polynomial} and Grover search~\cite{grover1996fast}, repeatedly invoke subroutines such as arithmetic circuits, incrementers, comparators, phase oracles, and reversible Boolean logic~\cite{cuccaro2004new,wang2025comprehensive}.
Their non-Clifford operations often occur in correlated patterns, including clusters of overlapping controlled-phase gates.
Decomposing each cluster into independently prepared canonical resources can discard this structure and introduce additional resource, scheduling, and routing overhead.
This observation motivates a particular design question: can a magic-state factory produce a correlated resource that is matched directly to a recurring algorithmic primitive or pattern, and if so, under what specific conditions is it favorable to utilize such states?

\input{includes/table_differences}

\emph{Synthillation}~\cite{campbell2017unified,campbell2017unifying} provides one answer to this question by directly synthesizing and distilling diagonal phase gates at the third level of the Clifford hierarchy, rather than assembling them from elementary resources.
In its current form, however, synthillation is formulated for the third level of the Clifford hierarchy and relies on a joint optimization of multi-qubit gate synthesis and $T$-count, which becomes harder as the distance and the size of the phase polynomial grow~\cite{campbell2017unified,campbell2017unifying}.

In this work, we introduce a complementary, code-centric perspective to synthillation using \emph{CSS-T-Product} (CSS-TP) codes.
Throughout, a \emph{transversal diagonal layer}, or simply a \emph{layer}, is a product of single-qubit phase gates $\mathrm{diag}(1,e^{i\theta_{j}})$, one on each physical qubit $j$ of a code, such as a pattern of $T$ and $T^{\dagger}$ gates; when it preserves the code space, it acts as a logical diagonal gate (\cref{section:gate_calculus} makes this precise).
A CSS-TP code is a CSS code for which a transversal diagonal layer induces a logical phase gate whose magic state is not a single canonical state (such as $\ket{T}$ or $\ket{CCZ}$) but rather a tensor product of several canonical states, or an \emph{irreducible} state that free Clifford unitaries cannot split into canonical ones.
Whereas synthillation begins with a logical phase operation and constructs a synthesis-and-distillation protocol for it, CSS-TP begins with a quantum code and derives the logical phase circuit generated by its transversal action.

The two approaches overlap: our smallest third-level constructions recover code-equivalent forms of known synthillation resources~\cite{campbell2017unified,campbell2017unifying}, as established in \cref{section:CSSTP_examples}.
Beyond this overlap, CSS-TP differs in its code-based generative construction, its use of logical basis transformations, its access to higher-level transversal phase circuits, and its potential extension to the \emph{XP} and \emph{non-Abelian} stabilizer formalisms, which tensor networks also represent efficiently~\cite{webster2022xp,Shen2025quantumlegoxp,kobayashi2026clifford}.
A comparative summary of synthillation and the CSS-TP formalism is provided in \cref{table:csstp_synthillation_comparison}.

We construct codes, determine the logical phase gates they implement, and evaluate factories for the corresponding resource states.
Tailored resources give substantial gains when their interaction pattern matches the target circuit.
At code distance $21$ and with one factory per family, these resources shorten the latency of a $GF(2^{8})$ multiplier by up to $48\%$ and of a shared-addend carry-lookahead adder by up to $38\%$, and the smallest tailored factory, \Mfac{12,5,2}, lowers latency and occupied volume together.
The size of the gain is set by how well the resource matches the circuit and by factory throughput, footprint and output quality; for circuits without matching structure, canonical states such as $\ket{CCZ}$ remain the better choice.
CSS-TP codes therefore widen the design space that links quantum code design, logical algorithm compilation, and magic-state factories for a given architecture.

\subsection{Summary of Main Results}

This work develops a code-and-compiler framework for producing and using magic states tailored to structured algorithmic primitives.
The main contributions are:

\begin{itemize}
    \item \textbf{A code framework for structured resource states.}
    We introduce CSS-T-Product (CSS-TP) codes, whose transversal diagonal layers implement products of logical phase gates rather than a single canonical ($T$ or $CCZ$) logical gate.

    \item \textbf{Phase-polynomial flow rules for quantum LEGO.}
    We derive the logical action of a transversal diagonal layer from the $X$-type generator matrix, and rules that track it exactly through the quantum LEGO operations used to build a code: Bell fusion of legs (conjoining and self-traces), stoppers, code shortening and concatenation (\cref{section:gate_calculus}).
    The rules extend the operator-matching picture of quantum LEGO~\cite{cao2022quantum,cao2026quantum} beyond matched legs: they also decide fusions of legs whose gates do not match, stoppers, and fusions that remove a logical qubit, and in each case give the resulting logical phase polynomial.
    The rules hold for diagonal gates at any level $\ell$ of the Clifford hierarchy, and the level of the resulting logical gate follows from its phase polynomial (\cref{cor:level}).

    \item \textbf{Code families from quantum LEGO.}
    Starting from quantum Reed--Muller and hypercube seed codes, we use quantum LEGO operations~\cite{cao2022quantum,pato2025hyper,cao2024quantum} to rederive the $\code{8,3,2}$ code and to build families of codes with diverse non-Clifford logical actions, including \emph{open} and \emph{closed} chains of $\code{8,3,2}$ codes, \emph{star-like} logical-hypergraph orientations of $CCZ$ and of $CCCZ$ gates that share one control qubit, and concatenated codes of distance $4$ and $6$ (\cref{section:CSSTP_examples,section:concatenation_distance_growth}), each with its logical action derived by these rules.
    The smallest members recover the $\code{12,5,2}$ and $\code{14,6,2}$ codes from Campbell and Howard~\cite{campbell2017unified}; the others give higher distance or fourth-level actions, such as the $\code{30,8,2}$ and $\code{40,10,2}$ codes.

    \item \textbf{A classification of the resources.}
    We classify resource states, up to the unitary operations a compiler applies, into three \emph{tiers}: \emph{canonical} (Tier 0), \emph{tensor products} of canonical states (Tier 1) and \emph{irreducible} states (Tier 2).
    The key concept for this classification is the \emph{stabilizer nullity} of Beverland \emph{et al.}~\cite{beverland2020lower}, which can be read off from the logical phase polynomial.
    Because the nullity is a Clifford-invariant magic monotone, the resulting test holds under the full Clifford group, and it bounds the number of $CCZ$ states from which a stabilizer protocol can prepare a resource; the stabilizer R\'enyi entropy gives a second test (\cref{section:tiers}).

    \item \textbf{A method for reshaping logical resources.}
    Logical basis transformations move a CSS-TP resource through its Clifford orbit; we give a search that returns a sequence of $CX$ moves to a logical basis with reduced non-Clifford support (\cref{section:CSSTP_gate_cleaning}).

    \item \textbf{A route from codes to factories.}
    We convert CSS-TP codes into Encode-$T$-Decode (ETD) protocols~\cite{litinski2019game,litinski2019magic} that prepare entangled many-qubit resource states and characterize their magic content, error suppression, and implementation costs.

    \item \textbf{An architecture-aware evaluation.}
    We compare tailored and canonical resources on standalone factories and arithmetic circuits at code distance $21$ and over cultivation capacity, accounting for cultivation supply, ETD retries, latency, occupied and installed volume, footprint and output error.
    With one factory per family and serial consumption, \Mfac{12,5,2} reduces latency by $16$--$38\%$ and occupied volume by $11$--$33\%$ on the three circuits with matching structure, and the largest stars are up to $48\%$ faster.
    On a single-lane carry-lookahead adder at code distance $d=21$, for example, one \Mfac{12,5,2} factory is $16\%$ faster than one \Mfac{8,3,2} factory, and a mixed portfolio of one \Mfac{12,5,2} and one \Mfac{14,6,2} factory is $39\%$ faster, with $24\%$ less occupied volume.
    We also cost two fourth-level resources and state the factory and $\sqrt T$-source conditions under which they would trade installed volume for improved latency.
\end{itemize}

Together, the aforementioned contributions define a pipeline from code construction to architectural evaluation. 
The resulting resources complement canonical factories: on circuits with matching structure they reduce latency at a similar installed volume, and in general they expand the latency--volume tradeoffs available to a compiler.

\subsection{Outline}

\cref{section:background} reviews the basics of Pauli stabilizer codes, their description as quantum LEGO tensor networks, and notions of transversality used throughout the paper.
\cref{section:CSSTP} introduces CSS-TP codes, develops the generative protocol and logical basis transformations, and converts the resulting codes into tailored Encode-$T$-Decode factories.
\cref{section:tailored_algorithmic_primitives} evaluates these factories on standalone and arithmetic benchmarks, as well as on several $CCCZ$ gate logical-hypergraph orientations for fourth-level resource states (\cref{section:fourth_level}).
\cref{section:discussion} summarizes the resulting tradeoffs and directions for future work.
The supporting proofs, constructions, algorithms, and resource-model details are collected in the appendices.

%% file: includes/table_differences.tex
\begin{table*}
\centering
\caption{Comparison of synthillation and the CSS-T-Product (CSS-TP)
formalism. Here \Tofs denotes two $CCZ$ gates that share one qubit and \twoCCZ two disjoint $CCZ$ gates.}
\label{table:csstp_synthillation_comparison}
\def\tdc#1{\begin{tabular}[c]{@{}c@{}}\rule{0pt}{1.6em}#1\rule[-0.95em]{0pt}{0pt}\end{tabular}}%
\widetable{%
\renewcommand{\arraystretch}{1}%
\setlength{\extrarowheight}{0pt}%
\setlength{\tabcolsep}{8pt}%
\begin{tabular}{c | c | c}
\hline
\tdc{\textbf{Feature}}
& \tdc{\textbf{Synthillation}}
& \tdc{\textbf{CSS-TP formalism}} \\
\hline\hline
\tdc{Construction/ Technique}
& \tdc{Joint $T$-count/ multi-qubit gate optimization}
& \tdc{Conjoining seed codes (QLEGO);\\ logical $CX$ moves to a representative} \\
\hline
\tdc{Logical Output}
& \tdc{Third-level Clifford hierarchy}
& \tdc{Diagonal gates at level $\ell$ from\\ transversal level-$\ell$ phase gates} \\
\hline
\tdc{Logical Predictability}
& \tdc{The target gate is the input}
& \tdc{Derived from the code; predicted for\\ conjoined seeds by \cref{section:gate_calculus}} \\
\hline
\tdc{Resource classification}
& \tdc{--}
& \tdc{Tensor-product (Tier 1) or irreducible (Tier 2),\\ \cref{def:tiers}} \\
\hline
\tdc{Overlap}
& \multicolumn{2}{c}{\tdc{The $\code{12,5,2}$ and $\code{14,6,2}$ codes coincide with the \Tofs and \twoCCZ\\ synthillation codes of Ref.~\cite{campbell2017unified}, with the same $A_2$}} \\
\hline
\tdc{Potential for Extendability}
& \tdc{Potential for higher levels\\ of Clifford hierarchy}
& \tdc{Extendible to the XP stabilizer formalism\\ and to non-Abelian topological codes} \\
\hline
\end{tabular}%
}
\end{table*}

%% file: sections/02_background.tex
\section{Background}\label{section:background}

\subsection{Pauli Stabilizer Codes}\label{section:background_stabs}

We shall work with $\code{n,k}$ and $\code{n,k,d}$ Pauli stabilizer codes, in which $k$ logical qubits are encoded into $n$ physical qubits with code distance $d$ (if specified).
The code $\mathsf{C}$ is the simultaneous $+1$ eigenspace of an abelian \emph{stabilizer group} $\mathcal{S} \subset \mathsf{P}_{n}$, where $\mathsf{P}_{n}$ denotes the $n$-qubit Pauli group:
\begin{align}
\mathsf{C} = \{ \ket{\psi} \in \mathcal{H}_{n}: S\ket{\psi} = \ket{\psi},\; \forall\, S \in \mathcal{S} \}~,
\end{align}
where $\mathcal{H}_{n}$ is the Hilbert space of $n$ qubits.
Each stabilizer generator $S \in \mathcal{S}$ can be compactly represented as a binary vector of length $2n$ via the \emph{binary symplectic representation}: the first $n$ bits record positions at which $X$ acts, and the second $n$ bits record positions at which $Z$ acts.
Stacking $r$ vectors of a basis for $\mathcal{S}$ yields a \emph{parity check matrix} (PCM) $\mathsf{H} \in \mathbb{F}_{2}^{r \times 2n}$, which provides a succinct classical description of the code.
The \emph{normalizer} of $\mathcal{S}$ is the set of Pauli strings that commute with $\mathcal{S}$:
\begin{align}
\mathcal{N}(\mathcal{S}) = \{ P \in \mathsf{P}_{n} : P\mathcal{S}P^{\dagger} = \mathcal{S} \}~,
\end{align}
and the logical Pauli operators are elements of $\mathcal{N}(\mathcal{S})/\mathcal{S}$: Pauli operators that commute with every stabilizer (thus preserving the code) but are not themselves stabilizers, and therefore act non-trivially on the encoded information.

\subsection{Quantum LEGO Codes}\label{section:background_QLEGO}

Tensor networks (TNs) provide a diagrammatic and algebraic language for assembling large quantum states, operators, or linear maps from smaller local pieces~\cite{berezutskii2025tensor,orus2014practical}.
In this language, a tensor is a multidimensional array whose indices label Hilbert-space degrees of freedom.
Diagrammatically, tensors are drawn as nodes and their indices as legs.
Contracting two legs means identifying the corresponding indices and summing over them, in direct analogy with matrix multiplication.
Legs that are not contracted remain as external indices, and therefore specify the input or output Hilbert spaces of the object represented by the full network.
Thus, a complicated global map can be encoded by a pattern of local tensors and contractions.

For a quantum error-correcting code, the relevant tensor is the \emph{encoding isometry}.
An $\code{n,k}$ code defines a map
\begin{align}
E:\mathcal{H}_{2}^{\otimes k}\to \mathcal{H}_{2}^{\otimes n}~,
\end{align}
and can therefore be viewed as an $(n+k)$-leg tensor: $k$ legs label logical input degrees of freedom, while $n$ legs label physical output degrees of freedom.
The tensor entries are the amplitudes of encoded logical basis states in the physical computational basis.
From this perspective, building larger codes by tensor contraction amounts to joining smaller encoding maps along selected Hilbert-space factors.
The \emph{quantum LEGO} code formalism~\cite{cao2022quantum,cao2024quantum,cao2025growing,cao2026quantum,farrelly2021tensor,farrelly2022local} makes this idea systematic: small \emph{seed codes} are treated as reusable tensor blocks, and larger codes are obtained by connecting their legs while tracking the induced stabilizers, logical operators, and transversal gates.
This framework has been used, for example, to construct \emph{holographic} codes with high rate and distance~\cite{fan2024lego_hqec,steinberg2025far,steinberg2025universal,cao2023quantum,farrelly2021tensor,farrelly2022local,steinberg2026towards}\footnote{It is often convenient, though not necessary, for the seed tensor to represent
a \emph{multipartite maximally-entangled} (MMS) state. In a $k$-uniform state every
subsystem $A$ with $|A|\le k$ has reduced state $\sigma_A\propto\mathbb I_{|A|}$;
absolutely maximally-entangled (AME) states are the case $k=\lfloor m/2\rfloor$, for
which this holds for every bipartition~\cite{rajchel2025absolutely,goyeneche2014genuinely}.}.

The central operation in the quantum LEGO formalism is called \emph{conjoining}: the contraction of indices between seed tensors to form a new code~\cite{cao2022quantum}.
Here, Ordinary concatenation joins physical output legs to logical input legs. 
Conjoining also permits logical–logical and physical–physical contractions; these are shown in (\cref{fig:TN_basics}(c) and \cref{fig:code_modification}(b)).
Operationally, contracting legs $i_{1}$ and $j_{1}$ of two seed tensors $\mathsf{T}_{i_{1}i_{2}i_{3}i_{4}}$ and $\mathsf{T}'_{j_{1}j_{2}j_{3}j_{4}}$ corresponds to projecting the two associated Hilbert spaces onto the maximally entangled state $\ket{\Phi^{+}}=\sum_{r} \ket{rr}/\sqrt{d}$ by Bell measurement and postselection~\cite{cao2022quantum}.
Algebraically, the resulting codespace is described by the following tensor:
\begin{align}
W_{i_{2}i_{3}i_{4}j_{2}j_{3}j_{4}}
=
\sum_{r}
\mathsf{T}_{r i_{2}i_{3}i_{4}}\,
\mathsf{T}'_{r j_{2}j_{3}j_{4}}~.
\end{align}
We depict this operation diagrammatically in \cref{fig:TN_basics}(c).
Ref.~\cite{cao2026quantum} also calls this contraction a \emph{tensor trace} (or \emph{Bell fusion}) of the two legs.
The same contraction applied to two legs of a single tensor is called a \emph{self-trace}~\cite{cao2022quantum}.
We write \emph{conjoining} for a contraction between two codes and \emph{self-trace} for one within a single code, and use \emph{Bell fusion} for either when the Bell projector $\bra{\Phi^{+}}$ itself matters, as in the rules of \cref{section:gate_calculus}.
A \emph{physical leg} of an encoding tensor is the index of one physical qubit, and we label the physical qubits of a code by its physical legs.
We also use the convention that all indices, whether logical or physical, are written in lower position and summed when repeated or explicitly contracted.

The stabilizers and logical operators of the conjoined code are obtained by \emph{operator matching}~\cite{cao2022quantum}.
Intuitively, an operator supported on an internal leg must be matched by the dual operator on the other side of the contraction, so that the two actions cancel on the contracted Hilbert space and leave a well-defined operator on the external legs.
More concretely, suppose an operator $\overline{\mathcal{O}}$ of the first seed has support $\mathcal{O}_{i_{1},i_{2},i_{3}}$, including the conjoined leg $i_{1}$.
If the second seed has an operator $(\overline{\mathcal{O}}')^{*}$ whose support on the shared leg $j_{1}$ matches $\mathcal{O}_{i_{1}}$ under the contraction, then the product pushes through the internal edge and becomes an operator on the remaining legs, such as $\mathcal{O}_{i_{2},i_{3}}\cdot(\mathcal{O}')^{*}_{j_{3}}$.
This mechanism, depicted in \cref{fig:TN_basics}(d), generalizes to arbitrary networks of seed tensors, as well as any unitary $\mathcal{O}$; indeed, this mechanism is the basic reason that transversal-gate and logical-operator structures can be inherited by quantum LEGO constructions.
The same push-through mechanism underlies the gate-inheritance guarantees of~\cref{con:protocol}.

Ref.~\cite{cao2026quantum} phrases this mechanism in terms of symmetries of the encoding tensor.
A product operator $U=\bigotimes_{j}U_{j}$ of single-leg unitaries is a \emph{symmetry} of a tensor if it leaves the tensor, viewed as a state on all of its legs, invariant up to a phase.
Splitting the legs into logical and physical ones, a symmetry $Q\otimes O$ with $Q$ on the logical legs and $O$ on the physical legs means that $O$ implements the logical operator $Q^{*}$~\cite[Prop.~2.3]{cao2026quantum}; a transversal logical gate is such a symmetry in which $O$ is a product of single-leg unitaries.
If two legs carrying $U_{i}$ and $U_{j}$ are joined by a Bell fusion and $U_{j}=U_{i}^{*}$, the \emph{Matching Lemma}~\cite{cao2022quantum,cao2024quantum}\cite[Lemma~3.2]{cao2026quantum} states that the rest of $U$ remains a symmetry of the resulting tensor, which is how transversal gates are inherited by quantum LEGO constructions.
A transversal gate is \emph{strongly transversal} if it applies the same single-qubit gate to every physical leg, and \emph{weakly transversal} if the single-leg gates differ~\cite[Def.~2.4]{cao2026quantum}; the layers of $T$ and $T^{\dagger}$ gates used throughout this work are weakly transversal.
\input{includes/TN_operator_matching}
In \cref{section:conjoining_pcm}, we revisit conjoining from the complementary perspective of tensor contraction at the level of parity-check matrices.

\subsection{Codes Featured in this Work}\label{section:background_featured_codes}

All of the quantum codes in this work are of the \emph{Calderbank--Shor--Steane} (CSS) variety ~\cite{steane1996multiple,nielsen2010quantum}.
We use the convention
\begin{align}
\mathrm{CSS}(X:C_{2},Z:C_{1}^{\perp})~,
\qquad
C_{2}\subseteq C_{1}\subseteq\mathbb{F}_{2}^{n}~,
\label{eq:main_css_convention}
\end{align}
where $C_{2}$ generates the $X$-stabilizer space and $C_{1}^{\perp}$ generates the $Z$-stabilizer space.
The inclusion $C_{2}\subseteq C_{1}$ is equivalent to the CSS commutation condition $C_{2}\perp C_{1}^{\perp}$.
Logical-$X$ representatives may be chosen from the quotient $C_{1}/C_{2}$, while logical-$Z$ representatives may be chosen from $C_{2}^{\perp}/C_{1}^{\perp}$.
CSS codes are among the most heavily studied quantum-code classes because of their simple structure, straightforward decoding strategies, and comparatively simple fault-tolerant state-preparation circuits~\cite{gottesman1997stabilizer}.
In this work, we mainly consider generalized triorthogonal (GTO), $\Delta$-divisible ($\Delta$D), and CSS-T codes, described below.

\emph{Triorthogonal} codes were introduced by Bravyi and Haah~\cite{bravyi2012magic}.
A binary $m\times n$ matrix $\mathsf{G}$ is triorthogonal if
\begin{align}
\sum_{j=1}^{n} \mathsf{G}_{a,j}\mathsf{G}_{b,j} &= 0 \text{ mod } 2~, \\
\sum_{j=1}^{n} \mathsf{G}_{a,j}\mathsf{G}_{b,j}\mathsf{G}_{c,j} &= 0 \text{ mod } 2~,
\end{align}
for all pairs $1 \leq a < b \leq m$ and triples $1 \leq a < b < c \leq m$ of rows; its
odd-weight rows then label logical qubits on which transversal $T$ acts as logical $T$,
up to Clifford corrections.
We use the generalization of Haah and Hastings~\cite{haah2018codes}, which also admits
logical $CS$ and $CCZ$ gates.
\begin{definition}[Generalized triorthogonality~\cite{haah2018codes}]\label[definition]{def:gto}
A binary matrix $\mathsf G$ with $n$ columns is \emph{generalized triorthogonal} if its
rows can be ordered as $\mathsf G=(\mathsf G_{T};\mathsf G_{CS};\mathsf G_{CCZ};\mathsf G_{0})$,
where $\mathsf G_{T}$ has $k_{T}$ rows, $\mathsf G_{CS}$ consists of $k_{CS}$ designated
pairs of rows and $\mathsf G_{CCZ}$ of $k_{CCZ}$ designated triples of rows, such that
for all rows $a\le b\le c$
\begin{equation}
\sum_{j=1}^{n}\mathsf G_{a,j}\mathsf G_{b,j}\mathsf G_{c,j}\equiv
\begin{cases}
1 & \{a,b,c\}\in\mathcal D~,\\
0 & \text{otherwise}
\end{cases}
\pmod 2~,
\label{eq:gto}
\end{equation}
where $\mathcal D$ consists of the one-element sets $\{r\}$ for rows $r$ of
$\mathsf G_{T}$, the designated pairs of $\mathsf G_{CS}$ and the designated
triples of $\mathsf G_{CCZ}$.
\end{definition}
Because $a\le b\le c$ may repeat rows, \cref{eq:gto} fixes the parity of every row
weight, every pair overlap and every triple overlap.
Write $\mathsf G_{1}=(\mathsf G_{T};\mathsf G_{CS};\mathsf G_{CCZ})$, let
$\mathcal G_{0},\mathcal G\subseteq\mathbb F_{2}^{n}$ be the row spaces of
$\mathsf G_{0}$ and $\mathsf G$, and let $\mathcal G^{\perp}$ be the orthogonal
complement of $\mathcal G$.
The \emph{GTO code} is CSS$(X:\mathcal G_{0},Z:\mathcal G^{\perp})$; it has
$k=k_{T}+2k_{CS}+3k_{CCZ}$ logical qubits, whose logical-$X$ representatives are the
rows of $\mathsf G_{1}$.
Transversal $T$ on this code implements, up to Clifford corrections, logical
$\overline T$ on each qubit of $\mathsf G_{T}$, $\overline{CS}$ on each designated pair
and $\overline{CCZ}$ on each designated triple~\cite{haah2018codes}; see also
Refs.~\cite{campbell2017unified,campbell2017unifying}.
Ordinary triorthogonality is the case $k_{CS}=k_{CCZ}=0$.

For example, the $\code{8,3,2}$ code, with qubits on the vertices of a cube, is generalized triorthogonal with $k_{T}=k_{CS}=0$ and $k_{CCZ}=1$.
Its $\mathsf G_{0}=(1^{8})$ is the $X$ stabilizer, and its designated triple $\mathsf G_{CCZ}$ consists of the three faces that meet at one vertex.
Each face has weight $4$, two faces share an edge of weight $2$, and the three faces share a single vertex, so the only odd entry of \cref{eq:gto} is the designated triple.
The code is therefore not triorthogonal in the ordinary sense: no row has odd weight, yet the three faces have an odd triple overlap.
\Cref{eq:gto} constrains only parities, and accordingly fixes the logical action only up to Clifford corrections.
For the $\code{8,3,2}$ code, uniform $T^{\otimes8}$ thus gives $\overline{CCZ}$ together with $\overline{CZ}$ on every pair and $\overline Z$ on every qubit.
The CSS-TP codes of this work use signed and non-uniform layers and read the logical action from the exact coefficients $c_{S}$ rather than from their parities (\cref{section:gate_calculus}); they therefore need not be generalized triorthogonal, as the closed chains of \cref{section:closed_chains} show (\cref{lem:fusion_structure}).

\emph{$\Delta$-divisible} ($\Delta$D) codes are defined by the divisibility of each $X$ stabilizer by $\Delta$. 4-divisible and 8-divisible codes are known as \emph{doubly even} (DE) and \emph{triply even} (TE) codes, respectively, and this divisibility is closely related to the logical action induced on the code~\cite{jain2025transversal}.
For a \emph{uniform} transversal layer $T^{\otimes n}$ (resp.\ $S^{\otimes n}$) with trivial character vector, $8$- (resp.\ $4$-) divisibility of the $X$-stabilizer weights is necessary~\cite{rengaswamy2020optimality}; for non-uniform layers such as $T^{\otimes a}T^{\dagger\otimes b}$ this is no longer required (e.g.\ the $\code{6,4,2}$ code with $S^{\otimes3}S^{\dagger\otimes3}$).
More generally, for uniform layers, divisibility of all $X$-stabilizer weights by $2^{\ell}$ is a necessary ingredient for transversal physical $T_{\ell}=\mathrm{diag}(1,\exp(2i\pi/2^{\ell}))$ to induce a logical gate at level $\ell$ of the Clifford hierarchy; for example, $T_{3}=T$ and $T_{4}=\sqrt T$.
Sufficiency additionally requires the appropriate higher-order overlap and phase-compatibility conditions.
The possibility of optimal TE codes implementing logical $\overline{T}$ has been discussed in Refs.~\cite{jain2025transversal,bravyi2015doubled}; these are regarded as \emph{weakly triply even} codes.
For the purposes of this work, however, we focus more broadly on $\Delta$D codes in their full generality.

\emph{CSS-T} codes are the most general class of codes for which transversal application of $T_{\ell}$ or $T_{\ell}^{\dagger}$ gates preserves the code space, by way of performing some logical phase gate ranging from $\overline{\mathbb{I}}$ to any logical phase gate, Clifford or non-Clifford~\cite{rengaswamy2020optimality}.
There exist several ways to define the class of CSS-T codes, one of which involves the condition
\begin{align}
|X| - 2|X*Z| = 0 \text{ mod } 8~,
\end{align}
where $X \in C_{2}$ and $Z \in C_{1} + \mathbf{y}$.
Here $|\cdot|$ represents the Hamming weight, $X*Z$ represents the elementwise product of $X$ and $Z$, and $\mathbf{y}$ is known as the \emph{character vector}, and determines the sign of each $Z$ stabilizer~\cite{jain2025transversal}.
If the sign of the character vector is positive only, then the definition further simplifies as
\begin{align}
|X| &= 0 \text{ mod } 8 \text{ , } \forall X \in C_{2}~, \nonumber \\
|X*Z| &= 0 \text{ mod } 4 \text{ , } \forall X \in C_{2} \text{ , } Z \in C_{1}~.
\end{align}
Additionally, one may define a CSS-T code using pairs of classical codes called \emph{CSS-T pairs}~\cite{jain2025transversal}, and is given by a set of conditions on the CSS-T pair.
Namely, if $C_{1}$ and $C_{2}$ are two classical codes, then the following statements are equivalent:
\begin{enumerate}
    \item $(C_{1}, C_{2})$ is a CSS-T pair.
  \item $C_{2} \subseteq C_{1}$, and $C_{2}$ is even-weighted.
  Moreover, for any $X \in C_{2}$, the punctured code $C_{1}|_{\mathrm{supp}(X)}$ is self-orthogonal.
    \item $C_{2} \subset C_{1} \cap (C_{1}^{*2})^{\perp}$.
    \item $C_{1}^{\perp} + C_{1}^{*2} \subset C_{2}^{\perp}$.
\end{enumerate}
Here, $C_{1}^{*2} = \{ X*Y \mid X,Y \in C_{1}\}$, $C_{1}|_{\mathrm{supp}(X)} = \{Y|_{\mathrm{supp}(X)} \mid Y \in C_{1}\}$, and $\mathrm{supp}(X) = \{i \mid X_{i} \neq 0\}$.
Thus $C_{1}|_{\mathrm{supp}(X)}$ is the projection of $C_{1}$ onto the positions where $X$ is non-zero; the condition follows from $X\cdot(Y*Z)=\langle Y|_{\mathrm{supp}(X)},Z|_{\mathrm{supp}(X)}\rangle$.

We shall refer to \emph{CSS-T codes} broadly as those codes in which transversal physical $T_{\ell}$ or $T_{\ell}^{\dagger}$ implements some logical gate.
A subclass of these codes, the \emph{CSS-T-Identity} (CSS-TI) codes, we reserve for those codes for which $T^{\otimes n}_{\ell} = \overline{\mathbb{I}}^{\otimes k}$ and therefore contains transversal $T_{\ell}$ as a member of the \emph{logical identity group}.
In \cref{section:CSSTP_generative_protocol} we show that CSS-T, CSS-TI, and \emph{CSS-T-S-Gate} (CSS-TS) codes appear naturally as a byproduct of many stopper operations.
CSS-TS codes are codes for which $T^{\otimes n}_{\ell} = \overline{S}^{\otimes k}$ or $S^{\otimes n} = \overline{S}^{\otimes k}$~\cite{rengaswamy2020optimality}\footnote{Our definition also includes codes for which $(T^{\dagger}_{\ell})^{\otimes n} = (\overline{S^{\dagger}})^{\otimes k}$ or $(S^{\dagger})^{\otimes n} = (\overline{S}^{\dagger})^{\otimes k}$.}; we explain in \cref{section:concatenation_distance_growth} why they do not yield new transversal gates under concatenation.
In \cref{section:further_examples_CSStp}, we list a few of the examples of novel code families found during the course of this work.

Since we defined the CSS-T condition as transversal $T_{\ell}$ giving rise to identity or \emph{any} logical phase operator, one particular possibility exists in the transversal application of $T_{\ell}$ or $T_{\ell}^{\dagger}$ giving rise to an \emph{entire logical circuit} of phase gates.
Several code families with this property have been identified in the literature~\cite{rengaswamy2020optimality,vuillot2022quantum,campbell2017unified,campbell2017unifying,haah2018codes}, mostly individually rather than as a class to be generated and costed systematically.

In this work, we focus on such codes, expand the class and analyze their potential for practical use cases.
Specifically, for a code with $k$ logical qubits, the transversal action takes the form
\begin{align}\label{eq:css_tp_action}
T^{\otimes n}_{\ell} = \prod_{j=1}^{m} \prod_{\mathbf{s} \in \mathcal{A}_{j}} \overline{\mathcal{M}^{(j)}}_{\mathbf{s}} = \overline{\mathcal{M}}~,
\end{align}
where $\overline{\mathcal{M}^{(j)}}_{\mathbf{s}}$ is a diagonal gate at the $j$-th level of the Clifford hierarchy acting on the logical qubit subset $\mathbf{s} \subseteq [k]$, and $\mathcal{A}_{j}$ collects all such subsets at level $j$~\cite{campbell2017unifying,campbell2017unified}.
The structure described encompasses Clifford and non-Clifford logical phase-gate circuits alike, including \emph{canonical} non-Clifford gates such as $\overline{T}$, $\overline{CS}$, $\overline{CS}^{\dagger}$, $\overline{C^{\ell-1}Z}$ (at level $\ell$), and products thereof.
Although the formalism accommodates arbitrary hierarchy levels, we restrict attention mainly to gates in $\mathrm{Cl}^{(3)}_{k}$ throughout this work for simplicity.
Every transversal diagonal gate can be written in the form \cref{eq:css_tp_action}, including a single $\overline{CCZ}$ up to Clifford terms, so the equation alone does not single out a class of codes.
We therefore classify codes by the magic state that their transversal gate prepares, up to the operations that the compiler applies for free.

\begin{definition}[Resource state, free operations, and tiers]\label[definition]{def:tiers}
Let a CSS-T code with $k$ logical qubits have transversal logical gate $\overline{\mathcal M}=\sum_{u}\omega^{\phi(u)}\ket{\bar u}\!\bra{\bar u}$, where $\omega=e^{2\pi i/2^{\ell}}$ and $\phi$ is its logical phase polynomial (\cref{eq:logical_poly}).
\begin{enumerate}
\item The \emph{resource state} is $\ket{\overline{\mathcal M}}=\overline{\mathcal M}\ket{+}^{\otimes k}$.
It is a weighted hypergraph state with one hyperedge for each monomial of $\phi$, and an ordinary hypergraph state when $\overline{\mathcal M}=(-1)^{f(u)}$ with $f$ a Boolean polynomial.
\item The \emph{free group} $\mathcal F$ is generated by logical $CX$ gates and qubit permutations, Pauli $X$ and $Z$ gates, and the diagonal Clifford gates $S$ and $CZ$.
These are the operations that the compiler applies without non-Clifford cost when a resource is consumed (\cref{section:CSSTP_gate_cleaning,section:comparison_framework}).
\item A state is \emph{canonical} if it is equivalent under the free group $\mathcal F$ to a single non-Clifford controlled phase applied to $\ket{+}^{\otimes q}$, tensored with $\ket{+}^{\otimes(k-q)}$; examples are $\ket{T}$, $\ket{CS}$, $\ket{CCZ}$ and $\ket{CCCZ}$.
\end{enumerate}
A non-stabilizer resource state is \emph{Tier 0} if it is canonical, \emph{Tier 1} (tensor-product) if it is equivalent under $\mathcal F$ to a tensor product of at least two canonical states on disjoint sets of qubits, and \emph{Tier 2} (irreducible) otherwise.
A code is \emph{CSS-T-Product} (CSS-TP) if its resource state is Tier 1 or Tier 2, and \emph{irreducible CSS-TP} if it is Tier 2.
\end{definition}

Before proceeding, we should clarify that the states $\ket{T}$ and $\ket{CCZ}$ are the currency that standard factories already produce~\cite{litinski2019magic,gidney2019efficient,gidney2024magic}.
If a resource state is Tier 0 or Tier 1, a portfolio of canonical factories can deliver a state equivalent under $\mathcal F$, and any advantage of the code can only come from producing several canonical states in one factory, with a different footprint, throughput or error rate.
In contrast, a Tier-2 state is not equivalent to any product of canonical states, so it cannot be assembled from the outputs of canonical factories by free unitaries.
By this estimation, a single Tier-2 magic factory output replaces several canonical outputs and is consumed in one step; therefore, it is reasonable to expect that, at the cost of higher installed volume, latency improvements for select algorithmic applications should be within reach.
With Pauli measurements and classical feed-forward, gate teleportation prepares every Boolean cubic Tier-2 state from finitely many $\ket{CCZ}$ states, at the price of additional consumptions; \cref{prop:nullity} gives the number needed.
Both tiers matter for the benchmarks of \cref{section:tailored_algorithmic_primitives}.
\cref{section:tiers} gives criteria for the tiers and classifies every code featured in this work.
For expediency, we will henceforth refer to this transversal non-Clifford gate as $\overline{\mathcal{M}}$ for CSS-TP codes.

%% file: includes/TN_operator_matching.tex
\begin{figure*}
\centering
\includegraphics[width=\textwidth]{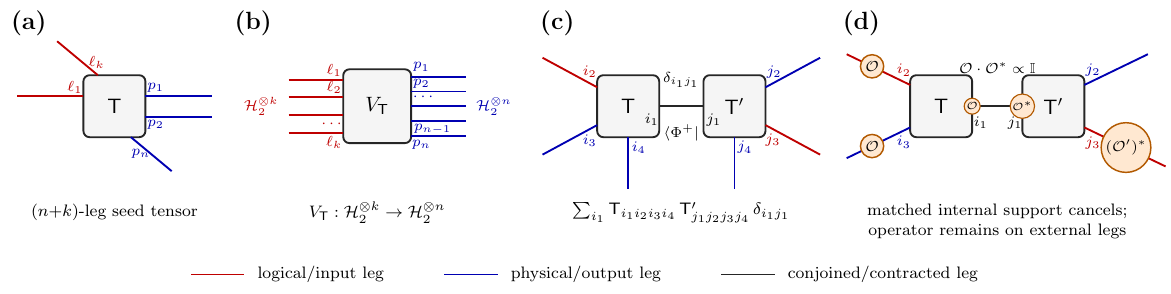}
\caption{
(a) A seed code is represented by an $(n+k)$-leg tensor.
(b) Choosing $k$ logical/input legs and $n$ physical/output legs defines the
encoding isometry $V_{\mathsf{T}}:\mathcal{H}_{2}^{\otimes k}\to
\mathcal{H}_{2}^{\otimes n}$.
(c) Conjoining contracts chosen legs between seed tensors; the contracted pair
may be logical--physical, physical--physical, or logical--logical.
(d) Operator pushing/matching: the operator components on the conjoined leg
cancel, leaving an induced operator on the external legs.
}
\label{fig:TN_basics}
\end{figure*}

%% file: sections/03_csstp_codes.tex
\section{CSS-T-Product Codes}\label{section:CSSTP}

CSS-TP codes are the subclass defined by \cref{eq:css_tp_action}, for which a transversal diagonal layer induces a product of canonical logical phase gates.
In this section, we develop a generative construction for these codes, analyze their logical actions, reshape their resources through logical basis transformations, and convert them into Encode-$T$-Decode protocols.
Preservation guarantees for the construction are proved in \cref{section:generative_protocol_proofs}.

Finally, we briefly mention some aspects of the \emph{Clifford hierarchy} here.\label{section:clifford_hierarchy} It is well-known that QEC codes with distance $d \geq 2$ cannot possess a set of transversal logical operators sufficient for \emph{universal logical quantum computation}~\cite{eastin2009restrictions}.
As such, it is necessary to identify and analyze codes that are able to access higher levels of the \emph{Clifford hierarchy}~\cite{gottesman1999quantum}, and to understand how to utilize them for the goal of attaining universal quantum logic.
Simply put, the Clifford hierarchy for $n$ qubits is a sequence of nested subsets of operators, taking the form
\begin{align}
\mathrm{Cl}^{(1)}_{n} \subset \mathrm{Cl}^{(2)}_{n} \subset \mathrm{Cl}^{(3)}_{n} \subset \cdots \subset \mathcal{U}(2^{n})~,
\end{align}
$\mathrm{Cl}^{(1)}_{n}$ is the $n$-qubit Pauli group $\mathsf{P}_{n}$, $\mathrm{Cl}^{(2)}_{n}$ is the $n$-qubit Clifford group, and generally $\mathrm{Cl}^{(\ell)}_{n}$ is the set of $n$-qubit unitaries that map $\mathsf{P}_{n}$ into $\mathrm{Cl}^{(\ell-1)}_{n}$ by conjugation~\cite{gottesman1999quantum}:
\begin{align}
\mathrm{Cl}^{(\ell)}_{n} = \{ U \in \mathcal{U}(2^{n}) \mid UPU^{\dagger} \in \mathrm{Cl}^{(\ell-1)}_{n} ~\forall P \in \mathsf{P}_{n} \}~.
\end{align}
Here, we focus on CSS codes that are equipped with transversal non-Clifford logical gates, i.e., gates that appear at the third level (or higher) of the Clifford hierarchy.
As an example, we may utilize a gate set such as the Clifford group $\mathrm{Cl}^{(2)}_{n}$ together with the $T$ gate, providing the universal gate set $(\mathrm{Cl}^{(2)}_{n}, T)$~\cite{nielsen2010quantum}.

\subsection{Modifying Quantum Codes}\label{section:background_code_modification}

Operations for generating new Pauli stabilizer codes from code modification techniques have been explored since the beginning of QEC as a field~\cite{gottesman1997stabilizer,steane1996multiple}.
One such operation is known as \emph{quantum code shortening} (QCS).
Given a QEC code $\mathsf{Q}$ with code parameters $\code{n,k,d}$, QCS permits the formation of a new code $\mathsf{Q}'$ with parameters $\code{n-1,k+1,d'}$, where $d' \leq d$~\cite{gottesman1997stabilizer}.
This procedure is realized by first choosing the $(n-k)$ stabilizer generators of $\mathsf{Q}$ such that $S_{1}^{X}, S_{1}^{Z}$ have $X$ or $Z$ support on qubit $l \in [1,n]$, and all other generators have support $\mathbb{I}$ on qubit $l$ (this condition is enforced by multiplying $S_{1}^{X}$ and $S_{1}^{Z}$ into other stabilizer generators as necessary).
Next, we remove the two stabilizers with support on qubit $l$.
Finally, the new code $\mathsf{Q}'$ has a stabilizer group defined from the $(n-k-2)$ remaining generators (\cref{fig:code_modification}(a)).
\input{includes/quantum_code_modifications}
We also use the conjoining operation defined in \cref{section:background_QLEGO}.
In the constructions below, conjoining is applied primarily between physical legs of seed-code tensors.
Whether a transversal diagonal layer survives a conjoining, and what it then implements, is decided by \cref{lem:operations} below.

Finally, we use \emph{stoppers}~\cite{pato2025hyper}: single-leg tensors given by a single-qubit state that is stabilized by one Pauli operator.
The \emph{$Z$ stopper} is $\ket{0}$, stabilized by $Z$, and the \emph{$X$ stopper} is $\ket{+}$, stabilized by $X$.
Contracting a stopper with physical leg $i$ removes that leg; the stabilizers of the new code are those stabilizers of the old code that act on leg $i$ as $\mathbb I$ or as the stopper's own Pauli operator, restricted to the remaining legs~\cite{pato2025hyper}.
Operationally, a stopper can be thought of as a single-qubit projective measurement in the $Z$ or $X$ basis, which removes one physical or logical qubit from the code.
The projection can be written as
\begin{align}
\Pi^{(i)}_{P} = \frac{\mathbb{I} + P_{i}}{2}~,
\end{align}
for $P \in \{X,Z\}$, where $P_{i}$ denotes $P$ acting on physical leg $i$, followed by removal of that leg.

\subsection{Transversal phase symmetries in quantum LEGO operations}\label{section:gate_calculus}

The codes in this work are built by starting from a code with a known transversal gate and modifying it step by step.
In the language of \cref{section:background_QLEGO}, the transversal diagonal layer of the introduction, made precise in \cref{eq:diag_layer} below and from here on simply the \emph{layer}, is a symmetry of the encoding tensor, and each step is a quantum LEGO operation on that tensor: a Bell fusion of two legs, a stopper on one leg, or the shortening of one leg.
After every step we need to know two things: whether the layer, restricted to the remaining legs, is still a symmetry, i.e.\ still acts as a logical gate, and if so, which gate it implements.
The Matching Lemma answers the first question for fusions of legs whose gates are conjugate~\cite{cao2022quantum,cao2026quantum}.
For diagonal layers on CSS codes, both answers follow in every case from a short calculation on the $X$-type generator matrix of the code, so we never have to simulate the code itself.
The calculation has three parts:
\begin{enumerate}[label=(\alph*)]
\item Write the phase that the layer puts on an encoded basis state as a polynomial in the logical and stabilizer bits and check that the stabilizer bits drop out (\cref{lem:logical_poly}).
\item Read off the logical gate and its level in the Clifford hierarchy (\cref{cor:level}).
\item Track how the polynomial changes under each operation (\cref{lem:operations}).
\end{enumerate}
The notation used in this section is summarized in \cref{tab:notation}.

\begin{table}
\centering
\caption{Notation for transversal diagonal layers on CSS codes.}
\label{tab:notation}
\renewcommand{\arraystretch}{1.15}
\small
\begin{tabular}{@{}ll@{}}
\toprule
Symbol & Meaning\\
\midrule
$n$, $k$ & physical and logical qubits\\
$m$ & independent $X$ stabilizers\\
$\ell$, $\omega$ & level of the layer, $\omega=e^{2\pi i/2^{\ell}}$\\
$a\in\mathbb Z_{2^{\ell}}^{n}$ & layer exponents; leg $j$ carries $\mathrm{diag}(1,\omega^{a_{j}})$\\
$L_{X}$ & $k\times n$ matrix of logical-$X$ rows\\
$C_{X}$ & $m\times n$ matrix of $X$-stabilizer rows\\
$G$ & $[L_{X};C_{X}]$, the $X$-type generator matrix\\
$g_{j}$ & column $j$ of $L_{X}$\\
$u\in\mathbb F_{2}^{k}$ & logical bits of a basis state $\ket{\bar u}$\\
$t\in\mathbb F_{2}^{m}$ & stabilizer coordinates\\
$y=uL_{X}+tC_{X}$ & $X$-type codeword, with bit $y_{j}$ at leg $j$\\
$f(u,t)$ & phase exponent $\sum_{j} a_{j} y_{j}$ on $\ket{y}$\\
$\phi(u)$ & logical phase polynomial, $f(u,0)$\\
$z=(u,t)$ & combined row labels of $G$\\
$c_{S}$ & coefficient of $\prod_{r\in S}z_{r}$ in $f$\\
$v_{2}(c)$ & largest $p$ with $2^{p}$ dividing $c$\\
$\alpha$ & edge residue (\cref{tab:operations})\\
\bottomrule
\end{tabular}
\end{table}

We begin with part (a), the phase that the layer puts on an encoded basis state, and write the layer as
\begin{align}
V=\bigotimes_{j=1}^{n}\mathrm{diag}\bigl(1,\omega^{a_{j}}\bigr)~,
&&
\omega=e^{2\pi i/2^{\ell}}~,
&&
a\in\mathbb Z_{2^{\ell}}^{n}~,
\label{eq:diag_layer}
\end{align}
so that each leg carries a power of a single phase gate.
At level $\ell=3$, for example, $T$, $T^{\dagger}$, $S$ and $Z$ correspond to $a_{j}=1$, $-1$, $2$ and $4$.
With every $Z$ stabilizer set to $+1$, the encoded basis state with logical bits $u$ is
\begin{align}
\ket{\bar u}\propto\sum_{t\in\mathbb F_{2}^{m}}\ket{uL_{X}+tC_{X}}~,
\end{align}
where $t$ runs over all combinations of $X$ stabilizers.
The physical transversal gates multiply a basis state $\ket{y}$ by $\omega^{\sum_{j} a_{j} y_{j}}$, so the codeword $uL_{X}+tC_{X}$ picks up the phase $\omega^{f(u,t)}$ with $f(u,t)=\sum_{j} a_{j}\,(uL_{X}+tC_{X})_{j}$.
$V$ is a logical gate exactly when every term of $\ket{\bar u}$ picks up the same phase, i.e.\ when $f$ does not depend on $t$.
It then acts as $\overline{\mathcal M}=\sum_{u}\omega^{\phi(u)}\ket{\bar u}\!\bra{\bar u}$ with $\phi(u)=f(u,0)$, which we call the \emph{logical phase polynomial}~\cite{amy2014polynomial}.
If $f$ does depend on $t$, then $V$ does not preserve the code space.

It remains to decide whether $f$ depends on $t$.
Because each bit $y_{j}$ is a parity, whereas the phase adds the bits as integers, we first rewrite $f$ as an ordinary polynomial.
Let $G=[L_{X};C_{X}]$ be the $(k+m)\times n$ binary $X$-type generator matrix, whose first $k$ rows are the logical-$X$ representatives and whose last $m$ rows are the $X$-stabilizer generators; $G_{rj}\in\{0,1\}$ indicates whether row $r$ acts on leg $j$.
With $z=(u,t)\in\mathbb F_{2}^{k+m}$, i.e.\ $z_{r}=u_{r}$ for $r\le k$ and $z_{r}=t_{r-k}$ for $r>k$, the bit at leg $j$ is the parity $y_{j}=\bigoplus_{r} z_{r}G_{rj}$, whereas the phase adds the $y_{j}$ as ordinary integers.

Applying the identity $b\oplus b'=b+b'-2bb'$ one bit at a time rewrites a parity of several bits as an ordinary polynomial with integer coefficients.
As an example, for three bits, $b_{1}\oplus b_{2}\oplus b_{3}=(b_{1}+b_{2}+b_{3})-2(b_{1}b_{2}+b_{1}b_{3}+b_{2}b_{3})+4b_{1}b_{2}b_{3}$.
In general,
\begin{align}
b_{1}\oplus\dots\oplus b_{p}=\sum_{\emptyset\ne S\subseteq[p]}(-2)^{|S|-1}\prod_{r\in S}b_{r}~.
\label{eq:parity_expansion}
\end{align}
Summing over legs with weights $a_{j}$ gives $f(z)=\sum_{\emptyset\ne S\subseteq[k+m]}c_{S}\prod_{r\in S}z_{r}$ with
\begin{align}
c_{S}=(-2)^{|S|-1}\sum_{j=1}^{n}a_{j}\prod_{r\in S}G_{rj}\pmod{2^{\ell}}~.
\label{eq:cS}
\end{align}
The sum is the $a$-weighted size of the common support of the rows in $S$, and the prefactor vanishes modulo $2^{\ell}$ whenever $|S|>\ell$.

For every nonempty set $S$ of rows of $G$, the coefficient $c_{S}$ is then computed in four steps: (i) find the legs on which all rows in $S$ act, (ii) add up the exponents $a_{j}$ on those legs, (iii) multiply by $(-2)^{|S|-1}$, and (iv) reduce modulo $2^{\ell}$.
The coefficients with a stabilizer row in $S$ decide whether the layer is logical, and those with only logical rows give the logical gate.

\begin{lemma}[Logical phase polynomial]\label[lemma]{lem:logical_poly}
$V$ from \cref{eq:diag_layer} is logical iff $c_{S}\equiv0$ for every $S$ with $|S|\le\ell$ that contains at least one stabilizer row.
Its logical phase polynomial is then
\begin{align}
\phi(u)=\sum_{\emptyset\neq S\subseteq[k]}c_{S}\prod_{r\in S}u_{r}\pmod{2^{\ell}}~,
\label{eq:logical_poly}
\end{align}
where $u_{r}$ is the value of logical qubit $r$ in $\ket{\bar u}$.
\end{lemma}

\begin{proof}
The multilinear expansion of a function $\mathbb F_{2}^{k+m}\to\mathbb Z_{2^{\ell}}$ is unique, so $f$ is independent of $t$ exactly when every $c_{S}$ whose set $S$ contains a stabilizer row vanishes; those with $|S|>\ell$ vanish automatically.
Setting $t=0$ leaves the sets $S\subseteq[k]$ of logical rows.
\end{proof}

Each term of $\phi$ corresponds to a familiar gate: a term $c_{S}\prod_{r\in S}u_{r}$ with $|S|=q$ is the controlled phase $C^{q-1}P(2\pi c_{S}/2^{\ell})$ on the logical qubits in $S$.
At $\ell=3$, for instance, $4u_{1}u_{2}u_{3}$ is a $CCZ$ gate, $4u_{1}u_{2}$ a $CZ$ gate and $2u_{1}u_{2}$ a controlled-$S$ gate.

To locate the gate in the Clifford hierarchy, let $v_{2}(c)$, for nonzero $c\in\mathbb Z_{2^{\ell}}$, be the largest integer $p$ such that $2^{p}$ divides $c$; e.g.\ $v_{2}(1)=0$, $v_{2}(2)=v_{2}(6)=1$ and $v_{2}(4)=2$.

\begin{corollary}[Level in the Clifford hierarchy]\label[corollary]{cor:level}
If $V$ is logical, it implements a gate at level
\begin{align}
\mathrm{lev}(\overline{\mathcal M})=\max_{\emptyset\ne S\subseteq[k]:\,c_{S}\not\equiv0}\bigl(\ell-v_{2}(c_{S})+|S|-1\bigr)
\label{eq:level}
\end{align}
of the Clifford hierarchy, or at level $1$ if every such $c_{S}$ vanishes.
This level is at most $\ell$.
\end{corollary}
For example, at $\ell=3$ the terms $4u_{1}u_{2}$ ($CZ$) and $2u_{1}$ ($S$) have level $3-2+1=3-1+0=2$, whereas $2u_{1}u_{2}$ (controlled-$S$) and $4u_{1}u_{2}u_{3}$ ($CCZ$) have level $3-1+1=3-2+2=3$; the gate $4u_{1}u_{2}u_{3}+4u_{1}u_{2}+2u_{3}$ therefore lies at level $3$.

The term in \cref{eq:level} is the level of a single controlled phase~\cite{cui2017diagonal}, so the level of the whole gate is the highest level among its terms.
Because $c_{S}$ always contains the factor $2^{|S|-1}$ from \cref{eq:cS}, the logical gate can never lie higher in the hierarchy than the physical layer, and it reaches level $\ell$ when some $c_{S}$ has no further factors of two.
The hypercube codes of \cref{section:cccz_star} reach this bound: their layer at level $\ell=D$ implements $\overline{C^{D-1}Z}$, which also lies at level $D$ (\cref{eq:cube_phase}).
We prove \cref{cor:level} in \cref{section:operation_rules_proof}.

\subsubsection{Example codes}\label{section:gate_calculus_examples}

Here we illustrate steps (i)--(iv) on the two smallest hypercube codes.
For the $\code{4,2,2}$ code, take $L_{X}$ with rows $(1,1,0,0)$ and $(1,0,1,0)$, $C_{X}=(1,1,1,1)$, and the layer $S\otimes S^{\dagger}\otimes S^{\dagger}\otimes S$, i.e.\ $\ell=2$ and $a=(1,3,3,1)$.
Call the rows $u_{0}$, $u_{1}$ and $t$, so that $G$ has the two logical rows $u_{0},u_{1}$ and the single stabilizer row $t$.
Equivalently, place the qubits on the corners $v=(v_{0},v_{1})$ of a square, with legs $1,2,3,4$ at $v=11,10,01,00$: then $u_{0}$ and $u_{1}$ are the edges $v_{0}=1$ and $v_{1}=1$, and the layer applies $S$ on the even-weight and $S^{\dagger}$ on the odd-weight corners (\cref{fig:hypercube_examples}(a)).
Since $\ell=2$, the phase is $\omega^{f}$ with $\omega=i$, and every coefficient only matters modulo $2^{\ell}=4$.

We apply steps (i)--(iv) to each of the seven nonempty sets of rows.
For instance, $u_{0}$ and $t$ both act on legs $1$ and $2$, so $c_{u_{0}t}=(-2)^{1}(a_{1}+a_{2})=-2\cdot4=-8\equiv0$.
\cref{tab:422_example} lists all seven coefficients.

\begin{figure}[t]
\centering
\begin{tikzpicture}[
  evenv/.style={circle,draw=darkblue,fill=darkblue,inner sep=0pt,minimum size=5pt},
  oddv/.style={circle,draw=darkred,fill=white,thick,inner sep=0pt,minimum size=5pt},
  vlab/.style={font=\tiny,inner sep=3pt},
  cedge/.style={gray!70},
  hedge/.style={gray!70,dashed},
  legend/.style={font=\scriptsize,anchor=west,inner sep=0pt},
  plab/.style={font=\bfseries\small,anchor=north west}]
\foreach \pn/\sx in {sa/0.35,sb/4.1}{
  \foreach \xa/\xb in {0/0,1/0,0/1,1/1}{
    \coordinate (\pn\xa\xb) at ({\sx+1.6*\xa},{4.1+1.6*\xb});
  }
}
\draw[blueBg,line width=6pt,line cap=round] (sb10)--(sb11)--(sb01);
\foreach \pn in {sa,sb}{
  \draw[cedge] (\pn00)--(\pn10)--(\pn11)--(\pn01)--cycle;
}
\draw[darkred,thick] (sb11) circle (5pt);
\foreach \pn in {sa,sb}{
  \foreach \vx in {00,11}{\node[evenv] at (\pn\vx) {};}
  \foreach \vx in {10,01}{\node[oddv] at (\pn\vx) {};}
}
\foreach \vx/\st/\an in {00/4/north east,10/2/north west,01/3/south east,11/1/south west}{
  \node[vlab,anchor=\an] at (sa\vx) {$\vx$\,(\st)};
}
\node[plab] at (-0.5,6.6) {(a)};
\node[plab] at (3.3,6.6) {(b)};
\node[evenv] at (-0.25,3.5) {};
\node[legend] at (-0.05,3.5) {$S$:\ $a_{v}=+1$, $|v|$ even};
\node[oddv] at (-0.25,3.15) {};
\node[legend] at (-0.05,3.15) {$S^{\dagger}$:\ $a_{v}=-1\equiv3$, $|v|$ odd};
\draw[blueBg,line width=6pt] (3.4,3.5)--(3.65,3.5);
\node[legend] at (3.8,3.5) {edges $u_{0}$, $u_{1}$: $\sum a_{v}\equiv0$};
\draw[darkred,thick] (3.525,3.15) circle (4pt);
\node[legend] at (3.8,3.15) {vertex $11$ ($u_{0}u_{1}$): $a_{11}=+1$};
\foreach \pn/\sx in {a/0,b/3.75}{
  \foreach \xa/\xb/\xc in {0/0/0,1/0/0,0/1/0,1/1/0,0/0/1,1/0/1,0/1/1,1/1/1}{
    \coordinate (\pn\xa\xb\xc) at ({\sx+1.6*\xa+0.7*\xc},{1.6*\xb+0.55*\xc});
  }
}
\fill[blueBg] (b100)--(b110)--(b111)--(b101)--cycle;
\foreach \pn in {a,b}{
  \foreach \vs/\ve in {000/100,000/010,100/110,010/110,100/101,110/111,010/011,101/111,011/111}{
    \draw[cedge] (\pn\vs)--(\pn\ve);
  }
  \foreach \vs/\ve in {000/001,001/101,001/011}{
    \draw[hedge] (\pn\vs)--(\pn\ve);
  }
}
\draw[darkred,line width=1.6pt] (b110)--(b111);
\draw[darkred,thick] (b111) circle (5pt);
\foreach \pn in {a,b}{
  \foreach \vx in {000,110,101,011}{\node[evenv] at (\pn\vx) {};}
  \foreach \vx in {100,010,001,111}{\node[oddv] at (\pn\vx) {};}
}
\foreach \vx/\an in {000/north east,100/north,010/east,110/north east,001/south west,101/west,011/south,111/south west}{
  \node[vlab,anchor=\an] at (a\vx) {$\vx$};
}
\node[plab] at (-0.5,2.75) {(c)};
\node[plab] at (3.3,2.75) {(d)};
\node[evenv] at (-0.25,-0.6) {};
\node[legend] at (-0.05,-0.6) {$T$:\ $a_{v}=+1$, $|v|$ even};
\node[oddv] at (-0.25,-0.95) {};
\node[legend] at (-0.05,-0.95) {$T^{\dagger}$:\ $a_{v}=-1$, $|v|$ odd};
\fill[blueBg] (3.4,-0.68) rectangle (3.65,-0.52);
\node[legend] at (3.8,-0.6) {face $v_{0}=1$ ($u_{0}$): $\sum a_{v}=0$};
\draw[darkred,line width=1.6pt] (3.4,-0.95)--(3.65,-0.95);
\node[legend] at (3.8,-0.95) {edge ($u_{0}u_{1}$): $\sum a_{v}=0$};
\draw[darkred,thick] (3.525,-1.3) circle (4pt);
\node[legend] at (3.8,-1.3) {vertex $111$ ($u_{0}u_{1}u_{2}$): $a_{111}=-1$};
\end{tikzpicture}
\caption{The two worked examples as hypercube codes: the $\code{4,2,2}$ code on a square (top) and the $\code{8,3,2}$ code on a cube (bottom).
Qubits sit on the vertices $v$, with $v_{0}$ increasing to the right, $v_{1}$ upward and $v_{2}$ into the page, and the logical row $u_{r}$ is the face $v_{r}=1$.
(a),\,(c)~The layer: $S$ (resp.\ $T$) on even-weight and $S^{\dagger}$ (resp.\ $T^{\dagger}$) on odd-weight vertices; the numbers in parentheses in (a) are the legs of \cref{tab:422_example}.
(b),\,(d)~Common supports of the row sets $S$ of \cref{tab:422_example,tab:832_example}.}
\label{fig:hypercube_examples}
\end{figure}

\begin{table}[t]
\centering
\caption{Coefficients $c_{S}$ of \cref{eq:cS} for the $\code{4,2,2}$ code with the layer $S\otimes S^{\dagger}\otimes S^{\dagger}\otimes S$ ($\ell=2$, $a=(1,3,3,1)$).
The upper block contains the stabilizer row $t$ and must vanish modulo $4$ for the layer to be logical; the lower block gives the logical phase polynomial $\phi$.}
\label{tab:422_example}
\begin{tabular}{lccccc}
\toprule
$S$ & common legs & $\sum a_{j}$ & $(-2)^{|S|-1}$ & $c_{S}$ & $c_{S}\bmod 4$\\
\midrule
$\{t\}$         & $1,2,3,4$ & $8$ & $1$  & $8$  & $0$\\
$\{u_{0},t\}$     & $1,2$     & $4$ & $-2$ & $-8$ & $0$\\
$\{u_{1},t\}$     & $1,3$     & $4$ & $-2$ & $-8$ & $0$\\
$\{u_{0},u_{1},t\}$ & $1$       & $1$ & $4$  & $4$  & $0$\\
\midrule
$\{u_{0}\}$       & $1,2$     & $4$ & $1$  & $4$  & $0$\\
$\{u_{1}\}$       & $1,3$     & $4$ & $1$  & $4$  & $0$\\
$\{u_{0},u_{1}\}$   & $1$       & $1$ & $-2$ & $-2$ & $2$\\
\bottomrule
\end{tabular}
\end{table}

The upper block of \cref{tab:422_example} checks that the layer is logical.
Every coefficient that involves $t$ is a multiple of $4$, so $f(u,t)$ does not depend on $t$: both codewords in each $\ket{\bar u}$ pick up the same phase, and the layer preserves the code space.
The three-row set $\{u_{0},u_{1},t\}$ has $|S|>\ell$ and vanishes automatically, because its prefactor $(-2)^{2}=4$ is already zero modulo $4$.
The two-row sets vanish because the prefactor $-2$ supplies one factor of two and the overlaps $a_{1}+a_{2}=a_{1}+a_{3}=4$ supply the other.

The lower block gives the logical action.
The single-row coefficients $c_{u_{0}}$ and $c_{u_{1}}$ vanish, because each edge $u_{r}$ carries one $S$ and one $S^{\dagger}$, so there are no single-qubit logical phases, and only $c_{u_{0}u_{1}}\equiv2$ survives, from the single shared corner $11$ (\cref{fig:hypercube_examples}(b)).
Hence $\phi(u)=2u_{0}u_{1}$, i.e.\ the phase $i^{2u_{0}u_{1}}=(-1)^{u_{0}u_{1}}$, and the layer implements $\overline{CZ}$.
As a direct check, $u=(1,1)$ gives the codeword $(0,1,1,0)$ with phase exponent $a_{2}+a_{3}=6\equiv2$, i.e.\ a sign of $-1$, whereas $u=(1,0)$ and $u=(0,1)$ give $a_{1}+a_{2}=a_{1}+a_{3}=4\equiv0$.

Moving up one dimension to the $\code{8,3,2}$ code, we place the qubits on the vertices $v=(v_{0},v_{1},v_{2})\in\{0,1\}^{3}$ of a cube.
The logical rows $u_{0},u_{1},u_{2}$ of $L_{X}$ are the three faces $v_{r}=1$, and the single stabilizer row $t$ is the $X$ stabilizer, so the codeword with labels $(u,t)$ is $y_{v}=t\oplus u\cdot v$.
Apply $T$ on the even-weight vertices and $T^{\dagger}$ on the odd-weight ones, i.e.\ $\ell=3$ and $a_{v}=(-1)^{|v|}$, so that coefficients now matter modulo $2^{\ell}=8$ (\cref{fig:hypercube_examples}(c)).

The geometry of the cube makes steps (i)--(iv) particularly transparent.
The common support of a set $S$ of rows is always a subcube: one logical row gives a face, two give an edge, and all three give the single vertex $111$ (\cref{fig:hypercube_examples}(d)); adding $t$ does not change the support, because $t$ covers every vertex.
\cref{tab:832_example} lists all $15$ coefficients, grouped by this geometry.

\begin{table}[t]
\centering
\caption{Coefficients $c_{S}$ of \cref{eq:cS} for the $\code{8,3,2}$ code with $T$ on even and $T^{\dagger}$ on odd vertices ($\ell=3$, $a_{v}=(-1)^{|v|}$).
Rows labeled by $r$ or by $r\ne s$ stand for three sets each.
The upper block contains the stabilizer row $t$ and must vanish modulo $8$; the lower block gives the logical phase polynomial $\phi$.}
\label{tab:832_example}
\begin{tabular}{lccccc}
\toprule
$S$ & common legs & $\sum a_{v}$ & $(-2)^{|S|-1}$ & $c_{S}$ & $c_{S}\bmod 8$\\
\midrule
$\{t\}$             & cube   & $0$  & $1$  & $0$  & $0$\\
$\{u_{r},t\}$         & face   & $0$  & $-2$ & $0$  & $0$\\
$\{u_{r},u_{s},t\}$     & edge   & $0$  & $4$  & $0$  & $0$\\
$\{u_{0},u_{1},u_{2},t\}$ & $111$  & $-1$ & $-8$ & $8$  & $0$\\
\midrule
$\{u_{r}\}$           & face   & $0$  & $1$  & $0$  & $0$\\
$\{u_{r},u_{s}\}$       & edge   & $0$  & $-2$ & $0$  & $0$\\
$\{u_{0},u_{1},u_{2}\}$   & $111$  & $-1$ & $4$  & $-4$ & $4$\\
\bottomrule
\end{tabular}
\end{table}

The key observation is that the signs $a_{v}=(-1)^{|v|}$ alternate along every edge of the cube.
The whole cube, each face and each edge therefore contain equally many $T$ and $T^{\dagger}$ legs, and their weighted sizes $\sum a_{v}$ vanish; only the single vertex $111$ is unbalanced, with $a_{111}=-1$.
Every coefficient therefore vanishes except for the two sets whose common support is this vertex.
For $S=\{u_{0},u_{1},u_{2},t\}$ we have $|S|>\ell$, and the prefactor $(-2)^{3}=-8$ is already zero modulo $8$.
For $S=\{u_{0},u_{1},u_{2}\}$ the prefactor is $(-2)^{2}=4$, so $c_{u_{0}u_{1}u_{2}}=-4\equiv4$.

The upper block of \cref{tab:832_example} shows that the layer is logical, and the lower block gives $\phi(u)=4u_{0}u_{1}u_{2}$, i.e.\ the phase $\omega^{4u_{0}u_{1}u_{2}}=(-1)^{u_{0}u_{1}u_{2}}$ with $\omega=e^{2\pi i/8}$.
The layer thus implements $\overline{CCZ}$, and \cref{cor:level} places it at level $\ell-v_{2}(4)+|S|-1=3-2+2=3$, as expected for $CCZ$.
As a direct check, $u=(1,1,1)$ and $t=0$ give the codeword on the four odd-weight vertices, with phase exponent $4\cdot(-1)\equiv4$, and $t=1$ gives the four even-weight vertices, with exponent $4\cdot(+1)=4$; both terms of $\ket{\bar u}$ thus pick up the sign $-1$.
Every other $u\ne0$ gives codewords with two even-weight and two odd-weight vertices, and hence no phase.

\subsubsection{Tracking the logical gate with quantum LEGO}\label{section:gate_calculus_operations}

Every operation of \cref{section:background_code_modification} removes one leg $i$, or two legs $i$ and $j$ for a contraction, and the new layer $V'$ is precisely the old layer $V$ restricted to the remaining legs.
In the case of conjoining two codes $\mathsf C_{1}$ and $\mathsf C_{2}$, we obtain a contraction of the single code $\mathsf C=\mathsf C_{1}\otimes\mathsf C_{2}$, whose legs, $X$ stabilizers and logical rows are those of both codes.
The layer $V=V_{1}\otimes V_{2}$ is logical on $\mathsf C$ with $\phi(u)=\phi_{1}(u^{(1)})+\phi_{2}(u^{(2)})$, where $u=(u^{(1)},u^{(2)})$ collects the logical bits of both codes; a self-trace is the same contraction with $i$ and $j$ in a single code.
Since $V$ is logical on $\mathsf C$, every codeword $y$ with logical label $u$ picks up the phase $\omega^{\phi(u)}$.
Removing leg $i$ removes the term $a_{i}y_{i}$ from this phase.
In contrast, a Bell fusion $\bra{\Phi^{+}}_{ij}$ keeps only codewords with $y_{i}=y_{j}$ and removes $a_{i}y_{i}+a_{j}y_{j}=(a_{i}+a_{j})y_{i}$.
In either case, the remaining legs therefore carry the phase $\phi(u)-\alpha\,y_{i}$, where the \emph{edge residue} $\alpha$ is $a_{i}+a_{j}$ for a fusion and $a_{i}$ otherwise.
The residue $\alpha$ is the exponent of the phase gate $\mathrm{diag}(1,\omega^{\alpha})$ that is left on the fused edge, or on the stopped or shortened leg, once the layer has been pushed onto it.
For a fusion, $\alpha\equiv0$ is equivalent to the condition $V_{j}=V_{i}^{*}$ of the quantum LEGO \emph{Matching Lemma}~\cite{cao2022quantum,cao2026quantum}, since $\mathrm{diag}(1,\omega^{a})^{*}=\mathrm{diag}(1,\omega^{-a})$.
The first term depends only on the logical bits.
The second does as well, unless $\alpha\not\equiv0$ and some $X$ stabilizer of the new code can flip $y_{i}$.

\cref{tab:operations} lists $\alpha$ for each operation, together with the constraint that the operation imposes on the codewords and the value of $y_{i}$ on the new code.
An $X$ stabilizer of $\mathsf C$ \emph{survives} the operation if it satisfies the constraint, i.e.\ if it matches the Bell state on the fused legs or the stopper on the stopped leg~\cite{pato2025hyper}; restricted to the kept legs, the surviving stabilizers generate the $X$ stabilizers of the new code $\mathsf C'$.
QCS at $i$ first clears leg $i$ as in \cref{lem:QCS}: it picks an $X$-stabilizer generator $p$ with $p_{i}=1$, the \emph{pivot}, and adds it to every other stabilizer and logical row that acts on $i$, so that afterwards only the pivot acts on $i$.
Deleting leg $i$ then turns the punctured pivot into a new logical-$X$ row with logical bit $u_{k+1}$.
The pivot is therefore not a surviving stabilizer, every codeword has $y_{i}=u_{k+1}$, and none is discarded.
In quantum LEGO terms, QCS \emph{dualizes} physical leg $i$ into a new logical leg~\cite[Sec.~5.3]{cao2026quantum}.

\begin{table}[t]
\centering
\caption{Code operations and their effect on a transversal diagonal layer.
Each operation removes leg $i$ (and $j$ for a fusion), keeps the codewords that satisfy the constraint, and changes the logical phase polynomial by $-\alpha\,y_{i}$ (\cref{lem:operations}).
The last column gives $y_{i}$ in the logical bits of the new code whenever the layer stays logical; $g_{i}$ is column $i$ of $L_{X}$, and $u_{k+1}$ is the logical bit created by QCS.}
\label{tab:operations}
\renewcommand{\arraystretch}{1.15}
\small
\begin{tabular}{@{}lccc@{}}
\toprule
Operation & $\alpha$ & Constraint & $y_{i}$\\
\midrule
Bell fusion of $i$, $j$ & $a_{i}+a_{j}$ & $y_{i}=y_{j}$ & logical parity\\
QCS at $i$ & $a_{i}$ & none & $u_{k+1}$\\
$Z$ stopper at $i$ & $a_{i}$ & $y_{i}=0$ & $0$\\
$X$ stopper at $i$ & $a_{i}$ & none & $u\cdot g_{i}$\\
\bottomrule
\end{tabular}
\end{table}

\begin{lemma}[Logical action under code operations]\label[lemma]{lem:operations}
Let $V$ be logical on $\mathsf C$ with logical phase polynomial $\phi$, and let $\mathsf C'$ be obtained by one of the operations of \cref{tab:operations}, with edge residue $\alpha$.
Then $V'$ is logical on $\mathsf C'$ iff $\alpha\equiv0\pmod{2^{\ell}}$ or no surviving $X$ stabilizer acts on leg $i$.
Its logical phase polynomial is then
\begin{align}
\phi'=\phi-\alpha\,y_{i}\pmod{2^{\ell}}~,
\label{eq:op_phase}
\end{align}
evaluated on the logical labels that satisfy the constraint, with $y_{i}$ as in \cref{tab:operations}.
For the individual operations this means:
\begin{enumerate}
\item[(a)] \emph{Bell fusion of legs $i$ and $j$} (conjoining or self-trace).
$V'$ is logical iff $a_{i}+a_{j}\equiv0$ or no surviving $X$ stabilizer acts on $i$.
If $a_{i}+a_{j}\equiv0$, $\phi'$ is the restriction of $\phi$ to the constraint.
\item[(b)] \emph{QCS at $i$.}
$V'$ is always logical, and $\phi'(u,u_{k+1})=\phi(u)-a_{i}\,u_{k+1}$.
\item[(c)] \emph{Stopper at $i$.}
After a $Z$ stopper, $V'$ is always logical and $\phi'$ is the restriction of $\phi$ to $y_{i}=0$.
After an $X$ stopper, $V'$ is logical iff $a_{i}\equiv0$ or no $X$ stabilizer acts on $i$, and then $\phi'(u)=\phi(u)-a_{i}\,(u\cdot g_{i})$.
\end{enumerate}
\end{lemma}

Removing a leg therefore costs nothing when its edge residue vanishes, and is otherwise allowed only where no $X$ stabilizer can flip the removed bit.
In particular, fusing two legs with opposite exponents is always allowed, and QCS and $Z$ stoppers never destroy the gate, because no surviving stabilizer acts on the removed leg.
Only two steps do more than restrict the gate: QCS, which adds a single-qubit phase on the new logical qubit, and the $X$ stopper, which subtracts the parity $u\cdot g_{i}$.
Expanding this parity with \cref{eq:parity_expansion} produces entangling terms.
\cref{fig:code_modification} annotates each operation with its effect on $\phi$.

As an illustrative example, at $\ell=3$ an $X$ stopper on a $T$ leg ($a_{i}=1$) on which no $X$ stabilizer and three logical rows act ($g_{i}=(1,1,1)$) gives
\begin{align}
\phi'(u)&=\phi(u)-(u_{1}\oplus u_{2}\oplus u_{3})\nonumber\\
&=\phi(u)-(u_{1}+u_{2}+u_{3})\nonumber\\
&\quad+2(u_{1}u_{2}+u_{1}u_{3}+u_{2}u_{3})-4u_{1}u_{2}u_{3}~,
\label{eq:xtrace_example}
\end{align}
i.e.\ the stopper multiplies the logical gate by a $T^{\dagger}$ gate on each of the three qubits, a controlled-$S$ gate on each pair and, since $-4\equiv4\pmod 8$, a $CCZ$ gate.
This mechanism is responsible for how the $X$ stoppers in the derivation of the $\code{8,3,2}$ code (\cref{section:family_832}) turn four single-qubit gates into a $CCZ$ gate.

In our formalism, the constraint decides which logical qubits survive.
For a fusion there are two cases, depending on whether a stabilizer coordinate can absorb the constraint $y_{i}=y_{j}$.

\begin{corollary}[Conjoining and self-traces]\label[corollary]{cor:conjoin_loop}
Contract legs $i$ and $j$ with $a_{i}+a_{j}\equiv0$, and assume that no $X$-type codeword of $\mathsf C$ is supported exactly on $\{i,j\}$.
\begin{enumerate}
\item If some $X$ stabilizer takes different values at $i$ and $j$, the constraint fixes a stabilizer coordinate, so $k'=k$ and $\phi'=\phi$.
This is the typical case when two codes are conjoined on legs covered by $X$ stabilizers, and then $\phi'=\phi_{1}(u^{(1)})+\phi_{2}(u^{(2)})$.
\item If every $X$ stabilizer takes equal values at $i$ and $j$, the constraint acts on the logical bits, $u\cdot(g_{i}+g_{j})=0$.
For $g_{i}\ne g_{j}$ one logical qubit is removed, $k'=k-1$, and $\phi'$ is the restriction of $\phi$ to this hyperplane.
\end{enumerate}
\end{corollary}

Having stated the rules in terms of generator matrices, we now relate them to the quantum LEGO framework of \cref{section:background_QLEGO}.
They specialize the quantum LEGO description of transversal gates to diagonal layers on CSS codes, and extend it beyond matched legs.
First, \cref{lem:operations}(a) with $\alpha\equiv0$ is the Matching Lemma~\cite{cao2022quantum,cao2024quantum}\cite[Lemma~3.2]{cao2026quantum} for diagonal layers (\cref{lem:gate_compatible_conjoining}): the conjugate gates cancel on the fused edge, and the rest of the layer remains a symmetry.
The lemma also decides the unmatched case $\alpha\not\equiv0$: the residue $\mathrm{diag}(1,\omega^{\alpha})$ left on the edge is harmless exactly when no surviving $X$ stabilizer can flip the edge bit, and it then adds the term $-\alpha\,y_{i}$ to the logical phase polynomial.
Second, \cref{lem:operations}(b) is an instance of \cite[Prop.~2.3]{cao2026quantum}. 
When leg $i$ is dualized into a logical leg, the phase gate $\mathrm{diag}(1,\omega^{a_{i}})$ on that leg becomes a logical gate on the new qubit, conjugated, which is the term $-a_{i}u_{k+1}$.
Third, an $X$ stopper on a leg with $a_{i}\not\equiv0$ applies the phase gadget~\cite{cowtan2020phase} $\mathrm{diag}(1,\omega^{-a_{i}})$ to the parity $u\cdot g_{i}$ of the logical bits; its parity expansion, as in \cref{eq:xtrace_example}, gives the entangling terms.
Fourth, case~2 of \cref{cor:conjoin_loop} is a fusion whose logical map has a kernel in the sense of \cite[Cor.~3.5]{cao2026quantum}: the logical states with $u\cdot(g_{i}+g_{j})=1$ are annihilated, and the layer implements the restriction of $\phi$ on the remaining logical subspace.
Finally, the $T$/$T^{\dagger}$ layers of this work are \emph{weakly transversal}~\cite{cao2026quantum}, and their alternation is what makes fused pairs match, $a_{i}+a_{j}\equiv0$.
Ref.~\cite{cao2026quantum} obtains strongly transversal layers from Clifford-deformed fusions instead: since $XTX\propto T^{\dagger}$, a fusion with $\ket{\Phi_{X}}=(\mathbb I\otimes X)\ket{\Phi^{+}}$ matches $T$ with $T$~\cite[Lemma~3.7]{cao2026quantum}, and the two descriptions differ by $X$ gates on the legs that carry $T^{\dagger}$ in ours.

Concatenation follows a similar rule (\cref{lem:concat}): the codewords of the upper code are substituted into the logical phase polynomial of the lower one.
We prove \cref{lem:operations,cor:conjoin_loop} in \cref{section:operation_rules_proof} and \cref{lem:concat} in \cref{section:concatenation_distance_growth}.
\cref{section:CSSTP_examples} applies these rules to every code family of this work.

\subsection{CSS-T-Product Codes from Quantum LEGO}\label{section:CSSTP_generative_protocol}

\begin{figure*}
\centering
\includegraphics[width=\textwidth]{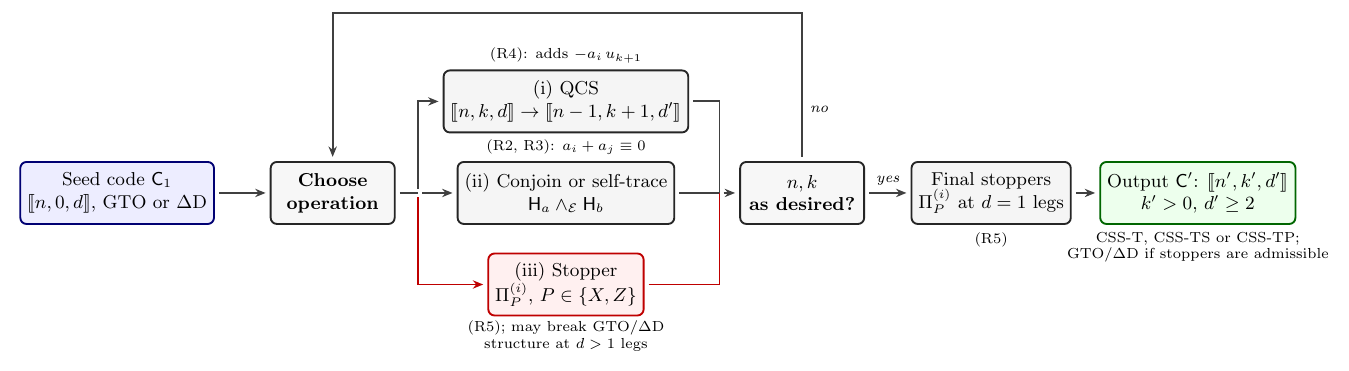}
\caption{Flowchart of the code-discovery protocol of \cref{con:protocol}, annotated with the rules of \cref{lem:operations} that govern the transversal layer at each operation; red marks stoppers that can break the GTO or $\Delta$D structure; Bell fusions can break it too (\cref{lem:fusion_structure}).}
\label{fig:generative_protocol}
\end{figure*}

We are now in a position to introduce the generative protocol used in this work.
In particular, \cref{fig:generative_protocol} introduces a method for discovering CSS-TP codes: starting from a seed code, it applies QCS, conjoining or a self-trace, or a stopper until $n$ and $k$ are as desired, and finishes with stoppers at legs of distance one.
The annotations in the flowchart name the rule of \cref{lem:operations} that decides whether the transversal layer survives each operation and what it then implements.
Unrestricted use of variant (iii) (shown in red), i.e.\ stoppers at legs where the code distance exceeds one, may break the stronger algebraic structure of the resulting codes (\cref{rem:trace_rules}) and thereby allows additional CSS and CSS-T codes to be discovered.
We treat these nuances in what follows.

\begin{construction}[CSS-TP Generative Protocol: Basic GTO/$\Delta$D Code Discovery]\label[construction]{con:protocol}
Given a GTO or $\Delta$D code $\mathsf{C}_{1}$ with parameters $\code{n,0,d}$, the following procedure generates candidate CSS codes.
The effect of every step on the layer is given by \cref{lem:operations}, and \cref{thm:protocol} states what this guarantees for the output.
\begin{enumerate}
    \item Initialize the GTO / $\Delta$D code $\mathsf{C}_{1}$ with parameters $\code{n,0,d}$.
    \item Choose either of the following:
    \begin{enumerate}
        \item[$(i)$] Perform QCS on $\mathsf{C}_{1}$ (cf.~\cref{section:background_code_modification}).
        This procedure works only inasmuch as the new code has $X$ stabilizers.
        For example, the $\code{16,0,4}$ QRM code\footnote{For $k=0$ we quote $d$ as the minimum weight of a non-trivial stabilizer.} has five $X$ stabilizers; therefore, we can execute quantum code shortening up to five times.
        \item[$(ii)$] Conjoin any pair of physical legs between $\mathsf{C}_{1}$ and another GTO or $\Delta$D code, which we shall call $\mathsf{C}_{2}$, as described in \cref{section:background_code_modification}. This procedure keeps GTO structure if no logical qubit is lost, but keeps $\Delta$-divisibility for $\Delta\ge4$ only if one of the two legs carries no $X$ stabilizer (\cref{lem:fusion_structure}).
        \item[$(iii)$] Apply an $X$ or $Z$ stopper $\Pi^{(i)}_{P}$ to a physical leg of $\mathsf{C}_{1}$.
    \end{enumerate}
    \item Repeat step~$(2)$ until $n,k$ are as desired or no acceptable operation remains.
    Because QCS and unrestricted stoppers can reduce the distance, recompute the stabilizer rank, logical operators, and distance after every operation.
    \item Inspect the stabilizer PCM and logical operators.
    At a leg where the relevant stabilizer support is absent, apply the appropriate $X$ or $Z$ stopper using the standard stabilizer-measurement update, remove the measured qubit, and recompute the stabilizers, logical operators, and distance.
    Repeat this procedure until either the code has $k>0$ and $d\geq2$, or no acceptable stopper remains.
    By \cref{lem:operations}(c) such stoppers keep the layer logical and give its new logical action.
    To keep GTO or $\Delta$D structure as well, require every stopper to be structure-admissible (\cref{def:structure_admissible_trace}).
\end{enumerate}
\end{construction}

  \noindent Allowing stoppers as intermediate operations substantially enlarges the search space, permitting the discovery of CSS, CSS-T, CSS-TS, and CSS-TP codes in addition to GTO and $\Delta$D codes.
  A stopper always preserves the CSS stabilizer form, but an unrestricted stopper need not preserve generalized triorthogonality, $\Delta$-divisibility, or a transversal non-Clifford logical action.
  Bell fusions preserve the first two only under the conditions of \cref{lem:fusion_structure}.
  Structure is checked through \cref{def:structure_admissible_trace,lem:fusion_structure}; the logical action is decided exactly by \cref{lem:operations}, for every operation.

We now state what the lemmas guarantee along the protocol.

  \begin{theorem}[Logical action along the generative protocol]\label[theorem]{thm:protocol}
  Let a seed code carry a layer $V$ as in \cref{eq:diag_layer} that is logical on it, with logical phase polynomial $\phi$ given by \cref{lem:logical_poly}.
  Apply a finite sequence of QCS steps, Bell fusions (conjoining or self-traces) and stoppers, and restrict the layer to the surviving legs after each step.
  \begin{enumerate}
    \item The restricted layer remains logical at every step iff each step meets the condition of \cref{lem:operations}; QCS steps and $Z$ stoppers always do.
    \item The logical phase polynomial of the output is obtained by composing the updates of \cref{lem:operations}, and its level in the Clifford hierarchy follows from \cref{cor:level}.
    The construction applies at every level $\ell$ of the physical layer, and the output gate never lies above level $\ell$; it is non-Clifford iff its level is at least $3$.
    \item If in addition the seed and every conjoined block are GTO, every QCS step clears
    the deleted coordinate as in \cref{lem:QCS}, every stopper is structure-admissible
    (\cref{def:structure_admissible_trace}) and every Bell fusion keeps all logical qubits,
    the output is GTO. 
    If instead the seed and every conjoined block are $\Delta$-divisible, the same QCS and stopper conditions hold, and at every Bell fusion with $\Delta\nmid2$ one of the two fused legs
    carries no $X$ stabilizer, the output is $\Delta$-divisible (\cref{lem:fusion_structure}).
  \end{enumerate}
  \end{theorem}

  Items 1 and 2 apply \cref{lem:operations,cor:level} step by step; item 3 adds the structural guarantee.
  Item 3 is not needed for the logical action: the $\code{14,6,2}$ code is GTO but only $2$-divisible, and the closed chains are neither GTO nor $8$-divisible, yet all carry the predicted gates.
  We prove \cref{thm:protocol} in \cref{section:generative_protocol_proofs}.
  Distance is not controlled by the theorem; we compute it for every output.

\subsection{Tensor-product and irreducible resources}\label{section:tiers}

\cref{def:tiers} asks whether the free operations $\mathcal F$ can split a resource state into canonical pieces: a Tier-0 state is a single canonical state, a Tier-1 state is a tensor product of several, and a Tier-2 state admits no such split.
Here we give a simple test for the tier, so that each code in \cref{section:CSSTP_examples} can be classified.
We restrict to the Boolean cubic resources that make up most of this work, $\overline{\mathcal M}=(-1)^{f(u)}$ with $f$ a cubic polynomial in the logical bits $u\in\mathbb F_{2}^{k}$; the canonical piece is then $\ket{CCZ}$.
As logical $CX$ networks replace $u$ by $Au$ for an invertible binary matrix $A$, a logical $X$ gate may flip bits of $u$; additionally, diagonal Clifford gates $Z$, $S$ and $CZ$ only add terms of degree at most two to $f$.
Only the cubic terms of $f$ therefore matter, and only up to a change of logical basis.
The tier asks whether some basis splits these cubic terms into $CCZ$ gates on disjoint triples of qubits.
We find that two numbers determine the tier, effectively: the \emph{stabilizer nullity} of the resource state~\cite{beverland2020lower} and its \emph{$CCZ$-rank}.

We start with the first of these two numbers.
For a pure $k$-qubit state $\ket\psi$, let
\begin{align}\label{eq:stab_group}
\mathrm{Stab}(\psi)=\bigl\{P\in\{\pm1\}\times\{I,X,Y,Z\}^{\otimes k}\;:\;P\ket\psi=\ket\psi\bigr\}
\end{align}
be the set of signed Pauli operators that have $\ket\psi$ as a $+1$ eigenvector.
$\mathrm{Stab}(\psi)$ is a group whose elements commute and square to $I$, and it does not contain $-I$, so $\mathrm{Stab}(\psi)\cong\mathbb F_{2}^{s}$ for some $0\le s\le k$.
The \emph{stabilizer nullity} of Beverland \emph{et al.}~\cite{beverland2020lower} is
\begin{align}\label{eq:nullity}
\nu(\psi)=k-\log_{2}\bigl|\mathrm{Stab}(\psi)\bigr|=k-s~.
\end{align}
The nullity $\nu$ vanishes on stabilizer states ($s=k$), is invariant under Clifford unitaries, is additive, $\nu(\psi\otimes\phi)=\nu(\psi)+\nu(\phi)$, and cannot increase under stabilizer protocols that prepare a pure state: Clifford unitaries, adjoining stabilizer states, Pauli measurements with classical feed-forward, and discarding qubits~\cite{beverland2020lower}.
For our resources we write $\nu(f)=\nu(\psi_{f})$, where $\ket{\psi_{f}}=\overline{\mathcal M}\ket{+}^{\otimes k}=2^{-k/2}\sum_{u}(-1)^{f(u)}\ket u$.
A single $\ket{CCZ}$ state has $\nu=3$ (shown below), so by additivity $r$ $CCZ$ states on disjoint triples have $\nu=3r$, and by Clifford invariance so does every state equivalent to them.
A resource whose stabilizer nullity is not a multiple of three can therefore never be split into $CCZ$ states.

To compute $\nu(f)$ without enumerating Pauli operators, we instead ask which logical $X$ operators the gate maps back to Pauli operators.
Let $\overline X^{w}$ flip the logical bits marked by $w\in\mathbb F_{2}^{k}$.
Then
\begin{align}\label{eq:pauli_direction}
\overline{\mathcal M}\,\overline X^{w}\,\overline{\mathcal M}^{\dagger}=\overline X^{w}\,(-1)^{f(u+w)+f(u)}~,
\end{align}
and the cubic terms cancel in $f(u+w)+f(u)$, which therefore has degree at most two.
If it has degree at most one, the right-hand side is a Pauli operator, and we call $w$ a \emph{Pauli direction} of $f$.
For example, flipping $u_{1}$ in $f=u_{1}u_{2}u_{3}$ changes $f$ by $u_{2}u_{3}$, and no nonzero $w$ is a Pauli direction of this $f$; flipping a qubit that no cubic term touches is always one.
The Pauli directions form a subspace $N(f)\subseteq\mathbb F_{2}^{k}$ (\cref{section:tier_proofs}).

First, we note that every Pauli direction gives a stabilizer.
For $w\in N(f)$ write $f(u+w)+f(u)=c_{w}+\ell_{w}\cdot u$ with $c_{w}\in\mathbb F_{2}$ and $\ell_{w}\in\mathbb F_{2}^{k}$ being the two parts of the change in $f$ when bits in $w$ are flipped.
Then the right-hand side of \cref{eq:pauli_direction} is the signed Pauli operator $P_{w}=(-1)^{c_{w}}\,\overline X^{w}\,\overline Z^{\ell_{w}}$.
Since $\overline X^{w}$ stabilizes $\ket{+}^{\otimes k}$, $P_{w}=\overline{\mathcal M}\,\overline X^{w}\,\overline{\mathcal M}^{\dagger}$ stabilizes $\ket{\psi_{f}}$.
Conversely, every stabilizer comes from a Pauli direction.
Up to sign, any Pauli operator is $\overline X^{w}\overline Z^{w'}$, and $\overline X^{w}\overline Z^{w'}\ket{\psi_{f}}\propto\sum_{u}(-1)^{f(u+w)+w'\cdot u}\ket u$.
The right-hand side equals $\pm\ket{\psi_{f}}$ iff $f(u+w)+f(u)+w'\cdot u$ is constant, i.e.\ iff $w\in N(f)$ and $w'=\ell_{w}$.

So $w\mapsto P_{w}$ is a bijection from $N(f)$ onto $\mathrm{Stab}(\psi_{f})$.
This bijection is also a group isomorphism, because conjugation by $\overline{\mathcal M}$ preserves products.
Hence $|\mathrm{Stab}(\psi_{f})|=2^{\dim N(f)}$, and
\begin{align}\label{eq:ess_equals_nullity}
\nu(f)=k-\dim N(f)~.
\end{align}
For example, $f=u_{1}u_{2}u_{3}$ has $N(f)=\{0\}$, so $\ket{CCZ}$ is stabilized only by $I$ and $\nu(\ket{CCZ})=3$.
The number $k-\dim N(f)$ also has a direct meaning: it is the smallest number of logical qubits that the cubic terms of $f$ act on, over all choices of logical basis (\cref{lem:ess_pauli}).

Whereas the stabilizer nullity counts the qubits on which a resource acts, the second number measures what it costs.
The \emph{CCZ-rank} $r(f)$ is the smallest number of $CCZ$ gates that implement $\overline{\mathcal M}$ together with $CX$ networks and diagonal Clifford gates.
Equivalently, it is the smallest $r$ for which $f$ is a sum of $r$ products of three linear combinations of the logical bits, up to terms of degree at most two.
Each such product involves at most three independent directions, so $\nu(f)\le3r(f)$; a split into disjoint $CCZ$ gates needs equality.

\begin{lemma}[Deciding the tier]\label[lemma]{lem:tiers}
Let $f$ and $g$ be Boolean cubic polynomials in $k$ variables.
\begin{enumerate}
\item[(i)] The states $(-1)^{f}\ket{+}^{\otimes k}$ and $(-1)^{g}\ket{+}^{\otimes k}$ are equivalent under $\mathcal F$ iff $g(u)=f(Au+b)$ up to terms of degree at most two, for some invertible binary matrix $A$ and some $b\in\mathbb F_{2}^{k}$.
\item[(ii)] The state $(-1)^{f}\ket{+}^{\otimes k}$ is equivalent under $\mathcal F$ to $\ket{CCZ}^{\otimes r}\otimes\ket{+}^{\otimes(k-3r)}$ iff $\nu(f)=3r$ and $r(f)=r$.
\end{enumerate}
Consequently, $(-1)^{f}\ket{+}^{\otimes k}$ is Tier 0 if $\nu(f)=3$, Tier 1 if $\nu(f)=3r(f)\ge6$, and Tier 2 otherwise.
In particular, it is Tier 2 whenever $\nu(f)$ is not divisible by $3$.
\end{lemma}

Part (i) states precisely what the free operations do to $f$.
Part (ii) is the test: a resource splits into $r$ $CCZ$ states exactly when its cubic terms need $3r$ qubits and only $r$ $CCZ$ gates.
Divisibility by three is the quickest way to recognize an irreducible resource, but it is not sufficient on its own: some polynomials with $\nu=6$ have CCZ-rank $3$ and are Tier 2 (\cref{section:tier_proofs}).
We prove the lemma in \cref{section:tier_proofs}.

To see the test at work, consider the $\code{12,5,2}$ code, which implements $f=c\,a_{1}b_{1}+c\,a_{2}b_{2}$ on $u=(c,a_{1},b_{1},a_{2},b_{2})$ (\cref{eq:1252_action}).
Flipping $c$ changes $f$ by $a_{1}b_{1}+a_{2}b_{2}$, and flipping $a_{1}$, $b_{1}$, $a_{2}$ or $b_{2}$ changes it by $cb_{1}$, $ca_{1}$, $cb_{2}$ or $ca_{2}$, plus terms of degree at most one.
When several bits are flipped at once, these quadratic changes add up, and since they involve different monomials they never cancel.
No nonzero $w$ is a Pauli direction, so $\nu(f)=5$.
Five is not divisible by three, so the resource is Tier 2: no free operation turns it into a product of $CCZ$ states.
For the $\code{14,6,2}$ code, by contrast, $f=u_{0}u_{1}u_{2}+u'_{0}u'_{1}u'_{2}$ has $\nu=6$ and $r(f)=2$, so that resource is Tier 1.

Beyond deciding the tier, the stabilizer nullity also bounds the cost of a resource, under the full Clifford group and not only under the free operations.

\begin{proposition}[Stabilizer nullity of Boolean cubic resources]\label[proposition]{prop:nullity}
For every Boolean cubic $f$ in $k$ variables, $\nu(f)=k-\dim N(f)$, and this is the smallest number of logical qubits that the cubic terms of $f$ act on, over all choices of logical basis.
Consequently:
\begin{enumerate}
\item[(i)] if $\nu(f)$ is not divisible by $3$, then $\ket{\psi_{f}}$ is not equivalent to $\ket{CCZ}^{\otimes r}\otimes\ket{+}^{\otimes(k-3r)}$ for any $r$, even under the full Clifford group, since the latter has $\nu=3r$;
\item[(ii)] a stabilizer protocol that prepares $\ket{\psi_{f}}$ exactly from $\ket{CCZ}^{\otimes m}$ and stabilizer ancillas needs $m\ge\lceil\nu(f)/3\rceil$, and $m=r(f)$ copies suffice.
\end{enumerate}
\end{proposition}

We prove the proposition in \cref{section:tier_proofs}.
Part (i) means that the divisibility test rules out a split into $CCZ$ states under the full Clifford group, not only under the free operations.
Part (ii) says how many $CCZ$ states an irreducible resource is worth.
For the $\code{12,5,2}$ code ($\nu=5$, $r=2$) and the $R=3$ star ($\nu=7$, $r=3$) the lower and upper bounds coincide: each resource is worth exactly two or three $CCZ$ states assembled by teleportation, and its advantage is that it arrives as a single output and is consumed in a single step.
The stabilizer R\'enyi entropy $M_{2}$ gives a further test, which also covers canonical states other than $\ket{CCZ}$ (\cref{section:sre_certificate}).

\subsection{Example Code Families}\label{section:CSSTP_examples}

This subsection works through five examples of \cref{con:protocol}.
The first rederives a known code and shows each rule of \cref{lem:operations} in action; the other four are infinite families.
In every case the logical action follows from \cref{lem:logical_poly,lem:operations}, and we also verified it numerically, with \texttt{CSSLO}~\cite{webster2023transversal} and by direct enumeration of the codewords.
The tiers of \cref{section:tiers} sort the resulting resources, and \cref{tab:classification} at the end of the subsection summarizes them.

\subsubsection{The \texorpdfstring{$\code{8,3,2}$}{[[8,3,2]]} code from the QRM code}\label{section:family_832}

We derive the $\code{8,3,2}$ color code~\cite{kubica2015unfolding,hangleiter2025fault,vasmer2022morphing} from the $\code{16,0,4}$ quantum Reed--Muller (QRM) code in three steps (\cref{fig:832_derivation}).
\begin{enumerate}
\item \emph{Seed.}
The QRM code is triply even and has five $X$ stabilizers.
The layer $T^{\otimes16}$ is logical, with trivial action since $k=0$.
\item \emph{Four QCS steps.}
In the standard Reed--Muller ordering we shorten on qubits $12$--$15$.
Each step removes a qubit and promotes one $X$ stabilizer to a logical operator, which gives a $\code{12,4,1}$ code.
By \cref{lem:operations}(b) each step adds $-u_{\mathrm{new}}$ to the logical phase polynomial, so $\phi=-(u_{1}+u_{2}+u_{3}+u_{4})$ and the layer implements $\overline{T}^{\dagger\otimes4}$.
\item \emph{Four $X$ stoppers.}
No $X$ stabilizer of the $\code{12,4,1}$ code covers qubits $0$, $1$, $2$ and $11$, so $X$ stoppers there keep the layer logical (\cref{lem:operations}(c)).
At the first three legs $X_{i}$ anticommutes with a $Z$ stabilizer, which the stopper removes together with the qubit, and $k$ is unchanged.
At the last leg $X_{i}$ is a logical operator, and the stopper removes a logical qubit.
Each stopper subtracts $u\cdot g_{i}$, and expanding these parities as in \cref{eq:xtrace_example} turns the single-qubit phases into quadratic and cubic terms.
\end{enumerate}
The output is an $\code{8,3,2}$ code with the stabilizers and logical operators of the standard one~\cite{ray2024implementing}.
The inherited layer $T^{\otimes8}$ implements $\overline{CCZ}$ times diagonal Clifford gates, and the balanced layer implements exactly
\begin{align}
\overline{\mathcal{M}}_{\code{8,3,2}} = \overline{CCZ}_{(0,1,2)} = T_{0,1,2,3}\, T^{\dagger}_{4,5,6,7}~.
\end{align}
The stoppers therefore do not preserve the gate of the $\code{12,4,1}$ code: they turn four single-qubit gates into an entangling gate, as \cref{lem:operations} predicts.
\input{includes/derivation_832}

\subsubsection{Open chains}\label{section:open_chains}
\input{includes/open_chain.tex}
Consider a sequence of $\code{8,3,2}$ codes $\mathsf{C}^{(1)}, \ldots, \mathsf{C}^{(m+2)}$ arranged in a linear chain, in which adjacent codes are conjoined along a single physical index via $\mathsf{H}^{(j)} \wedge_{\mathcal{E}} \mathsf{H}^{(j+1)}$ for $j = 1, \ldots, m+1$; some of these are shown in \cref{fig:832_openchain}.
Each conjoining operation identifies one physical qubit from each code and contracts the resulting tensor network (cf.~\cref{section:background_code_modification}).
The resulting code family\footnote{This is the same family as the $\code{6(L+1)+2,\,3(L+1),\,2}$ chains of $\code{8,3,2}$ tiles of Cao and Lackey~\cite{cao2026quantum}, with $L=m+1$ and the same $\overline{CCZ}^{\otimes(m+2)}$ action.
What we add for this family is the analytic derivation of its logical action and the closed-chain and star variants below.} has parameters
\begin{align}
\code{6m+14,\, 3m+6,\, 2}~, \quad m \geq 0~.
\end{align}
The first representative, at $m=0$, is a $\code{14,6,2}$ code formed by conjoining two copies of $\code{8,3,2}$, and we display its stabilizers and logical operators in \cref{section:further_examples_CSStp}.
\input{includes/logical_gates_unclean}
Using the numerical tool \texttt{CSSLO}~\cite{webster2023transversal}, we verify the transversal non-Clifford logical gate, which is depicted in \cref{fig:1462_uncleaned_gate}.
Here, the first line expresses the logical action on six logical qubits (with subscript pairs and triples indicating the logical qubit support of each $\overline{CZ}$ and $\overline{CCZ}$ gate, respectively), and the second line denotes the transversal physical implementation in terms of $T$, $T^{\dagger}$, and $Z$ gates.
The $\overline{CZ}$ terms depend on the choice of logical basis and carry no structure.
Up to free operations the $\code{14,6,2}$ code implements $\overline{CCZ}\otimes\overline{CCZ}$, the product of its two cubes' gates (\cref{cor:conjoin_loop}).
The $\code{14,6,2}$ resource is therefore a Tier-1, tensor-product CSS-TP resource (\cref{def:tiers}): $\nu=6$, CCZ-rank $2$ and $M_{2}=2M_{2}(CCZ)$ (\cref{section:tiers}).

The same action also follows analytically from the rules of \cref{section:gate_calculus_operations}, without numerical tools.
To see this, label the qubits of an $\code{8,3,2}$ cube by its vertices $v=(v_{0},v_{1},v_{2})\in\{0,1\}^{3}$ as in the worked example of \cref{section:gate_calculus_examples}, so that its $X$-type codewords are $y_{v}=t\oplus u\cdot v$, with logical coordinates $u\in\mathbb F_{2}^{3}$ and stabilizer coordinate $t\in\mathbb F_{2}$, and the layer $a_{v}=(-1)^{|v|}$ gives $\sum_{v}a_{v}y_{v}\equiv4u_{0}u_{1}u_{2}\pmod 8$ for both values of $t$.
Take two cubes with coordinates $(t,u)$ and $(t',u')$, and conjoin vertex $h$ of the first with vertex $s$ of the second, where $|h|$ and $|s|$ have opposite parity, so that $a_{h}+a'_{s}\equiv0$.
The conjoined code keeps the codeword pairs with $y_{h}=y'_{s}$, i.e.\ $t'=t\oplus u\cdot h\oplus u'\cdot s$.
The constraint fixes the second stabilizer coordinate and leaves $(t,u,u')$ free (\cref{cor:conjoin_loop}(1)), so $n=14$ and $k=6$, and the two $X$-checks merge into one of weight $14$, so the code is GTO with $k_{CCZ}=2$ but only $2$-divisible (\cref{lem:fusion_structure}).
The phase on the $14$ kept legs is
\begin{align}
&\sum_{v\ne h}a_{v}y_{v}+\sum_{v\ne s}a'_{v}y'_{v}\nonumber\\
&\quad=4u_{0}u_{1}u_{2}+4u'_{0}u'_{1}u'_{2}-(a_{h}+a'_{s})\,y_{h}\nonumber\\
&\quad\equiv4\,(u_{0}u_{1}u_{2}+u'_{0}u'_{1}u'_{2})\pmod 8~,
\end{align}
so in the basis of the two cubes
\begin{align}
\overline{\mathcal M}_{\code{14,6,2}}=\overline{CCZ}(u_{0},u_{1},u_{2})\,\overline{CCZ}(u'_{0},u'_{1},u'_{2})
\label{eq:1462_action}
\end{align}
exactly, with no Clifford terms.
The $\overline{CZ}$ terms of \cref{fig:1462_uncleaned_gate} come from a different choice of logical representatives.
The same argument for two $\code{16,4,2}$ codes ($\ell=4$, $c_{S}=8$) gives a $\code{30,8,2}$ code with $\overline{CCCZ}\otimes\overline{CCCZ}$; the $\code{30,8,2}$ code of \cref{section:further_examples_CSStp} is this code up to a permutation of the qubits.

\subsubsection{Closed chains}\label{section:closed_chains}
A further family is obtained by a self-trace of the first and last physical legs of the open chain, which closes the chain.
The two closing legs must carry opposite layer values: the merged $X$ stabilizer covers both, so by \cref{lem:operations} any other choice destroys the transversal gate.
Because the merged stabilizer covers both legs, the self-trace removes one logical qubit, and the result is not GTO (\cref{lem:fusion_structure}).
The self-trace then restricts the product of $\overline{CCZ}$ gates to a hyperplane that couples the factors (\cref{cor:conjoin_loop}).
The self-trace thus yields a code family with parameters
\begin{align}
\code{6m+12,\, 3m+5,\, 2}~, \quad m \geq 0~.
\end{align}
\cref{fig:832_openchain}(b) draws the $m=1$ member, $\code{18,8,2}$.
The transversal logical non-Clifford gate is shown in \cref{fig:1252_uncleaned_gate}.
As for the open chain, this action can be derived by hand, now through case~2 of \cref{cor:conjoin_loop}.
We close the two-cube chain ($m=0$) by conjoining a second pair, $h\oplus e_{0}$ with $s\oplus e_{0}$, where $h$ and $s$ are the vertices conjoined above and $e_{0}=(1,0,0)$; that is, the edge $\{h,h\oplus e_{0}\}$ of the first cube is conjoined to the edge $\{s,s\oplus e_{0}\}$ of the second cube.
Both parities flip, so again $a_{h\oplus e_{0}}+a'_{s\oplus e_{0}}\equiv0$.
The new constraint is $y_{h\oplus e_{0}}=y'_{s\oplus e_{0}}$.
Since $y_{h\oplus e_{0}}=y_{h}\oplus u_{0}$, $y'_{s\oplus e_{0}}=y'_{s}\oplus u'_{0}$ and $y_{h}=y'_{s}$, it reads $u_{0}=u'_{0}$.
The merged $X$-check takes the value $1$ at both legs, so no stabilizer coordinate absorbs the constraint, and one logical qubit is removed (\cref{cor:conjoin_loop}(2)): $n=12$, $k=5$.
The phase is that of \cref{eq:1462_action} restricted to $u_{0}=u'_{0}=:c$,
\begin{align}
\overline{\mathcal M}_{\code{12,5,2}}=\overline{CCZ}(c,u_{1},u_{2})\,\overline{CCZ}(c,u'_{1},u'_{2})~,
\label{eq:1252_action}
\end{align}
two $\overline{CCZ}$ gates that share the qubit $c$, again with no Clifford terms.
We checked both derivations by enumerating all $2^{k+1}$ $X$-type codewords of the conjoined codes; the codes obtained are the $\code{14,6,2}$ and $\code{12,5,2}$ codes of \cref{section:further_examples_CSStp} up to a permutation of the physical qubits.
The star family below repeats the second step with $R-1$ spokes.
Both families satisfy $n=2k+2$, so the closed chain offers no block-length advantage at fixed $k$.
The closed chain's logical action differs: the two $\overline{CCZ}$ factors of the $\code{12,5,2}$ code share one logical qubit, whereas the open chain yields disjoint factors.
No free operation separates them: for $m=0,1,2$ the closed chains have $\nu=5,8,11$, which is not divisible by $3$, so by \cref{lem:tiers} they are Tier 2, irreducible CSS-TP resources.

\subsubsection{Stars of \texorpdfstring{$CCZ$}{CCZ} gates}\label{section:star_family}

The closed chain generalizes to $R$ $\overline{CCZ}$ gates that share one logical qubit.
Label the qubits of each $\code{8,3,2}$ cube by $v\in\{0,1\}^{3}$ as above, so that logical qubit $u_{i}$ has $\bar X_{i}$ on the face $v_{i}=1$ and $\bar Z_{i}$ on an edge parallel to direction $i$.
Take a hub cube and $R-1$ spoke cubes, all with the layer $a_{v}=(-1)^{|v|}$.
\emph{Glue} each spoke to the hub, i.e.\ conjoin it twice, as in the closed chain: the endpoints of a hub edge $\{h,h\oplus e_{0}\}$ to the endpoints of a spoke edge $\{s,s\oplus e_{0}\}$, with $|h|$ and $|s|$ of opposite parity (for example $h=(0,1,0)$ and $s=(0,0,0)$), and a different hub edge for each spoke.
The hub has four direction-$0$ edges, so $R\le5$.
The result is a $\code{4R+4,\,2R+1,\,2}$ code with logical action
\begin{align}
\overline{\mathcal M}=\prod_{r=1}^{R}\overline{CCZ}\bigl(c,u^{(r)}_{1},u^{(r)}_{2}\bigr)~,
\label{eq:star_action}
\end{align}
where $c=u_{0}$ is the hub's direction-$0$ logical qubit, $u^{(1)}_{1,2}=u_{1,2}$ are the hub's other two logical qubits, and $u^{(r)}_{1,2}$ for $r\ge2$ are the direction-$1$ and direction-$2$ logical qubits of spoke $r$.
The $R=2$ member is the $\code{12,5,2}$ closed chain, and \cref{fig:832_openchain}(c) draws the $R=3$ member, $\code{16,7,2}$, with $a_{r}=u^{(r)}_{1}$ and $b_{r}=u^{(r)}_{2}$.

The derivation is that of the $\code{12,5,2}$ code, repeated for each spoke.
The first conjoining of a spoke fixes the spoke's stabilizer coordinate (\cref{cor:conjoin_loop}(1)).
The second identifies the spoke's direction-$0$ logical qubit with $c$ (\cref{cor:conjoin_loop}(2)), which removes one logical qubit.
Every conjoined pair joins vertices of opposite parity, so the removed legs contribute no phase, and each cube contributes $4$ times the product of its three logical bits.
Summing gives \cref{eq:star_action}, with $n=8+4(R-1)$, $k=3+2(R-1)$ and no Clifford terms.
\cref{prop:glued_stars} below proves the same statement for hypercubes of any dimension.
For general $R$ the same action is obtained by a logical merge: take $R$ disjoint cubes and add the $Z$-type stabilizers $\bar Z_{0}^{(1)}\bar Z_{0}^{(r)}$ for $r=2,\dots,R$.
This merge gives $\code{8R,\,2R+1,\,2}$, which uses more qubits than the glued star and has $R$ $X$ checks; we do not use it in the benchmarks.

The shared qubit changes the resource.
Every star is Tier 2 (\cref{section:tiers}): $\nu=2R+1$, which is not divisible by $3$ for $R=2,3,5$, and for $R=4$ the rank histogram of \cref{section:tier_proofs} differs from that of three disjoint $CCZ$ gates.
The stabilizer R\'enyi entropy of the $R=3$ star is $M_{2}=2.850$, against $4.622$ for three disjoint $CCZ$ gates, so by \cref{prop:sre_certificate} the star is not even Clifford-equivalent to three independent $CCZ$ gates.

\subsubsection{Stars of \texorpdfstring{$CCCZ$}{CCCZ} gates}\label{section:cccz_star}
The same gluing works one level up, and for hypercubes of any dimension.
Label the qubits of the $\code{2^{m},m,2}$ hypercube code~\cite{vasmer2022morphing,hangleiter2025fault} by $v\in\{0,1\}^{m}$.
Its logical-$X$ rows are the faces $v_{i}=1$ and its $X$ stabilizer is the all-ones vector, so its $X$-type codewords are $y_{v}=t\oplus u\cdot v$.
The signed layer $a_{v}=(-1)^{|v|}$ at level $\ell=m$ ($T_{m}$ on even and $T_{m}^{\dagger}$ on odd vertices; for $m=4$ the signed $\sqrt T$ layer) gives
\begin{align}
\sum_{v}a_{v}\,y_{v}\equiv2^{m-1}\,u_{0}u_{1}\cdots u_{m-1}\pmod{2^{m}}
\label{eq:cube_phase}
\end{align}
for both values of $t$.
If some $u_{i}=0$, the sum over $v_{i}$ of $(-1)^{v_{i}}$ vanishes and so does the whole sum.
If every $u_{i}=1$, then $y_{v}=t\oplus|v|$ and the sum is $\mp2^{m-1}$ for $t=0,1$, which is $2^{m-1}$ modulo $2^{m}$ in both cases.
The hypercube therefore implements $\overline{C^{m-1}Z}$.

\begin{proposition}[Glued $C^{m-1}Z$ stars]\label[proposition]{prop:glued_stars}
Let $m\ge3$ and $2\le R\le2^{m-1}+1$.
Glue $R-1$ spoke hypercubes to a hub hypercube, each spoke twice, joining a hub edge $\{h,h\oplus e_{0}\}$ to a spoke edge $\{s,s\oplus e_{0}\}$ with $|h|$ and $|s|$ of opposite parity, and a different hub edge for each spoke.
With the signed layer on every hypercube, the result is an $\code{(2^{m}-4)R+4,\,(m-1)R+1}$ code on which the layer implements
\begin{align}
\overline{\mathcal M}=\prod_{r=1}^{R}\overline{C^{m-1}Z}\bigl(c,u^{(r)}_{1},\dots,u^{(r)}_{m-1}\bigr)
\label{eq:glued_star_action}
\end{align}
in the hub/spoke basis, with no Clifford terms.
Here $c=u_{0}$ and $u^{(1)}_{i}=u_{i}$ are the hub's logical qubits, and $u^{(r)}_{i}$ for $r\ge2$ and $i\ge1$ are those of spoke $r$.
\end{proposition}

\begin{proof}
Give spoke $r$ the coordinates $(t_{r},u^{(r)})$.
The contraction of $h$ with $s$ imposes $t_{r}=t\oplus u\cdot h\oplus u^{(r)}\cdot s$, which fixes the spoke's stabilizer coordinate (\cref{cor:conjoin_loop}(1)).
Given this, the contraction of $h\oplus e_{0}$ with $s\oplus e_{0}$ imposes $u^{(r)}_{0}=u_{0}=c$ (\cref{cor:conjoin_loop}(2)) and removes one logical qubit.
Every contracted pair joins vertices of opposite parity, so $a_{h}+a'_{s}\equiv0$ and the deleted legs contribute nothing to \cref{eq:op_phase}.
The phase on the kept legs is therefore the sum of the hypercube phases \cref{eq:cube_phase}, $2^{m-1}\sum_{r=1}^{R}c\,u^{(r)}_{1}\cdots u^{(r)}_{m-1}$, which is \cref{eq:glued_star_action}.
Each spoke adds $2^{m}-4$ qubits and $m-1$ logical qubits, and the hub has $2^{m-1}$ direction-$0$ edges, which bounds $R$.
\end{proof}

For $m=4$, the proposition gives $\code{12R+4,\,3R+1,\,2}$ codes with
\begin{align}
\overline{\mathcal M}=\prod_{r=1}^{R}\overline{CCCZ}\bigl(c,u^{(r)}_{1},u^{(r)}_{2},u^{(r)}_{3}\bigr)~,
\end{align}
for example $\code{28,7,2}$, $\code{40,10,2}$ and $\code{52,13,2}$ for $R=2,3,4$.
As for the other glued codes, $A_{2}=\binom n2$ ($378$, $780$ and $1326$).
These codes provide $CCCZ$ gates that share a control, which \Mfac{30,8,2} cannot.
We cost the $R=3$ member as a factory in \cref{section:fourth_level}.
The $m=5$ and $m=6$ members, stars of $C^{4}Z$ and $C^{5}Z$ gates, are given in \cref{section:higher_stars}.

We also note that up to physical-qubit permutations, changes of logical representatives, and invertible row operations, the $\code{12,5,2}$ and $\code{14,6,2}$ codes are equivalent to the \Tofs and \twoCCZ synthillation codes of Ref.~\cite{campbell2017unified}, where $\Tofs=CCZ_{123}CCZ_{345}$ is a pair of $CCZ$ gates sharing one qubit and $\twoCCZ=CCZ_{123}CCZ_{456}$ is a disjoint pair.
We computed the column permutation and invertible row transformation for each code (\cref{section:further_examples_CSStp}); the leading error coefficients also agree, $A_{2}=66$ and $91$.
The star family has the same block length $4R+4$, the same logical action and the same $A_{2}=\binom{n}{2}$ as the $U_{N\#}$ family of Ref.~\cite{campbell2017unified} with $N=R$.
Campbell and Howard give the generator matrix explicitly only for $N=2$, so for $R\ge3$ we establish the correspondence through parameters, action and $A_{2}$ rather than an explicit permutation.
Recovering these protocols provides a consistency check for the CSS-TP construction.

In their own right, Gong et al.~\cite{anqiwills2026} reach the same states from a different direction: they contain multi-qubit gates such as $CS$, $CCZ$ and \Tofs, the \Mfac{12,5,2} action, into single-qudit gates over binary extension fields $\mathbb F_{2^{s}}$, and obtain compact qubit distillation protocols from algebraic-geometry codes over those fields.
Their construction starts from the desired gate, as synthillation does, whereas ours starts from conjoined codes and derives the action.
Our broader contribution then is the generative code framework, the logical basis transformation method of \cref{section:CSSTP_gate_cleaning}, and the extension to additional code families and higher levels of the Clifford hierarchy.

Moving past the works of \cite{campbell2017unified,campbell2017unifying,anqiwills2026}, our formalism also extends beyond the third level of the Clifford hierarchy.
In Refs.~\cite{vasmer2022morphing,hangleiter2025fault}, the class of $\code{2^{\mathcal{D}},\mathcal{D},2}$ \emph{hypercube codes} was shown to be equipped with transversal logical $\overline{C^{\mathcal{D}-1}Z}$ gates.
Setting $\mathcal{D} = 5$ yields a code with transversal logical $\overline{CCCCZ}$ gate, which is amenable to the conjoining procedure of ~\cref{con:protocol}.
We provide more details about this code in \cref{section:higher_stars}.

Beyond the families above, we include additional ones in \cref{section:further_examples_CSStp}, several of which have distance greater than two.
The codes presented here can also be concatenated to achieve higher distance, as discussed in \cref{section:concatenation_distance_growth}.

With all families in hand, we can now sort them by the tiers of \cref{section:tiers}.
\cref{tab:classification} lists the tier of every code in this work, the method used to establish it, and the stabilizer R\'enyi entropy $M_{2}$ of its resource state.
Most of our codes are Tier 1.
The open chains, the $\code{30,8,2}$ code and the distance-three QRM chains exhibit two or more canonical states in one factory; this can reduce cost (\cref{section:tailored_algorithmic_primitives}) but does not create a new resource.
The closed chains ($m\le2$), the stars, and the concatenated $\code{64,9,4}$, $\code{96,15,4}$ and $\code{180,5,6}$ codes are Tier 2, i.e.\ irreducible CSS-TP: each delivers in one output what canonical factories deliver only with additional teleportations.
Their $M_{2}$ is not an integer multiple of $M_{2}(CCZ)=1.541$.

\input{includes/classification_table}

\subsection{Logical Basis Transformations and Reduced Forms}\label{section:CSSTP_gate_cleaning}

Any diagonal gate on $n$ qubits can be represented as a \emph{phase polynomial} over $\mathbb{Z}_{2^{\ell}}$~\cite{amy2014polynomial}.
That is, for a diagonal unitary defined by
\begin{align}
U\ket{u} = \omega^{\phi(u)}\ket{u}~,
\end{align}
$\phi(u)$ can be written as
\begin{align}
\phi(u) = \sum_{S \subseteq \{1,\ldots,n\}} c_{S} \prod_{j \in S} u_{j} \pmod{2^{\ell}}~,
\end{align}
where $c_{S} \in \mathbb{Z}_{2^{\ell}}$, as in \cref{eq:logical_poly}.
For $\ell=3$, i.e.\ $\omega=e^{i\pi/4}$, the standard diagonal gates contribute
\begin{align}
T_{i} &= u_{i}~, \nonumber\\
S_{i} &= 2u_{i}~, \nonumber\\
Z_{i} &= 4u_{i}~, \nonumber\\
CS_{(i,j)} &= 2u_{i}u_{j}~, \nonumber\\
CZ_{(i,j)} &= 4u_{i}u_{j}~, \nonumber\\
CCZ_{(i,j,k)} &= 4u_{i}u_{j}u_{k}~.
\end{align}

As discussed previously, the resource state produced by a given CSS-TP code encodes a fixed product of logical non-Clifford gates specified by its transversal gate $\overline{\mathcal{M}}$.
However, for practical applications one often requires a different arrangement of the same non-Clifford gates.
For instance, overlapping $\overline{CCZ}$ gates that share one control, or a specific correlation pattern matching a subroutine in a quantum circuit.
Here we show that the logical non-Clifford content of a resource state can be reshaped by conjugation with logical Clifford gates.

The main observation for altering the resource states of CSS-TP codes lies in the fact that conjugation of $\overline{\mathcal{M}}$ by a sequence of logical $\overline{CX}$ gates ($A_{\overline{CX}})$ induces the substitution
\begin{align}
u_{i} \mapsto \sum_{j} A_{ij} u_{j} ~\text{ (mod 2)}~,
\end{align}
for some $A \in GL(k, \mathbb{F}_{2})$, thereby producing a new diagonal gate $\overline{\mathcal{M}}' = A_{\overline{CX}}\,\overline{\mathcal{M}}\,A_{\overline{CX}}^{\dagger}$ with a different logical phase polynomial $\phi'(u)$.
This alteration simply changes the logical phase-gate pattern while preserving Clifford-invariant properties of the resource state..
Any residual Clifford terms ($\overline{CZ}$, $\overline{S}$, $\overline{Z}$) must still be reduced by logical Clifford corrections or handled in a rigorously supported frame; our resource estimates charge these terms explicitly~\cite{amy2014polynomial,litinski2019game}.

In this work, we shall refer to this procedure of altering $\overline{\mathcal{M}}$ as a \emph{logical basis transformation}.
By employing $\overline{CX}$ conjugations, we effectively follow a particular orbit of $\phi(u)$ in the group action of $GL(k, \mathbb{F}_{2})$, residing in the quotient space $\mathrm{RM}(3,k)/\mathrm{RM}(2,k)$~\cite{amy2019t}.
For \Mfac{12,5,2}, the $\mathrm{GL}(5,2)$ orbit of the cubic part modulo $\mathrm{RM}(2,5)$ (i.e.\ up to diagonal Cliffords) has $868$ elements, and the exact gate orbit has $5{,}208$; for \Mfac{14,6,2} the corresponding cubic orbit has $357{,}120$ elements.
Neither \Mfac{12,5,2} nor \Mfac{14,6,2} is $CX$-equivalent to a single $CCZ$: the minimum number of cubic monomials over each orbit is $2$.
Of the two, only \Mfac{14,6,2} is $CX$-equivalent to two disjoint $CCZ$ gates (\cref{section:tiers}).
To find a representative with a prescribed shape in this orbit, we use a Metropolis walk over $\overline{CX}$ conjugations whose energy is the number of monomials; it returns the move sequence, which we replay to check the result.
We detail this algorithm in \cref{section_random_walk_alg}.

Although conjugation by $\overline{CX}$ has found wide application in quantum compilation as a technique for reducing $T$-count in $(\mathrm{Cl}^{(2)}_{n}, T)$ circuits~\cite{amy2013meet,amy2014polynomial}, we shall nevertheless work out a small example for the benefit of those who are unfamiliar with the technique.
Consider the phase polynomial
\begin{align}
\phi(u) = 4(u_{0}u_{1}u_{2} + u_{0}u_{1}u_{3})~,
\end{align}
which corresponds to two $CCZ$ gates.
Applying $CX_{32}$ (control on qubit 2, target on qubit 3) induces the substitution $u_{3} \mapsto u_{3} \oplus u_{2}$.
Using the identity $a \oplus b = a + b - 2ab$ for bits, we obtain:
\begin{align}
\phi'(u) &= 4\bigl(u_{0}u_{1}u_{2} + u_{0}u_{1}(u_{3} \oplus u_{2})\bigr) \nonumber\\
&= 4\bigl(u_{0}u_{1}u_{2} + u_{0}u_{1}u_{3} + u_{0}u_{1}u_{2} - 2u_{0}u_{1}u_{2}u_{3}\bigr) \nonumber\\
&= 4\bigl(2u_{0}u_{1}u_{2} + u_{0}u_{1}u_{3} - 2u_{0}u_{1}u_{2}u_{3}\bigr) \nonumber\\
&= 8u_{0}u_{1}u_{2} + 4u_{0}u_{1}u_{3} - 8u_{0}u_{1}u_{2}u_{3} \nonumber\\
&= 4u_{0}u_{1}u_{3} \pmod{2^{\ell}}~.
\end{align}
Thus, a single $CX$ conjugation has reduced the polynomial from two $CCZ$-equivalent terms to one, demonstrating the utility of this technique for non-Clifford gate synthesis.
The key feature is the overlap of the two $CCZ$ terms on two qubits.
Equivalently, the original pair can be represented as one $CCZ$ operation in a conjugated logical basis; in our ETD accounting this basis choice changes measurement axes rather than adding physical $CX$ gates.

Taking into account the transformation depicted, one can immediately start to simplify the $\overline{\mathcal{M}}_{\code{n,k,d}}$ logical gates from \cref{section:CSSTP_examples}.
Here we briefly showcase a few results from the codes described previously.
For the $\code{14,6,2}$ code, applying logical basis transformations to $\overline{\mathcal{M}}_{\code{14,6,2}}$ via a $\overline{CX}$ sequence results in the new gate $\overline{\mathcal{M}}'_{\code{14,6,2}}$, as shown in \cref{fig:1462_gate_cleaned}.
We can apply similar techniques to any of the other CSS-TP codes detailed here as well.
Applying conjugation sequences can simplify both $\overline{\mathcal{M}}_{\code{12,5,2}}$ and $\overline{\mathcal{M}}_{\code{30,8,2}}$.
The effect is largest for the $\code{16,7,2}$ star code: in the logical basis of its construction its gate has one $\overline{Z}$, six $\overline{CZ}$ and $25$ $\overline{CCZ}$ terms (\cref{fig:1672_uncleaned_gate}), and seven logical $CX$ gates reduce it to three $\overline{CCZ}$ gates sharing one qubit, with no Clifford terms (\cref{fig:1672_gate_cleaned}).
\input{includes/logical_gates_clean}

We call a representative $\overline{\mathcal{M}}'$ of this kind, chosen to reduce the support of the non-Clifford terms, the \emph{reduced form} of the resource.
The exact conjugation steps for these transformations are in \cref{fig:cx_conjugation_sequence}, and the reduced forms $\overline{\mathcal{M}}'_{\code{12,5,2}}$, $\overline{\mathcal{M}}'_{\code{14,6,2}}$ and $\overline{\mathcal{M}}'_{\code{16,7,2}}$ are in \cref{fig:cleaned_logical_gates}.

We have mentioned that the conversion steps above are contained within $GL(k,\mathbb{F}_{2})$.
However, in practice, it is not necessary to be able to reach every distinct phase polynomial in the group.

The compiler need not search the entire orbit; rather, it needs a representative whose phase-gate pattern matches the target circuit.
In what follows, we will characterize how to construct efficient magic state distillation protocols using CSS-TP codes.

\subsection{Constructing New Distillation Protocols}\label{section:CSSTP_distillation_protocols}

We now describe how CSS-TP codes are converted into magic-state distillation protocols.
The construction is a direct application of Litinski's Encode-$T$-Decode protocol~\cite{litinski2019game,litinski2019magic} (see also Figures 14 and 15 of \cite{litinski2019magic}).
Let $\mathsf{C}$ be a CSS-TP code with parameters $\code{n,k,d}$, encoding isometry $E_{\mathsf{C}}$, and layer $V$ (\cref{eq:diag_layer}).
We assume that $V$ consists of single-qubit $T_{\ell}$ and $T_{\ell}^{\dagger}$ gates ($T$ for third-level resources, $\sqrt T$ for fourth-level ones), and possibly single-qubit Clifford phase gates, and that its induced logical action is
\begin{align}
V \cdot E_{\mathsf{C}} = E_{\mathsf{C}} \cdot \overline{\mathcal{M}}~,
\end{align}
where $\overline{\mathcal{M}}$ is the CSS-TP logical product gate introduced in \cref{eq:css_tp_action}.
The corresponding output state is
\begin{align}
\ket{\overline{\mathcal{M}}} = \overline{\mathcal{M}}\ket{+}^{\otimes k}~.
\end{align}

In the idealized Encode-$T$-Decode picture, the protocol proceeds as follows.
One begins with $k$ logical input qubits initialized in $\ket{+}^{\otimes k}$, encodes them into the quantum code $\mathsf{C}$ using the Clifford encoder $E_{\mathsf{C}}$, applies the layer $V$ by magic state injection, and then decodes using $E_{\mathsf{C}}^{\dagger}$.
The remaining $n-k$ qubits are measured to extract the error-detection syndrome.
A run is accepted only if the trivial syndrome is observed.
Conditioned on acceptance, the $k$ output qubits are projected onto the distilled resource state $\ket{\overline{\mathcal{M}}}$; the byproducts that arise when this state is consumed are treated at the end of this subsection.
Since $E_{\mathsf{C}}$ and $E_{\mathsf{C}}^{\dagger}$ are Clifford circuits, Litinski's rearrangement commutes them through the non-Clifford injections.
The resulting implementation does not require the quantum code to exist as an encoded block; instead, one prepares the noisy single-qubit magic states required by $V$, measures the Pauli-product checks specified by the encoder and decoder, postselects on the trivial syndrome, and tracks any Clifford byproducts in software.
In a surface-code architecture, these Pauli-product measurements are implemented by lattice surgery, exactly as in the original ETD construction~\cite{litinski2019game,litinski2019magic}.
If the code is instead held as an encoded block, $E_{\mathsf{C}}$ must be compiled as a physical circuit whose two-qubit gate count and depth enter the cost directly, in which case it is known that automated encoder synthesis techniques for stabilizer codes can reduce both substantially~\cite{peham2026synthesis}.

This treatment gives the following CSS-TP distillation primitive:
\begin{align}
\mathcal{D}_{\mathsf{C}}: \bigotimes_{i=1}^{n} \ket{A_{i}}_{\mathrm{noisy}} \longrightarrow \ket{\overline{\mathcal{M}}}_{\mathrm{distilled}}~,
\end{align}
where each $\ket{A_{i}}$ is either a noisy $\ket{T_{\ell}}$ or $\ket{T^{\dagger}_{\ell}}$ state, depending on the sign of the layer on qubit $i$.
Physical Clifford phases appearing in $V$ do not require magic states and can be absorbed into the surrounding Clifford circuit.
Thus an $\code{n,k,d}$ CSS-TP code defines an $n$-to-$1$ distillation routine whose single output is a $k$-qubit entangled resource state rather than $k$ independent magic states.

The error-detection behavior is inherited from the quantum code.
In the standard input-error model, where each noisy injected magic state suffers an independent error with probability $p$, any error pattern of weight below the distance $d$ is detected.
Consequently, the leading accepted output error scales as
\begin{align}
p_{\mathrm{out}} = A_{\mathsf{C}} p^{d} + O(p^{d+1})~,
\end{align}
where $A_{\mathsf{C}}$ counts the accepted weight-$d$ input fault patterns that induce a nontrivial logical error on the distilled resource.
For the distance-two codes emphasized in this work the output error is quadratically suppressed, and $A_{2}$ can be extracted from the quantum weight enumerator of the code~\cite{rains2002quantum,cao2023quantum,cao2024quantum}.
For each code used here, every pair of input faults is undetected and malignant, so $A_{2}=\binom{n}{2}$:
\begin{align}
p_{\mathrm{out}}^{\code{8,3,2}}
&=28p^{2}+O(p^{3})~,
\nonumber\\
p_{\mathrm{out}}^{\code{12,5,2}}
&=66p^{2}+O(p^{3})~,
\nonumber\\
p_{\mathrm{out}}^{\code{14,6,2}}
&=91p^{2}+O(p^{3})~,
\nonumber\\
p_{\mathrm{out}}^{\code{16,7,2}}
&=120p^{2}+O(p^{3})~.
\label{eq:third_level_output_errors}
\end{align}
Thus these CSS-TP factories have the same quadratic suppression order as the $\code{8,3,2}$ ETD factory, but with a higher leading malignant-pair coefficient under the same input-noise convention.
Higher-distance CSS-TP codes, or concatenated CSS-TP constructions such as those described in \cref{section:concatenation_distance_growth}, yield correspondingly higher-order suppression.
We emphasize though that this is not necessarily required for practically relevant suppression.
The quantity $p$ is the error of the raw magic state and is distinct from the hardware circuit error $p_{\rm phys}$.
Cultivation relates these quantities through a cultivation setting, i.e.\ a choice of protocol and post-selection threshold, rather than a universal power law~\cite{gidney2024magic,chen2026efficient}.
The numerical study therefore takes cultivated $T$ and $\sqrt T$ errors directly from the selected cultivation settings and separately schedules their acceptance losses, escape costs, and ETD consumption.
The formulas above describe only the ETD stage conditional on those cultivated inputs.

The logical basis transformation that we introduced in \cref{section:CSSTP_gate_cleaning} is naturally compatible with this generalized ETD protocol.
Suppose a sequence of logical Clifford gates $A_{\overline{CX}}$ transforms the native CSS-TP gate into a more useful representative
\begin{align}
\overline{\mathcal{M}}' = A_{\overline{CX}} \overline{\mathcal{M}} A_{\overline{CX}}^{\dagger}~.
\end{align}
Then the distilled state can be converted from $\ket{\overline{\mathcal{M}}}$ to $\ket{\overline{\mathcal{M}}'}$ by applying the same logical basis transformation to the output.
Equivalently, $A_{\overline{CX}}$ and $A_{\overline{CX}}^{\dagger}$ can be absorbed into the ETD encoder and decoder, changing only the Pauli-product measurement pattern and amounting to a virtual $\overline{CX}$ sequence by simply choosing a more convenient logical basis.
Thus the distillation cost is governed by the same code, while the logical resource can be chosen from the Clifford orbit explored in \cref{section:CSSTP_gate_cleaning}.
This virtual implementation does not remove lower-degree terms induced in the chosen phase-polynomial representative; any remaining Clifford terms are still included as explicit Clifford corrections in the resource estimates.

We now record several concrete CSS-TP factories that will be used as primitive resources in \cref{section:tailored_algorithmic_primitives}.
Applying the above construction to the reduced forms $\overline{\mathcal{M}}'$ of the $\code{12,5,2}$, $\code{14,6,2}$ and $\code{16,7,2}$ codes gives factories that consume $12$, $14$ and $16$ noisy $T$-type states and output one $5$-, $6$- or $7$-qubit resource, with the output errors of \cref{eq:third_level_output_errors}.
Each input count denotes the full signed layer, including the appropriate choices of $T$, $T^{\dagger}$, and any additional single-qubit Clifford deformations.
The protocols for $\overline{\mathcal{M}}'_{\code{12,5,2}}$ and $\overline{\mathcal{M}}'_{\code{14,6,2}}$ are shown in \cref{fig:1252_etd_factory,fig:1462_etd_factory}, and the protocol of the $\code{16,7,2}$ star code, which outputs three $CCZ$ gates sharing one qubit, in \cref{fig:1672_etd_factory}; we shall make use of these resources in several application-level comparisons below.
\cref{fig:concrete_csstp_etd_factories} draws them in the layout of Ref.~\cite{litinski2019game}, next to the $8$-to-$CCZ$ protocol of that reference.
The drawings use the logical basis of \cref{section:further_examples_CSStp} for the $\code{12,5,2}$ and $\code{14,6,2}$ codes and that of \cref{section:star_family} for the star code, so each outputs $\overline{\mathcal M}$ in the drawn basis, which the logical $CX$ networks of \cref{fig:cx_conjugation_sequence} relate to the reduced form.
Each column is a Pauli rotation $P_\phi=e^{-i\phi P}$ in the convention of Ref.~\cite{litinski2019game}, with the signed angle of the verified physical layer; $P_{5\pi/8}=P_{\pi/8}P_{\pi/2}$, and the Pauli part is tracked in the frame.
The $X$-stabilizer patch, drawn at the bottom, is measured in the $X$ basis at the end.
The $\code{12,5,2}$ and $\code{14,6,2}$ codes are equivalent to the \Tofs and \twoCCZ codes of Ref.~\cite{campbell2017unified}.
\input{includes/3082_uncleaned_and_cleaned}
\input{includes/litinski_factories}

Finally, the distilled state may be consumed by standard gate teleportation, or by the corresponding lattice-surgery-based injection procedures when available \cite{horsman2012surface,litinski2019game}.
Given $\ket{\overline{\mathcal{M}}}=\overline{\mathcal{M}}\ket{+}^{\otimes k}$, one teleports the data through the resource state.
The Bell-measurement outcomes produce Pauli byproducts.
If $\overline{\mathcal{M}}\in \mathrm{Cl}^{(\ell)}_{k}$, then conjugating such Pauli byproducts through $\overline{\mathcal{M}}$ yields corrections one level lower in the Clifford hierarchy:
\begin{align}
\overline{\mathcal{M}} P \overline{\mathcal{M}}^{\dagger} \in \mathrm{Cl}^{(\ell-1)}_{k}~, \quad P \in \mathsf{P}_{k}~.
\end{align}
Therefore, for the third-level CSS-TP resources emphasized in most of our examples, the required corrections are Clifford operations and can be applied by lattice surgery or tracked in a Clifford frame.
As an example, consider a Pauli byproduct $X_{1}$ arising when injecting a $CCZ_{123}$ resource.
Since $CCZ$ is diagonal and self-adjoint, its conjugation action can be evaluated on computational-basis states:
\begin{align}
CCZ_{123} X_{1} CCZ_{123}^{\dagger}\ket{a,b,c} &= (-1)^{abc+(a\oplus 1)bc}\ket{a\oplus 1,b,c} \nonumber \\
&= (-1)^{bc}\ket{a\oplus 1,b,c}~.
\end{align}
Thus
\begin{align}
CCZ_{123} X_{1} CCZ_{123}^{\dagger} = X_{1} CZ_{23}~.
\end{align}
An $X$ byproduct on one leg of a $CCZ$ therefore induces a $CZ$ correction on the other two legs, together with the original Pauli byproduct, which may be tracked in the Pauli frame.

The same ETD construction also applies to CSS-TP codes whose transversal action lies above the third level of the Clifford hierarchy.
For the $\code{30,8,2}$ code, whose transversal action contains fourth-level components such as $\overline{CCCZ}$, the corresponding factory is
\begin{align}
30 \cdot \ket{\sqrt{T}}_{\mathrm{noisy}} &\longrightarrow \ket{\overline{\mathcal{M}}'_{\code{30,8,2}}}~,
\end{align}
with $p_{\mathrm{out}} = 435p^{2}+O(p^{3})$, where $p$ is now the error of the $\sqrt T$ inputs and again every pair of input faults is malignant, $435=\binom{30}{2}$.
We provide all details regarding the $\code{30,8,2}$ code in \cref{section:further_examples_CSStp}.
By the consumption rule above, its injection requires third-level corrections, here $CCZ$ gates arising from $X$ byproducts on a $CCCZ$ component.
\cref{section:fourth_level} costs such a resource, including its correction burden, and finds that it does not beat $CCZ$-only compilation.
The resulting ETD factory is shown in \cref{fig:3082_etd_factory}, where every angle is $\pm\pi/16$ up to a Clifford rotation $P_{m\pi/4}$, and its logical gate and reduced form are in \cref{fig:3082_gates}.

%% file: includes/quantum_code_modifications.tex
\begin{figure*}
\centering
\includegraphics[width=\textwidth]{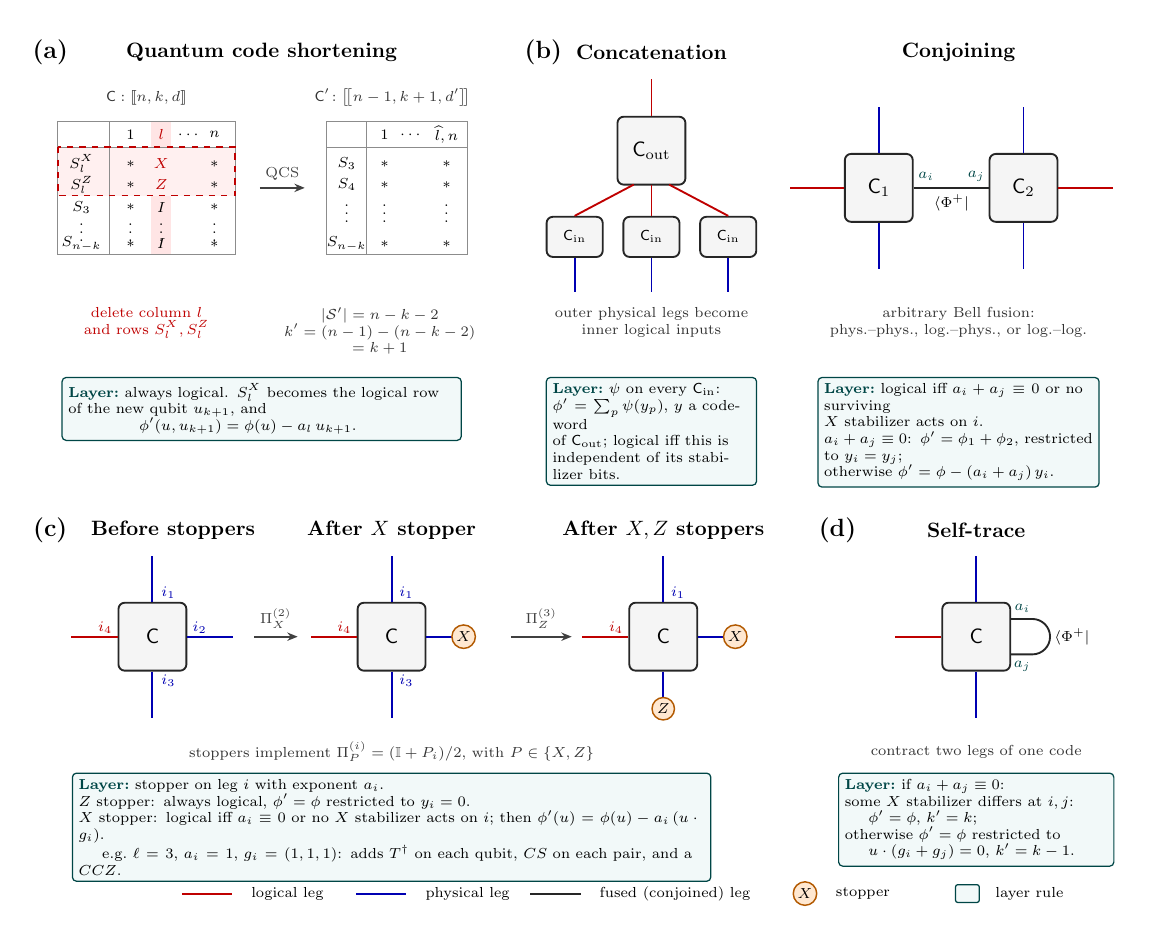}
\caption{Code modification operations of \cref{section:background_code_modification} and their effect on a transversal diagonal layer with logical phase polynomial $\phi$ (teal boxes), as given by \cref{lem:operations} and \cref{tab:operations} (notation in \cref{tab:notation}). (a)~QCS on qubit $l$. (b)~Concatenation and conjoining, drawn without index notation (\cref{lem:concat} and \cref{lem:operations}(a)). (c)~$Z$ and $X$ stoppers~\cite{pato2025hyper}. (d)~A self-trace (\cref{cor:conjoin_loop}).}
\label{fig:code_modification}
\end{figure*}

%% file: includes/derivation_832.tex
\begin{figure*}
\centering
\includegraphics[width=0.85\textwidth]{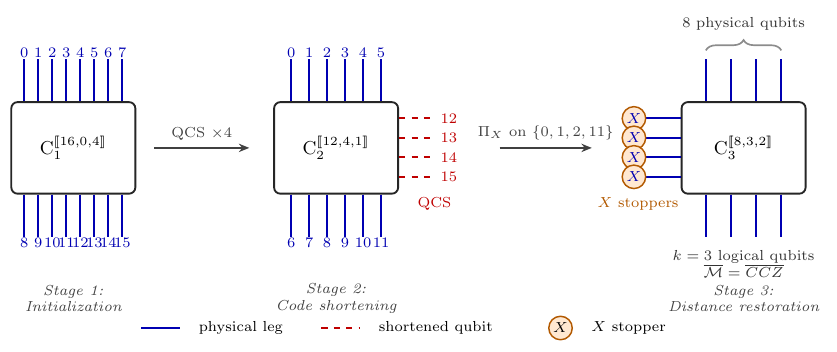}
\caption{Derivation of the $\code{8,3,2}$ code from the $\code{16,0,4}$ QRM code via \cref{con:protocol}: four QCS steps (stage~2) followed by four $X$ stoppers (stage~3), with the code parameters after each step.}
\label{fig:832_derivation}
\end{figure*}

%% file: includes/open_chain.tex
\begin{figure}
\centering
\includegraphics[width=\columnwidth]{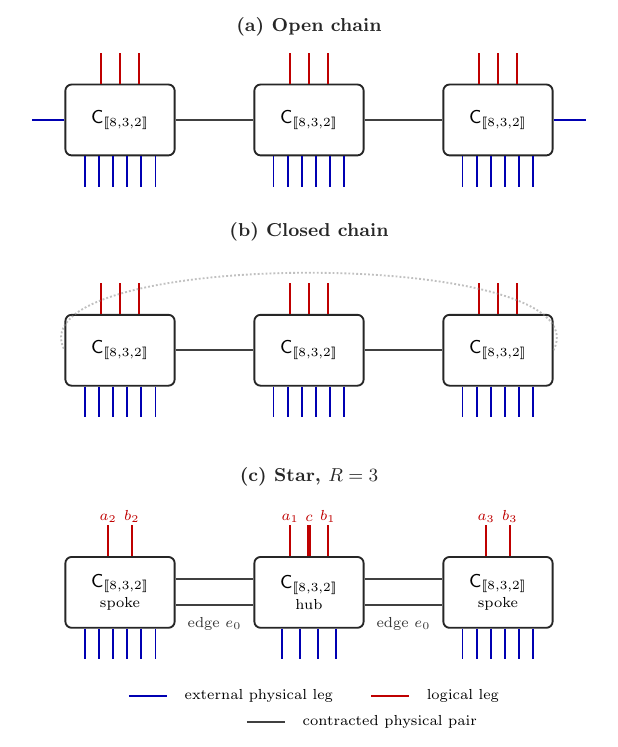}
\caption{CSS-TP code families built by conjoining $\code{8,3,2}$ codes. (a)~Open chain, $\code{20,9,2}$ (\cref{section:open_chains}). (b)~Closed chain, $\code{18,8,2}$; the closing self-trace removes one of the nine drawn logical legs (\cref{section:closed_chains}). (c)~Star with $R=3$, $\code{16,7,2}$, implementing $\prod_{r}\overline{CCZ}(c,a_r,b_r)$ (\cref{section:star_family}).}
\label{fig:832_openchain}
\end{figure}

%% file: includes/logical_gates_unclean.tex
\begin{figure*}
\centering

\begin{minipage}[t]{0.47\textwidth}
\centering
\panelimage[
    width=\linewidth,
    trim=0pt 0pt 0pt 0pt,
    clip
]{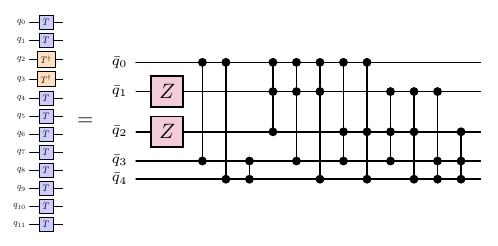}{(a)}
\end{minipage}
\hfill
\begin{minipage}[t]{0.45\textwidth}
\centering
\panelimage[
    width=\linewidth,
    trim=0pt 0pt 0pt 0pt,
    clip
]{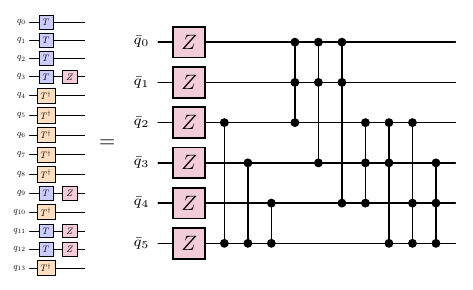}{(b)}
\end{minipage}

\vspace{0.6em}
\begin{minipage}[t]{0.9\textwidth}
\centering
\panelimage[
    width=\linewidth,
    trim=0pt 0pt 0pt 0pt,
    clip
]{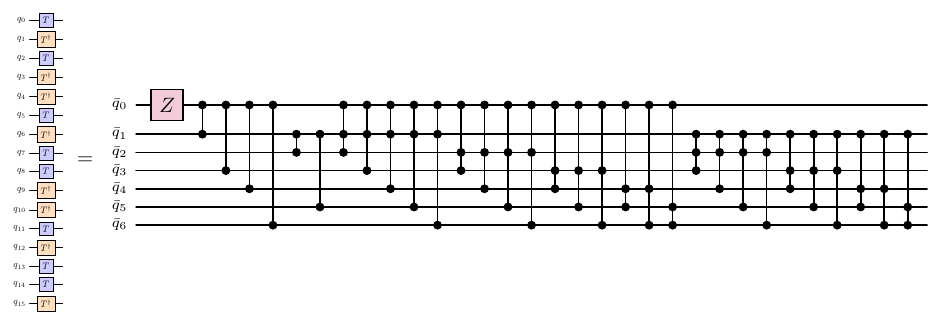}{(c)}
\end{minipage}

\caption{
Transversal non-Clifford logical gates for the
$\code{12,5,2}$, $\code{14,6,2}$ and $\code{16,7,2}$ CSS-TP codes. Panels
(a), (b) and (c) show
$\overline{\mathcal{M}}_{\code{12,5,2}}$,
$\overline{\mathcal{M}}_{\code{14,6,2}}$ and
$\overline{\mathcal{M}}_{\code{16,7,2}}$, respectively, each in the logical basis of its construction (\cref{section:CSSTP_examples}).
}
\label{fig:uncleaned_logical_gates}

\setcounter{subfigure}{0}
\refstepcounter{subfigure}\label{fig:1252_uncleaned_gate}
\refstepcounter{subfigure}\label{fig:1462_uncleaned_gate}
\refstepcounter{subfigure}\label{fig:1672_uncleaned_gate}

\end{figure*}

%% file: includes/classification_table.tex
\begin{table*}[t]
\centering
\caption{Tier of the resource state $\ket{\overline{\mathcal M}}$ of every code in this work (\cref{def:tiers}), with the method used to establish it and the stabilizer R\'enyi entropy $M_2$.
``Walk'' is an explicit sequence of logical $CX$ (and $X$) moves to disjoint canonical hyperedges, replayed and re-checked; ``scan'' is an exhaustive scan of $GL(k,2)$ and all $X$ frames; $\nu$ (the stabilizer nullity) and ``hist.'' (the rank histogram) are the invariants of \cref{section:tiers,section:tier_proofs}.
For Tier 0 and 1 the last column gives the canonical factors, whose $M_2$ values add up to the computed $M_2$ to within $10^{-5}$.
Canonical values: $M_2(T)=0.415$, $M_2(\sqrt T)=0.193$, $M_2(CS)=1.193$, $M_2(CCZ)=1.541$, $M_2(CCCZ)=1.378$.}
\label{tab:classification}
\renewcommand{\arraystretch}{1.12}
\widetable{%
\begin{tabular}{@{}llcllc@{}}
\toprule
Code & Family & Tier & Method & $M_2$ & Canonical factors \\
\midrule
$\code{8,3,2}$, $\code{120,3,6}$ & seed; $\code{8,3,2}\circ\code{15,1,3}$ & 0 & $\nu=3$ & $1.541$ & $CCZ$ \\
$\code{16,4,2}$ & seed & 0 & walk & $1.378$ & $CCCZ$ \\
$\code{15,1,3}$, $\code{31,1,3}$ & seeds & 0 & scan & $0.415$, $0.193$ & $T$, $\sqrt T$ \\
\midrule
$\code{14,6,2}$, $\code{20,9,2}$, $\code{26,12,2}$ & open $\code{8,3,2}$ chain & 1 & walk ($\nu=6,9,12$) & $3.081$, $4.622$, $6.162$ & $2,3,4\times CCZ$ \\
$\code{30,8,2}$, $\code{44,12,2}$ & open $\code{16,4,2}$ chain & 1 & walk & $2.756$, $4.134$ & $2,3\times CCCZ$ \\
$\code{28,2,3}$, $\code{41,3,3}$, $\code{54,4,3}$ & QRM$_{15}$ chain & 1 & scan & $0.830$, $1.245$, $1.660$ & $k\times T$ \\
$\code{26,2,2}$, $\code{39,3,3}$, $\code{52,4,3}$ & QRM$_{15}$ variant & 1 & scan & $0.830$, $1.245$, $1.660$ & $k\times T$ \\
$\code{60,2,3}$, $\code{89,3,3}$ & QRM$_{31}$ chain & 1 & scan & $0.385$, $0.578$ & $k\times\sqrt T$ \\
\midrule
$\code{12,5,2}$, $\code{18,8,2}$, $\code{24,11,2}$ & closed $\code{8,3,2}$ chain & 2 & $\nu=5,8,11$ & $2.476$, $4.430$, $6.107$ & -- \\
$\code{16,7,2}$, $\code{20,9,2}$, $\code{24,11,2}$ & star, $R=3,4,5$ & 2 & $\nu=7$, hist., $\nu=11$ & $2.850$, $2.961$, $2.990$ & -- \\
$\code{180,5,6}$ & $\code{12,5,2}\circ\code{15,1,3}$ & 2 & $\nu=5$ & $2.476$ & -- \\
$\code{64,9,4}$, $\code{96,15,4}$ & $\code{8,3,2}\circ\code{8,3,2}$, $\code{8,3,2}\circ\code{12,5,2}$ & 2 & hist. & $5.781$, $8.660$ & -- \\
\bottomrule
\end{tabular}%
}
\end{table*}

%% file: includes/logical_gates_clean.tex
\begin{figure*}
\centering

\begin{minipage}[t]{0.36\textwidth}
\centering
\panelimage[
    width=\linewidth,
    trim=0pt 0pt 0pt 0pt,
    clip
]{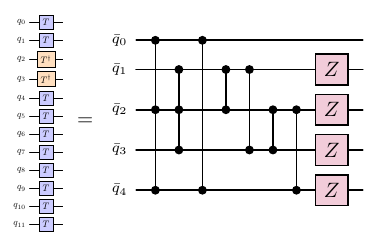}{(a)}
\end{minipage}
\hfill
\begin{minipage}[t]{0.32\textwidth}
\centering
\panelimage[
    width=\linewidth,
    trim=0pt 0pt 0pt 0pt,
    clip
]{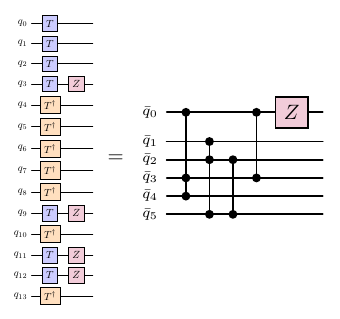}{(b)}
\end{minipage}
\hfill
\begin{minipage}[t]{0.23\textwidth}
\centering
\panelimage[
    width=\linewidth,
    trim=0pt 0pt 0pt 0pt,
    clip
]{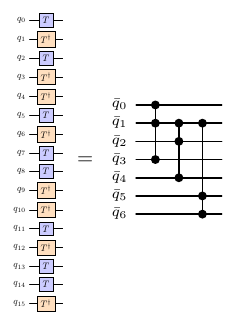}{(c)}
\end{minipage}

\caption{
Reduced forms, after a logical basis transformation by $\overline{CX}$ gates, of the logical gates of the $\code{12,5,2}$
, $\code{14,6,2}$ and $\code{16,7,2}$ CSS-TP codes. Panels (a), (b) and (c)
show $\overline{\mathcal{M}}'_{\code{12,5,2}}$,
$\overline{\mathcal{M}}'_{\code{14,6,2}}$ and $\overline{\mathcal{M}}'_{\code{16,7,2}}$, respectively.
}
\label{fig:cleaned_logical_gates}

\setcounter{subfigure}{0}
\refstepcounter{subfigure}\label{fig:1252_gate_cleaned}
\refstepcounter{subfigure}\label{fig:1462_gate_cleaned}
\refstepcounter{subfigure}\label{fig:1672_gate_cleaned}

\end{figure*}

%% file: includes/3082_uncleaned_and_cleaned.tex
\begin{figure*}[!t]
\centering
\edef\figtenheightval{\the\dimexpr 0.45\textwidth\relax}%

\begin{minipage}[t]{0.50\textwidth}
\centering
\panelimage[
    height=\figtenheightval,
    trim=0pt 0pt 0pt 0pt,
    clip
]{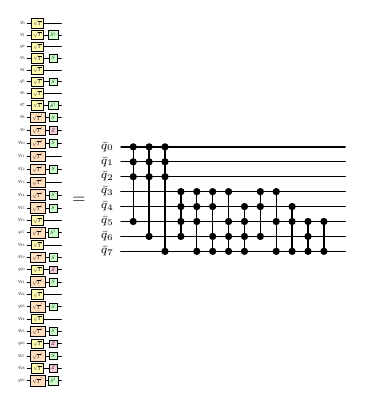}{(a)}
\end{minipage}
\hspace{2em}
\begin{minipage}[t]{0.28\textwidth}
\centering
\panelimage[
    height=\figtenheightval,
    trim=0pt 0pt 0pt 0pt,
    clip
]{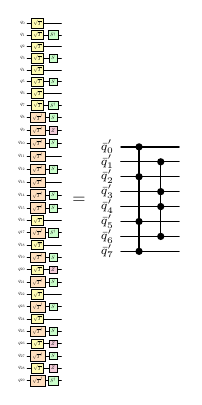}{(b)}
\end{minipage}

\caption{
Transversal non-Clifford logical gate of the $\code{30,8,2}$ CSS-TP code, for the same physical layer of $15$ $\sqrt T$ and $15$ $\sqrt T^{\dagger}$ gates times Clifford phases.
(a)~In the logical basis of \cref{section:further_examples_CSStp}.
(b)~In the reduced form (\cref{section:fourth_level}).
}
\label{fig:3082_gates}
\setcounter{subfigure}{0}
\refstepcounter{subfigure}\label{fig:3082_uncleaned_gate}
\refstepcounter{subfigure}\label{fig:3082_cleaned_gate}

\end{figure*}

%% file: includes/litinski_factories.tex
\begin{figure*}
\centering
\begin{minipage}[t]{0.50\textwidth}
\centering
\panelimage[
    width=\linewidth,
    trim=0pt 0pt 0pt 0pt,
    clip
]{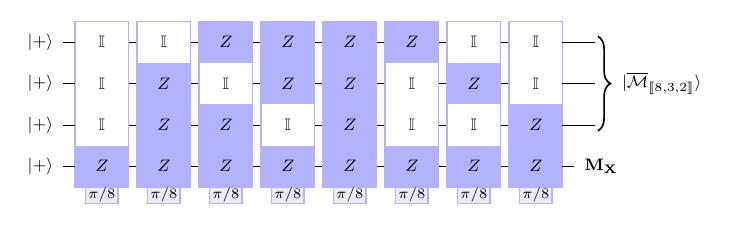}{(a)}
\end{minipage}
\vspace{0.15em}

\begin{minipage}[t]{0.675\textwidth}
\centering
\panelimage[
    width=\linewidth,
    trim=0pt 0pt 0pt 0pt,
    clip
]{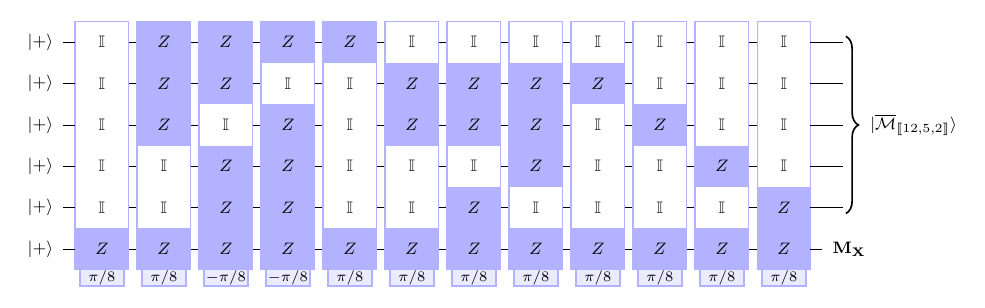}{(b)}
\end{minipage}
\vspace{0.15em}

\begin{minipage}[t]{0.753\textwidth}
\centering
\panelimage[
    width=\linewidth,
    trim=0pt 0pt 0pt 0pt,
    clip
]{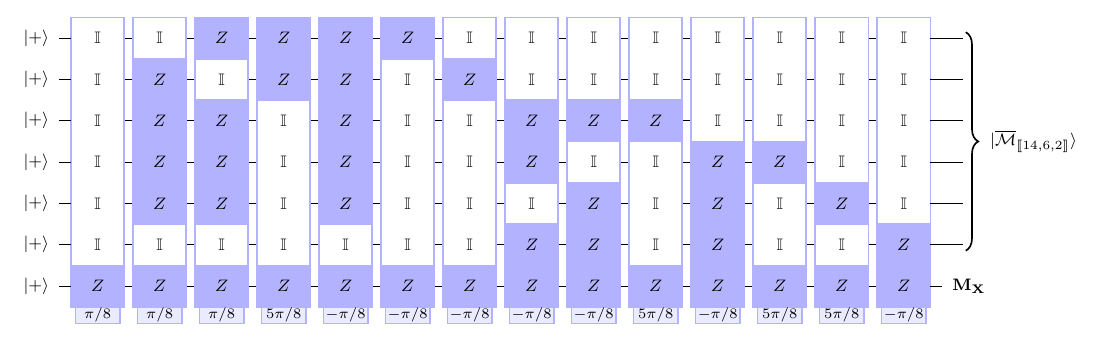}{(c)}
\end{minipage}
\vspace{0.15em}

\begin{minipage}[t]{0.845\textwidth}
\centering
\panelimage[
    width=\linewidth,
    trim=0pt 0pt 0pt 0pt,
    clip
]{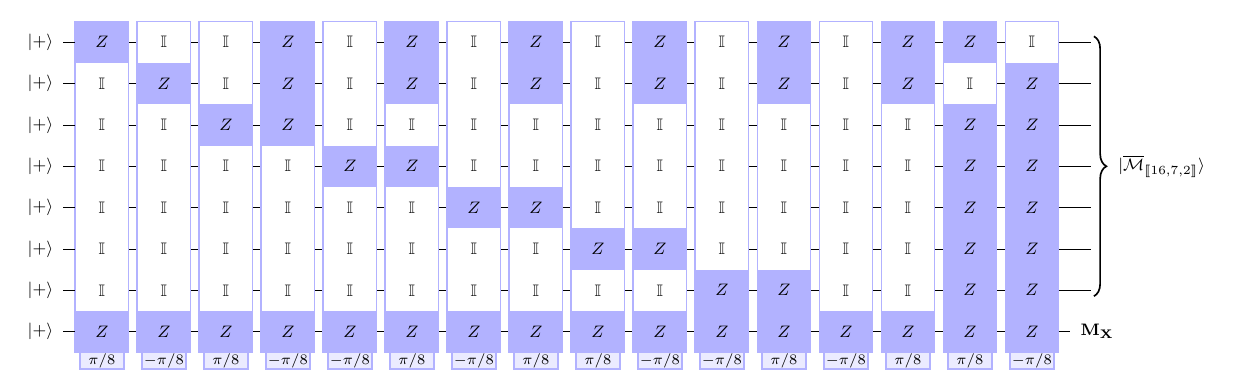}{(d)}
\end{minipage}
\vspace{0.15em}

\begin{minipage}[t]{0.90\textwidth}
\centering
\panelimage[
    width=\linewidth,
    trim=0pt 0pt 0pt 0pt,
    clip
]{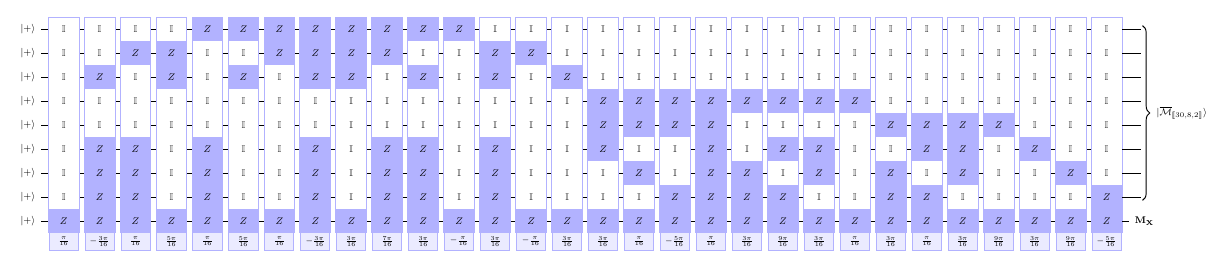}{(e)}
\end{minipage}

\caption{
ETD factories in the layout of Ref.~\cite{litinski2019game}: (a)~the $8$-to-$CCZ$ protocol of that reference, and the CSS-TP factories of (b)~the $\code{12,5,2}$ code, (c)~the $\code{14,6,2}$ code, (d)~the $\code{16,7,2}$ star code and (e)~the $\code{30,8,2}$ code (\cref{section:CSSTP_distillation_protocols}).
Rows are the logical patches and, at the bottom, the $X$-stabilizer patch; each column is a Pauli rotation $P_\phi=e^{-i\phi P}$.
}
\label{fig:concrete_csstp_etd_factories}
\setcounter{subfigure}{0}
\refstepcounter{subfigure}\label{fig:litinski832}
\refstepcounter{subfigure}\label{fig:1252_etd_factory}
\refstepcounter{subfigure}\label{fig:1462_etd_factory}
\refstepcounter{subfigure}\label{fig:1672_etd_factory}
\refstepcounter{subfigure}\label{fig:3082_etd_factory}
\end{figure*}

%% file: sections/04_tailored_primitives.tex
\section{Tailored Magic States for Algorithmic Primitives}\label{section:tailored_algorithmic_primitives}

The preceding sections construct CSS-TP resource states that carry several correlated $CCZ$ interactions in one output.
We now ask when a factory for such a state is cheaper than a factory for single $CCZ$ states, for the same circuit, the same code distance and the same cultivation supply.
We state the resource model in \cref{section:comparison_framework} and apply it to small standalone targets in \cref{section:standalone}.
We then evaluate five complete arithmetic circuits: a $GF(2^{8})$ multiplier (\cref{section:finitefield_multiplication}), a single-lane carry-lookahead adder and a single-lane ripple-carry adder (\cref{section:single_lane_adders}), and a carry-lookahead adder and a ripple-carry adder that each perform two additions sharing one addend (\cref{section:carry_lookahead_addition}).
All are evaluated in a fixed configuration: one factory per resource family, $\apxlanes$ cultivation lanes and code distance $d=21$.

\subsection{Resource model}
\label{section:comparison_framework}

All benchmarks use one model, which \cref{section:spacetime_model} derives step by step with a worked example.
\begin{itemize}
\item \emph{Units.}
The physical error rate is $p=10^{-3}$.
A distance-$d$ surface-code tile holds $2d^{2}$ physical qubits and fails with probability $p_{L}(d)=0.1(p/0.01)^{(d+1)/2}$ per round; a logical step lasts $d$ rounds~\cite{litinski2019game}.
\item \emph{$T$ supply.}
Raw $\ket T$ states come from magic-state cultivation~\cite{gidney2024magic}, which grows a $T$ state inside a small color code of distance $d_{1}$, then into a surface code, and keeps it only if a final post-selection test passes; the threshold of this test is called the gap, and a larger gap keeps fewer attempts, each with a lower error.
We simulated end-to-end cultivation with $d_{1}=3$ grown into a $d=21$ surface code, and use a gap of $70$\,dB, with $p_{T}=\apxpT$ (see \cref{fig:appendix_cultivation}).
A cultivation \emph{lane} reserves $\apxlaneq$ qubits and runs attempts back to back.
Each attempt lasts $\apxkeptlen$ rounds and is kept with probability $\apxaccept$, so a lane delivers a kept state every $\apxmeankept$ rounds on average, with a geometric tail.
Unless stated otherwise there are $\apxlanes$ lanes.
\item \emph{Factories.}
An encode-$T$-decode (ETD) factory~\cite{litinski2019game,litinski2019magic} on an $\code{n,k,2}$ code with $m_{x}$ $X$-stabilizers runs for $\tau=n-m_{x}$ logical steps on $F=1.5(k+m_{x})+4$ tiles and outputs one $k$-qubit resource.
A faulty input $T$ state is modeled as a $Z$ error with probability $p_{T}$, a form that Clifford twirling enforces (\cref{section:appendix_model_factory}), and the accepted-output error is $\epsilon_{\rm ETD}\approx A_{2}p_{T}^{2}$, with $A_{2}$ the number of undetected weight-two input errors (\cref{tab:appendix_factories}).
Applied to Litinski's 15-to-1 protocol, the two rules give $11$ steps and $11.5$ tiles, against his $11$ and $11$~\cite{litinski2019game}.
We label each factory by the resource it outputs, $\overline{\mathcal M}_{\code{n,k,d}}$ for the ETD protocol on the $\code{n,k,d}$ code.
The factory families are \Mfac{8,3,2} (one $CCZ$), \Mfac{12,5,2} (two $CCZ$ sharing one qubit), \Mfac{14,6,2} (two disjoint $CCZ$) and the stars \Mfac{4R+4,2R+1,2} ($R$ $CCZ$ sharing one qubit; \cref{section:star_family}), of which we use $R=3$, $4$ and $5$.
A factory that is asked for a resource smaller than its native output, such as a single $CCZ$ from a star factory, runs the smallest ETD protocol producing it, on the same tiles.
We compare strategies at a fixed number of factories $N_{\rm fac}$.
\item \emph{Consumption.}
The algorithm's logical qubits sit in a \emph{data block} of surface-code patches laid out as in Ref.~\cite{litinski2019game}. Logical Clifford operations, including lattice-surgery corrections, are free.
The circuit is compiled into Pauli-based computation~\cite{litinski2019game}, and consuming one resource is one Pauli-product measurement on the data block, lasting $\delta=9$, $5$ or $1$ logical steps in the compact, intermediate and fast blocks (\cref{tab:appendix_regimes}).
Under \emph{serial} consumption, the default, one consumption runs on a block at a time, as in Ref.~\cite{litinski2019game}.
Under \emph{parallel} consumption, any number of consumptions run at once provided their logical supports are pairwise disjoint; each still lasts $\delta$, and no additional routing area is charged.
Serial consumption is pessimistic for a block that can route several measurements at once, and parallel consumption without a routing cost is optimistic; the adder tables report both.
Logical basis transformations of a resource or of a target are therefore free for canonical and tailored factories alike.
\item \emph{Cover.}
Before scheduling, the $CCZ$ gates are partitioned into factory outputs whose gates the dependencies of the circuit allow to be consumed together.
We use the partition with the fewest consumptions, computed by a mixed-integer program.
Among the minimum partitions, we use the one with the shortest chain of dependent consumptions, so that the choice is reproducible; all of these covers are proven optimal in both respects.
The one solve that does not finish, \Mfac{16,7,2} on the $GF(2^{8})$ multiplier, stops at the $300$~s limit with $22=\lceil64/3\rceil$ consumptions and no dependencies, which is optimal by counting.
\item \emph{Schedule.}
An event-driven scheduler runs the lanes, the raw-state buffers, the factories (with sampled ETD rejections), the output storage and the consumptions from a cold start, and respects the dependencies of the circuit.
The latency is the time until the last consumption ends, averaged over random seeds; tables also give its 95th percentile.
\item \emph{Cost.}
We report two spacetime volumes.
The occupied volume integrates busy physical qubits over time (qubit-rounds).
The footprint counts all installed qubits, including a raw-state buffer of two ETD batches ($2n$ tiles) per factory, and the installed volume is footprint $\times$ latency.
\item \emph{Error.}
The error per circuit is the union bound $E_{1}+E_{2}+E_{3}$.
Here $E_{1}$ comes from cultivated states consumed directly (only the $\sqrt T$ rotations of \cref{section:fourth_level}), and $E_{2}$ is the sum of $\epsilon_{\rm ETD}$ over consumed outputs.
$E_{3}$ is the topological error, the probability that some surface-code patch suffers a logical error during the circuit, i.e.\ $p_{L}(d)$ times the number of tile-rounds.
To leading order $E_{2}=p_{T}^{2}\sum_{r}A_{2}^{(r)}N_{r}$, with $N_{r}$ consumed outputs of type $r$ and $A_{2}^{(r)}=28$, $66$, $91$, $120$ and $190$ for \Mfac{8,3,2}, \Mfac{12,5,2}, \Mfac{14,6,2}, \Mfac{16,7,2} and \Mfac{20,9,2} (\cref{eq:third_level_output_errors} and \cref{tab:appendix_factories}).
\item \emph{Tables.}
Unless a caption states otherwise, a table uses one factory per strategy, $\apxlanes$ lanes and serial consumption.
Latencies are averaged over $200$ seeds ($400$ in \cref{tab:standalone_factory_costs}).
``Cons.'' is the number of consumptions in the cover, ``Rounds'' the latency, ``Qubit-rounds'' the occupied volume and ``Qubits'' the footprint.
The error columns are $\epsilon_{\rm inj}=E_{1}$, $\epsilon_{\rm ETD}=E_{2}$, $\epsilon_{\rm topo}=E_{3}$ and their sum $\epsilon_{\rm total}$.
Rows labeled fast$^{\parallel}$ use parallel consumption in the fast block.
\end{itemize}
In the tables, a latency printed in \textcolor{darkred}{dark red} is shorter than that of \Mfac{8,3,2} in the same configuration, and an occupied volume printed in \textcolor{darkblue}{dark blue} is smaller.

At $d=21$, the distance of the cultivated states, $E_{3}$ exceeds $E_{2}$ by one to three orders of magnitude for every third-level strategy below, so $\epsilon_{\rm total}$ follows the occupied volume and lies between $5.9\times10^{-8}$ and $2.4\times10^{-6}$ per arithmetic circuit. 
Our comparisons are therefore at equal code distance, not at
equal error, and a strategy with a smaller occupied volume is also the more accurate one.

\subsection{Standalone resources}
\label{section:standalone}

\Mfac{12,5,2} outputs two $CCZ$ gates whose supports share one qubit (\cref{fig:1252_gate_cleaned}); \Mfac{14,6,2} outputs two $CCZ$ gates on disjoint supports (\cref{fig:1462_gate_cleaned}); \Mfac{16,7,2} outputs three $CCZ$ gates sharing one qubit, the $R=3$ star (\cref{fig:1672_gate_cleaned}).
\cref{tab:standalone_factory_costs} compares one such output with the same gates from \Mfac{8,3,2} outputs made one after the other by one \Mfac{8,3,2} factory.
At $d=21$ with $\apxlanes$ lanes, \Mfac{12,5,2} makes the shared pair in $299$ rounds, against $361$ for \Mfac{8,3,2}, with a smaller occupied volume: $3.73\times10^{6}$ qubit-rounds, against $4.69\times10^{6}$.
\Mfac{14,6,2} makes the disjoint pair in $341$ rounds with about the same volume as \Mfac{8,3,2}.
The star of three gains most: \Mfac{16,7,2} makes it in $383$ rounds, against $507$ for three \Mfac{8,3,2} runs, with $5.90\times10^{6}$ qubit-rounds against $7.90\times10^{6}$.
One tailored run replaces two or three \Mfac{8,3,2} runs, so at equal factory count the tailored factory is faster.
The worked example of \cref{section:appendix_model_worked} derives these latencies by hand.

One level higher in the Clifford hierarchy, however, the comparison reverses.
The last two blocks of \cref{tab:standalone_factory_costs} concern two fourth-level targets: two disjoint $CCCZ$ gates, the reduced form of the output of the $\code{30,8,2}$ factory \Mfac{30,8,2} (\cref{fig:3082_etd_factory}), and three $CCCZ$ gates sharing one control qubit, the action of the $R=3$ glued $CCCZ$ star \Mfac{40,10,2} of \cref{section:cccz_star}.
Both factories consume $\sqrt T$ states, and consuming their outputs leaves byproducts that need $CCZ$ corrections, so every row installs a second factory.
In the fast block at $d=21$, $CCZ$-only compilation with measurement-based uncomputation applies the two disjoint $CCCZ$ gates in $372$ rounds and the star in $517$, against $1{,}172$ and $1{,}649$ rounds for \Mfac{30,8,2} and \Mfac{40,10,2}.
The fourth-level factories lose because their ETD runs are long, $29$ and $39$ logical steps plus about half as many again for the $T$-state fix-ups of the $\sqrt T$ injections, and because their error floors, $1.1\times10^{-7}$ and $1.9\times10^{-7}$, lie above a target of $10^{-7}$.
\cref{section:fourth_level} gives the full costing and the optimistic parameters under which a fourth-level factory would trade installed volume for latency.

\input{includes/generated/standalone.tex}

\subsection{Finite-Field Multiplication}
\label{section:finitefield_multiplication}

Multiplication over $GF(2^{m})$ is a canonical source of structured non-Clifford cost in fault-tolerant arithmetic, for example in elliptic-curve cryptanalysis~\cite{wang2025comprehensive,shor1999polynomial,maslov2009depth}.
It multiplies binary polynomials and then applies a linear reduction by an irreducible polynomial $p(x)$, so it should not be confused with integer modular multiplication $ab\bmod N$.
In a reversible implementation the reduction is a $CX$ network, while the bilinear products need $CCX$ (equivalently $CCZ$) gates.

We use the complete out-of-place multiplier of Refs.~\cite{cheung2008design,maslov2009depth}, $\ket{a,b,0}\mapsto\ket{a,b,ab}$, with $p(x)=x^{8}+x^{4}+x^{3}+x+1$.
It acts on $N=24$ logical qubits and contains $28$ $CCX$ gates for the high-degree products, a reduction network of $23$ $CX$ gates, and $36$ $CCX$ gates for the low-degree products, $64$ $CCX$ gates in total (\cref{fig:gf256_full} shows the complete circuit).
We verified it by simulation on all $2^{16}$ inputs $(a,b)$, for every one of the $30$ irreducible polynomials of degree eight.
Conjugating the output register by Hadamards turns the circuit into a $CX$ network followed by the diagonal gate
\begin{align}
\prod_{i,j=0}^{7}CCZ\!\left(a_{i},\,b_{j},\,\ell_{i+j}\cdot c\right)~,
\label{eq:gf_phase_form}
\end{align}
where $\ell_{i+j}$ is the coefficient vector of $x^{i+j}\bmod p(x)$, so $\ell_{i+j}\cdot c$ is a parity of output qubits (the single qubit $c_{i+j}$ when $i+j<8$).
All $64$ gates are diagonal and commute.
Gates that share one argument form an $R$-star resource state (\cref{eq:star_action}) up to a $CX$ basis transformation, provided their other arguments are linearly independent: the $CX$ network then maps each of those arguments to its own qubit.
For the gates sharing $a_{i}$, the other arguments are $b_{0},\dots,b_{7}$ and the parities $\ell_{i}\cdot c,\dots,\ell_{i+7}\cdot c$.
These parities are independent because $x^{i},\dots,x^{i+7}$ are the images of $1,x,\dots,x^{7}$ under multiplication by $x^{i}$, which is invertible modulo $p(x)$.
The same argument applies to the gates sharing $b_{j}$.
Gates with the same value of $i+j$ share the parity $\ell_{i+j}\cdot c$ and act on distinct pairs $(a_{i},b_{j})$.
The multiplier is therefore eight stars of eight gates centered on $a_{0},\dots,a_{7}$, and the minimum covers also use stars centered on $b_{j}$ and on output parities.
$32$ \Mfac{12,5,2} outputs, $22$ \Mfac{16,7,2} outputs ($20$ three-$CCZ$ stars and two pairs) or $16$ \Mfac{20,9,2} outputs replace $64$ \Mfac{8,3,2} outputs.
Twenty-two is the fewest possible, $\lceil 64/3\rceil$; stars centered only on the $a_{i}$ would need $24$.
The $64$-gate schoolbook decomposition is a fixed benchmark rather than an optimal compilation of the primitive: recursive Karatsuba multiplication~\cite{vanhoof2020space} writes the same product with $27$ $CCZ$ gates on parities of the inputs and outputs, keeping the reduction modulo $p(x)$ as a free $CX$ network, and these $27$ gates group
into $13$ pairs that are disjoint up to a basis change plus one single gate.
All comparisons below refer to the $64$-gate circuit; recompiling both canonical and tailored strategies on the Karatsuba form is left to future work.

\input{includes/finite_field_full_circuit}

\cref{tab:finite_field_mult_litinski_regimes} evaluates one factory of each family with $\apxlanes$ lanes in the three data blocks.
With a single factory, the factory almost always sets the pace: the latency is close to the number of consumptions times the ETD run length $\tau$, so the ETD time per $CCZ$, $\tau/R$, decides the order.
Every tailored factory is therefore faster than \Mfac{8,3,2} in every block, and larger stars are faster: in the fast block the latency falls from $9{,}475$ rounds for \Mfac{8,3,2} ($\tau/R=7$) to $7{,}459$ for \Mfac{12,5,2} ($5.5$) and $6{,}265$ for \Mfac{24,11,2} ($4.6$).
The gain is largest in the compact block, where $64$ consumptions of $9$ logical steps each make the data block, not the factory, the bottleneck for \Mfac{8,3,2}; halving the number of consumptions removes that bottleneck, and the latency drops by $38$--$48\%$.
Volume favors the smallest tailored factory.
\Mfac{12,5,2} has the lowest occupied volume of all single-factory strategies in every block, $11$--$26\%$ below \Mfac{8,3,2}, while larger stars pay for their speed with a larger footprint: \Mfac{24,11,2} saves only $5\%$ in the fast block.
\Mfac{14,6,2} applies the same two gates as \Mfac{12,5,2} with a run two steps longer, so it is slower than \Mfac{12,5,2} and uses more volume than \Mfac{8,3,2} in the intermediate and fast blocks.
Running \Mfac{12,5,2} and \Mfac{14,6,2} side by side gives the shortest latency in the intermediate and fast blocks, $41\%$ below \Mfac{8,3,2}, but installs $50\%$ more qubits.
All rows have $\epsilon_{\rm total}$ between $5.7\times10^{-7}$ and $9.0\times10^{-7}$; the topological term dominates, so the error follows the occupied volume.
In short, the largest star wins on latency, \Mfac{12,5,2} wins on volume, and both gain most when the data block is slow.

\input{includes/generated/finite_field.tex}

\subsection{Single-lane adders}
\label{section:single_lane_adders}

Addition is a harder test for tailored resources, because its non-Clifford gates form long dependency chains rather than clusters.
The ripple-carry adder is the extreme case, since $CCZ$ gates are sparse in the circuit~\cite{cuccaro2004new,gidney2018halving}.
On the other hand, prefix and carry-lookahead adders have more parallelism, but must still respect propagate, generate, prefix, carry and sum dependencies~\cite{gossett1998quantum,draper2004logarithmic,wang2025comprehensive}.

All adders in this and the next section (\cref{section:carry_lookahead_addition}) compute the sum out of place, $\ket{A,B,0}\mapsto\ket{A,B,A+B}$, and use the same garbage-free construction~\cite{gidney2018halving}.
Each Toffoli gate is written as a temporary logical AND into a fresh ancilla, followed by $CX$ gates.
Each ancilla is uncomputed by an $X$-basis measurement and a classically controlled $CZ$, which costs no non-Clifford resource but needs one reaction time.
The non-Clifford cost of an adder is therefore its number of AND gates.

We first add two 4-bit numbers with a carry-in.
The carry-lookahead adder~\cite{draper2004logarithmic} has $16$ AND gates on $N=42$ logical qubits; its AND gates have the same control functions as the $CCX$ gates of the carry-lookahead adder of the preprint.
The ripple-carry adder of Ref.~\cite{gidney2018halving} has $4$ AND gates on $N=30$ qubits.
Both are verified exhaustively on all inputs, including the return of every ancilla to $\ket0$, and are drawn in \cref{fig:adder_ripple_single,fig:adder_cla_single}.
Within one lane, the carry-lookahead adder contains AND gates that share a control, which suit \Mfac{12,5,2}, and AND gates on disjoint supports, which suit \Mfac{14,6,2}.

\cref{table:cla_single} compares both adders in the fast block at $d=21$; \cref{table:cla_single_cla,table:cla_single_ripple} give all three data blocks.
Tailored resources help only when the circuit contains AND gates that can be packed together.
The carry-lookahead adder does, and every tailored strategy beats \Mfac{8,3,2} on latency in every block.
With one factory the latency again tracks the total ETD time, so \Mfac{12,5,2}, which packs $12$ of the $16$ AND gates into shared-control pairs, wins among single factories: it is $16$--$28\%$ faster than \Mfac{8,3,2} and uses $11$--$22\%$ less occupied volume.
\Mfac{14,6,2} packs all $16$ gates into disjoint pairs, but its longer ETD run leaves it slower than \Mfac{12,5,2}, and it saves volume only in the compact block.
Both gain most in the compact block, where the $16$ consumptions of \Mfac{8,3,2} make the data block the bottleneck.
The two pair types are complementary, so running \Mfac{12,5,2} and \Mfac{14,6,2} side by side is fastest in every block, $35$--$39\%$ below \Mfac{8,3,2}, and has the lowest occupied volume in the intermediate and fast blocks, at the price of $37$--$48\%$ more installed qubits.
The ripple-carry adder is the opposite case: its four AND gates form one dependent chain, no cover packs two of them, and every strategy needs four consumptions, with latencies within $4\%$ and volumes within $1\%$ of each other.
Simultaneous consumption of disjoint-support resources (rows fast$^{\parallel}$) changes none of these numbers, because the factories, not the data block, limit the schedule.

\input{includes/adders_figure}

\input{includes/generated/adder_cla_summary.tex}

\subsection{Shared-addend adders}
\label{section:carry_lookahead_addition}

We next consider $R$ additions that share one addend $A$, computed out of place with the same garbage-free AND construction: $\ket{A,B^{(r)},0}\mapsto\ket{A,B^{(r)},A+B^{(r)}}$ for $r=1,\dots,R$.
At bit position $i$ the leaf-generate ANDs $g_{i}^{(r)}=a_{i}b_{i}^{(r)}$ of the $R$ lanes share the control $a_{i}$, i.e.\ they form an $R$-star centered on $a_{i}$, while the prefix ANDs of different lanes act on disjoint supports and suit \Mfac{14,6,2}.
We build the Sklansky carry-lookahead version~\cite{sklansky1960conditional} with $R=2$ and width $8$: $50$ AND gates on $N=132$ logical qubits.
As a baseline with little parallelism we also build the ripple-carry version of Ref.~\cite{gidney2018halving}: $16$ AND gates on $N=100$ qubits.
Both are verified exhaustively on all inputs as in the other arithmetic examples, including the return of every ancilla to $\ket0$.
\cref{fig:adder_ripple_shared,fig:adder_cla_shared} in \cref{section:extended_tables} show their width-4 instances.

\cref{table:batched_addition} compares both shared-addend adders in the fast block at $d=21$; \cref{table:batched_addition_cla,table:batched_addition_ripple} give all three data blocks.
Similarly to the single-lane adders, these two adders favor different resources, because they offer different kinds of pairs.
In the carry-lookahead adder, many AND gates share a control, so the star family wins.
\Mfac{12,5,2} covers the $50$ AND gates with $26$ consumptions and is $20\%$ faster than \Mfac{8,3,2} in the fast and intermediate blocks and $36\%$ faster in the compact block, with $17$--$33\%$ less occupied volume.
\Mfac{16,7,2} saves two more consumptions and is about $2\%$ faster still; larger stars find no further packing and only add footprint.
\Mfac{14,6,2} is the weakest tailored choice here: it needs about as many consumptions as \Mfac{12,5,2}, but its longer ETD run cuts the latency by only $7\%$ outside the compact block, and adding it to \Mfac{12,5,2} makes the schedule slower than \Mfac{12,5,2} alone.
The ripple-carry adder is the opposite case.
Its AND gates act on the carry-dependent inputs of each lane, so the shared addend yields only one shared-control pair, and \Mfac{12,5,2} and the star factories stay within $3\%$ of \Mfac{8,3,2}.
The carry chains of the two lanes are independent, however, so AND gates of different lanes pair on disjoint supports: \Mfac{14,6,2} needs $9$ consumptions instead of $16$ and is $7$--$23\%$ faster, and the mixed \Mfac{12,5,2}+\Mfac{14,6,2} portfolio, which installs two factories, is fastest in every block, $16$--$29\%$ below \Mfac{8,3,2}, with the lowest occupied volume.
In both adders the gain is largest in the compact block, where the $50$ or $16$ consumptions of \Mfac{8,3,2} make the data block the bottleneck, and $\epsilon_{\rm total}$ follows the occupied volume.
As in the single-lane case, simultaneous consumption of disjoint-support resources (rows fast$^{\parallel}$) changes none of these numbers, because the factories limit the schedule.

\input{includes/generated/adder_shared_summary.tex}

%% file: includes/generated/standalone.tex
\begin{table*}\centering
\caption{Standalone comparison of one tailored output against the same gates made from \Mfac{8,3,2} outputs at $d=21$. The last two blocks concern the fourth-level targets of \cref{section:fourth_level} in the fast block; there every row installs two factories, and the CCZ-only row uses measurement-based uncomputation. Settings, columns and colors are those of \cref{section:comparison_framework}; in the fourth-level blocks the colors compare with the CCZ-only row.}
\label{tab:standalone_factory_costs}
\widetable{\begin{tabular}{rll|rr|rr|rrr}\toprule
$d$ & Target & Strategy & Rounds & 95\% & Qubit-rounds & Qubits & $\epsilon_{\rm ETD}$ & $\epsilon_{\rm topo}$ & $\epsilon_{\rm total}$ \\\midrule
21 & Shared pair & \Mfac{8,3,2} & 361 & 387 & 4,691,194 & 53,930 & $1.2\times10^{-10}$ & $4.2\times10^{-9}$ & $4.3\times10^{-9}$ \\
21 & Shared pair & \Mfac{12,5,2} & \textcolor{darkred}{299} & 324 & \textcolor{darkblue}{3,732,647} & 65,396 & $1.4\times10^{-10}$ & $3.3\times10^{-9}$ & $3.5\times10^{-9}$ \\
\midrule
21 & Disjoint pair & \Mfac{8,3,2} & 361 & 387 & 4,691,194 & 53,930 & $1.2\times10^{-10}$ & $4.2\times10^{-9}$ & $4.3\times10^{-9}$ \\
21 & Disjoint pair & \Mfac{14,6,2} & \textcolor{darkred}{341} & 366 & 4,760,258 & 71,129 & $1.9\times10^{-10}$ & $4.3\times10^{-9}$ & $4.5\times10^{-9}$ \\
\midrule
21 & Star of three & \Mfac{8,3,2} & 507 & 534 & 7,901,704 & 53,930 & $1.8\times10^{-10}$ & $7.3\times10^{-9}$ & $7.5\times10^{-9}$ \\
21 & Star of three & \Mfac{16,7,2} & \textcolor{darkred}{383} & 408 & \textcolor{darkblue}{5,903,145} & 76,862 & $2.5\times10^{-10}$ & $5.5\times10^{-9}$ & $5.7\times10^{-9}$ \\
\midrule
21 & 2 disjoint $CCCZ$ & CCZ-only (\ensuremath{2{\times}}\Mfac{8,3,2}) & 372 & 382 & 19,012,573 & 105,919 & $2.4\times10^{-10}$ & $1.9\times10^{-8}$ & $2.0\times10^{-8}$ \\
21 & 2 disjoint $CCCZ$ & \Mfac{30,8,2} + \Mfac{8,3,2} & 1,172 & 1,266 & 59,133,960 & 176,969 & $1.1\times10^{-7}$ & $6.3\times10^{-8}$ & $1.7\times10^{-7}$ \\
\midrule
21 & 3 $CCCZ$, shared control & CCZ-only (\ensuremath{2{\times}}\Mfac{8,3,2}) & 517 & 529 & 32,484,442 & 112,317 & $3.5\times10^{-10}$ & $3.4\times10^{-8}$ & $3.4\times10^{-8}$ \\
21 & 3 $CCCZ$, shared control & \Mfac{40,10,2} + \Mfac{8,3,2} & 1,649 & 1,754 & 102,229,285 & 212,200 & $1.9\times10^{-7}$ & $1.1\times10^{-7}$ & $3.0\times10^{-7}$ \\
\bottomrule\end{tabular}}\end{table*}

%% file: includes/finite_field_full_circuit.tex
\begin{figure*}
\centering
\includegraphics[width=\textwidth]{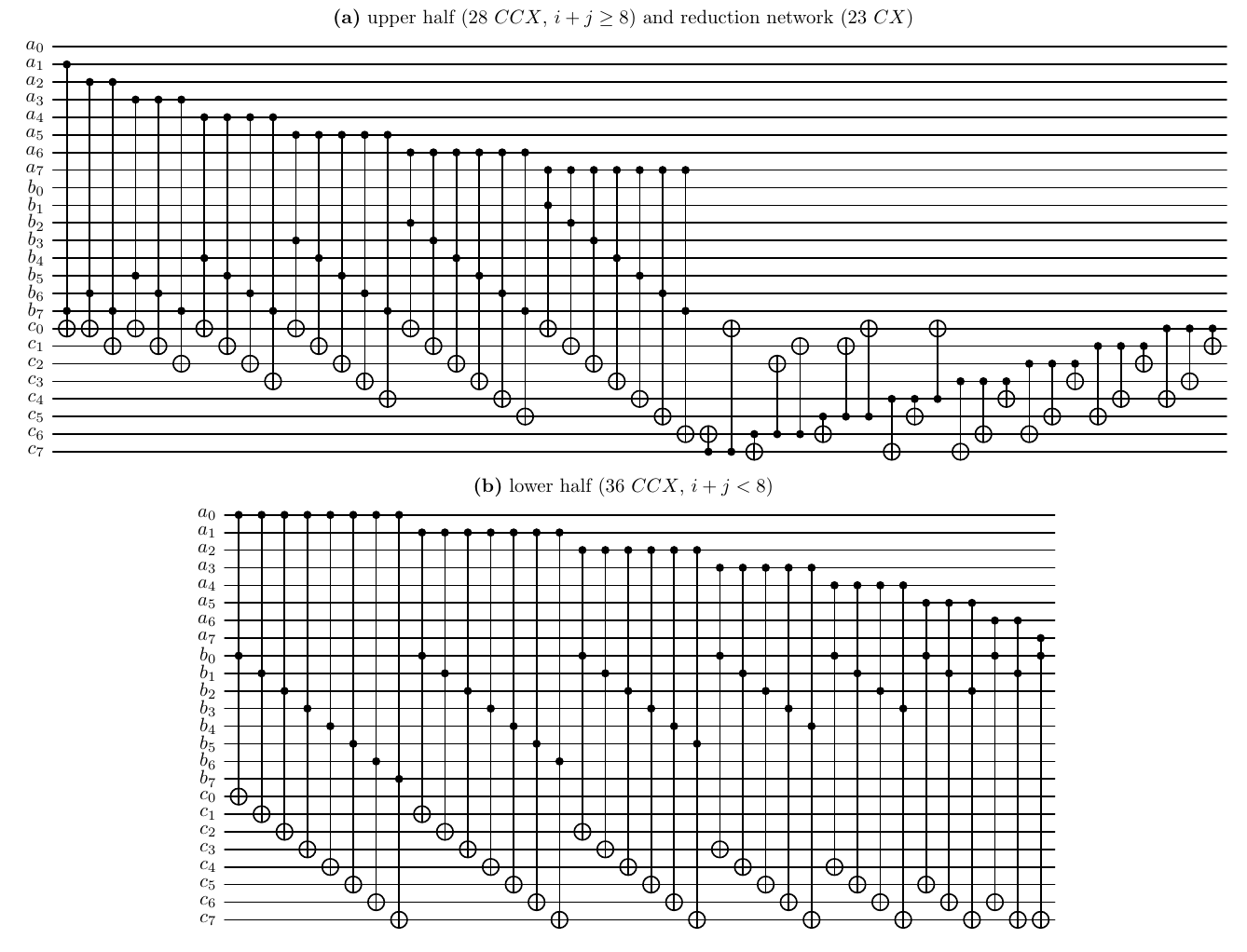}
\caption{
The reversible $GF(2^{8})$ multiplier $\ket{a,b,0}\mapsto\ket{a,b,ab}$ with $p(x)=x^8+x^4+x^3+x+1$ (\cref{section:finitefield_multiplication}).
(a)~$CCX(a_i,b_j;c_{i+j-8})$ for $i+j\ge8$, then the $CX$ network $c\mapsto Mc$ with $Me_k=x^{8+k}\bmod p(x)$.
(b)~$CCX(a_i,b_j;c_{i+j})$ for $i+j<8$.
}
\label{fig:gf256_full}
\end{figure*}

%% file: includes/generated/finite_field.tex
\begin{table*}\centering
\caption{Multiplication over $GF(2^8)$ with $p(x)=x^8+x^4+x^3+x+1$ by the Cheung--Maslov multiplier~\cite{cheung2008design,maslov2009depth} (64 $CCX$ gates, $N=24$ logical qubits) in all data blocks at $d=21$. Settings, columns and colors are those of \cref{section:comparison_framework}.}
\label{tab:finite_field_mult_litinski_regimes}
\widetable{\begin{tabular}{llrr|rr|rr|rrr}\toprule
Strategy & Block & $d$ & Cons. & Rounds & 95\% & Qubit-rounds & Qubits & $\epsilon_{\rm ETD}$ & $\epsilon_{\rm topo}$ & $\epsilon_{\rm total}$ \\\midrule
\Mfac{8,3,2} & compact & 21 & 64 & 12,289 & 12,315 & 711,616,276 & 88,328 & $3.8\times10^{-9}$ & $7.7\times10^{-7}$ & $7.8\times10^{-7}$ \\
 & intermediate & 21 & 64 & 9,559 & 9,585 & 669,675,449 & 99,794 & $3.8\times10^{-9}$ & $7.3\times10^{-7}$ & $7.3\times10^{-7}$ \\
 & fast & 21 & 64 & 9,475 & 9,501 & 742,323,465 & 109,369 & $3.8\times10^{-9}$ & $8.1\times10^{-7}$ & $8.1\times10^{-7}$ \\
\midrule
\Mfac{12,5,2} & compact & 21 & 32 & \textcolor{darkred}{7,627} & 7,635 & \textcolor{darkblue}{524,012,756} & 99,794 & $4.4\times10^{-9}$ & $5.7\times10^{-7}$ & $5.7\times10^{-7}$ \\
 & intermediate & 21 & 32 & \textcolor{darkred}{7,543} & 7,551 & \textcolor{darkblue}{595,755,104} & 111,260 & $4.4\times10^{-9}$ & $6.5\times10^{-7}$ & $6.6\times10^{-7}$ \\
 & fast & 21 & 32 & \textcolor{darkred}{7,459} & 7,467 & \textcolor{darkblue}{651,469,168} & 120,835 & $4.4\times10^{-9}$ & $7.1\times10^{-7}$ & $7.2\times10^{-7}$ \\
\midrule
\Mfac{14,6,2} & compact & 21 & 32 & \textcolor{darkred}{8,970} & 8,979 & \textcolor{darkblue}{659,167,156} & 105,527 & $6.1\times10^{-9}$ & $7.2\times10^{-7}$ & $7.2\times10^{-7}$ \\
 & intermediate & 21 & 32 & \textcolor{darkred}{8,886} & 8,895 & 743,941,768 & 116,993 & $6.1\times10^{-9}$ & $8.1\times10^{-7}$ & $8.2\times10^{-7}$ \\
 & fast & 21 & 32 & \textcolor{darkred}{8,802} & 8,811 & 810,148,255 & 126,568 & $6.1\times10^{-9}$ & $8.9\times10^{-7}$ & $9.0\times10^{-7}$ \\
\midrule
\Mfac{12,5,2}+\Mfac{14,6,2} (mixed) & compact & 21 & 32 & \textcolor{darkred}{6,985} & 6,987 & \textcolor{darkblue}{666,794,373} & 142,571 & $5.5\times10^{-9}$ & $7.3\times10^{-7}$ & $7.3\times10^{-7}$ \\
 & intermediate & 21 & 32 & \textcolor{darkred}{5,662} & 5,664 & \textcolor{darkblue}{604,117,037} & 154,037 & $5.5\times10^{-9}$ & $6.6\times10^{-7}$ & $6.6\times10^{-7}$ \\
 & fast & 21 & 32 & \textcolor{darkred}{5,544} & 5,559 & \textcolor{darkblue}{635,072,193} & 163,612 & $5.5\times10^{-9}$ & $6.9\times10^{-7}$ & $7.0\times10^{-7}$ \\
\midrule
\Mfac{16,7,2} & compact & 21 & 22 & \textcolor{darkred}{6,986} & 7,005 & \textcolor{darkblue}{538,097,779} & 111,260 & $5.3\times10^{-9}$ & $5.9\times10^{-7}$ & $5.9\times10^{-7}$ \\
 & intermediate & 21 & 22 & \textcolor{darkred}{6,902} & 6,921 & \textcolor{darkblue}{603,229,465} & 122,726 & $5.3\times10^{-9}$ & $6.6\times10^{-7}$ & $6.7\times10^{-7}$ \\
 & fast & 21 & 22 & \textcolor{darkred}{6,818} & 6,837 & \textcolor{darkblue}{653,545,078} & 132,301 & $5.3\times10^{-9}$ & $7.2\times10^{-7}$ & $7.2\times10^{-7}$ \\
\midrule
\Mfac{20,9,2} & compact & 21 & 16 & \textcolor{darkred}{6,623} & 6,645 & \textcolor{darkblue}{564,561,392} & 122,726 & $6.4\times10^{-9}$ & $6.2\times10^{-7}$ & $6.3\times10^{-7}$ \\
 & intermediate & 21 & 16 & \textcolor{darkred}{6,539} & 6,561 & \textcolor{darkblue}{625,981,985} & 134,192 & $6.4\times10^{-9}$ & $6.9\times10^{-7}$ & $7.0\times10^{-7}$ \\
 & fast & 21 & 16 & \textcolor{darkred}{6,455} & 6,477 & \textcolor{darkblue}{673,271,731} & 143,767 & $6.4\times10^{-9}$ & $7.4\times10^{-7}$ & $7.5\times10^{-7}$ \\
\midrule
\Mfac{24,11,2} & compact & 21 & 13 & \textcolor{darkred}{6,433} & 6,438 & \textcolor{darkblue}{597,119,093} & 134,192 & $7.4\times10^{-9}$ & $6.6\times10^{-7}$ & $6.6\times10^{-7}$ \\
 & intermediate & 21 & 13 & \textcolor{darkred}{6,349} & 6,354 & \textcolor{darkblue}{656,585,588} & 145,658 & $7.4\times10^{-9}$ & $7.2\times10^{-7}$ & $7.3\times10^{-7}$ \\
 & fast & 21 & 13 & \textcolor{darkred}{6,265} & 6,270 & \textcolor{darkblue}{702,280,101} & 155,233 & $7.4\times10^{-9}$ & $7.8\times10^{-7}$ & $7.8\times10^{-7}$ \\
\bottomrule\end{tabular}}\end{table*}

%% file: includes/adders_figure.tex
\begin{figure*}[tp]
\centering
\panelimage[width=0.397\textwidth]{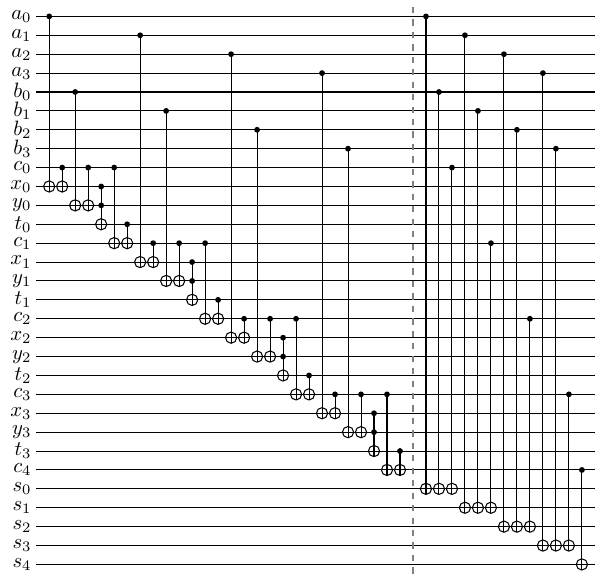}{(a)}%
\hspace{0.06\textwidth}%
\panelimage[width=0.482\textwidth]{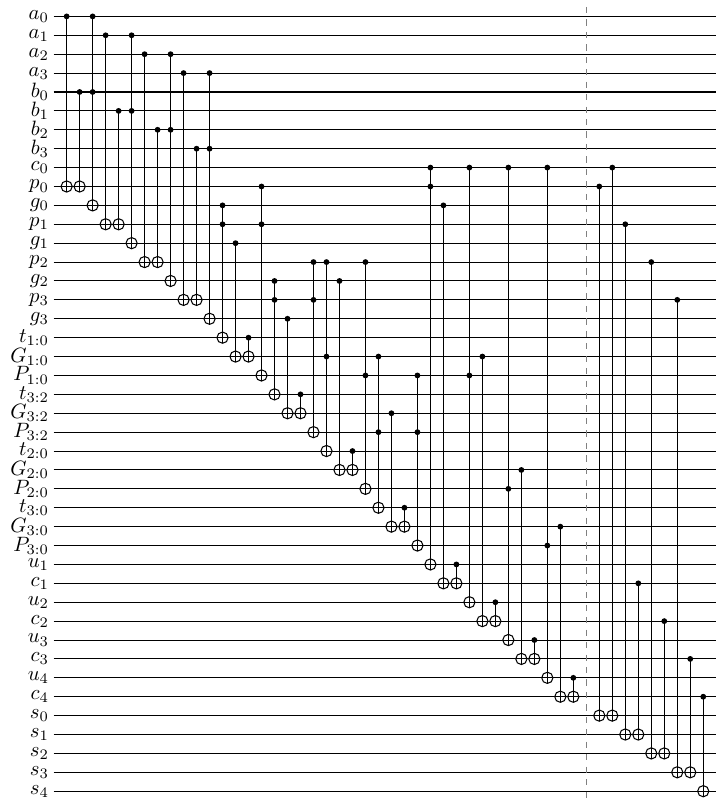}{(b)}
\caption{The two single-lane adders of \cref{section:single_lane_adders}, drawn from the verified and benchmarked circuits.
Each three-qubit gate is a temporary logical AND into a fresh ancilla; every wire other than $a$, $b$ and $c_0$ starts in $\ket0$.
Gates left of the dashed line compute the carries, and gates right of it copy the sum out; the uncomputation is not drawn.
(a)~Single-lane 4-bit ripple-carry adder with carry-in~\cite{gidney2018halving} ($4$ AND gates, $N=30$): $x_i=a_i\oplus c_i$, $y_i=b_i\oplus c_i$, $t_i=x_iy_i$ and $c_{i+1}=c_i\oplus t_i$.
(b)~Single-lane 4-bit carry-lookahead adder with carry-in~\cite{draper2004logarithmic} ($16$ AND gates, $N=42$): $p_i=a_i\oplus b_i$, $g_i=a_ib_i$, $G_{i:j}=G_{i:k}\oplus t_{i:j}$ with $t_{i:j}=P_{i:k}G_{k-1:j}$, $P_{i:j}=P_{i:k}P_{k-1:j}$, and $c_{i}=G_{i-1:0}\oplus u_{i}$ with $u_{i}=P_{i-1:0}c_0$.}
\label{fig:adders}
\setcounter{subfigure}{0}
\refstepcounter{subfigure}\label{fig:adder_ripple_single}
\refstepcounter{subfigure}\label{fig:adder_cla_single}
\end{figure*}

%% file: includes/generated/adder_cla_summary.tex
\begin{table*}\centering
\caption{Single-lane 4-bit carry-lookahead~\cite{draper2004logarithmic} and ripple-carry~\cite{gidney2018halving} adders with carry-in in the fast block at $d=21$; \cref{table:cla_single_cla,table:cla_single_ripple} give all data blocks. Settings, columns and colors are those of \cref{section:comparison_framework}.}
\label{table:cla_single}
\widetable{\begin{tabular}{l|rrrr|rrrr}\toprule
 & \multicolumn{4}{c}{Carry-lookahead (16 AND, $N=42$)} & \multicolumn{4}{c}{Ripple-carry (4 AND, $N=30$)} \\
\cmidrule(lr){2-5}\cmidrule(lr){6-9}
Strategy & Cons. & Rounds & Qubit-rounds & $\epsilon_{\rm total}$ & Cons. & Rounds & Qubit-rounds & $\epsilon_{\rm total}$ \\\midrule
\Mfac{8,3,2} & 16 & 2,418 & 272,779,765 & $3.0\times10^{-7}$ & 4 & 653 & 55,365,397 & $6.1\times10^{-8}$ \\
\Mfac{12,5,2} & 10 & \textcolor{darkred}{2,025} & \textcolor{darkblue}{241,503,444} & $2.7\times10^{-7}$ & 4 & \textcolor{darkred}{639} & 55,481,243 & $6.1\times10^{-8}$ \\
\Mfac{14,6,2} & 8 & \textcolor{darkred}{2,250} & 278,920,783 & $3.1\times10^{-7}$ & 4 & \textcolor{darkred}{635} & 55,463,290 & $6.1\times10^{-8}$ \\
\Mfac{12,5,2}+\Mfac{14,6,2} (mixed) & 8 & \textcolor{darkred}{1,468} & \textcolor{darkblue}{206,612,387} & $2.3\times10^{-7}$ & 4 & \textcolor{darkred}{632} & 55,771,921 & $6.1\times10^{-8}$ \\
\bottomrule\end{tabular}}\end{table*}

%% file: includes/generated/adder_shared_summary.tex
\begin{table*}\centering
\caption{2-lane shared-addend Sklansky carry-lookahead and ripple-carry~\cite{gidney2018halving} adders of width 8 in the fast block at $d=21$; \cref{table:batched_addition_cla,table:batched_addition_ripple} give all data blocks. Settings, columns and colors are those of \cref{section:comparison_framework}.}
\label{table:batched_addition}
\widetable{\begin{tabular}{l|rrrr|rrrr}\toprule
 & \multicolumn{4}{c}{Carry-lookahead (50 AND, $N=132$)} & \multicolumn{4}{c}{Ripple-carry (16 AND, $N=100$)} \\
\cmidrule(lr){2-5}\cmidrule(lr){6-9}
Strategy & Cons. & Rounds & Qubit-rounds & $\epsilon_{\rm total}$ & Cons. & Rounds & Qubit-rounds & $\epsilon_{\rm total}$ \\\midrule
\Mfac{8,3,2} & 50 & 7,416 & 2,114,811,602 & $2.4\times10^{-6}$ & 16 & 2,418 & 541,358,778 & $6.1\times10^{-7}$ \\
\Mfac{12,5,2} & 26 & \textcolor{darkred}{5,906} & \textcolor{darkblue}{1,735,635,326} & $2.0\times10^{-6}$ & 15 & \textcolor{darkred}{2,356} & 541,479,877 & $6.1\times10^{-7}$ \\
\Mfac{14,6,2} & 25 & \textcolor{darkred}{6,890} & \textcolor{darkblue}{2,057,677,263} & $2.3\times10^{-6}$ & 9 & \textcolor{darkred}{2,252} & \textcolor{darkblue}{528,319,663} & $5.9\times10^{-7}$ \\
\Mfac{12,5,2}+\Mfac{14,6,2} (mixed) & 25 & \textcolor{darkred}{6,087} & \textcolor{darkblue}{1,949,893,658} & $2.2\times10^{-6}$ & 8 & \textcolor{darkred}{1,993} & \textcolor{darkblue}{490,600,017} & $5.5\times10^{-7}$ \\
\Mfac{16,7,2} & 24 & \textcolor{darkred}{5,768} & \textcolor{darkblue}{1,732,675,848} & $1.9\times10^{-6}$ & 15 & \textcolor{darkred}{2,345} & 551,236,691 & $6.2\times10^{-7}$ \\
\Mfac{20,9,2} & 24 & \textcolor{darkred}{5,767} & \textcolor{darkblue}{1,765,933,521} & $2.0\times10^{-6}$ & 15 & \textcolor{darkred}{2,344} & 561,995,691 & $6.3\times10^{-7}$ \\
\Mfac{24,11,2} & 24 & \textcolor{darkred}{5,767} & \textcolor{darkblue}{1,798,989,033} & $2.0\times10^{-6}$ & 15 & \textcolor{darkred}{2,344} & 571,764,644 & $6.4\times10^{-7}$ \\
\bottomrule\end{tabular}}\end{table*}

%% file: sections/05_discussion.tex
\section{Discussion}
\label{section:discussion}

CSS-T-Product codes provide a code-theoretic route to multi-qubit magic state resources.
Their smallest third-level representatives overlap with known synthillation resources, while the broader framework supplies generative quantum LEGO constructions, explicit stabilizer descriptions, logical basis orbits, higher-distance families, and transversal actions above the third level of the Clifford hierarchy.
Our principal contribution therefore has been a method for discovering, organizing and costing structured resource states, which carries known resources to higher distance and to higher levels of the Clifford hierarchy; this method we consider complementary to synthillation.
The organization is made precise by the tiers of \cref{def:tiers}.
Most of the codes we construct produce several canonical states in one factory, which can still pay off architecturally; some examples include the closed-chain and star resources, such as \Mfac{12,5,2} and \Mfac{16,7,2}, which are irreducible.
Gate teleportation could still assemble these from two and three $CCZ$ states, respectively, so an irreducible resource saves teleportations and consumptions rather than $CCZ$ states (\cref{prop:nullity}).
Whether irreducible resources can be matched to circuits more efficiently than tensor-product ones is a question our benchmarks, which use tensor-product (Tier-1) and irreducible (Tier-2) resources side by side, only begin to address.

Across the benchmarks, the decisive quantity is not simply the number of non-Clifford interactions contained in an output.
A tailored state pays off when its logical gate pattern can be matched to non-Clifford patterns in the target circuit without introducing excessive factory occupancy, waiting between dependent consumptions, output error, or unused resource components.
Tailored factories then reduce the number of consumption events; the price is that a larger accepted output has a larger leading error coefficient.
Latency (the run time of the circuit), occupied volume (busy qubits integrated over time), installed volume (all installed qubits times latency), footprint, and output quality consequently define distinct optimization objectives.

The benchmarks of \cref{section:tailored_algorithmic_primitives} locate where tailored third-level resources win.
In our model the algorithm's logical qubits sit in a \emph{data block}, factories prepare resource states from cultivated $T$ states, and each use of a resource state (a \emph{consumption}) is one lattice-surgery measurement on the data block of fixed duration; a simulation of all three gives the latency.
Two conditions must hold together.
First, the circuit must contain the matching structure, i.e.\ groups of $CCZ$ gates with the pattern of the resource: shared controls for \Mfac{12,5,2} and the stars (the schoolbook $GF(2^{8})$ multiplier is eight stars of eight $CCZ$ gates) or disjoint pairs for \Mfac{14,6,2}.
On a circuit with little matching structure, such as the single-lane ripple-carry adder, a tailored factory changes the latency by at most $3\%$ and needs $7$--$16\%$ more installed volume.
Second, the gain depends on what limits the latency.
With one factory per family the factory sets the pace, and strategies are ordered by
the factory run time per $CCZ$, $\tau/R$, from $7$ logical steps for \Mfac{8,3,2} down
to $4.6$ for \Mfac{24,11,2}; the gain is largest in the compact block, where the many
consumptions of \Mfac{8,3,2} make the data block the bottleneck.
When $T$ states are scarce, the order follows the raw $T$ states per $CCZ$ ($8$ for
\Mfac{8,3,2} down to $5$ for \Mfac{20,9,2}), and with many factories and lanes the
number of consumptions limits the latency (\cref{fig:cultivation_capacity}).
When both hold, the gain is mainly in latency: at $d=21$ with one factory per family, \Mfac{12,5,2} is $21$--$38\%$ faster than \Mfac{8,3,2} on the multiplier and $20$--$36\%$ faster on the shared-addend carry-lookahead adder, with $11$--$33\%$ less occupied volume, and the largest stars reach $48\%$ and $38\%$, respectively, at a larger footprint.

For fourth-level resources a third condition must be added: a factory that applies its $\sqrt T$ rotations in parallel, and a $\sqrt T$ source about as accurate as $T$ cultivation whose injections need no $T$-state fix-ups (\cref{section:fourth_level_optimistic}).
In addition, CSS-TP resource states above the third level introduce a separate limitation: consuming them requires adaptive corrections one level lower in the Clifford hierarchy, which for \Mfac{30,8,2} are zero, one or two canonical $CCZ$ corrections with multiplicities $1$, $30$ and $225$.
With these corrections, the $T$-state fix-ups of the $\sqrt T$ injections and the $\sqrt T$ supply included in complete schedules (\cref{section:fourth_level}), \Mfac{30,8,2} and the glued $CCCZ$ star \Mfac{40,10,2} lose to $CCZ$-only compilation at the cultivation settings considered, and their error floors lie above a target of $10^{-7}$.
Only with both parts of the third condition does \Mfac{30,8,2} become $5$--$24\%$ faster than a $CCZ$-only construction given the same parallel factory layout, and then only in the two slower data blocks and at $26$--$54\%$ more installed volume.
Better preparation and injection of fourth-level resources, in particular a parallel-injection factory, are therefore a natural subject for further study.

Several theoretical and architectural questions remain open.
Our rules predict the logical action of a code built by quantum LEGO operations, and its level in the Clifford hierarchy, for single-qubit diagonal transversal layers on CSS codes; within this scope they also give the logical action after a fusion whose logical map has a kernel, which Ref.~\cite{cao2026quantum} identifies as an open question for general symmetries (\cref{cor:conjoin_loop}).
Extending them to physical layers of multi-qubit diagonal gates, such as a transversal $CCZ$ between (or inside) code blocks, appears direct; non-diagonal or fold-transversal gates are not covered~\cite{cao2026quantum}.
Our prediction of the logical phase polynomial from overlaps among stabilizers and logical operators can also be related to cup-product methods for homological codes~\cite{breuckmann2026cups}, which could give CSS-TP families of distance greater than two.
The inverse problem, constructing a CSS-TP code with a prescribed transversal action, is open, although partially answered by synthillation.
On the compilation side, the Clifford orbit of \cref{section:CSSTP_gate_cleaning} could serve directly in phase-oracle compilation, by matching representatives of an available resource to blocks of the oracle's phase polynomial, for example in the phase separators of variational algorithms.
Our model also does not lay out how a resource patch is injected into the data block: logical Clifford operations, including the lattice-surgery basis changes and Pauli corrections around a consumption, are free, and every consumption costs the same time $\delta$ regardless of how many qubits the resource acts on. 
A more detailed injection model, in which this cost may grow with the support of a tailored resource, could narrow the latency gains reported here.
More detailed placement, routing, and correction models are also needed to determine how the present logical-patch tradeoffs change in specific near- or medium-term architectures, such as those involving surface code, color code, or QLDPC codes~\cite{breuckmann2021quantum,williamson2026low,swaroop2024universal,cross2024improved,cowtan2026parallel,he2025extractors,perlin2026quantum}.
Finally, quantum LEGO, and with it the CSS-TP formalism, could be extended beyond Pauli stabilizer codes, to XP codes and codes with non-commuting stabilizers~\cite{Shen2025quantumlegoxp,kobayashi2026clifford,huang2025hybrid,zhu2023non,ni2015non,webster2022xp}; quantum LEGO constructions of such codes exist~\cite{schotte2022quantum,Shen2025quantumlegoxp}, but how non-Clifford logical gates carry over while preserving code properties remains to be worked out.

CSS-TP codes connect transversal-gate code design, structured magic-state preparation, and architecture-aware compilation.
We gave phase-polynomial flow rules for quantum LEGO that predict transversal logical actions and their level in the Clifford hierarchy, for diagonal layers at any level, a classification of the resulting resources into tensor-product and irreducible ones, and code families of each kind.
Principally we found that the value of such magic resources lies in enlarging the set of available resource patterns: on circuits with matching structure they open new tradeoffs between installed hardware occupancy, latency, and circuit structure that a compiler can exploit.

%% file: sections/06_backmatter.tex
\section*{Acknowledgments}

We thank Dylan Herman, Anuj Apte, Enrico Fontana, Sivaprasad Omanakuttan, and Jacob Watkins for help regarding potential exotic magic state applications for algorithmic primitives.
We thank Anqi Gong and Adam Wills for bringing the related work \cite{anqiwills2026} to our attention during QEC26. 
We thank Ruslan Shaydulin and Rob Otter for executive support of this work.

\section*{Author Contributions}

MS devised the CSS-TP framework, the code-generation protocol, and developed the logical basis transformation technique.
CC and MS investigated magic properties of CSS-TP codes.
MS selected the set of applications and performed all of the spacetime volume comparisons, with guidance from JS and MP.
CC, JS, and MP provided project guidance and supervision.

\section*{Software Usage}

The following software libraries were utilized in this manuscript: \texttt{qLDPC}~\cite{perlin2023qldpc}; \texttt{MQT QECC}~\cite{mqt}; \texttt{PyZX}~\cite{kissinger2020Pyzx}; \texttt{PlanqTN}~\cite{pato_2025_planqtn}; and \texttt{CSSLO}~\cite{webster2023transversal}.
Beyond these libraries, artificial-intelligence assistants, including \texttt{ChatGPT}~\cite{openai_chatgpt_2026}, \texttt{Claude Opus 5.5}~\cite{anthropic_claude_2026}, and GitHub Copilot, assisted with code generation, debugging, and manuscript review.
MS directed the scientific development and is responsible for the generated code, verifying results, and approving the scientific claims.
The authors take responsibility for the final content.

\section*{Disclaimer}

This paper was prepared for informational purposes with contributions from the Global Technology Applied Research center of JPMorgan Chase \& Co.
This paper is not a product of the Research Department of JPMorgan Chase \& Co. or its affiliates.
Neither JPMorgan Chase \& Co. nor any of its affiliates makes any explicit or implied representation or warranty and none of them accept any liability in connection with this paper, including, without limitation, with respect to the completeness, accuracy, or reliability of the information contained herein and the potential legal, compliance, tax, or accounting effects thereof.
This document is not intended as investment research or investment advice, or as a recommendation, offer, or solicitation for the purchase or sale of any security, financial instrument, financial product or service, or to be used in any way for evaluating the merits of participating in any transaction.

%% file: sections/A_conjoining_pcm.tex
\section{Conjoining Operations at the Parity-Check-Matrix Level}\label{section:conjoining_pcm}

Conjoining was defined conceptually in \cref{section:background_QLEGO} as a tensor contraction between code tensors.
This appendix gives the corresponding parity-check-matrix implementation.
The implementation does not introduce a separate code operation.
Consider two qubit stabilizer codes $\mathsf{Q}_{1}$ and $\mathsf{Q}_{2}$ with PCMs $\mathsf{H}_{1}$ and $\mathsf{H}_{2}$.
What is depicted diagrammatically as a conjoining operation can be represented at the level of PCMs:
\begin{align}
\mathsf{H}_{1} \wedge_{\mathcal{E}} \mathsf{H}_{2} = \wedge_{\mathcal E} (H_{1}\oplus H_{2})~,
\end{align}
where $\mathcal{E}$ represents the pairwise set of edges to be tensor contracted together.
These edges correspond to physical qubits at the level of the PCM, and are thus represented as columns of $H_{1}\oplus H_{2}$.
In this way, the operation of conjoining can be considered a mapping that inputs two PCMs and outputs a single PCM, corresponding to the conjoined code.

More generally, a conjoining operation is defined by \emph{self-trace} operations~\cite{cao2022quantum} applied to a PCM, where the above \emph{single-trace} operation forms a special subclass that corresponds to a self-trace on a block diagonal PCM.
In this work we use conjoining, i.e.\ single traces between two codes, and self-traces of two legs of one code, contracting one pair of legs at a time; this keeps the logical gate predictable~\cite{cao2026quantum} (\cref{lem:operations,cor:conjoin_loop}).
Conjoining can reduce the distance below $\min(d_{1},d_{2})$ (we observe this in $24$ of $308$ random conjoins), so all distances quoted here are verified by computer.
In order to execute a single trace, one must first identify the columns (qubits) of the PCM which must be tensor contracted (traced) over.
Let us assume for simplicity that we wish to conjoin the first qubit of each code; we can represent these PCMs in the following way if each code is error detecting:
\begin{align}
\mathsf{H}_{1} = \left[
\begin{array}{cc|cc}
1 & \alpha_{1}^{\intercal} & 0 & \zeta_{1}^{\intercal}  \\
0 & \beta_{1}^{\intercal} & 1 & \eta_{1}^{\intercal}  \\
\mathbf{0} & \mathbf{A}_{1} & \mathbf{0} & \mathbf{B}_{1} \\
\end{array}
\right]
~,~\quad
\mathsf{H}_{2} = \left[
\begin{array}{cc|cc}
1 & \alpha_{2}^{\intercal} & 0 & \zeta_{2}^{\intercal}  \\
0 & \beta_{2}^{\intercal} & 1 & \eta_{2}^{\intercal}  \\
\mathbf{0} & \mathbf{A}_{2} & \mathbf{0} & \mathbf{B}_{2} \\
\end{array}
\right]~,
\end{align}
where the vertical line in each PCM separates the $X$ and $Z$ sectors.
$\alpha_{1,2}^{\intercal}, \beta_{1,2}^{\intercal}, \zeta_{1,2}^{\intercal}, \eta_{1,2}^{\intercal}$ represent additional support for $X$ and $Z$ stabilizers on a given row; we depict this extra physical support accordingly, since they do not participate in the contraction.
$\mathbf{0}, \mathbf{A}_{1,2}, \mathbf{B}_{1,2}$ signify the zero matrix, as well as extraneous matrices of $X$ and $Z$ stabilizer content that do not participate in the conjoining, respectively.

The resultant PCM for a tensor product of $\mathsf{Q}_{1}$ and $\mathsf{Q}_{2}$ is given as the sum of PCMs:
\begin{align}
\mathsf{H}_{\oplus} = \mathsf{H}_{1} \oplus \mathsf{H}_{2} =
\left[
\begin{array}{G c G c | B c B c}
1 & \alpha_{1}^{\intercal} & 0 & 0 & 0 & \zeta_{1}^{\intercal} & 0 & 0 \\
0 & \beta_{1}^{\intercal} & 0 & 0 & 1 & \eta_{1}^{\intercal} & 0 & 0 \\
\mathbf{0} & \mathbf{A}_{1} & \mathbf{0} & \mathbf{0} & \mathbf{0} & \mathbf{B}_{1} & \mathbf{0} & \mathbf{0} \\
0 & 0 & 1 & \alpha_{2}^{\intercal} & 0 & 0 & 0 & \zeta_{2}^{\intercal} \\
0 & 0 & 0 & \beta_{2}^{\intercal} & 0 & 0 & 1 & \eta_{2}^{\intercal} \\
\mathbf{0} & \mathbf{0} & \mathbf{0} & \mathbf{A}_{2} & \mathbf{0} & \mathbf{0} & \mathbf{0} & \mathbf{B}_{2} \\
\end{array}
\right]
~.
\label{eq:TNcontraction}
\end{align}
Therefore, performing the conjoining of the first qubits of both $\mathsf{Q}_{1}$ and $\mathsf{Q}_{2}$ corresponds to contracting the light green and light blue columns in \cref{eq:TNcontraction}.
Continuing, the conjoined PCM takes the form
\begin{align}
\mathsf{H}' = \mathsf{H}_{1} \wedge_{\mathcal{E}} \mathsf{H}_{2} = \left[
\begin{array}{cc|cc}
\alpha_{1}^{\intercal} & \alpha_{2}^{\intercal} & \zeta_{1}^{\intercal} & \zeta_{2}^{\intercal}  \\
\beta_{1}^{\intercal} & \beta_{2}^{\intercal} & \eta_{1}^{\intercal} & \eta_{2}^{\intercal} \\
\mathbf{A}_{1} & \mathbf{0} & \mathbf{B}_{1} & \mathbf{0} \\
\mathbf{0} & \mathbf{A}_{2} & \mathbf{0} & \mathbf{B}_{2} \\
\end{array}
\right]~,
\end{align}
where the uncolored row entries from the two blocks are concatenated if their respective entries in the green and blue columns are matching.
More generally, neither $H_{1}$ nor $H_{2}$ need to be error detecting and a similar matching procedure applies.

%% file: sections/B_protocol_proofs.tex
\section{Proofs Regarding the Generative CSS-TP Protocol}
\label{section:generative_protocol_proofs}

This appendix establishes the conditions under which the operations in \cref{con:protocol} preserve generalized triorthogonality, $\Delta$-divisibility, and a transversal logical phase action.
Two distinctions will be important throughout.

First, quantum code shortening is not arbitrary puncturing.
Before a physical coordinate is removed, binary symplectic row operations clear that coordinate from every stabilizer generator that survives the operation.
The subsequent deletion therefore does not alter the weights or overlap data of the surviving generators.

Second, a stopper always maps a CSS stabilizer code to another CSS stabilizer code, but preservation of a particular transversal logical gate is an additional condition.
That condition is given exactly by \cref{lem:operations}, which we prove in \cref{section:operation_rules_proof}.

\subsection{Quantum code shortening}

We begin with quantum code shortening.

\begin{lemma}[QCS preservation under stabilizer clearing]
\label[lemma]{lem:QCS}
Let $\mathsf{Q}$ be a GTO code represented by a generalized triorthogonal matrix
\begin{align}
\mathsf{G}
=
\begin{pmatrix}
\mathsf{G}_{0}\\
\mathsf{G}_{1}
\end{pmatrix}~,
\end{align}
where $\mathsf{G}_{0}$ contains the $X$-stabilizer rows and $\mathsf{G}_{1}=(\mathsf G_{T};\mathsf G_{CS};\mathsf G_{CCZ})$ the logical-$X$ representatives (\cref{def:gto}).
Choose an $X$-stabilizer row $p\in\mathcal{G}_{0}$ and a coordinate $i$ such that $p_{i}=1$.
For every other row $r$ with $r_{i}=1$, perform the binary row operation
\begin{align}
r\longmapsto \widetilde r=r+p~.
\label{eq:qcs_row_clearing}
\end{align}
Remove the stabilizer generated by $p$, delete coordinate $i$, and use the punctured row $p\setminus i$ as an additional logical-$X$ representative.
Then the resulting matrix is generalized triorthogonal, with $p\setminus i$ as an additional row of $\mathsf G_{T}$.

The analogous statement holds for a $\Delta$-divisible $X$-stabilizer space: after clearing coordinate $i$, deleting that coordinate preserves the divisibility of every surviving $X$ stabilizer.
\end{lemma}

\begin{proof}
The essential point is that QCS first clears the coordinate that will be deleted.
For any row $r\neq p$, define
\begin{align}
\widetilde r
=
r+r_{i} p~.
\label{eq:qcs_cleared_row}
\end{align}
Since $p_{i}=1$, every cleared row satisfies
\begin{align}
\widetilde r_{i}
=
r_{i}+r_{i} p_{i}
=
0
\pmod 2~.
\end{align}
Thus coordinate $i$ is zero on every row that survives as an $X$-stabilizer or pre-existing logical-$X$ representative.

We first verify that the clearing operations preserve generalized triorthogonality.
Since $p$ is a row of $\mathsf G_{0}$, its weight and all its pair and triple overlaps are even by \cref{def:gto}, so
\begin{align}
p\cdot \widetilde r
&=
p\cdot r+r_{i} |p|
=
0
\pmod 2~.
\end{align}
For any two cleared rows,
\begin{align}
\widetilde r\cdot\widetilde s
&=
(r+r_{i} p)\cdot(s+s_{i} p)
\nonumber\\
&=
r\cdot s
+r_{i}(p\cdot s)
+s_{i}(r\cdot p)
+r_{i} s_{i}|p|
\nonumber\\
&\equiv r\cdot s\pmod 2~.
\end{align}
Similarly, for three cleared rows,
\begin{align}
\sum_{j}
\widetilde r_{j}\widetilde s_{j}\widetilde t_{j}
&=
\sum_{j}
(r_{j}+r_{i} p_{j})
(s_{j}+s_{i} p_{j})
(t_{j}+t_{i} p_{j})~.
\end{align}
Expanding this expression produces only original triple overlaps, pairwise overlaps involving $p$, and the weight of $p$.
Every term other than the original triple overlap involves $p$ and is therefore even, so the triple overlap of the cleared rows equals that of the original rows modulo two; the same holds for weights and pair overlaps. Hence the clearing preserves every parity in \cref{eq:gto}.

After clearing, coordinate $i$ is zero on every surviving row.
Deleting that coordinate therefore changes neither the weight of a surviving row nor any pairwise or triple overlap among surviving rows.
The deleted pivot row $p$ has even weight and $p_{i}=1$, so its punctured representative $p\setminus i$ has odd weight.
Moreover, because all surviving rows have zero in coordinate $i$, the pairwise and triple overlaps involving $p\setminus i$ are identical to those involving $p$ before puncturing.
Consequently, $p\setminus i$ has odd weight and even pair and triple overlaps with every other row, which are the conditions on a row of $\mathsf G_{T}$; the resulting matrix is generalized triorthogonal with $k_{T}$ increased by one.

For a $\Delta$-divisible code, let $\mathcal{C}_{X}$ denote the complete $X$-stabilizer space.
Divisibility is a property of this linear space, not of one particular generating matrix.
Binary row operations leave $\mathcal{C}_{X}$ unchanged.
After the pivot is removed, the surviving space is the subspace
\begin{align}
\mathcal{C}_{X}^{(i=0)}
=
\{x\in\mathcal{C}_{X}:x_{i}=0\}~.
\end{align}
Every vector in this subspace was already $\Delta$-divisible before QCS, and deleting coordinate $i$ does not alter its weight because $x_{i}=0$.
Thus every surviving $X$ stabilizer remains $\Delta$-divisible.
The punctured pivot becomes logical and is not required to remain a $\Delta$-divisible stabilizer.
\end{proof}

\begin{remark}
The clearing step in \cref{lem:QCS} is essential.
Arbitrary deletion of a column can change an even pairwise or triple overlap into an odd one, and can change the weight of a divisible stabilizer by one.
The preservation claim therefore applies to QCS as defined in \cref{section:background_code_modification}, not to arbitrary puncturing.
\end{remark}

\subsection{Conjoining and the Matching Lemma}

We next record the condition under which conjoining preserves a transversal phase action.
This condition is the Matching Lemma of quantum LEGO~\cite{cao2022quantum,cao2024quantum}, in the form of Ref.~\cite[Lemma~3.2]{cao2026quantum}, applied to the fusion of two physical legs; we include the short proof for completeness.

\begin{lemma}[Matching Lemma~\cite{cao2022quantum,cao2026quantum}]
\label[lemma]{lem:gate_compatible_conjoining}
Let $E_{1}$ and $E_{2}$ be encoding tensors for two stabilizer codes, and suppose
\begin{align}
V_{a}E_{a}
=
E_{a}\overline{\mathcal{M}}_{a}~,
\qquad
a\in\{1,2\}~,
\label{eq:seed_intertwining}
\end{align}
for product layers $V_{1}=V_{1,i}\otimes V_{1,\mathrm{ext}}$ and $V_{2}=V_{2,j}\otimes V_{2,\mathrm{ext}}$ of single-leg unitaries.
Fuse physical legs $i$ and $j$ with $\bra{\Phi^{+}}_{ij}$, so that $E'=\bra{\Phi^{+}}_{ij}(E_{1}\otimes E_{2})$ encodes the conjoined code.
If $V_{2,j}=V_{1,i}^{*}$ up to a phase, then $V_{1,\mathrm{ext}}\otimes V_{2,\mathrm{ext}}$ implements $\overline{\mathcal{M}}_{1}\otimes\overline{\mathcal{M}}_{2}$ on the conjoined code, up to a global phase.
\end{lemma}

\begin{proof}
For every unitary $U$, $(U\otimes U^{*})\ket{\Phi^{+}}=(UU^{\dagger}\otimes\mathbb I)\ket{\Phi^{+}}=\ket{\Phi^{+}}$, and hence
\begin{align}
\bra{\Phi^{+}}_{ij}
\left(V_{1,i}\otimes V_{2,j}\right)
=
e^{i\theta}
\bra{\Phi^{+}}_{ij}
\label{eq:bell_gate_matching}
\end{align}
for some phase $e^{i\theta}$.
The edge factors can therefore be pushed through the fusion, and with \cref{eq:seed_intertwining},
\begin{align}
(V_{1,\mathrm{ext}}\otimes V_{2,\mathrm{ext}})\,E'
&=e^{-i\theta}\bra{\Phi^{+}}_{ij}(V_{1}\otimes V_{2})(E_{1}\otimes E_{2})\nonumber\\
&=e^{-i\theta}E'\left(\overline{\mathcal{M}}_{1}\otimes\overline{\mathcal{M}}_{2}\right)~.
\end{align}
The same argument applies to a self-trace, and edge by edge to larger networks~\cite{cao2026quantum}.
\end{proof}

\begin{remark}
Conjoining always produces a well-defined tensor-network code when the contraction is nonzero.
The Matching Lemma is sufficient for a layer to survive, but not necessary.
For diagonal layers the matching condition $V_{2,j}=V_{1,i}^{*}$ is $a_{i}+a_{j}\equiv0$, and \cref{lem:operations}(a) shows that it is necessary whenever a surviving $X$ stabilizer covers the fused legs; when none does, an unmatched fusion also keeps the layer logical and adds $-(a_{i}+a_{j})\,y_{i}$ to the logical phase polynomial.
In addition, the induced logical action must remain non-Clifford; it is possible in principle for non-Clifford factors to cancel after contraction.
\end{remark}

\begin{lemma}[Bell fusion and code structure]\label[lemma]{lem:fusion_structure}
Let $\mathsf C$ be a CSS code (for conjoining, the disjoint union of the two codes)
with $X$-stabilizer space $\mathcal S$. Fuse legs $i$ and $j$ with $\bra{\Phi^{+}}_{ij}$,
let $K$ be the kept legs and $\pi_K$ the restriction to them.
\begin{enumerate}
\item The fused code has $X$-stabilizer space
$\mathcal S'=\pi_K\{x\in\mathcal S:x_i=x_j\}$, and for $X$-type vectors $x,y,z$ with
$x_i=x_j$, $y_i=y_j$ and $z_i=z_j$,
\begin{align}
|\pi_Kx|&=|x|-2x_i~,\quad
|\pi_Kx\wedge\pi_Ky|=|x\wedge y|-2x_iy_i~,\nonumber\\
|\pi_Kx\wedge\pi_Ky\wedge\pi_Kz|&=|x\wedge y\wedge z|-2x_iy_iz_i~.
\end{align}
\item If $\mathcal S$ is $\Delta$-divisible, so is $\mathcal S'$ when $\Delta\mid2$.
For $\Delta\nmid2$, $\mathcal S'$ is $\Delta$-divisible iff no $x\in\mathcal S$ has
$x_i=x_j=1$, i.e.\ iff column $i$ or column $j$ of an $X$-stabilizer generator matrix
is zero.
\item If $\mathsf C$ is GTO (\cref{def:gto}) and the fusion keeps all logical qubits
(case~1 of \cref{cor:conjoin_loop}), the fused code is GTO with the same $k_T$,
$k_{CS}$ and $k_{CCZ}$.
\end{enumerate}
\end{lemma}
\begin{proof}
Item~1 for $\mathcal S'$ is shown in the proof of \cref{lem:operations}; the
displayed identities hold because $x_i=x_j$ and both legs are deleted.
For item~2, a surviving $x$ with $x_i=x_j=1$ has $|\pi_Kx|\equiv-2\not\equiv0
\pmod\Delta$, and every other surviving $x$ keeps its weight. With $v_i,v_j$ the two
columns, $x_i$ vanishes on $\ker(x_i+x_j)$ iff $v_i\in\operatorname{span}\{v_i+v_j\}$,
i.e.\ iff $v_i=0$ or $v_j=0$.
For item~3, in case~1 every logical label survives, so each row of $\mathsf G_{1}$ can
be replaced by its sum with a stabilizer that satisfies $x_i=x_j$. Adding a row of
$\mathsf G_{0}$ changes no parity in \cref{eq:gto}, since every overlap involving it
is even, and by item~1 the restriction to $K$ changes no parity either.
\end{proof}
Two $\code{8,3,2}$ cubes illustrate item~2: their surviving $X$ stabilizer $1^{16}$
becomes $1^{14}$, so the $\code{14,6,2}$ code is GTO with $k_{CCZ}=2$ but only
$2$-divisible, while it keeps the logical action $\overline{CCZ}\otimes\overline{CCZ}$
of \cref{eq:1462_action}. A self-trace in case~2 of \cref{cor:conjoin_loop} restricts
the logical labels to a hyperplane and need not preserve GTO: the closed chains of
\cref{section:closed_chains} are irreducible and therefore not GTO.

\subsection{Proof of the operation rules}\label{section:operation_rules_proof}

\begin{proof}[Proof of \cref{lem:operations}]
Let $K$ be the set of kept legs, and write $E\ket u\propto\sum_{t}\ket{uL_{X}+tC_{X}}$ for the encoding of $\mathsf C$.
On computational basis states, the contraction $\bra{\Phi^{+}}_{ij}$ maps $\ket y$ to $\ket{y|_{K}}$ if $y_{i}=y_{j}$ and to zero otherwise; $\bra0_{i}$ maps $\ket y$ to $\ket{y|_{K}}$ if $y_{i}=0$ and to zero otherwise; $\bra+_{i}$ maps $\ket y$ to $\ket{y|_{K}}/\sqrt2$; and QCS deletes coordinate $i$ after the clearing of \cref{lem:QCS}, with the punctured pivot promoted to a logical row.
In every case the operation keeps exactly the codewords that satisfy the constraint of \cref{tab:operations}.
Let $\mathcal S$ be the group generated by the surviving $X$ stabilizers; since the constraints are linear and homogeneous, $y+c$ satisfies the constraint whenever $y$ does and $c\in\mathcal S$.
The encoded states of $\mathsf C'$ are therefore proportional to $\sum_{y}\ket{y|_{K}}$, where $y$ runs over the codewords in one coset of $\mathcal S$ that satisfy the constraint (for QCS, the coset is labeled by $(u,u_{k+1})$).
The restriction map is linear, so two such cosets restrict to identical or disjoint sets; the encoded states are equal or orthogonal, and the $X$ stabilizers of $\mathsf C'$ are the restrictions of $\mathcal S$.

On a codeword $y$ that satisfies the constraint, the removed legs carry the phase $\alpha\,y_{i}$: for a contraction, $a_{i}y_{i}+a_{j}y_{j}=(a_{i}+a_{j})y_{i}$ because $y_{i}=y_{j}$.
Since $V$ is logical on $\mathsf C$, the phase on all legs is $\phi(u(y))$, so $V'$ multiplies $\ket{y|_{K}}$ by $\omega^{\phi(u(y))-\alpha y_{i}}$.
A diagonal operator maps the uniform superposition over a coset into the code space iff it multiplies all terms by the same phase.
The label $u(y)$ is constant on each coset, so $V'$ is logical iff $\alpha(y+c)_{i}\equiv\alpha y_{i}$ for every such $y$ and every $c\in\mathcal S$.
Since $(y+c)_{i}-y_{i}=(1-2y_{i})c_{i}$ and $1-2y_{i}=\pm1$, this holds iff $\alpha c_{i}\equiv0$ for every $c\in\mathcal S$, i.e.\ iff $\alpha\equiv0$ or no surviving $X$ stabilizer acts on $i$.
If no surviving $X$ stabilizer acts on $i$, the linear function $y\mapsto y_{i}$ is constant on each coset of $\mathcal S$ and is therefore a parity of the logical bits of $\mathsf C'$; the common phase is \cref{eq:op_phase}.
If $\alpha\equiv0$, the common phase is $\phi$ on the labels that satisfy the constraint, which is \cref{eq:op_phase} as well.

It remains to identify $y_{i}$ for each operation.
For QCS, after clearing, the surviving stabilizers and the old logical rows vanish at $i$ and the pivot equals $1$ there, so the condition always holds and $y_{i}=u_{k+1}$; this is item~(b).
For a $Z$ stopper, the surviving stabilizers vanish at $i$ and $y_{i}=0$ on every kept codeword, so the layer is always logical and $\phi'$ is the restriction of $\phi$.
For an $X$ stopper there is no constraint and every $X$ stabilizer survives; if none acts on $i$, then $y_{i}=(uL_{X})_{i}=u\cdot g_{i}$.
Together with the contraction case, these give items~(a) and~(c).
\end{proof}

\begin{proof}[Proof of \cref{cor:conjoin_loop}]
For $y=uL_{X}+tC_{X}$ the constraint $y_{i}=y_{j}$ reads $u\cdot(g_{i}+g_{j})+t\cdot\gamma=0$, where $\gamma$ collects, for each $X$-stabilizer generator, the sum of its entries at $i$ and $j$.
If $\gamma\ne0$, some $t$ satisfies the constraint for every $u$, so every logical label survives; if $\gamma=0$, the constraint is $u\cdot(g_{i}+g_{j})=0$.
On codewords that satisfy the constraint, the kernel of the restriction map is $\{0,e_{i}+e_{j}\}$ intersected with the $X$-type codewords, which is $\{0\}$ by assumption.
Therefore distinct surviving labels give orthogonal encoded states, and $k'=k$ in case~1 and $k'=k-1$ in case~2.
With $a_{i}+a_{j}\equiv0$, \cref{lem:operations} gives $\phi'=\phi$ on the surviving labels.
For two codes, an $X$ stabilizer of the first code that acts on $i$ vanishes at $j$, so $\gamma\ne0$.
\end{proof}

\begin{proof}[Proof of \cref{cor:level}]
For $\psi:\mathbb F_{2}^{k}\to\mathbb Z_{2^{\ell}}$ let $U_{\psi}=\sum_{u}\omega^{\psi(u)}\ket u\!\bra u$, and let $\lambda(\psi)$ be the right-hand side of \cref{eq:level} computed from the multilinear coefficients $c_{S}$, $S\ne\emptyset$, of $\psi$, with $\lambda(\psi)=1$ if all of them vanish.
For a single term, $\lambda(c\,u^{S})=\ell-v_{2}(c)+|S|-1$ with $u^{S}=\prod_{r\in S}u_{r}$.
Since $v_{2}(c+c')\ge\min(v_{2}(c),v_{2}(c'))$, $\lambda$ of a sum is at most the larger of the two values, and $\lambda(-\psi)=\lambda(\psi)$.
We show $\mathrm{lev}(U_{\psi})=\lambda(\psi)$ by induction on $\lambda(\psi)$.

(a) $U_{\psi}$ is a Pauli operator up to a phase iff $\psi(u)-\psi(0)\equiv2^{\ell-1}(z\cdot u\bmod2)$ for some $z$, i.e.\ iff the only nonzero coefficients are $c_{\{r\}}=2^{\ell-1}$.
This is the condition $\lambda(\psi)=1$: a term with $|S|=1$ has $\lambda\le1$ iff $c=2^{\ell-1}$, and a term with $|S|\ge2$ has $\lambda\ge2$.

(b) For a Pauli operator $P=X^{w}Z^{z}$, $U_{\psi} PU_{\psi}^{\dagger}=U_{-D_{w}\psi}P$ with $D_{w}\psi(u)=\psi(u+w)-\psi(u)$.
Multiplying by a Pauli operator or a phase preserves every level, so if $U_{\psi}$ is not a Pauli operator, $\mathrm{lev}(U_{\psi})=1+\max_{w\ne0}\mathrm{lev}(U_{-D_{w}\psi})$.

(c) We show $\max_{w\ne0}\lambda(D_{w}\psi)=\lambda(\psi)-1$ when $\lambda(\psi)\ge2$.
Over the integers $u_{r}+1\equiv1-u_{r}$, so the shift $u\mapsto u+w$ turns $c\,u^{S}$ into $\pm c$ times monomials $u^{T}$, $T\subseteq S$, and does not increase $\lambda$.
For a unit vector $e_{a}$, $D_{e_{a}}u^{S}=u^{S\setminus a}-2u^{S}$ if $a\in S$ and $0$ otherwise; both terms have $\lambda\le\lambda(c\,u^{S})-1$.
Since $D_{w+e_{a}}\psi(u)=D_{e_{a}}\psi(u+w)+D_{w}\psi(u)$, induction on the weight of $w$ gives $\lambda(D_{w}\psi)\le\lambda(\psi)-1$ for every $w\ne0$.
For equality, let $S^{*}$ attain $\lambda(\psi)$ and $a\in S^{*}$.
If $|S^{*}|\ge2$, the monomial $u^{S^{*}\setminus a}$ arises in $D_{e_{a}}\psi$ only from $S^{*}$, with coefficient $c_{S^{*}}$ and $\lambda=\lambda(\psi)-1$.
If $S^{*}=\{a\}$, the monomial $u_{a}$ has coefficient $-2c_{S^{*}}$, which is nonzero because $\lambda(\psi)\ge2$ forces $v_{2}(c_{S^{*}})\le\ell-2$, and again $\lambda=\lambda(\psi)-1$.

If $\lambda(\psi)=1$, then $\mathrm{lev}(U_{\psi})=1$ by (a).
Otherwise $U_{\psi}$ is not a Pauli operator, and (b), (c) and the induction hypothesis give $\mathrm{lev}(U_{\psi})=1+\max_{w\ne0}\lambda(D_{w}\psi)=\lambda(\psi)$.
Applied to $\psi=\phi$ this is \cref{eq:level}.
Finally, $c_{S}=2^{|S|-1}\cdot(\pm\sum_{j}a_{j}\prod_{r\in S}G_{rj})$ by \cref{eq:cS}, so $v_{2}(c_{S})\ge|S|-1$ and $\mathrm{lev}\le\ell$, with equality iff $v_{2}(c_{S})=|S|-1$ for some $S$.
\end{proof}

\subsection{Stoppers at distance-one legs}

A stopper consists of postselecting a physical qubit in a Pauli eigenstate and then removing that qubit.
It is useful to distinguish the projector from the corresponding map to the reduced Hilbert space.
Define
\begin{align}
R_{P,i}
=
\bra{P{+}}_{i}~,
\qquad
P\in\{X,Z\}~,
\label{eq:pauli_trace_map}
\end{align}
where $\ket{X{+}}=\ket{+}$ and $\ket{Z{+}}=\ket{0}$.
Up to normalization, $R_{P,i}$ implements $\Pi^{(i)}_{P}$ followed by deletion of qubit $i$.

\begin{proposition}[CSS structure under a stopper]
\label[proposition]{prop:trace_css}
Let $\mathsf{Q}$ be a CSS code with stabilizer group
\begin{align}
\mathcal{S}
=
\langle\mathcal{S}_{X},\mathcal{S}_{Z}\rangle~.
\end{align}
If $R_{P,i}E_{\mathsf{Q}}\neq0$, then a stopper $R_{P,i}$ produces another CSS stabilizer code.
Its stabilizer group is obtained by the standard Pauli-measurement update:
\begin{enumerate}
    \item generators commuting with $P_{i}$ are retained;
    \item if generators anticommute with $P_{i}$, one is chosen as a pivot and multiplied into the others to clear their support on qubit $i$;
    \item the measured Pauli $P_{i}$ is fixed to eigenvalue $+1$; and
    \item qubit $i$ and the resulting single-qubit stabilizer are removed.
\end{enumerate}
The resulting number of logical qubits is determined by the rank of the updated stabilizer group and is not fixed solely by whether the stopper is an $X$ or a $Z$ stopper.
\end{proposition}

\begin{proof}
Pauli measurement maps a stabilizer code to a stabilizer code by the standard generator-update rule.
If $P_{i}$ commutes with every stabilizer and is independent of the stabilizer group, it is added as a new generator.
If $P_{i}$ anticommutes with one or more generators, one anticommuting generator is replaced by $P_{i}$, and that generator is used to clear the anticommutation of all remaining generators.

Because $P_{i}$ is either purely $X$-type or purely $Z$-type, the updated generators can be chosen to remain purely $X$-type or purely $Z$-type.
The resulting code is therefore CSS.
After postselection, the measured qubit is stabilized by $P_{i}$ and factors from the remaining code, so it may be removed.

Depending on the support and rank of the original stabilizer generators, the measurement may add an independent constraint, replace an existing constraint, or remove a redundant physical degree of freedom.
Hence the change in $k$ must be computed from the updated stabilizer rank rather than inferred from the measurement basis alone.
\end{proof}

Distance-one legs can be identified directly from missing stabilizer support.

\begin{lemma}[Missing support and weight-one logical operators]
\label[lemma]{lem:missing_support_weight_one}
Let $\mathsf{Q}$ be a CSS code.

\begin{enumerate}
    \item If no $X$ stabilizer has support on qubit $i$, then $Z_{i}$ commutes with the complete stabilizer group.
    Unless $Z_{i}$ is itself a stabilizer, it is a weight-one logical $Z$ operator.

    \item If no $Z$ stabilizer has support on qubit $i$, then $X_{i}$ commutes with the complete stabilizer group.
    Unless $X_{i}$ is itself a stabilizer, it is a weight-one logical $X$ operator.
\end{enumerate}
\end{lemma}

\begin{proof}
The operator $Z_{i}$ commutes with every $Z$-type stabilizer.
It also commutes with every $X$-type stabilizer precisely when every $X$ stabilizer has zero support at coordinate $i$.
The first statement follows.
The second follows by exchanging $X$ and $Z$.
\end{proof}

Thus, if missing $X$ support produces a weight-one logical $Z_{i}$, an $X$ stopper is a natural way to remove that distance-one logical direction.
Likewise, a missing-$Z$ leg can be treated using a $Z$ stopper.
This observation explains the stopper choices used in the $\code{8,3,2}$ derivation.
The observation does not, by itself, prove preservation of a particular transversal logical gate.

\begin{definition}[Structure-admissible stopper]
\label[definition]{def:structure_admissible_trace}
A stopper is called \emph{structure-admissible} for a GTO or $\Delta$D code if, after the stabilizer-measurement update and removal of the measured qubit, the surviving $X$-stabilizer and logical-$X$ representatives satisfy the corresponding generalized triorthogonality or divisibility conditions.
\end{definition}

A simple sufficient condition for preserving the $X$-stabilizer divisibility data is that every surviving $X$ stabilizer has zero support on the removed coordinate before deletion.
This condition ensures that deleting the coordinate changes neither the weight of any surviving $X$ stabilizer nor its overlap data.
Generalized triorthogonality also requires compatible logical-$X$ representatives; these may be obtained by the same type of clearing procedure used in \cref{lem:QCS}, when such a pivot exists.

\begin{remark}[Stoppers and the logical action]\label[remark]{rem:trace_rules}
For a diagonal layer, \cref{lem:operations} decides every stopper exactly.
A $Z$ stopper always keeps the layer logical and restricts the logical phase polynomial to $y_{i}=0$; this may remove the logical qubits that carry the non-Clifford part, which the Clifford test detects.
An $X$ stopper keeps the layer logical iff $a_{i}\equiv0$ or no $X$ stabilizer acts on $i$.
At the distance-one legs of \cref{lem:missing_support_weight_one} with missing $X$ support, the second condition holds by construction, so such stoppers always keep the layer; they change the logical phase polynomial by $-a_{i}\,(u\cdot g_{i})$ and can turn single-qubit logical gates into entangling ones, as in the $\code{8,3,2}$ derivation.
Stoppers elsewhere follow the same rule, so no separate verification of the logical action is needed.
\end{remark}

\subsection{Guarantee of the generative protocol}

We now combine the preceding results into a proof of \cref{thm:protocol}.

\begin{proof}[Proof of \cref{thm:protocol}]
Induct on the number of steps.
If the layer is logical before a step, \cref{lem:operations} states that it is logical after the step iff the step meets the stated condition, and gives the new logical phase polynomial in terms of the old one.
Composing these updates gives item~2; the level follows from \cref{cor:level}.
Item~3 follows from \cref{lem:QCS} for QCS steps and from \cref{def:structure_admissible_trace} for stoppers.
Every step produces a CSS code (\cref{prop:trace_css}).
The number of logical qubits and the distance are not determined by these statements; we compute both for every output.
\end{proof}

\subsection{Reversibility of the construction operations}

We next consider whether the construction can be reversed: can one begin with a CSS code that is not GTO or $\Delta$D and obtain a code in one of those stricter classes?

\begin{remark}[No general converse guarantee]
\label[remark]{prop:nonreverse}
QCS, conjoining, and stoppers do not provide a general converse theorem from arbitrary CSS or CSS-T codes to GTO or $\Delta$D codes.
These operations may remove the rows or coordinates responsible for a failed orthogonality or divisibility condition, so special instances can enter a stricter code class.
Such behavior, however, is instance-dependent and must be verified after the operation.
\end{remark}

For QCS, the clearing operations preserve the relevant shortened row spaces, but they do not force an initially non-triorthogonal space to become triorthogonal.
Nevertheless, shortening can remove a coordinate or a pivot row responsible for a failed overlap condition.
It is therefore possible, but not guaranteed, that a particular QCS operation maps a non-GTO code to a GTO code.

Likewise, conjoining combines operator spaces through tensor contraction.
Without a compatible seed and an operator-matching relation, it does not generally impose triorthogonality or divisibility.
A special contraction may nevertheless remove offending degrees of freedom or inherit a useful transversal action from one of its inputs.
For example, conjoining a non-GTO CSS-T code with a GTO block can preserve part of the GTO block's logical non-Clifford action when the relevant operators match across the contracted edge.

Finally, a stopper restricts and punctures the constituent stabilizer spaces.
Such a restriction may accidentally eliminate a failed overlap or nondivisible vector and thereby produce a code in a stricter class.
Consequently, the correct statement is not that these operations can never produce GTO or $\Delta$D codes from more general CSS codes, but that no such upgrade is guaranteed without an explicit algebraic check.

\subsection{Relationship to classical puncturing and shortening}

The stabilizer updates above can be expressed precisely using classical puncturing and shortening.
To avoid ambiguity in the conventions for $C_{1}$ and $C_{2}$, we formulate the correspondence directly in terms of the $X$- and $Z$-stabilizer spaces.

Let
\begin{align}
\mathsf{Q}
=
\mathrm{CSS}(X:C_{X},Z:C_{Z})~,
\qquad
C_{X}\subseteq C_{Z}^{\perp}~,
\label{eq:css_cx_cz_convention}
\end{align}
where $C_{X},C_{Z}\subseteq\mathbb{F}_{2}^{n}$ are the row spaces of the $X$- and $Z$-stabilizer matrices.

For a classical linear code $C\subseteq\mathbb{F}_{2}^{n}$, define
\begin{align}
\mathrm{Punc}_{i}(C)
&=
\{c\setminus i:c\in C\}~,
\\
\mathrm{Short}_{i}(C)
&=
\{c\setminus i:c\in C,\ c_{i}=0\}~.
\end{align}
Puncturing deletes coordinate $i$ from every codeword.
Shortening first restricts to the subcode with zero in coordinate $i$ and then deletes that coordinate.
They satisfy the duality relations
\begin{align}
\mathrm{Punc}_{i}(C)^{\perp}
&=
\mathrm{Short}_{i}(C^{\perp})~,
\\
\mathrm{Short}_{i}(C)^{\perp}
&=
\mathrm{Punc}_{i}(C^{\perp})~.
\label{eq:puncture_shorten_duality}
\end{align}

\begin{proposition}[Quantum-to-classical correspondence]
\label[proposition]{prop:q_to_c}
Let $\mathsf{Q}=\mathrm{CSS}(X:C_{X},Z:C_{Z})$, and let $i$ be a physical coordinate.

\begin{enumerate}
    \item If QCS is performed using both an $X$-type and a $Z$-type pivot at coordinate $i$, then
    \begin{align}
    \mathrm{QCS}_{i}(\mathsf{Q})
    =
    \mathrm{CSS}\left(
    X:\mathrm{Short}_{i}(C_{X}),
    Z:\mathrm{Short}_{i}(C_{Z})
    \right)~.
    \label{eq:qcs_classical_correspondence}
    \end{align}

    \item A $Z$ stopper, corresponding to postselection onto $\ket{0}$, gives
    \begin{align}
    R_{Z,i}(\mathsf{Q})
    =
    \mathrm{CSS}\left(
    X:\mathrm{Short}_{i}(C_{X}),
    Z:\mathrm{Punc}_{i}(C_{Z})
    \right)~.
    \label{eq:z_trace_classical_correspondence}
    \end{align}

    \item An $X$ stopper, corresponding to postselection onto $\ket{+}$, gives
    \begin{align}
    R_{X,i}(\mathsf{Q})
    =
    \mathrm{CSS}\left(
    X:\mathrm{Punc}_{i}(C_{X}),
    Z:\mathrm{Short}_{i}(C_{Z})
    \right)~.
    \label{eq:x_trace_classical_correspondence}
    \end{align}
\end{enumerate}
In each expression, redundant rows are removed after the operation.
\end{proposition}

\begin{proof}
For QCS, choose an $X$-stabilizer pivot with a one in coordinate $i$ and use it to clear coordinate $i$ from every other $X$-stabilizer generator.
The surviving generators span exactly
\begin{align}
\{x\in C_{X}:x_{i}=0\}~.
\end{align}
Deleting coordinate $i$ therefore gives $\mathrm{Short}_{i}(C_{X})$.
Repeating the same binary symplectic clearing procedure with a $Z$-stabilizer pivot gives $\mathrm{Short}_{i}(C_{Z})$.
Both pivots are removed, so when the coordinate maps on $C_{X}$ and $C_{Z}$ are nonzero, the stabilizer rank decreases by two while the block length decreases by one.
Hence $k$ increases by one, as expected for QCS.

For a $Z$ stopper, every surviving $X$ stabilizer must commute with $Z_{i}$.
It must therefore have zero in coordinate $i$.
Restricting to such vectors and deleting the coordinate gives $\mathrm{Short}_{i}(C_{X})$.
The $Z$-stabilizers commute with the measurement and may be punctured directly, giving $\mathrm{Punc}_{i}(C_{Z})$.
The measured generator $Z_{i}$ factors with the removed qubit and does not appear in the reduced code.

The $X$-stopper case follows by exchanging $X$ and $Z$.
Existing $X$-stabilizers commute with $X_{i}$ and are punctured, whereas surviving $Z$-stabilizers must have zero in coordinate $i$ and are therefore shortened.
\end{proof}

The correspondence is summarized in \cref{table:classical_correspondence}.

\begin{table}
\centering
\caption{Classical-code form of the code-modification operations for $\mathrm{CSS}(X:C_{X},Z:C_{Z})$.
The formulas assume that dependent generators are removed after each operation.}
\label{table:classical_correspondence}
\begin{tabular}{@{}lcc@{}}
\toprule
Quantum operation
&
Action on $C_{X}$
&
Action on $C_{Z}$
\\
\midrule
QCS at $i$
&
$\mathrm{Short}_{i}$
&
$\mathrm{Short}_{i}$
\\
$Z$ stopper at $i$
&
$\mathrm{Short}_{i}$
&
$\mathrm{Punc}_{i}$
\\
$X$ stopper at $i$
&
$\mathrm{Punc}_{i}$
&
$\mathrm{Short}_{i}$
\\
\bottomrule
\end{tabular}
\end{table}

\begin{remark}[Distance-one elimination in classical form]
Suppose no $X$ stabilizer has support on coordinate $i$.
Then every vector in $C_{X}$ already has $x_{i}=0$, and therefore
\begin{align}
\mathrm{Short}_{i}(C_{X})
=
\mathrm{Punc}_{i}(C_{X})~.
\end{align}
The operator $Z_{i}$ is then in the normalizer, as described in \cref{lem:missing_support_weight_one}.
An $X$ stopper acts as
\begin{align}
C_{X}
&\longmapsto
\mathrm{Punc}_{i}(C_{X})~,
&
C_{Z}
&\longmapsto
\mathrm{Short}_{i}(C_{Z})~.
\end{align}
The dual statement applies when no $Z$ stabilizer has support and a $Z$ stopper is used.
These updates give an exact classical implementation of the weight-one logical elimination step, which is not, in general, equivalent to shortening both constituent stabilizer spaces.
\end{remark}

\begin{remark}[Logical-qubit count]
The change in $k$ under a stopper depends on the dimensions of the punctured and shortened spaces.
In particular, puncturing can reduce the dimension of a classical code if the code contains the unit vector $e_{i}$, while shortening reduces the dimension when the coordinate functional $c\mapsto c_{i}$ is nonzero on the code.
Therefore neither an $X$ stopper nor a $Z$ stopper has a universal effect on $k$.
The value of $k$ is recomputed after every operation from
\begin{align}
k'
=
(n-1)
-
\dim C_{X}'
-
\dim C_{Z}'~.
\end{align}
\end{remark}

\begin{remark}[Classical duality and $X/Z$ symmetry]
The exchange $X\leftrightarrow Z$ swaps $C_{X}$ and $C_{Z}$.
By \cref{eq:puncture_shorten_duality}, this also exchanges the roles of puncturing and shortening.
The $X$- and $Z$-stopper formulas in \cref{prop:q_to_c} therefore exhibit the expected Hadamard symmetry.
\end{remark}

\begin{remark}[Search through the puncture-and-shorten lattice]
The stochastic implementation of \cref{con:protocol} can be formulated entirely using the operations in \cref{prop:q_to_c}.
After every QCS or stopper, the code removes dependent stabilizer generators, recomputes the logical operators and distance, and verifies the induced transversal logical action.
Thus the search explores a puncture-and-shorten lattice of CSS stabilizer spaces, while the GTO, $\Delta$D, CSS-T, and CSS-TP conditions are properties checked at the resulting nodes rather than assumed for every unrestricted move.
\end{remark}

\subsection{Proofs for the resource tiers}\label{section:tier_proofs}

This subsection proves \cref{lem:tiers,prop:nullity} and explains how we establish the tier of each code.
Throughout, $f$ is a Boolean cubic polynomial in the logical bits $u\in\mathbb F_{2}^{k}$, and a \emph{linear function} is a sum $\lambda(u)=c_{1}u_{1}+\dots+c_{k}u_{k}$ of logical bits modulo two.

We first recall how a flip acts on $f$.
As in \cref{section:tiers}, flipping the logical bits marked by $w\in\mathbb F_{2}^{k}$ changes $f$ by $f(u+w)+f(u)$.
The cubic terms cancel in this change, so it has degree at most two, and $w$ is a Pauli direction when it has degree at most one.
For example, flipping $u_{1}$ in $u_{1}u_{2}u_{3}$ changes it by $u_{2}u_{3}$, and flipping a bit in a term of degree at most two changes that term by at most a linear function.
We write $\mathcal P_{f}\subseteq\mathbb F_{2}^{k}$ for the set of Pauli directions of $f$.

A change of logical basis acts on flips in a simple way.
Suppose $g(u)=f(Au+b)+h(u)$ with $A$ an invertible binary matrix, $b\in\mathbb F_{2}^{k}$, and $h$ of degree at most two.
Then
\begin{align}\label{eq:flip_relabel}
g(u+w)+g(u)=f\bigl(Au+b+Aw\bigr)+f\bigl(Au+b\bigr)+\bigl(h(u+w)+h(u)\bigr)~,
\end{align}
and the last bracket has degree at most one.
Flipping $w$ in $g$ is therefore the same as flipping $Aw$ in $f$, with the logical bits renamed by $u\mapsto Au+b$.
The renaming changes neither the degree nor the number of independent linear functions that a polynomial depends on.
In particular, $w$ is a Pauli direction of $g$ iff $Aw$ is a Pauli direction of $f$.

\begin{lemma}[Cubic support from Pauli directions]\label[lemma]{lem:ess_pauli}
The Pauli directions of $f$ form a subspace $\mathcal P_{f}$ of $\mathbb F_{2}^{k}$, and the smallest number of logical qubits that the cubic terms of $f$ act on, over all logical bases, is $k-\dim\mathcal P_{f}$.
\end{lemma}

\begin{proof}
If $w$ and $w'$ are Pauli directions, then $f(u+w+w')+f(u)=\bigl[f(u+w+w')+f(u+w)\bigr]+\bigl[f(u+w)+f(u)\bigr]$.
The first bracket is the change under flipping $w'$, evaluated at $u+w$, so both brackets have degree at most one, and $w+w'$ is a Pauli direction.

Flipping the single bit $u_{j}$ turns each cubic monomial $u_{j}u_{a}u_{b}$ into $u_{a}u_{b}$, plus terms of degree at most one, and turns every other term into terms of degree at most one.
Different cubic monomials containing $u_{j}$ give different quadratic monomials, which cannot cancel.
So the $j$-th unit vector $e_{j}$ is a Pauli direction iff no cubic monomial of $f$ contains $u_{j}$.

Now change the logical basis by $u=Av+b$.
The shift $b$ leaves the cubic terms unchanged.
By the previous paragraph and \cref{eq:flip_relabel}, the cubic terms of $f(Av+b)$ avoid $v_{j}$ iff the $j$-th column $Ae_{j}$ of $A$ is a Pauli direction of $f$.
The columns of $A$ are linearly independent, so at most $\dim\mathcal P_{f}$ of them lie in $\mathcal P_{f}$, and the cubic terms act on at least $k-\dim\mathcal P_{f}$ qubits in every basis.
Choosing $A$ whose last $\dim\mathcal P_{f}$ columns span $\mathcal P_{f}$ attains this number.
\end{proof}

With \cref{lem:ess_pauli} in hand, we turn to \cref{lem:tiers}(i), which states what the free operations do to $f$.
The proof of part (i) works for any phase gate, not only for Boolean ones, and Step 5 below needs this general form.
Let $\omega=e^{2\pi i/2^{\ell}}$ with $\ell\ge2$, so that $i=\omega^{2^{\ell-2}}$.
A product of $S$, $Z$ and $CZ$ gates multiplies $\ket u$ by $i^{q(u)}$ with
\begin{align}
q(u)=\sum_{j}a_{j}u_{j}+2\sum_{j<j'}b_{jj'}u_{j}u_{j'}
\end{align}
for integers $a_{j}$ (one for each $S$, two for each $Z$) and $b_{jj'}$ (one for each $CZ$).
For phase functions $\phi,\psi:\mathbb F_{2}^{k}\to\mathbb Z_{2^{\ell}}$, we show that $\omega^{\phi}\ket{+}^{\otimes k}$ and $\omega^{\psi}\ket{+}^{\otimes k}$ are equivalent under $\mathcal F$ iff
\begin{align}\label{eq:tiers_general}
\psi(u)\equiv\phi(Au+b)+2^{\ell-2}q(u)+c
\end{align}
for some invertible binary matrix $A$, some $b\in\mathbb F_{2}^{k}$, some constant $c$ and some $q$ of this form.

\begin{proof}[Proof of \cref{lem:tiers}(i)]
\emph{Step 1: every free operation has a standard form.}
Moving an $S$, $Z$ or $CZ$ gate past a $CX$ or $X$ gate turns it into another product of $S$, $Z$ and $CZ$ gates, up to a global phase.
Moving an $X$ gate past a $CX$ gate gives $X$ gates again.
Every element of $\mathcal F$ is therefore, up to a global phase, a $CX$ network, followed by $X$ gates on the bits marked by $b'$, followed by the diagonal gate $i^{q(u)}$.

\emph{Step 2: action on the state.}
The $CX$ network maps $\ket u$ to $\ket{Au}$ for an invertible $A$, and the $X$ gates map $\ket u$ to $\ket{u+b'}$.
Applied to $\sum_{u}\omega^{\phi(u)}\ket u$, the operation therefore gives $\sum_{u} i^{q(u)}\omega^{\phi(A^{-1}u+A^{-1}b')}\ket u$.
The resulting state equals $\omega^{\psi}\ket{+}^{\otimes k}$ up to a global phase iff $\psi$ has the form of \cref{eq:tiers_general}, with $A^{-1}$ and $A^{-1}b'$ in place of $A$ and $b$.

\emph{Step 3: the Boolean case.}
For $\phi=2^{\ell-1}f$ and $\psi=2^{\ell-1}g$, the term $2^{\ell-2}q(u)+c$ must be a multiple of $2^{\ell-1}$ for every $u$.
At $u=0$ this makes $c$ a multiple of $2^{\ell-1}$, and at the unit vectors it makes every $a_{j}$ even.
Then $2^{\ell-2}q(u)=2^{\ell-1}\bigl(\sum_{j}\tfrac{a_{j}}{2}u_{j}+\sum_{j<j'}b_{jj'}u_{j}u_{j'}\bigr)$, so $g(u)$ and $f(Au+b)$ differ by a Boolean polynomial of degree at most two.
\end{proof}

\begin{proof}[Proof of \cref{lem:tiers}(ii)]
\emph{Step 1: invariance.}
The free operations are Clifford unitaries, so they leave $\nu$ unchanged, and by part (i) they change $f$ only by a change of logical basis and terms of degree at most two, which leave the CCZ-rank unchanged.

\emph{Step 2: $\nu(f)\le3r(f)$.}
Suppose $f$ equals $\sum_{i=1}^{s}\lambda_{i1}\lambda_{i2}\lambda_{i3}$ up to terms of degree at most two, with linear functions $\lambda_{ij}$, and let $W$ be the span of these $3s$ functions.
If $\lambda_{ij}(w)=0$ for all $i,j$, then $\lambda_{ij}(u+w)=\lambda_{ij}(u)$, every product is unchanged by the flip, and $w$ is a Pauli direction.
These $w$ form a subspace of dimension at least $k-\dim W$, so by \cref{eq:ess_equals_nullity} $\nu(f)=k-\dim\mathcal P_{f}\le\dim W\le3s$.
Taking $s=r(f)$ gives $\nu(f)\le3r(f)$.

\emph{Step 3: the conditions are necessary.}
If the state is equivalent under $\mathcal F$ to $\ket{CCZ}^{\otimes r}\otimes\ket{+}^{\otimes(k-3r)}$, then by part (i) a change of basis brings $f$ to $\sum_{i=1}^{r}u_{3i-2}u_{3i-1}u_{3i}$ up to terms of degree at most two.
This polynomial has Pauli directions exactly the $w$ with $w_{1}=\dots=w_{3r}=0$, so $\nu(f)=3r$ by Step 1 and \cref{eq:ess_equals_nullity}.
The same expression gives $r(f)\le r$, and Step 2 gives $r(f)\ge\nu(f)/3=r$.

\emph{Step 4: the conditions are sufficient.}
Suppose $\nu(f)=3r$ and $f$ equals $\sum_{i=1}^{r}\lambda_{i1}\lambda_{i2}\lambda_{i3}$ up to terms of degree at most two.
By Step 2, $3r=\nu(f)\le\dim W\le3r$, so the $3r$ linear functions are linearly independent.
Using them as the first $3r$ logical bits of a new basis turns $f$ into $\sum_{i}u_{3i-2}u_{3i-1}u_{3i}$ up to terms of degree at most two, and part (i) gives the equivalence.

\emph{Step 5: the tiers.}
Suppose the state is Tier 0 or Tier 1, i.e.\ equivalent under $\mathcal F$ to a product of canonical states on disjoint qubits.
Writing that product as $\omega^{\psi}\ket{+}^{\otimes k}$ for $\ell$ large enough, $\psi$ is a sum of single non-Clifford terms on disjoint sets of qubits, and \cref{eq:tiers_general} gives $\psi(u)\equiv2^{\ell-1}f(Au+b)+2^{\ell-2}q(u)+c$.
A phase function has a unique expansion into products of bits, and on the right-hand side every term of degree at most two has a Clifford coefficient.
So every term of $\psi$ is cubic with coefficient $2^{\ell-1}$, every canonical factor is $\ket{CCZ}$, and Steps 3 and 4 apply.
A single factor ($\nu(f)=3$) is Tier 0, and two or more are Tier 1.
\end{proof}

\cref{lem:tiers} states what has to be checked; it remains to explain how we check it for each code.
When $\nu(f)=3r$, we establish Tier 1 by an explicit sequence of logical $CX$ and $X$ gates that brings $f$ to $r$ $CCZ$ terms on disjoint triples, up to terms of degree at most two.
The sequence is found by the random walk of \cref{section_random_walk_alg} and replayed to check the result.
It is a free operation, so by part (i) it proves the equivalence.

We establish Tier 2 in one of two ways.
The first is $\nu(f)$ not divisible by three.
When $\nu(f)=3r$ we compare a finer invariant with its value on $r$ disjoint $CCZ$ gates.
For each flip $w$, take the quadratic terms of the change $f(u+w)+f(u)$, and call their \emph{rank} the smallest number of linear functions in which they can be written, up to terms of degree at most one.
For example, flipping $c$ in the $\code{12,5,2}$ polynomial $c\,a_{1}b_{1}+c\,a_{2}b_{2}$ gives $a_{1}b_{1}+a_{2}b_{2}$, of rank $4$, and flipping $a_{1}$ gives $c\,b_{1}$, of rank $2$.
In practice the rank is the rank over $\mathbb F_{2}$ of the symmetric binary matrix of coefficients of these quadratic terms.
The \emph{rank histogram} counts how many flips give each rank.
By \cref{eq:flip_relabel}, a change of basis only relabels the flips and preserves every rank, so the histogram is invariant under the free operations.
Flips that differ by a Pauli direction give the same quadratic terms (as in the proof of \cref{lem:ess_pauli}), so we count each such class once, excluding the Pauli directions themselves.

For two disjoint $CCZ$ gates on $u_{1},\dots,u_{6}$, a flip of rank $2$ touches only one of the two triples ($7+7=14$ flips), and a flip of rank $4$ touches both ($7\times7=49$ flips).
The polynomial $u_{1}u_{2}u_{3}+u_{1}u_{4}u_{5}+u_{2}u_{4}u_{6}$ also has $\nu=6$, but its histogram is $7$ flips of rank $2$ and $56$ of rank $4$.
It is therefore not equivalent to two disjoint $CCZ$ gates; it is Tier 2, has CCZ-rank $3$, and shows that the condition $r(f)=r$ in \cref{lem:tiers}(ii) cannot be dropped.
Likewise, the $R=4$ star has $\nu=9$ and histogram $(255,256)$ in ranks $(2,8)$, whereas three disjoint $CCZ$ gates give $(21,147,343)$ in ranks $(2,4,6)$.

Two cases fall outside the Boolean cubic setting.
For the quartic Boolean polynomials of the $\code{16,4,2}$ chains we use the same explicit gate sequences, now up to terms of degree at most three.
For weighted polynomials with $k\le4$ we try every invertible $A$ and every $b$ in \cref{eq:tiers_general}.
We also checked \cref{lem:tiers}(ii) numerically on $4713$ random sums of $r\le3$ products of linear functions in $k\le10$ variables and on all $1023$ cubic forms in five variables.

\begin{proof}[Proof of \cref{prop:nullity}]
\emph{Step 1: stabilizing Pauli operators.}
Up to a phase, every Pauli operator is $X^{w}Z^{w'}$ with $w,w'\in\mathbb F_{2}^{k}$, and
\begin{align}
Z^{w'}X^{w}\,(-1)^{f}\ket{+}^{\otimes k}\propto\sum_{u}(-1)^{f(u+w)+w'\cdot u}\ket u~.
\end{align}
The resulting state is $\pm(-1)^{f}\ket{+}^{\otimes k}$ iff $f(u+w)+f(u)+w'\cdot u$ is constant, which happens iff $w$ is a Pauli direction and $w'\cdot u$ is the linear part of the change $f(u+w)+f(u)$, as in \cref{eq:pauli_direction}.
Each Pauli direction therefore gives exactly one stabilizing Pauli operator up to sign, and every other $w$ gives none.
So $\nu=k-\dim\mathcal P_{f}$, and by \cref{lem:ess_pauli} this is the smallest number of logical qubits that the cubic terms of $f$ act on.

\emph{Step 2: part (i).}
$u_{1}u_{2}u_{3}$ has no nonzero Pauli direction, so $\nu(\ket{CCZ})=3$, and by Clifford invariance and additivity every state Clifford-equivalent to $\ket{CCZ}^{\otimes r}\otimes\ket{+}^{\otimes(k-3r)}$ has $\nu=3r$.

\emph{Step 3: part (ii).}
Appending stabilizer ancillas to $\ket{CCZ}^{\otimes m}$ gives $\nu=3m$, which a stabilizer protocol cannot increase, so $3m\ge\nu(f)$.
Conversely, write $f$ as $\sum_{i=1}^{r(f)}\lambda_{i1}\lambda_{i2}\lambda_{i3}$ up to terms of degree at most two.
Each phase $(-1)^{\lambda_{i1}\lambda_{i2}\lambda_{i3}}$ is a $CCZ$ gate conjugated by a $CX$ network, applied by teleportation with one $\ket{CCZ}$ state and Clifford corrections.
The remaining terms of degree at most two are applied by $CZ$, $S$ and $Z$ gates.
\end{proof}

The rank histogram is invariant under $\mathcal F$ but not under the full Clifford group.
For the $R=4$ star and the concatenated $\code{64,9,4}$ and $\code{96,15,4}$ codes, whose tier it establishes, the SRE test of \cref{section:sre_certificate} provides the Clifford-group statement against the canonical states listed there.
We checked $\nu(f)=k-\dim\mathcal P_{f}$ against the stabilizer group computed from the state vector for $29$ cubic polynomials with $k\le7$, including linear and quadratic terms.

%% file: sections/C_csstp_examples.tex
\section{Selected Examples of CSS-T-Product Codes}\label{section:further_examples_CSStp}

To explore the space of CSS-TP codes beyond the manually constructed families above, we have implemented ~\cref{con:protocol} as a \emph{stochastic tree search} algorithm.
Given an input seed code, the algorithm performs iterative \emph{rollouts}: at each step, one of the three operations (QCS, $X$ stopper, or $Z$ stopper) is selected uniformly at random and applied to a randomly chosen qubit.
After each operation, dependent stabilizers are removed and the code parameters and induced logical action are recomputed.
If the resulting code has $k>0$ and $d\geq2$, it is retained as a candidate; otherwise, the weight-one logical-elimination procedure is applied iteratively until the code either satisfies these acceptance conditions or no acceptable stopper remains.
Only candidates whose transversal logical action is verified directly are recorded as CSS-TP codes.
Since all three operations reduce the block length $n$, each rollout terminates in a finite number of steps.
In this section, we report some of the codes and code families discovered using this technique.

We begin with the two smallest codes of the main text and show the stabilizers, logical operators, and $\overline{CX}$ circuits needed in order to conjugate the $\code{14,6,2}$ and $\code{12,5,2}$ codes into their reduced forms.
The commensurate stabilizer generators for the $\code{14,6,2}$ code are
\begin{align}
S_{X}^{\code{14,6,2}} =
\begin{pmatrix}
  1 & 1 & 1 & 1 & 1 & 1 & 1 & 1 & 1 & 1 & 1 & 1 & 1 & 1
\end{pmatrix}
&&
S_{Z}^{\code{14,6,2}} =
\begin{pmatrix}
  1 & 0 & 0 & 1 & 0 & 1 & 1 & 0 & 0 & 0 & 0 & 0 & 0 & 0 \\
  0 & 1 & 0 & 1 & 0 & 1 & 0 & 0 & 0 & 1 & 0 & 1 & 1 & 0 \\
  0 & 0 & 1 & 1 & 0 & 0 & 1 & 0 & 0 & 1 & 0 & 1 & 1 & 0 \\
  0 & 0 & 0 & 0 & 0 & 0 & 0 & 1 & 0 & 1 & 0 & 1 & 0 & 1 \\
  0 & 0 & 0 & 0 & 1 & 1 & 1 & 0 & 0 & 1 & 0 & 1 & 1 & 0 \\
  0 & 0 & 0 & 0 & 0 & 0 & 0 & 0 & 1 & 1 & 0 & 0 & 1 & 1 \\
  0 & 0 & 0 & 0 & 0 & 0 & 0 & 0 & 0 & 0 & 1 & 1 & 1 & 1
\end{pmatrix}
\end{align}
with logical operators
\begin{align}
L_{X}^{\code{14,6,2}} =
\begin{pmatrix}
  0 & 0 & 1 & 1 & 1 & 1 & 0 & 0 & 0 & 0 & 0 & 0 & 0 & 0 \\
  0 & 1 & 0 & 1 & 1 & 0 & 1 & 0 & 0 & 0 & 0 & 0 & 0 & 0 \\
  0 & 1 & 1 & 0 & 1 & 0 & 0 & 1 & 1 & 1 & 0 & 0 & 0 & 0 \\
  0 & 1 & 1 & 0 & 1 & 0 & 0 & 1 & 0 & 0 & 1 & 1 & 0 & 0 \\
  0 & 1 & 1 & 0 & 1 & 0 & 0 & 0 & 1 & 0 & 1 & 0 & 1 & 0 \\
  0 & 0 & 0 & 0 & 0 & 0 & 0 & 1 & 1 & 0 & 1 & 0 & 0 & 1
\end{pmatrix}
&&
L_{Z}^{\code{14,6,2}} =
\begin{pmatrix}
  1 & 0 & 0 & 0 & 0 & 1 & 0 & 0 & 0 & 0 & 0 & 0 & 0 & 0 \\
  1 & 0 & 0 & 0 & 0 & 0 & 1 & 0 & 0 & 0 & 0 & 0 & 0 & 0 \\
  1 & 0 & 0 & 0 & 0 & 0 & 0 & 0 & 0 & 1 & 0 & 0 & 0 & 0 \\
  1 & 0 & 0 & 0 & 0 & 0 & 0 & 0 & 0 & 0 & 0 & 1 & 0 & 0 \\
  1 & 0 & 0 & 0 & 0 & 0 & 0 & 0 & 0 & 0 & 0 & 0 & 1 & 0 \\
  1 & 0 & 0 & 0 & 0 & 0 & 0 & 0 & 0 & 0 & 0 & 0 & 0 & 1
\end{pmatrix}
\end{align}
The stabilizers and logical operators for the $\code{12,5,2}$ code are given by
\begin{align}
S_{X}^{\code{12,5,2}} =
\begin{pmatrix}
  1  &  1  &  1  &  1  &  1  &  1  &  1  &  1  &  1  &  1  &  1  &  1
\end{pmatrix}
&&
S_{Z}^{\code{12,5,2}} =
\begin{pmatrix}
  0  &  0  &  1  &  0  &  1  &  1  &  0  &  0  &  0  &  1  &  1  &  1 \\
  0  &  1  &  0  &  0  &  1  &  0  &  0  &  0  &  1  &  1  &  0  &  0 \\
  1  &  0  &  0  &  0  &  0  &  1  &  0  &  0  &  1  &  1  &  0  &  0 \\
  0  &  0  &  0  &  1  &  1  &  1  &  0  &  0  &  1  &  0  &  1  &  1 \\
  0  &  0  &  0  &  0  &  0  &  0  &  1  &  0  &  1  &  1  &  0  &  1 \\
  0  &  0  &  0  &  0  &  0  &  0  &  0  &  1  &  1  &  1  &  1  &  0
\end{pmatrix}
\end{align}
for the stabilizers, and
\begin{align}
L_{X}^{\code{12,5,2}} =
\begin{pmatrix}
  0 & 1 & 1 & 1 & 1 & 0 & 0 & 0 & 0 & 0 & 0 & 0 \\
  0 & 1 & 1 & 0 & 0 & 1 & 1 & 1 & 1 & 0 & 0 & 0 \\
  0 & 1 & 0 & 1 & 0 & 1 & 1 & 1 & 0 & 1 & 0 & 0 \\
  0 & 0 & 1 & 1 & 0 & 0 & 0 & 1 & 0 & 0 & 1 & 0 \\
  0 & 0 & 1 & 1 & 0 & 0 & 1 & 0 & 0 & 0 & 0 & 1
\end{pmatrix}
&&
L_{Z}^{\code{12,5,2}} =
\begin{pmatrix}
  1 & 0 & 0 & 0 & 1 & 0 & 0 & 0 & 0 & 0 & 0 & 0 \\
  1 & 0 & 0 & 0 & 0 & 0 & 0 & 0 & 1 & 0 & 0 & 0 \\
  1 & 0 & 0 & 0 & 0 & 0 & 0 & 0 & 0 & 1 & 0 & 0 \\
  1 & 0 & 0 & 0 & 0 & 0 & 0 & 0 & 0 & 0 & 1 & 0 \\
  1 & 0 & 0 & 0 & 0 & 0 & 0 & 0 & 0 & 0 & 0 & 1
\end{pmatrix}
\end{align}
for the logical operators.

With these matrices, we can make explicit the equivalence with the synthillation codes stated in \cref{section:CSSTP_examples}.
Let $G=[L_{X};S_{X}]$, with $S_{X}$ the all-ones row, and let $G_{:,\pi}$ be the matrix whose column $j$ is column $\pi(j)$ of $G$ (qubits and rows indexed from $0$).
For each code there is a permutation $\pi$ and an invertible $M$ with $M\,G_{:,\pi}=G_{\rm CH}$, where $G_{\rm CH}$ is the generator matrix of Ref.~\cite{campbell2017unified} ($G_{\Tofs}$ for \Tofs and $G_{\twoCCZ}$ for \twoCCZ):
\begin{align*}
\code{12,5,2}:&\quad \pi=(0,1,2,6,7,5,4,3,11,10,9,8)~,\\
&\quad M=(000110,\,000010,\,100110,\,111001,\,110111,\,000001)~,\\
\code{14,6,2}:&\quad \pi=(0,1,6,2,3,4,5,7,8,10,11,12,13,9)~,\\
&\quad M=(1111111,\,1011111,\,1100000,\,0010111,\,0011011,\,0000010,\,0000001)~,
\end{align*}
with the rows of $M$ written as bit strings.
In both cases the last row of $M$ maps the $X$ stabilizer to itself, and the $Z$ stabilizers are the full dual of $G$, as in synthillation.

\cref{fig:cx_conjugation_sequence} gives the logical $\overline{CX}$ networks that take the logical gates of \cref{fig:uncleaned_logical_gates,fig:3082_uncleaned_gate} to their reduced forms.
Writing $\overline{A}=\overline{CX}_1\cdots\overline{CX}_r$ for the drawn gates of panels (a)--(c), the reduced-form gate acts as $\overline{\mathcal M}'\ket{x}=(-1)^{f(\overline{A}x)}\ket{x}$, where $\overline{\mathcal M}\ket{y}=(-1)^{f(y)}\ket{y}$; equivalently, the logical representatives change as $L_X\mapsto \overline{A}^{\mathsf T}L_X$.
Each of these three networks takes the gate of \cref{fig:uncleaned_logical_gates} to that of \cref{fig:cleaned_logical_gates} exactly, including the single-qubit $\overline{Z}$ terms; for the star code it takes the $32$-term $\overline{\mathcal M}_{\code{16,7,2}}$ to $\overline{CCZ}(0,1,3)\,\overline{CCZ}(1,2,4)\,\overline{CCZ}(1,5,6)$.
In panel (d), $11$ logical $CX$ gates, an $X$ frame and the wire relabelling on the right define $U$ with $U^\dagger\,\overline{\mathcal M}_{\code{30,8,2}}\,U=\overline{CCCZ}(0,2,5,7)\,\overline{CCCZ}(1,3,4,6)$ exactly.
We checked all four identities on all $2^{k}$ logical basis states.

\input{includes/CX_conjugation}

In addition to the examples shown in the main text, one can directly combine QRM seed codes together in order to develop distance-three CSS-TP codes.
Two such examples include:
\begin{enumerate}[label=(\alph*)]
    \item An open-chain family with parameters $\code{13m+28,\, m+2,\, 3}$ for $m \geq 0$, and
    \item A variant family with parameters $\code{13m+26,\, m+2,\, d \geq 2}$ for $m \geq 0$, whose first representative at $m=0$ is a $\code{26,2,2}$ code (achieving distance $3$ for $m \geq 1$).
\end{enumerate}

As a concrete example of both families, we provide the examples of $m=1$ for both, resulting in $\code{41,3,3}$ and $\code{39,3,3}$ codes, respectively.
The logical action of the $\code{39,3,3}$ code in particular takes on the form shown in \cref{fig:css_logical_operators_multipanel}(a).

Many other examples exist which were found during the course of this study.
In particular we show a few CSS-TI and CSS-TS code, i.e., codes for whom the transversal action of $T_{\ell}$ allows for a logical gate that is either identity ($\overline{\mathbb{I}}$) or a phase gate ($\overline{S}$ or $\overline{S}^{\dagger}$).

A first example is of the $\code{5,1,2}$ CSS-TI code.
Its stabilizers are of the form
\begin{align}
S_{X} = 
\begin{pmatrix}
1,1,1,1,1
\end{pmatrix} 
&&
S_{Z} = 
\begin{pmatrix}
1,1,1,0,1 \\
0,0,1,1,0 \\ 
1,1,0,0,0
\end{pmatrix}~.
\end{align}
The logical operators are $\overline{X} = X_{2,3,4}$ and $\overline{Z} = Z_{0,4}$.

The logical identity operator for this code takes the form shown in \cref{fig:css_logical_operators_multipanel}(b).

An example of a CSS-TS code can be found in the example of the $\code{13,2,2}$ code, with stabilizers
\begin{align}
S_{X} =
\begin{pmatrix}
  0 & 1 & 0 & 1 & 0 & 1 & 1 & 0 & 1 & 0 & 1 & 0 & 1 \\
  1 & 1 & 0 & 0 & 1 & 1 & 0 & 1 & 1 & 0 & 0 & 1 & 1 \\
  0 & 0 & 1 & 1 & 1 & 1 & 0 & 0 & 0 & 1 & 1 & 1 & 1
\end{pmatrix}
&&
S_{Z} =
\begin{pmatrix}
  1 & 1 & 0 & 0 & 1 & 1 & 0 & 1 & 1 & 0 & 0 & 1 & 1 \\
  0 & 0 & 1 & 1 & 1 & 1 & 0 & 0 & 0 & 1 & 1 & 1 & 1 \\
  0 & 1 & 0 & 0 & 0 & 1 & 0 & 0 & 1 & 0 & 0 & 0 & 1 \\
  0 & 0 & 0 & 1 & 0 & 1 & 0 & 0 & 0 & 0 & 1 & 0 & 1 \\
  0 & 0 & 0 & 0 & 1 & 1 & 0 & 0 & 0 & 0 & 0 & 1 & 1 \\
  0 & 0 & 0 & 0 & 0 & 0 & 0 & 1 & 1 & 0 & 0 & 1 & 1 \\
  0 & 0 & 0 & 0 & 0 & 0 & 0 & 0 & 0 & 1 & 1 & 1 & 1 \\
  0 & 0 & 0 & 0 & 0 & 0 & 1 & 0 & 1 & 0 & 1 & 0 & 1
\end{pmatrix}
\end{align}
The logical gates implemented with this code are shown in \cref{fig:css_logical_operators_multipanel}(c,d).
\input{includes/css_ts}

We also report a code family that can distill $\sqrt{T}$ gates (one per logical qubit) at distance three.
This family follows the open-chain form of the other families above, and has parameters $\code{29m+60,\, m+2,\, 3}$, with $m \geq 0$ ($m+2$ copies of the $\code{31,1,3}$ code).
The first representative, at $m=0$, is the $\code{60,2,3}$ code.
It is built from two punctured $\code{31,1,3}$ QRM codes with layers $\sqrt{T}^{\otimes 31}$ and $\sqrt{T}^{\dagger\otimes 31}$, conjoined on one leg.
A computer check confirms $n=60$, $k=2$, commuting $X$- and $Z$-type generators, and exact distance $3$.
The inherited layer is logical at level $\ell=4$ and implements $\overline{\sqrt{T}}^{\dagger}\otimes\overline{\sqrt{T}}$.
No layer on this code reaches a two-qubit non-Clifford monomial, so it does not implement a $CT$ gate.
The complete generating set, with qubits labeled $1,\dots,60$, is
\par\smallskip
\input{includes/generated/pcm_60_2_3}
\smallskip

The non-Clifford transversal logical gate for this code is shown in \cref{fig:css_logical_operators_multipanel}(e).

Finally, we report on a code family that can access the fourth level of the Clifford hierarchy.
Iterating the open-chain procedure with the $\code{16,4,2}$ code yields a family with parameters
\begin{align}
\code{14m+30, 4m+8, 2}~, \quad m \geq 0~.
\end{align}
The stabilizers for the first representative (at $m=0$, two conjoined copies of $\code{16,4,2}$), the $\code{30,8,2}$ code, are given by
\begin{align}
S_{X}^{\code{30,8,2}} =
\begin{pmatrix}
  \mathtt{111111111111111111111111111111}
\end{pmatrix}
&&
S_{Z}^{\code{30,8,2}} =
\begin{pmatrix}
  \mathtt{000000000011001000000000000111} \\
  \mathtt{000100010001000000000000000111} \\
  \mathtt{000010010000011000000000000000} \\
  \mathtt{000001010000010000000000000111} \\
  \mathtt{000000110000001000000000000111} \\
  \mathtt{000000001001011000000000000000} \\
  \mathtt{000000000101010000000000000111} \\
  \mathtt{010000010001010000000000000000} \\
  \mathtt{000000000000111000000000000111} \\
  \mathtt{100000010001011000000000000111} \\
  \mathtt{000000000000000100000010001100} \\
  \mathtt{000000000000000010000010001010} \\
  \mathtt{000000000000000001000010001001} \\
  \mathtt{000000000000000000100010001111} \\
  \mathtt{000000000000000000010010000011} \\
  \mathtt{000000000000000000001010000101} \\
  \mathtt{000000000000000000000110000110} \\
  \mathtt{000000000000000000000001001011} \\
  \mathtt{000000000000000000000000101101} \\
  \mathtt{001000010001001000000000000000} \\
  \mathtt{000000000000000000000000011110}
\end{pmatrix}
\end{align}
and the logical operators are
\begin{align}
L_{X}^{\code{30,8,2}} =
\begin{pmatrix}
  \mathtt{000011111111000000000000000000} \\
  \mathtt{001100111100110000000000000000} \\
  \mathtt{010101011010101000000000000000} \\
  \mathtt{000000000000000111111110000000} \\
  \mathtt{000000000000000111100001111000} \\
  \mathtt{011010010110100100101100110100} \\
  \mathtt{011010010110100010110101010010} \\
  \mathtt{011010010110100001111001100001}
\end{pmatrix}
&&
L_{Z}^{\code{30,8,2}} =
\begin{pmatrix}
  \mathtt{100000000001000000000000000000} \\
  \mathtt{100000000000010000000000000000} \\
  \mathtt{100000000000001000000000000000} \\
  \mathtt{100000000000000000000010000000} \\
  \mathtt{100000000000000000000000001000} \\
  \mathtt{100000000000000000000000000100} \\
  \mathtt{100000000000000000000000000010} \\
  \mathtt{100000000000000000000000000001}
\end{pmatrix}
\end{align}
We verified numerically that the layer of $15$ $\sqrt T$ and $15$ $\sqrt T^{\dagger}$ gates times Clifford phases is logical on this code; \cref{fig:3082_gates} shows its logical action in this basis and in the reduced form.

\subsection{Stars of \texorpdfstring{$C^{4}Z$}{CCCCZ} and \texorpdfstring{$C^{5}Z$}{CCCCCZ} gates}\label{section:higher_stars}

\cref{prop:glued_stars} applies to hypercubes of any dimension $m$.
The hub and every spoke are $\code{2^{m},m,2}$ hypercube codes with the signed layer $T_{m}^{(-1)^{|v|}}$, where $T_{m}=\mathrm{diag}(1,e^{2\pi i/2^{m}})$, and each spoke is glued twice to a direction-$0$ edge of the hub as in \cref{section:cccz_star}.

For $m=5$, the glued stars are $\code{28R+4,\,4R+1,\,2}$ codes for $2\le R\le17$, and the layer $T_{5}^{(-1)^{|v|}}$ implements
\begin{align}
\overline{\mathcal M}=\prod_{r=1}^{R}\overline{C^{4}Z}\bigl(c,u^{(r)}_{1},u^{(r)}_{2},u^{(r)}_{3},u^{(r)}_{4}\bigr)
\end{align}
in the hub/spoke basis.
The first members are $\code{60,9,2}$, $\code{88,13,2}$ and $\code{116,17,2}$ for $R=2,3,4$.

Going one dimension higher, for $m=6$ the glued stars are $\code{60R+4,\,5R+1,\,2}$ codes for $2\le R\le33$, and the layer $T_{6}^{(-1)^{|v|}}$ implements
\begin{align}
\overline{\mathcal M}=\prod_{r=1}^{R}\overline{C^{5}Z}\bigl(c,u^{(r)}_{1},\dots,u^{(r)}_{5}\bigr)~.
\end{align}
The first members are $\code{124,11,2}$ and $\code{184,16,2}$ for $R=2,3$.

For each member listed we built the code with the generic contraction routine and checked that the layer is logical, that $d=2$, and that the phase of the codewords used in the proof of \cref{prop:glued_stars} matches the action above, on all codewords for $k\le15$ and on $4096$ random codewords otherwise.
The block length grows as $2^{m}$ per spoke, so these codes are of interest mainly where several multi-controlled gates share one control; we do not cost them as factories.

%% file: includes/CX_conjugation.tex
\begin{figure*}
\centering

\begin{minipage}[t]{0.17\textwidth}
\centering
\panelimage[
    width=\linewidth,
    height=0.55\textheight,
    keepaspectratio,
    trim=0pt 0pt 0pt 0pt,
    clip
]{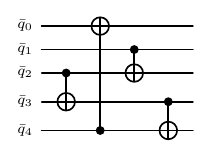}{(a)}
\end{minipage}
\hfill
\begin{minipage}[t]{0.5\textwidth}
\centering
\panelimage[
    width=\linewidth,
    height=0.45\textheight,
    keepaspectratio,
    trim=0pt 0pt 0pt 0pt,
    clip
]{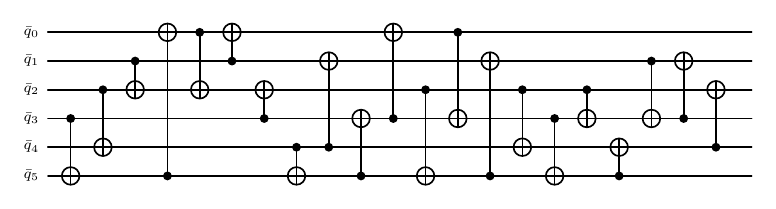}{(b)}
\end{minipage}
\hfill
\begin{minipage}[t]{0.24\textwidth}
\centering
\panelimage[
    width=\linewidth,
    height=0.55\textheight,
    keepaspectratio,
    trim=0pt 0pt 0pt 0pt,
    clip
]{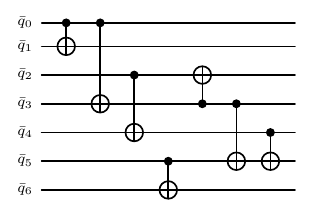}{(c)}
\end{minipage}

\vspace{0.8em}
\begin{minipage}[t]{0.4\textwidth}
\centering
\panelimage[
    width=\linewidth,
    height=0.55\textheight,
    keepaspectratio,
    trim=0pt 0pt 0pt 0pt,
    clip
]{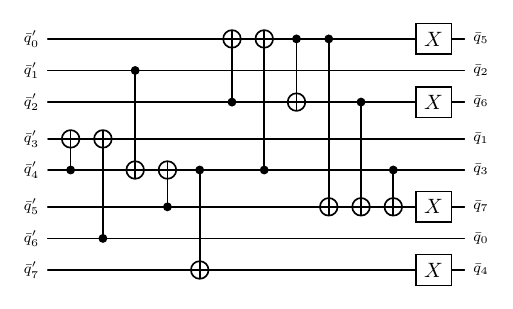}{(d)}
\end{minipage}

\caption[Logical CX conjugation sequences used to reach the reduced form]{
Logical $\overline{CX}$ conjugation sequences used to reach the reduced form of the
$\code{12,5,2}$, $\code{14,6,2}$, $\code{16,7,2}$ and $\code{30,8,2}$ CSS-TP codes. Panels
(a), (b), (c) and (d) show
$\overline{A}_{\code{12,5,2}}$,
$\overline{A}_{\code{14,6,2}}$, $\overline{A}_{\code{16,7,2}}$ and
$\overline{A}_{\code{30,8,2}}$, respectively.
}
\label{fig:cx_conjugation_sequence}

\setcounter{subfigure}{0}
\refstepcounter{subfigure}\label{fig:1252_conjugations}
\refstepcounter{subfigure}\label{fig:1462_conjugation}
\refstepcounter{subfigure}\label{fig:1672_conjugation}
\refstepcounter{subfigure}\label{fig:3082_conjugations}

\end{figure*}

%% file: includes/css_ts.tex
\begin{figure*}
    \centering

    \begin{minipage}[c]{0.325\textwidth}
        \centering
        \panelimage[width=0.95\linewidth]{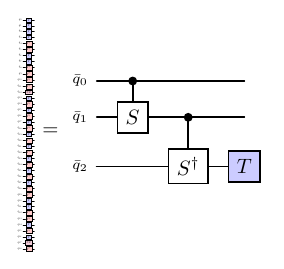}{(a)}

        \vspace{0.5em}

        \panelimage[width=\linewidth]{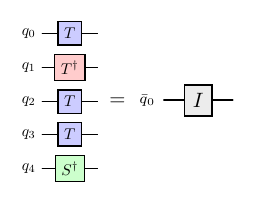}{(b)}
    \end{minipage}
    \hfill
    \begin{minipage}[c]{0.325\textwidth}
        \centering
        \panelimage[width=\linewidth]{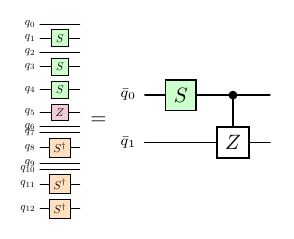}{(c)}

        \vspace{0.5em}

        \panelimage[width=\linewidth]{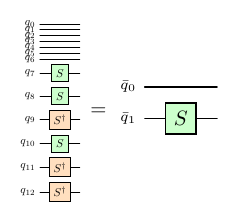}{(d)}
    \end{minipage}
    \hfill
    \begin{minipage}[c]{0.325\textwidth}
        \centering
        \panelimage[width=\linewidth]{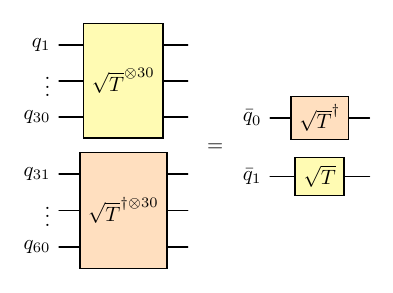}{(e)}
    \end{minipage}

    \caption{
        Logical gates and operators for several CSS codes.
        (a) The non-Clifford logical gate represented by the $\code{39,3,3}$ CSS-TP code.
        (b) The logical identity operator for the $\code{5,1,2}$ CSS-TI code.
        (c,d) The two logical $\overline{S}$ operators for the $\code{13,2,2}$ CSS-TS code.
        (e) The non-Clifford logical gate for the $\code{60,2,3}$ CSS-TP code.
    }
    \label{fig:css_logical_operators_multipanel}
\end{figure*}

%% file: includes/generated/pcm_60_2_3.tex
{\footnotesize\raggedright
\noindent $X$-type ($9$ generators): $X_{1}=\{$1,\allowbreak 4,\allowbreak 5,\allowbreak 8,\allowbreak 9,\allowbreak 12,\allowbreak 13,\allowbreak 16,\allowbreak 17,\allowbreak 20,\allowbreak 21,\allowbreak 24,\allowbreak 25,\allowbreak 28,\allowbreak 29,\allowbreak 32,\allowbreak 34,\allowbreak 36,\allowbreak 38,\allowbreak 40,\allowbreak 42,\allowbreak 44,\allowbreak 46,\allowbreak 48,\allowbreak 50,\allowbreak 52,\allowbreak 54,\allowbreak 56,\allowbreak 58,\allowbreak 60$\}$;\allowbreak\ $X_{2}=\{$2,\allowbreak 5,\allowbreak 6,\allowbreak 9,\allowbreak 10,\allowbreak 13,\allowbreak 14,\allowbreak 17,\allowbreak 18,\allowbreak 21,\allowbreak 22,\allowbreak 25,\allowbreak 26,\allowbreak 29,\allowbreak 30,\allowbreak 32,\allowbreak 34,\allowbreak 36,\allowbreak 38,\allowbreak 40,\allowbreak 42,\allowbreak 44,\allowbreak 46,\allowbreak 48,\allowbreak 50,\allowbreak 52,\allowbreak 54,\allowbreak 56,\allowbreak 58,\allowbreak 60$\}$;\allowbreak\ $X_{3}=\{$3,\allowbreak 4,\allowbreak 5,\allowbreak 6,\allowbreak 11,\allowbreak 12,\allowbreak 13,\allowbreak 14,\allowbreak 19,\allowbreak 20,\allowbreak 21,\allowbreak 22,\allowbreak 27,\allowbreak 28,\allowbreak 29,\allowbreak 30$\}$;\allowbreak\ $X_{4}=\{$7,\allowbreak 8,\allowbreak 9,\allowbreak 10,\allowbreak 11,\allowbreak 12,\allowbreak 13,\allowbreak 14,\allowbreak 23,\allowbreak 24,\allowbreak 25,\allowbreak 26,\allowbreak 27,\allowbreak 28,\allowbreak 29,\allowbreak 30$\}$;\allowbreak\ $X_{5}=\{$15,\allowbreak 16,\allowbreak 17,\allowbreak 18,\allowbreak 19,\allowbreak 20,\allowbreak 21,\allowbreak 22,\allowbreak 23,\allowbreak 24,\allowbreak 25,\allowbreak 26,\allowbreak 27,\allowbreak 28,\allowbreak 29,\allowbreak 30$\}$;\allowbreak\ $X_{6}=\{$31,\allowbreak 32,\allowbreak 35,\allowbreak 36,\allowbreak 39,\allowbreak 40,\allowbreak 43,\allowbreak 44,\allowbreak 47,\allowbreak 48,\allowbreak 51,\allowbreak 52,\allowbreak 55,\allowbreak 56,\allowbreak 59,\allowbreak 60$\}$;\allowbreak\ $X_{7}=\{$33,\allowbreak 34,\allowbreak 35,\allowbreak 36,\allowbreak 41,\allowbreak 42,\allowbreak 43,\allowbreak 44,\allowbreak 49,\allowbreak 50,\allowbreak 51,\allowbreak 52,\allowbreak 57,\allowbreak 58,\allowbreak 59,\allowbreak 60$\}$;\allowbreak\ $X_{8}=\{$37,\allowbreak 38,\allowbreak 39,\allowbreak 40,\allowbreak 41,\allowbreak 42,\allowbreak 43,\allowbreak 44,\allowbreak 53,\allowbreak 54,\allowbreak 55,\allowbreak 56,\allowbreak 57,\allowbreak 58,\allowbreak 59,\allowbreak 60$\}$;\allowbreak\ $X_{9}=\{$45,\allowbreak 46,\allowbreak 47,\allowbreak 48,\allowbreak 49,\allowbreak 50,\allowbreak 51,\allowbreak 52,\allowbreak 53,\allowbreak 54,\allowbreak 55,\allowbreak 56,\allowbreak 57,\allowbreak 58,\allowbreak 59,\allowbreak 60$\}$.\par
\noindent $Z$-type ($49$ generators): $Z_{1}=\{$1,\allowbreak 28,\allowbreak 29,\allowbreak 44,\allowbreak 52,\allowbreak 56,\allowbreak 58,\allowbreak 60$\}$;\allowbreak\ $Z_{2}=\{$2,\allowbreak 29,\allowbreak 30,\allowbreak 44,\allowbreak 52,\allowbreak 56,\allowbreak 58,\allowbreak 60$\}$;\allowbreak\ $Z_{3}=\{$3,\allowbreak 26,\allowbreak 28,\allowbreak 44,\allowbreak 52,\allowbreak 56,\allowbreak 58,\allowbreak 60$\}$;\allowbreak\ $Z_{4}=\{$4,\allowbreak 26,\allowbreak 28,\allowbreak 29,\allowbreak 30,\allowbreak 44,\allowbreak 52,\allowbreak 56,\allowbreak 58,\allowbreak 60$\}$;\allowbreak\ $Z_{5}=\{$5,\allowbreak 26,\allowbreak 30,\allowbreak 44,\allowbreak 52,\allowbreak 56,\allowbreak 58,\allowbreak 60$\}$;\allowbreak\ $Z_{6}=\{$6,\allowbreak 26,\allowbreak 29,\allowbreak 44,\allowbreak 52,\allowbreak 56,\allowbreak 58,\allowbreak 60$\}$;\allowbreak\ $Z_{7}=\{$7,\allowbreak 22,\allowbreak 28,\allowbreak 44,\allowbreak 52,\allowbreak 56,\allowbreak 58,\allowbreak 60$\}$;\allowbreak\ $Z_{8}=\{$8,\allowbreak 22,\allowbreak 28,\allowbreak 29,\allowbreak 30,\allowbreak 44,\allowbreak 52,\allowbreak 56,\allowbreak 58,\allowbreak 60$\}$;\allowbreak\ $Z_{9}=\{$9,\allowbreak 22,\allowbreak 30,\allowbreak 44,\allowbreak 52,\allowbreak 56,\allowbreak 58,\allowbreak 60$\}$;\allowbreak\ $Z_{10}=\{$10,\allowbreak 22,\allowbreak 29,\allowbreak 44,\allowbreak 52,\allowbreak 56,\allowbreak 58,\allowbreak 60$\}$;\allowbreak\ $Z_{11}=\{$11,\allowbreak 22,\allowbreak 26,\allowbreak 28,\allowbreak 30,\allowbreak 44,\allowbreak 52,\allowbreak 56,\allowbreak 58,\allowbreak 60$\}$;\allowbreak\ $Z_{12}=\{$12,\allowbreak 22,\allowbreak 26,\allowbreak 28,\allowbreak 29,\allowbreak 44,\allowbreak 52,\allowbreak 56,\allowbreak 58,\allowbreak 60$\}$;\allowbreak\ $Z_{13}=\{$13,\allowbreak 22,\allowbreak 26,\allowbreak 44,\allowbreak 52,\allowbreak 56,\allowbreak 58,\allowbreak 60$\}$;\allowbreak\ $Z_{14}=\{$14,\allowbreak 22,\allowbreak 26,\allowbreak 29,\allowbreak 30,\allowbreak 44,\allowbreak 52,\allowbreak 56,\allowbreak 58,\allowbreak 60$\}$;\allowbreak\ $Z_{15}=\{$15,\allowbreak 22,\allowbreak 26,\allowbreak 28,\allowbreak 29,\allowbreak 30$\}$;\allowbreak\ $Z_{16}=\{$16,\allowbreak 22,\allowbreak 26,\allowbreak 28$\}$;\allowbreak\ $Z_{17}=\{$17,\allowbreak 22,\allowbreak 26,\allowbreak 29$\}$;\allowbreak\ $Z_{18}=\{$18,\allowbreak 22,\allowbreak 26,\allowbreak 30$\}$;\allowbreak\ $Z_{19}=\{$19,\allowbreak 22,\allowbreak 28,\allowbreak 29$\}$;\allowbreak\ $Z_{20}=\{$20,\allowbreak 22,\allowbreak 28,\allowbreak 30$\}$;\allowbreak\ $Z_{21}=\{$21,\allowbreak 22,\allowbreak 29,\allowbreak 30$\}$;\allowbreak\ $Z_{22}=\{$23,\allowbreak 26,\allowbreak 28,\allowbreak 29$\}$;\allowbreak\ $Z_{23}=\{$24,\allowbreak 26,\allowbreak 28,\allowbreak 30$\}$;\allowbreak\ $Z_{24}=\{$25,\allowbreak 26,\allowbreak 29,\allowbreak 30$\}$;\allowbreak\ $Z_{25}=\{$27,\allowbreak 28,\allowbreak 29,\allowbreak 30$\}$;\allowbreak\ $Z_{26}=\{$31,\allowbreak 44,\allowbreak 52,\allowbreak 56,\allowbreak 59,\allowbreak 60$\}$;\allowbreak\ $Z_{27}=\{$32,\allowbreak 44,\allowbreak 52,\allowbreak 56$\}$;\allowbreak\ $Z_{28}=\{$33,\allowbreak 44,\allowbreak 52,\allowbreak 58,\allowbreak 59,\allowbreak 60$\}$;\allowbreak\ $Z_{29}=\{$34,\allowbreak 44,\allowbreak 52,\allowbreak 58$\}$;\allowbreak\ $Z_{30}=\{$35,\allowbreak 44,\allowbreak 52,\allowbreak 59$\}$;\allowbreak\ $Z_{31}=\{$36,\allowbreak 44,\allowbreak 52,\allowbreak 60$\}$;\allowbreak\ $Z_{32}=\{$37,\allowbreak 44,\allowbreak 56,\allowbreak 58,\allowbreak 59,\allowbreak 60$\}$;\allowbreak\ $Z_{33}=\{$38,\allowbreak 44,\allowbreak 56,\allowbreak 58$\}$;\allowbreak\ $Z_{34}=\{$39,\allowbreak 44,\allowbreak 56,\allowbreak 59$\}$;\allowbreak\ $Z_{35}=\{$40,\allowbreak 44,\allowbreak 56,\allowbreak 60$\}$;\allowbreak\ $Z_{36}=\{$41,\allowbreak 44,\allowbreak 58,\allowbreak 59$\}$;\allowbreak\ $Z_{37}=\{$42,\allowbreak 44,\allowbreak 58,\allowbreak 60$\}$;\allowbreak\ $Z_{38}=\{$43,\allowbreak 44,\allowbreak 59,\allowbreak 60$\}$;\allowbreak\ $Z_{39}=\{$45,\allowbreak 52,\allowbreak 56,\allowbreak 58,\allowbreak 59,\allowbreak 60$\}$;\allowbreak\ $Z_{40}=\{$46,\allowbreak 52,\allowbreak 56,\allowbreak 58$\}$;\allowbreak\ $Z_{41}=\{$47,\allowbreak 52,\allowbreak 56,\allowbreak 59$\}$;\allowbreak\ $Z_{42}=\{$48,\allowbreak 52,\allowbreak 56,\allowbreak 60$\}$;\allowbreak\ $Z_{43}=\{$49,\allowbreak 52,\allowbreak 58,\allowbreak 59$\}$;\allowbreak\ $Z_{44}=\{$50,\allowbreak 52,\allowbreak 58,\allowbreak 60$\}$;\allowbreak\ $Z_{45}=\{$51,\allowbreak 52,\allowbreak 59,\allowbreak 60$\}$;\allowbreak\ $Z_{46}=\{$53,\allowbreak 56,\allowbreak 58,\allowbreak 59$\}$;\allowbreak\ $Z_{47}=\{$54,\allowbreak 56,\allowbreak 58,\allowbreak 60$\}$;\allowbreak\ $Z_{48}=\{$55,\allowbreak 56,\allowbreak 59,\allowbreak 60$\}$;\allowbreak\ $Z_{49}=\{$57,\allowbreak 58,\allowbreak 59,\allowbreak 60$\}$.\par
\noindent Logical operators: $\bar X_{1}=\{$1,\allowbreak 5,\allowbreak 6,\allowbreak 9,\allowbreak 10,\allowbreak 11,\allowbreak 12,\allowbreak 17,\allowbreak 18,\allowbreak 19,\allowbreak 20,\allowbreak 23,\allowbreak 24,\allowbreak 29,\allowbreak 30$\}$;\allowbreak\ $\bar X_{2}=\{$2,\allowbreak 5,\allowbreak 6,\allowbreak 9,\allowbreak 10,\allowbreak 13,\allowbreak 14,\allowbreak 17,\allowbreak 18,\allowbreak 21,\allowbreak 22,\allowbreak 25,\allowbreak 26,\allowbreak 29,\allowbreak 30,\allowbreak 35,\allowbreak 36,\allowbreak 39,\allowbreak 40,\allowbreak 41,\allowbreak 42,\allowbreak 47,\allowbreak 48,\allowbreak 49,\allowbreak 50,\allowbreak 53,\allowbreak 54,\allowbreak 59,\allowbreak 60$\}$. $\bar Z_{1}=\{$1,\allowbreak 29,\allowbreak 30$\}$;\allowbreak\ $\bar Z_{2}=\{$2,\allowbreak 29,\allowbreak 30,\allowbreak 59,\allowbreak 60$\}$.\par
}

%% file: sections/D_magic_properties.tex
\section{Magic-Resource Properties of CSS-TP Codes}
\label{section:magic_resource_properties}

This appendix collects three complementary properties of the resources produced by CSS-TP codes: their realization as distillation protocols, their asymptotic yield, and their stabilizer R\'enyi entropy and synthesis costs.
These quantities describe different aspects of a resource and should not be interpreted as defining a common ordering.
In particular, stabilizer R\'enyi entropy measures a feature of the state's Pauli spectrum, whereas distillation yield, exact synthesis cost, and architectural cost are operational quantities determined by different resource theories.

\subsection{Distillation and asymptotic yield}
\label{section:distillation_etd}
\label{section:magic_yield}

The ETD construction, accepted output, and finite-size error model are developed in \cref{section:CSSTP_distillation_protocols}.
Here we consider only the asymptotic yield of the resulting code families.

The conventional asymptotic yield exponent of an $n$-to-$k$, distance-$d$ distillation protocol is
\begin{align}
\gamma
=
\frac{\log(n/k)}{\log d}
\label{eq:magic_yield}
\end{align}
\cite{bravyi2012magic}.
Several CSS-TP families satisfy $\gamma\to1$, as summarized in \cref{table:yields}.
This value is not a record: any family with fixed $d=2$ and $k/n\to1/2$ has $\gamma\to1$.
Constant and sub-constant overhead, $\gamma\to0$, has been achieved~\cite{hastings2018distillation,krishna2018towards,wills2025constant}.
We report $\gamma$ only as a bookkeeping quantity for the code families.

\input{includes/yield_table}

The exponent $\gamma$ records only the asymptotic relation among $n$, $k$, and $d$.
It does not include the accepted-output error prefactor, acceptance probability, input-state production cost, injection corrections, factory footprint, or spacetime volume.
The exponent is therefore a code-family descriptor rather than a complete measure of factory performance.

\subsection{Stabilizer R\'enyi entropy}
\label{section:unitary_stab_entropy}

For a pure $n$-qubit state $\ket{\psi}$, define its Pauli probability distribution by
\begin{align}
\Xi_{P}(\psi)
=
2^{-n}
\left|
\bra{\psi}P\ket{\psi}
\right|^{2}~,
\qquad
P\in\{I,X,Y,Z\}^{\otimes n}~.
\label{eq:pauli_probability_distribution}
\end{align}
The order-$\alpha$ stabilizer R\'enyi entropy is
\begin{align}
M_{\alpha}(\ket{\psi})
=
\frac{1}{1-\alpha}
\log_{2}
\left[
\sum_{P}
\Xi_{P}(\psi)^{\alpha}
\right]
-n~.
\label{eq:state_sre}
\end{align}
The stabilizer R\'enyi entropy vanishes on pure stabilizer states, is invariant under Clifford unitaries, and is additive under tensor products \cite{leone2022stabilizer,leone2024stabilizer}.
We use $\alpha=2$ throughout and write the resulting quantity as $M_{2}$, as in \cref{eq:M2}; the main text uses it as a test for the tiers of \cref{def:tiers} (\cref{prop:sre_certificate}).

The stabilizer R\'enyi entropy provides a computable description of the delocalization of a state's Pauli spectrum.
It is useful for detecting and comparing non-stabilizerness according to this particular functional, but it is not a complete operational ordering of magic resources.
In particular, a larger value does not by itself imply: a higher level of the Clifford hierarchy; a larger exact Clifford+$T$ synthesis cost; a larger distillation or preparation cost; a more useful resource for a given algorithm; or convertibility between two states under a specified set of free operations.

A unitary version of the entropy can be defined using the Choi state~\cite{hou2025stabilizer}
\begin{align}
\ket{U}
=
(U\otimes I)\ket{\Phi^{+}}^{\otimes n}~,
\qquad
\ket{\Phi^{+}}
=
\frac{\ket{00}+\ket{11}}{\sqrt{2}}~,
\end{align}
through
\begin{align}
H_{\alpha}(U)
=
M_{\alpha}(\ket{U})~.
\end{align}
For the diagonal operations considered here, this quantity agrees with the state entropy of the corresponding resource state.

\begin{lemma}
If $U$ is diagonal, then
\begin{align}
H_{\alpha}(U)
=
M_{\alpha}
\left(U\ket{+}^{\otimes n}\right)~.
\label{eq:diagonal_unitary_state_sre}
\end{align}
\end{lemma}

\begin{proof}
Write
\begin{align}
U
=
\sum_{x\in\{0,1\}^{n}}
e^{i\theta_{x}}\ket{x}\bra{x}~.
\end{align}
Its Choi state is
\begin{align}
\ket{U}
=
2^{-n/2}
\sum_{x} e^{i\theta_{x}}\ket{x}\ket{x}~.
\end{align}
Applying transversal $CX$ gates from the first register to the second maps this state to
\begin{align}
U\ket{+}^{\otimes n}
\otimes
\ket{0}^{\otimes n}~.
\end{align}
The result follows from Clifford invariance, tensor-product additivity, and the vanishing stabilizer R\'enyi entropy of $\ket{0}^{\otimes n}$.
\end{proof}

\cref{eq:diagonal_unitary_state_sre} is an identity for this entropy functional.
The identity does not imply that $U\ket{+}^{\otimes n}$ captures every operational cost associated with implementing $U$.

\subsection{The stabilizer R\'enyi entropy as a tier test}\label{section:sre_certificate}
The stabilizer R\'enyi entropy of a pure $n$-qubit state,
\begin{align}
M_{2}(\psi)=-\log_{2}\Bigl(2^{-n}\sum_{P\in\{I,X,Y,Z\}^{\otimes n}}\bra\psi P\ket\psi^{4}\Bigr)~,
\label{eq:M2}
\end{align}
is invariant under Clifford unitaries, additive under tensor products, and zero on stabilizer states~\cite{leone2022stabilizer,leone2024stabilizer}; it is the order-$2$ member of the family $M_{\alpha}$ of \cref{section:unitary_stab_entropy}.

\begin{proposition}[SRE test]\label[proposition]{prop:sre_certificate}
If a resource state is Clifford-equivalent to $\ket{c_{1}}\otimes\dots\otimes\ket{c_{m}}\otimes\ket{+}^{\otimes k'}$, then $M_{2}(\ket{\overline{\mathcal M}})=\sum_{i=1}^{m}M_{2}(\ket{c_{i}})$.
In particular this holds for every Tier-0 and Tier-1 state, with $\ket{c_{i}}$ its canonical factors.
\end{proposition}

\begin{proof}
$\mathcal F$ is a subgroup of the Clifford group, and $M_{2}(\ket+)=0$.
\end{proof}

A value of $M_{2}$ that matches no sum of canonical values therefore shows that the state is not equivalent to such a product even under the full Clifford group, which contains Hadamard gates and is larger than $\mathcal F$; the test covers only the canonical states whose values are compared.
The converse fails, so a match is only a consistency check.
\cref{tab:classification} lists the result for every code.
Every Tier-0 and Tier-1 state matches the sum of its canonical factors to within $10^{-5}$.
For every Tier-2 state, $M_{2}/M_{2}(CCZ)$ is not an integer, so none is Clifford-equivalent to $\ket{CCZ}^{\otimes r}\otimes\ket{+}^{\otimes(k-3r)}$; each $M_{2}$ also differs by at least $10^{-3}$ from every sum of the values of $\ket T$, $\ket{\sqrt T}$, $\ket{CS}$, $\ket{CT}$, $\ket{CCZ}$, $\ket{CCS}$ and $\ket{CCCZ}$ whose supports fit into $k$ qubits.

Beyond this test, $M_{2}$ also gives conversion bounds.
Leone and Bittel showed that $M_{2}$ cannot increase under a stabilizer protocol (Clifford gates, Pauli measurements, stabilizer ancillas and classical feed-forward) that maps a pure state to a pure state with certainty~\cite{leone2024stabilizer}.
Any such protocol that prepares the $\code{12,5,2}$ resource state from $m$ copies of $\ket{CCZ}$ therefore needs $m\ge\lceil2.476/1.541\rceil=2$, the same bound as \cref{prop:nullity}(ii), which two $CCZ$ gate teleportations attain; from $\ket T$ states it needs at least $\lceil2.476/0.415\rceil=6$, one more than the stabilizer-nullity bound $\nu(f)=5$.

However, $M_{2}$ is not a cost measure.
$M_{2}(CCCZ)=1.378$ is smaller than $M_{2}(CCZ)=1.541$ although $CCCZ$ lies one level higher in the Clifford hierarchy, and the $M_{2}$ of the $R$-star stays below $3$ as $R$ grows ($2.476$, $2.850$, $2.961$ and $2.990$ for $R=2,\dots,5$), whereas $R$ disjoint $CCZ$ gates give $1.541R$.
A larger $M_{2}$ therefore does not indicate a more valuable or more expensive resource.

\subsection{Numerical resource diagnostics}
\label{section:magic_properties}

For each selected CSS-TP code, we compute
\begin{align}
M_{2}
\left(\ket{\overline{\mathcal M}}\right)~,
\qquad
\ket{\overline{\mathcal M}}
=
\overline{\mathcal M}\ket{+}^{\otimes k}~.
\end{align}
We also report an upper bound on the number of non-Clifford phase rotations needed for the logical unitary without ancillas, the smaller of a \texttt{PyZX}~\cite{kissinger2020Pyzx} circuit and $7$ (or $15$) rotations per $CCZ$-type (or $CCCZ$-type) factor of an explicit representative; the latter is smaller in every multi-factor row (\texttt{PyZX} finds $19$, $17$, $41$, $41$ and $27$ rotations for the last five rows of \cref{table:SRE_results}).
The two columns are independent diagnostics: the first describes the Pauli spectrum of the resource state, while the second records the size of one explicit circuit for its logical unitary.
\input{includes/SRE_values}
The nonzero values in \cref{table:SRE_results} show that the listed states are non-stabilizer resources and show that their Pauli spectra vary substantially with the logical interaction structure.
They do not imply that a state with larger entropy is more expensive to synthesize, distill, or consume.
For example,
\begin{align}
M_{2}(\ket{CCCZ})
<
M_{2}(\ket{CCZ})~,
\end{align}
even though $CCCZ$ lies one level higher in the Clifford hierarchy.
The Clifford-hierarchy level records how conjugation acts on Pauli operators; the stabilizer R\'{e}nyi entropy instead records the distribution of Pauli expectation values.
These are distinct properties.
As a case in point, the same distinction explains why
\begin{align}
M_{2}(\ket{T})
&\approx 0.415~,
&
M_{2}(\ket{\sqrt T})
&\approx 0.193~.
\end{align}
The state $\ket{\sqrt T}=\sqrt T\ket{+}$ is closer to the stabilizer state $\ket{+}$ according to this Pauli-spectrum functional, despite the fact that $\sqrt T$ lies at a higher level of the Clifford hierarchy and is not exactly representable by a finite Clifford+$T$ circuit.

The final column of \cref{table:SRE_results} gives a separate upper bound from explicit synthesis.
This bound should be interpreted only as the size of the compiled logical-unitary circuit used in the calculation.

Beyond showing that the states are non-stabilizer, the values serve as the tier test of \cref{section:tiers}: Clifford invariance means that a logical basis transformation changes the shape of a resource but not its $M_{2}$, and additivity means that a tensor-product resource has exactly the sum of its canonical values.
Whether a particular state is practically useful is determined by the complete distillation, injection, compilation, and architectural resource model developed in \cref{section:CSSTP_distillation_protocols,section:tailored_algorithmic_primitives}.

%% file: includes/yield_table.tex
\begin{table}
\centering
\caption{Magic state yields $\gamma$ obtained for various code families and examples in this work.}
\label{table:yields}
\begin{tabular}{@{} | c | c | c | @{}}
\toprule
Asymptotic Parameters & Example Codes & $\gamma$ \\
\midrule
$\code{6m + 14, 3m + 6, 2}$ & - & $\mathbf{1.00}$ \\
- & $\code{14,6,2}_{m=0}$ & $1.22$ \\
- & $\code{20,9,2}_{m=1}$ & $1.15$ \\
- & $\code{26,12,2}_{m=2}$ & $1.12$ \\
$\code{6m+12, 3m+5, 2}$ & - & $\mathbf{1.00}$ \\
- & $\code{12,5,2}_{m=0}$ & $1.26$ \\
- & $\code{18,8,2}_{m=1}$ & $1.17$ \\
- & $\code{24,11,2}_{m=2}$ & $1.13$ \\
$\code{13m + 28, m + 2, 3}$ & - & $2.33$ \\
- & $\code{28,2,3}_{m=0}$ & $2.40$ \\
- & $\code{41,3,3}_{m=1}$ & $2.38$ \\
$\code{13m + 26, m + 2, d \geq 2}$ & - & $2.33$ \\
- & $\code{26,2,2}_{m=0}$ & $3.70$ \\
- & $\code{39,3,3}_{m=1}$ & $2.33$ \\
$\code{14m + 30, 4m + 8, 2}$ & - & $1.81$ \\
- & $\code{30,8,2}_{m=0}$ & $1.91$ \\
- & $\code{44,12,2}_{m=1}$ & $1.87$ \\
\bottomrule
\end{tabular}
\end{table}

%% file: includes/SRE_values.tex
\begin{table}
\centering
\caption{Non-stabilizerness and synthesis diagnostics for selected CSS-TP
resources. Here $\ell$ is the highest Clifford-hierarchy level occurring
in the logical phase operation, and $N_T^{\mathrm{UB}}$ is the upper bound of \cref{section:magic_properties} on the number of non-Clifford phase rotations in an ancilla-free circuit for the logical unitary. The tier of each state, and the use of $M_2$ as a tier test, are given in \cref{tab:classification}. The $\code{20,9,2}$ row is the open chain with $m=1$.}
\label{table:SRE_results}
\renewcommand{\arraystretch}{1.15}
\resizebox{0.25\textwidth}{!}{%
\begin{tabular}{@{}lcccc@{}}
\toprule
Resource
& $k$
& $\ell$
& $M_2(\ket{\overline{\mathcal M}})$
& $N_T^{\mathrm{UB}}$
\\
\midrule
$\ket{CS}$
& $2$ & $3$ & $1.193$ & $3$
\\
$\ket{CCZ}$
& $3$ & $3$ & $1.541$ & $7$
\\
$\ket{CCCZ}$
& $4$ & $4$ & $1.378$ & $15$
\\
$\ket{\mathcal M}_{\code{12,5,2}}$
& $5$ & $3$ & $2.476$ & $14$
\\
$\ket{\mathcal M}_{\code{14,6,2}}$
& $6$ & $3$ & $3.081$ & $14$
\\
$\ket{\mathcal M}_{\code{18,8,2}}$
& $8$ & $3$ & $4.430$ & $21$
\\
$\ket{\mathcal M}_{\code{30,8,2}}$
& $8$ & $4$ & $2.756$ & $30$
\\
$\ket{\mathcal M}_{\code{20,9,2}}$
& $9$ & $3$ & $4.622$ & $21$
\\
\bottomrule
\end{tabular}%
}
\end{table}

%% file: sections/E_distance_growth.tex
\section{Growing the Distance}\label{section:concatenation_distance_growth}

\begin{figure*}
\centering
\includegraphics[width=\textwidth]{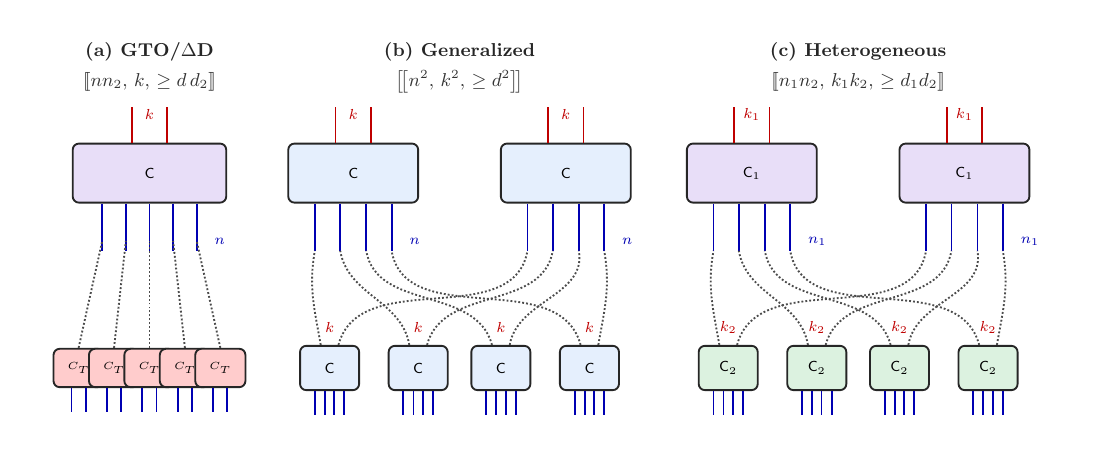}
\caption{
Three concatenation strategies for growing the distance of CSS-TP codes (\cref{section:concatenation_distance_growth}).
\textbf{(a)}~GTO/$\Delta$D concatenation: the outer code $\mathsf{C}$ (purple) with an inner $T$-type code $C_{T}$ (red) on every physical leg.
\textbf{(b)}~Generalized self-concatenation of $\mathsf{C}$; dotted lines show the cross-wiring.
\textbf{(c)}~Heterogeneous concatenation of $\mathsf{C}_{1}$ and $\mathsf{C}_{2}$ with the same wiring.
}
\label{fig:concatenation_types}
\end{figure*}
Concatenating CSS-TP codes provides a method for increasing code distance.
We identify several distinct strategies, illustrated in \cref{fig:concatenation_types}, each yielding higher-distance codes with distinct structural properties and trade-offs.
To start, each physical qubit of a CSS-TP code $\mathsf{C}$ can be encoded into an inner code satisfying $T^{\otimes n_{2}} = \overline{T}$~\cite{jain2025transversal}; this is shown in \cref{fig:concatenation_types}(a).
The resulting code has parameters $\code{n\,n_{2}, k, \geq d\,d_{2}}$ and preserves the logical action of $\mathsf{C}$.
This approach is particularly natural for \emph{Bravyi--Haah} magic state distillation protocols~\cite{bravyi2012magic,haah2018codes}.
As a concrete example, concatenating the $\code{12,5,2}$ code with the $\code{49,1,5}$ GTO code (for which $T^{\otimes 49} = \overline{T}$) yields a code with parameters $\code{588,5,\geq10}$ (the distance is the product lower bound) that preserves the logical gate $\overline{\mathcal{M}}_{\code{12,5,2}}$.

Generalized code concatenation takes in a CSS-TP code $\mathsf{C}$ and produces an $\code{n^{2},k^{2},\geq d^{2}}$ CSS-TP code by using $k$ copies of $\mathsf{C}$ at the first concatenation radius (the upper codes) and $n$ copies of the same code at the second (the lower, or inner, codes)~\cite{grassl2009generalized}; this is shown in \cref{fig:concatenation_types}(b).
Each inner code's $k$ logical legs connect to one physical leg of each upper code, as depicted by the dotted lines in the figure.
This scheme naturally gives rise to CSS-TP codes with unconventional resource states.
Consider the case $\mathsf{C} = \code{8,3,2}$.
Self-concatenation yields a $\code{64,9,4}$ code whose transversal $T$/$T^{\dagger}$ layer (\cref{fig:6494_uncleaned_gate}) implements a disjoint logical circuit of overlapping $\overline{CCZ}$ gates:
\input{includes/6494_uncleaned}
The structure of this gate follows from a concatenation rule.

\begin{lemma}[Logical action under concatenation]\label[lemma]{lem:concat}
Let the lower code carry a layer that is logical on it, with logical phase polynomial $\psi(v)=\sum_{Q}d_{Q}\prod_{c\in Q}v_{c}$ on its $K$ logical qubits, and apply that layer to each of the $n_{1}$ lower blocks of the generalized concatenation with $K$ upper copies.
On the encoded state whose upper copies carry the $X$-type codewords $y^{(1)},\dots,y^{(K)}$, the layer multiplies by $\omega^{F}$ with
\begin{align}
F=\sum_{p=1}^{n_{1}}\psi\bigl(y^{(1)}_{p},\dots,y^{(K)}_{p}\bigr)=\sum_{Q} d_{Q}\sum_{p=1}^{n_{1}}\prod_{c\in Q}y^{(c)}_{p}~.
\label{eq:concat_phase}
\end{align}
The layer is logical iff $F$ does not depend on the upper stabilizer coordinates, and the logical phase polynomial is then $F$ evaluated at $y^{(c)}=u^{(c)}L_{X}$.
\end{lemma}

\begin{proof}
The encoded state is a sum over upper stabilizer coordinates of $\bigotimes_{p}E_{2}\ket{y^{(1)}_{p}\cdots y^{(K)}_{p}}$, where $E_{2}$ encodes the lower code.
The layer on each lower block multiplies $E_{2}\ket v$ by $\omega^{\psi(v)}$, and summing over the blocks gives $F$.
The logical phase polynomial criterion follows as in \cref{lem:logical_poly}.
\end{proof}

For $\code{8,3,2}\circ\code{8,3,2}$ the lower polynomial is $\psi=4v_{1}v_{2}v_{3}$, and the upper codewords are affine functions $y^{(c)}_{p}=t_{c}\oplus\sum_{r}u^{(c)}_{r}p_{r}$ of the vertex $p\in\mathbb F_{2}^{3}$.
Modulo $2$, the sum of a function over $\mathbb F_{2}^{3}$ equals its coefficient of $p_{0}p_{1}p_{2}$.
For the product of three affine functions this coefficient is the permanent, equal modulo $2$ to the determinant, of the matrix $U=(u^{(c)}_{r})$, and it does not depend on $t$.
Hence
\begin{align}
\phi_{\code{64,9,4}}(U)=4\det U\pmod 8~,
\end{align}
which consists of the six $\overline{CCZ}$ gates on the triples $\{u^{(1)}_{\sigma(1)},u^{(2)}_{\sigma(2)},u^{(3)}_{\sigma(3)}\}$ for permutations $\sigma$; each logical qubit lies in two of them.
The lower $\code{8,3,2}$ code has no transversal $\overline T$: its layer acts as an entangling $\overline{CCZ}$ across the three upper copies, and that is why the concatenated gate is a structured product of $\overline{CCZ}$ gates rather than a copy of the upper gate.

One may generalize the previous strategy by using distinct codes at each concatenation radius; this is shown in \cref{fig:concatenation_types}(c).
The first radius consists of $k_{2}$ copies of a code $\mathsf{C}_{1}$ with parameters $\code{n_{1}, k_{1}, d_{1}}$ and the second of $n_{1}$ copies of a different code $\mathsf{C}_{2}$ with parameters $\code{n_{2}, k_{2}, d_{2}}$.
This \emph{heterogeneous code concatenation} produces a code with parameters $\code{n_{1}n_{2},k_{1}k_{2},\geq d_{1}d_{2}}$~\cite{steinberg2025universal,jochym2014using}.
For example, taking $\mathsf{C}_{1} = \code{8,3,2}$ and $\mathsf{C}_{2} = \code{12,5,2}$ yields a concatenated code with parameters $\code{96,15,4}$ and transversal logical non-Clifford gate shown in \cref{fig:9614_uncleaned_gate}
\input{includes/96154_uncleaned}
where $\mathcal{S}$ is the set of 54 triples $(a,b,c)$ with $a \in \{0,\ldots,4\}$, $b \in \{5,\ldots,9\}$, $c \in \{10,\ldots,14\}$.
By \cref{lem:concat}, each of the nine cubic terms of the $\code{12,5,2}$ polynomial, in the logical basis used for the lower code, contributes the six triples of a $3\times3$ determinant, giving $9\times6=54$.
The lower terms of degree at most two have coefficient $4$ and drop out, because a function of degree at most two sums to an even number over $\mathbb F_{2}^{3}$.

To combine the $\code{8,3,2}$ code with inner codes of higher distance, one might try to absorb a diagonal unitary into the encoding; however, this cannot create transversal diagonal gates.
If $W$ is diagonal and $\mathsf C'=W\mathsf C$, then a diagonal $D$ satisfies $D\,W\Pi_{\mathsf C}W^{\dagger}=W\,D\,\Pi_{\mathsf C}W^{\dagger}$, so $D$ is logical on $\mathsf C'$ exactly when it is logical on $\mathsf C$, with the same action.
An exact search over $\mathbb Z_{8}$ layers confirms this for the two cases one might try: neither $\code{8,3,2}\circ\code{7,1,3}=\code{56,3,6}$ nor $\code{8,3,2}\circ C_{6}=\code{48,6,4}$ (two outer copies) has a non-Clifford transversal diagonal gate at the $T$ level.

The working alternative is the strategy of \cref{fig:concatenation_types}(a) with a $T$-type inner code.
Concatenating the $\code{8,3,2}$ code with the $\code{15,1,3}$ punctured QRM code, and applying $T^{\otimes15}$ or $T^{\dagger\otimes15}$ to each inner block according to the sign of the $\code{8,3,2}$ layer on that leg, gives a $\code{120,3,6}$ code on which the layer implements exactly $\overline{CCZ}$.
The same construction on the $\code{12,5,2}$ code gives a $\code{180,5,6}$ code with exactly the \Mfac{12,5,2} logical phase polynomial.
Both distances were computed exactly and meet the lower bound $d\,d_{2}$.

%% file: includes/6494_uncleaned.tex
\begin{figure}
\centering
\includegraphics[width=0.5\textwidth]{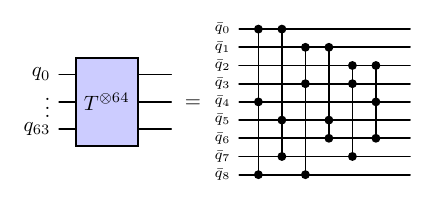}
\caption{The transversal non-Clifford logical gate $\overline{\mathcal{M}}_{\code{64,9,4}}$. The box on the left is schematic: the physical layer is the signed $\code{8,3,2}$ layer, $T$ on the even-weight and $T^\dagger$ on the odd-weight vertices of each inner cube, repeated on all eight inner blocks ($32$ $T$ and $32$ $T^\dagger$); we checked that this layer is logical and that its action is the six-term cubic phase shown.}
\label{fig:6494_uncleaned_gate}
\end{figure}

%% file: includes/96154_uncleaned.tex
\begin{figure*}
\centering
\includegraphics[width=0.75\textwidth]{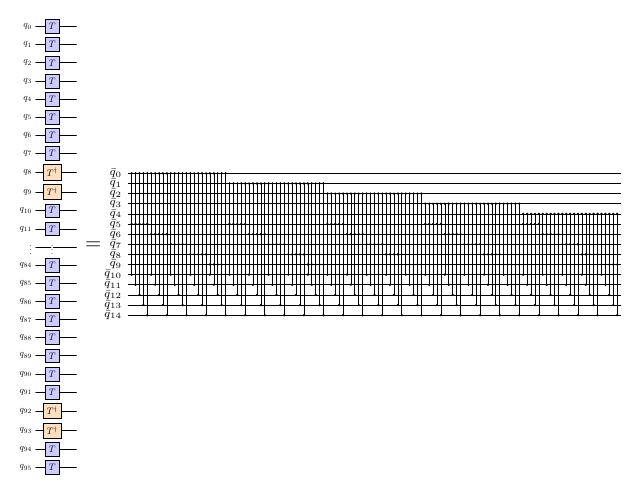}
\caption{The transversal non-Clifford logical gate $\overline{\mathcal{M}}_{\code{96,15,4}}$. The physical layer is shown schematically: it is a layer of six $T$ and six $T^\dagger$ gates that is logical on the $\code{12,5,2}$ code, repeated on all eight inner blocks ($48$ $T$ and $48$ $T^\dagger$); we checked that this layer is logical and that its action is the $54$-term cubic phase described in the text.}
\label{fig:9614_uncleaned_gate}
\end{figure*}

%% file: sections/F_random_walk.tex
\section{Random Walk Algorithm in Phase Polynomial Space}\label{section_random_walk_alg}

The algorithm performs a random walk over phase polynomials representing diagonal quantum gates on $k$ logical qubits.
The initial state is encoded as a set of polynomial terms, where each term is a subset of qubit indices.
Degree-three terms correspond to $\mathrm{CCZ}$ gates, degree-two terms correspond to $\mathrm{CZ}$ gates, and degree-one terms correspond to $\mathrm{Z}$ gates.
Terms from higher levels of the Clifford hierarchy need the $\mathbb Z_{2^{\ell}}$ update described below.
Since the polynomial is defined over $\mathbb{F}_{2}$, duplicate terms cancel modulo two; computationally, this is implemented using the symmetric difference identity we discussed in \cref{section:CSSTP_gate_cleaning}.

At each step, the algorithm randomly selects a control qubit $c$ and a target qubit $t$, then applies the action of conjugation by a $\mathrm{CX}_{c,t}$ gate.
Algebraically, this is equivalent to the Boolean substitution
\begin{align}
x_{t} \mapsto x_{t} \oplus x_{c}~.
\end{align}
Any polynomial term containing $x_{t}$ is therefore expanded into two contributions: one retaining $x_{t}$ and one replacing it with $x_{c}$.
Duplicate terms cancel modulo two, and overlapping variables can naturally lower the degree of a term.
This $\mathbb F_{2}$ update is exact for gates of the form $(-1)^{f(u)}$.
For gates with $\mathbb Z_{2^{\ell}}$ coefficients ($T$, $S$, $CS$, $\sqrt T$), $CX$ substitution generates additional cross terms, e.g.\ $CX_{c,t}\,T_{t}\,CX_{c,t}=T_{t}T_{c}\,CS^{\dagger}_{tc}$, and the update must be carried out in $\mathbb Z_{2^{\ell}}$ with $a\oplus b=a+b-2ab$.
Repeating this process samples the orbit of the initial diagonal phase resource under randomly chosen $\mathrm{CX}$ transformations.
An unbiased walk only samples the orbit, which for \Mfac{14,6,2} already has $357{,}120$ cubic classes.
\cref{alg:random_walk_phase_polynomials} therefore biases the walk: the energy is the number of monomials (modulo terms of degree below $\delta$), moves are accepted with the Metropolis rule under an increasing inverse temperature, the walk returns to the best state found after $W$ steps without improvement, and it stops as soon as a target predicate holds or after $N_{\max}$ steps.
With $X$ moves it also absorbs lower-degree terms of quartic polynomials.
Every move is recorded, and the returned sequence is replayed to check the result.
Our tests start from random orbit points (obtained by $40$ to $60$ random $CX$ moves) and recover two disjoint $CCZ$ gates for \Mfac{14,6,2}, the two-$CCZ$ star for \Mfac{12,5,2} (all $10$ seeds each, median $1{,}294$ and $17$ steps), the literal star for $R=3,4,5$, and two disjoint $CCCZ$ gates from the $21$-monomial action of the $\code{30,8,2}$ code.
The same routine produces the Tier-1 move sequences of \cref{tab:classification}.
Below we detail in \cref{alg:random_walk_phase_polynomials} a high-level summary of the algorithm.

\input{includes/random_walk_pseudocode}

%% file: includes/random_walk_pseudocode.tex
\begin{center}
\vspace{1em}
\end{center}

\begin{algorithm}[H]
\caption{Metropolis walk on the $CX$ (and $X$) orbit of a Boolean phase polynomial, modulo terms of degree below $\delta$}
\label{alg:random_walk_phase_polynomials}

\KwIn{ANF $P$ of $f$ (set of monomials) on $k$ qubits; cut-off degree $\delta$ ($3$ for cubic classes mod $RM(2)$); target predicate $\mathcal T$ (e.g.\ $r$ disjoint monomials, a literal star); schedule $\beta_0\to\beta_1$; step budget $N_{\max}$; patience $W$; $X$-move rate $q$; seed}
\KwOut{found flag, representative $P^*$, move list $\mathcal L^*$ with $P^*=\mathcal L^*(P)$}
$\mathcal L\gets()$;\ $(P^*,\mathcal L^*)\gets(P,())$;\ $w\gets0$\;
\For{$i=0,\dots,N_{\max}-1$}{
    \lIf{$\mathcal T(P)$}{\Return (true, $P$, $\mathcal L$)}
    \eIf{$X$ moves allowed and $\mathrm{rand}()<q$}{
        draw $j$; $Q\gets P(x_j\to x_j\oplus1)$, dropping monomials of degree $<\delta$; $m\gets X_j$\;
    }{
        draw $c\ne t$; $Q\gets P(x_t\to x_t\oplus x_c)$, dropping monomials of degree $<\delta$; $m\gets CX_{c,t}$\;
    }
    $\Delta E\gets|Q|-|P|$;\ $\beta\gets\beta_0+(\beta_1-\beta_0)\,i/N_{\max}$\;
    \If{$\Delta E\le0$ or $\mathrm{rand}()<e^{-\beta\Delta E}$}{
        $P\gets Q$;\ append $m$ to $\mathcal L$\;
        \lIf{$|P|<|P^*|$}{$(P^*,\mathcal L^*)\gets(P,\mathcal L)$;\ $w\gets0$}
    }
    $w\gets w+1$;\ \lIf{$w>W$}{$(P,\mathcal L)\gets(P^*,\mathcal L^*)$;\ $w\gets0$}
}
\Return ($\mathcal T(P^*)$, $P^*$, $\mathcal L^*$)\;
\BlankLine
\end{algorithm}

%% file: sections/G_resource_model.tex
\section{Resource Model}\label{section:spacetime_model}

This appendix explains the resource model behind every comparison in \cref{section:tailored_algorithmic_primitives}, for a reader who knows what a qubit and an error-correcting code are but has not worked with lattice surgery.
We first introduce the units, then follow a magic state from its preparation to its consumption, and finally work one small example end to end.

\subsection{Units}\label{section:appendix_model_units}
A logical qubit is stored in a distance-$d$ rotated surface-code patch: $d^2$ data qubits and $d^2-1$ measurement qubits.
Following Ref.~\cite{litinski2019game}, the layout is a grid of \emph{tiles}; a tile holds one patch, or an equal area of routing space used for lattice surgery between patches.
We charge
\begin{align}
q_{\rm tile}=2d^2
\end{align}
physical qubits per tile, i.e.\ $\apxtile$ qubits at $d=\apxd$.
Areas (data blocks, factories, storage) are counted in tiles.

Time, in turn, is counted in cycles of syndrome extraction: one \emph{round} is one such cycle on every patch.
A lattice-surgery operation (a multi-patch Pauli-product measurement) must be repeated for $d$ rounds to be fault tolerant, so one \emph{logical step} lasts $d$ rounds ($\apxstep$ at $d=\apxd$).
Durations are reported in rounds.

Every tile is a source of error: each tile fails with probability
\begin{align}
p_L(d)=0.1\left(\frac{p}{p_{\rm th}}\right)^{(d+1)/2}\label{eq:appendix_pL}
\end{align}
per round~\cite{litinski2019game}, with physical error rate $p=\apxpphys$ and threshold $p_{\rm th}=\apxpth$.
This gives $p_L(\apxd)=\apxpLd$ and $p_L(23)=\apxpLtwentythree$.

Combining area and time, the spacetime volume of a schedule is the integral of the occupied physical qubits over its duration, in qubit-rounds.
A tile that is occupied for $r$ rounds contributes $r$ tile-rounds, i.e.\ $q_{\rm tile}\,r$ qubit-rounds.
The \emph{footprint} is the number of physical qubits installed, whether or not they are busy.
The \emph{latency} $T$ of a circuit is the number of rounds until its last consumption ends, averaged over the random $T$ supply, and its \emph{installed volume} is footprint $\times$ latency.
Volume without qualification means occupied volume.

\subsection{Where a \texorpdfstring{$T$}{T} state comes from: cultivation}\label{section:appendix_model_cultivation}

Every factory in this paper is fed with $\ket{T}$ states prepared by magic-state cultivation~\cite{gidney2024magic}.
Cultivation grows a $T$ state inside a small color code of distance $d_1$, checks it repeatedly, and finally grows it into a surface code of distance $d_2$ and decodes; an attempt is kept only if the decoder's complementary gap, a measure of its confidence, exceeds a chosen threshold.
A larger gap threshold keeps fewer attempts, each with a lower error.
We simulated the end-to-end cultivation circuits of Ref.~\cite{gidney2024magic} with $d_1=3$, grown directly into the $d_2=\apxd$ surface code of the data, at physical error rate $p=10^{-3}$, and use a gap threshold of $70$\,dB for all factory inputs.
A \emph{lane} is a reserved region of $\apxlaneq$ physical qubits, the footprint of one cultivation attempt, that runs attempts one after another.

We now follow a single lane.
Each attempt lasts $\apxkeptlen$ rounds and is kept with probability $a=\apxaccept$; a kept state has error $p_T=\apxpT$.
The kept fraction is the fraction of attempts that pass the gap threshold, $0.241$, divided by the fraction that pass the initial color-code encoding, $0.862$: the $d_1=3$ color code occupies about $2\%$ of the lane, so a failed encoding is detected and restarted at negligible cost and is not charged.
We charge every attempt its full $\apxkeptlen$ rounds, so on average $1/a=\apxattemptsperkept$ attempts are needed per kept state.
The time a lane needs to deliver one kept state is therefore random: it is at least $\apxkeptmin$ rounds (probability $\apxkeptpfirst\%$, the first attempt is kept), its median is $\apxkeptmedian$ rounds, its 90th percentile $\apxkeptpninety$ rounds, and its mean
\begin{align}
\bar t_T=\frac{\apxkeptlen}{a}=\apxmeankept\ \text{rounds}.
\end{align}
The $\apxlanes$ lanes of our default configuration thus supply about $\apxrate$ states per round.

As for its cost, we charge the lane's full $\apxlaneq$ qubits for every round it is running, i.e.\ $\apxlaneq\times\apxmeankept=\apxqrperkept$ qubit-rounds per kept state, because the region cannot be reused while an attempt may still grow into it.
Counting only the qubits activated at each stage of an attempt, as in Ref.~\cite{gidney2024magic}, gives about $\apxgidneyqr$ qubit-rounds per kept state; we report it as a diagnostic only.
For comparison, a deterministic lane of $\apxvtwoq$ qubits delivering one state every $\apxvtwoperiod$ rounds in lockstep would be about $\apxsupplyratio\times$ slower per lane than the lane above, which is larger and random.

\begin{figure*}[t]\centering
\panelimage[width=0.49\textwidth]{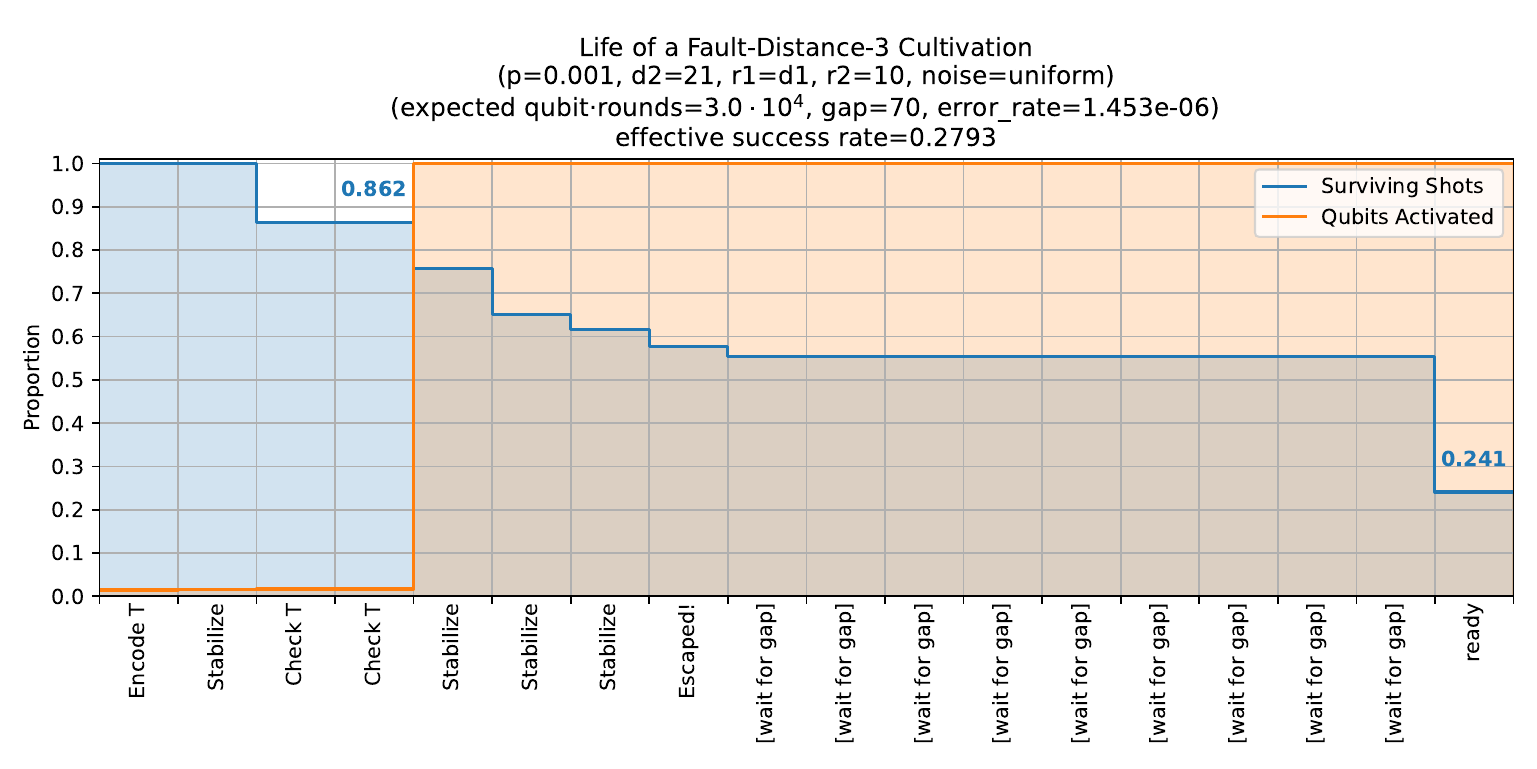}{(a)}\hfill
\panelimage[width=0.49\textwidth]{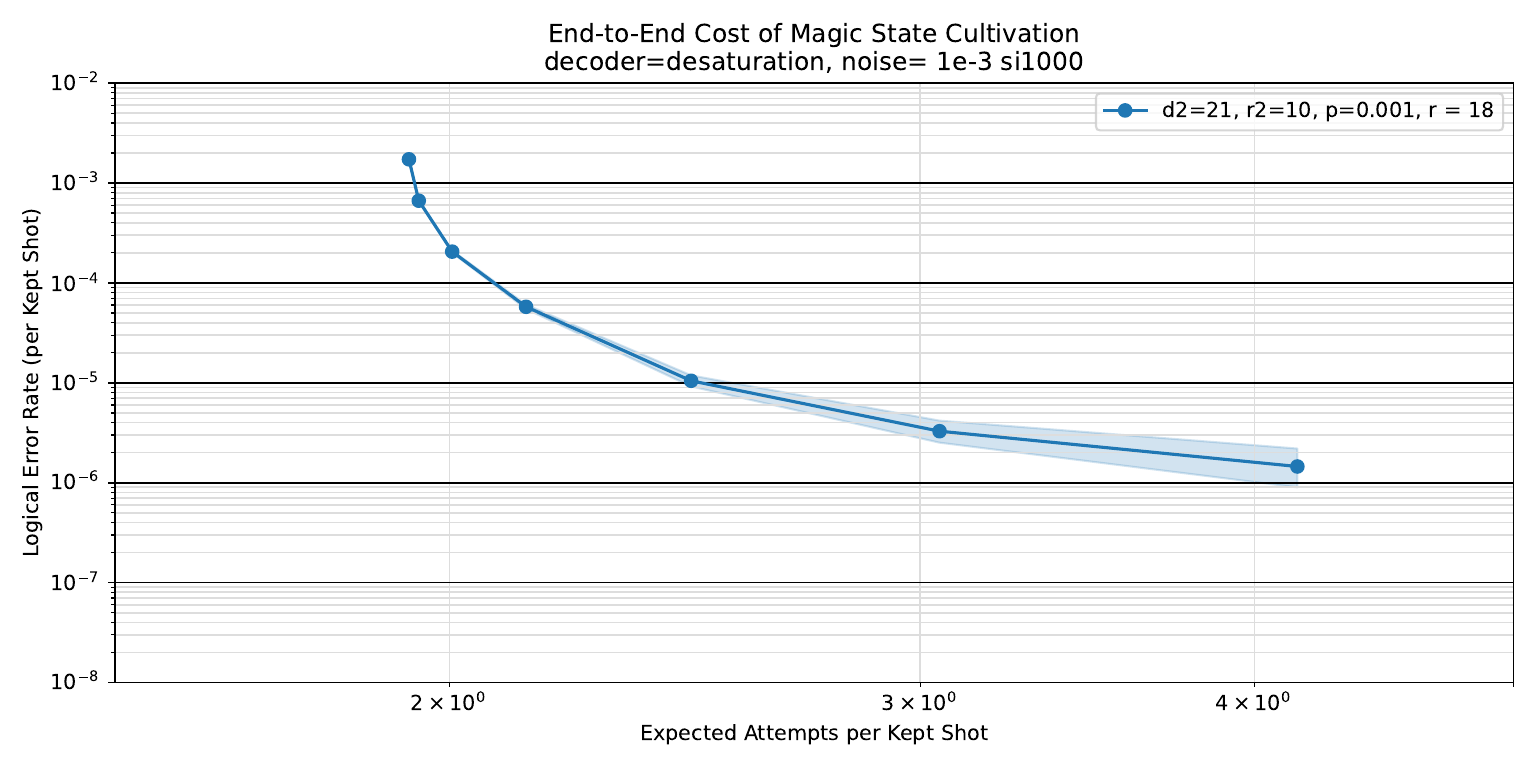}{(b)}
\caption{End-to-end simulation of magic-state cultivation with $d_1=3$ grown into a $d_2=21$ surface code ($p=10^{-3}$, SI1000 noise, $r=18$ rounds of which $r_2=10$ after the escape). (a)~Fraction of surviving shots (blue) and of active qubits (orange) after each round of one attempt, labeled by circuit stage. (b)~Logical error rate of a kept shot against the expected number of attempts per kept shot, one point per gap threshold.}\label{fig:appendix_cultivation}
\end{figure*}

These lane parameters come from our own simulation of cultivation.
\cref{fig:appendix_cultivation} shows its two outputs, from which $p_T$, $a$, the attempt length and the lane size are read off.
The circuits are the end-to-end cultivation circuits of Ref.~\cite{gidney2024magic} with $d_1=3$, $d_2=21$, the $CZ$ gate set and the SI1000 noise model at $p=10^{-3}$, decoded with the desaturation decoder of that reference.

Panel~(a) follows the shots of one attempt through its $18$ rounds; each column is one round, labeled by the stage of the circuit it executes.
The blue curve is the fraction of shots that have not yet been discarded, and the orange curve the fraction of the lane's qubits that are active.
In the first seven rounds the $T$ state is encoded into the $d_1=3$ color code, checked twice and stabilized; any detection event discards the shot, and $0.862$ of the shots survive this stage.
The state then escapes into the $d_2=21$ surface code, at which point all $\apxlaneq$ qubits of the lane become active, and the remaining rounds collect the syndrome data from which the decoder computes the complementary gap.
At the last round (\emph{ready}) the shots with a gap below $70$\,dB are discarded, leaving $0.241$.
Our kept fraction $a=0.241/0.862$ conditions on surviving the color-code stage, as explained above; the title of the panel gives the same ratio, $0.2793$, computed from the unrounded survivals, together with the error rate of the kept shots and the expected number of active qubit-rounds per kept shot, the diagnostic $\apxgidneyqr$ quoted above.
The small active footprint before the escape is why our reserved-lane accounting, which charges all $\apxlaneq$ qubits for all $18$ rounds, is conservative.

Panel~(b) shows the trade-off that the gap threshold controls.
Each point is one threshold: the horizontal axis is the expected number of attempts per kept shot and the vertical axis the logical error rate of a kept shot.
Raising the threshold discards more shots and lowers the error of the ones kept.
We operate at $70$\,dB, where the kept error is $p_T=\apxpT$, low enough that the distillation error $\epsilon_{\rm ETD}\approx A_2p_T^2$ of every third-level factory in this paper is one to three orders of magnitude below the topological error of the circuits at $d=21$ (\cref{section:comparison_framework}).

\subsection{What a distillation factory does}\label{section:appendix_model_factory}

An Encode-$T$-Decode (ETD) factory~\cite{litinski2019game,litinski2019magic} on an $\code{n,k,2}$ CSS-TP code with $m_x$ $X$-stabilizers consumes $n$ raw $T$ states and, if no error is detected, outputs one $k$-qubit resource state.
We use the \Mfac{8,3,2} factory, on the $\code{8,3,2}$ code, as the worked example: $\apxseightn$ inputs make one $\ket{CCZ}$ on $k=\apxseightk$ qubits, and there is $m_x=\apxseightmx$ $X$-stabilizer.

We first determine how long one run takes.
The $n$ transversal $T$ rotations are applied as Pauli-product rotations on the $k+m_x$ encoded patches, one logical step each; $m_x$ of them are absorbed into the $\ket{+}$ preparation of the $X$-stabilizer patches.
The run therefore takes
\begin{align}
\tau=n-m_x
\end{align}
logical steps, $\tau=\apxseighttau$ steps $=\apxseighttaur$ rounds for \Mfac{8,3,2}.

During the run, the encoded patches sit in a compact block ($1.5N+3$ tiles for $N=k+m_x$ patches, \cref{section:appendix_model_blocks}) plus one tile where the raw $T$ states are injected:
\begin{align}
F=1.5(k+m_x)+4,
\end{align}
$F=\apxseightF$ tiles for \Mfac{8,3,2}.
The output waits for consumption in a storage area of $k$ tiles.
As a check, the same two rules applied to Litinski's 15-to-1 protocol ($n=15$, $k=1$, $m_x=4$) give $\apxlitsteps$ steps and $\apxlittiles$ tiles, against his $11$ steps and $11$ tiles~\cite{litinski2019game}.

Duration and area fix the cost of a run; its benefit is error suppression, for which we first need an error model of the inputs.
A faulty raw $T$ state can be in any state close to $\ket T$.
Applying, at random, either the identity or the Clifford $e^{-i\pi/4}SX$, which leaves the ideal $\ket T$ unchanged, turns every such fault into a $Z$ error with some probability; this is \emph{Clifford twirling}~\cite{bravyi2005universal}.
We therefore model each raw $T$ state as the ideal state followed by a $Z$ error with probability $p_T$.

With this model, a faulty raw $T$ state carries a $Z$ error on its input.
An odd number of $Z$ errors anticommutes with the weight-$8$ $X$-stabilizer and is detected; the run is then discarded and repeated with fresh inputs.
The run is accepted with probability $q=(1-p_T)^n+O(p_T^2)$, here $q=\apxseightq$ (rejection $\apxseightrej$).
Errors of even weight are not detected.
Let $A_w$ be the number of weight-$w$ $Z$ patterns that are undetected and act non-trivially on the output.
For $\code{8,3,2}$, $A_2=\binom82=\apxseightAtwo$ and $A_4=\binom84-14=\apxseightAfour$, since the 14 weight-4 $Z$-stabilizers act trivially.
The output error per accepted output is
\begin{align}
\epsilon_{\rm ETD}=\frac{1}{q}\sum_w A_w\,p_T^w(1-p_T)^{n-w}\approx A_2p_T^2,
\end{align}
here $\apxseighteE$ (leading term $\apxseightlead$).
\cref{tab:appendix_factories} lists these quantities for all factories used.

Distillation is not the only source of error, however.
While it runs, the factory occupies $F\tau d=\apxseightrunTR$ tile-rounds, each failing with probability $p_L(d)$.
Lattice surgery during the run therefore adds $\apxseightrunEthree$ per run at $d=\apxd$, more than an order of magnitude above $\epsilon_{\rm ETD}$, so at $d=\apxd$ the lattice-surgery error, not the distillation error, dominates.

\input{includes/generated/appendix_factories.tex}

\subsection{How a resource is consumed: data blocks}\label{section:appendix_model_blocks}

The algorithm's $N$ logical data qubits sit in a data block.
Consuming a CCZ (or a tailored resource) teleports its gate into the data by one Pauli-product measurement between the stored resource and the data block, which lasts $\delta$ logical steps.
The three data blocks of Ref.~\cite{litinski2019game} trade area against $\delta$ (\cref{tab:appendix_regimes}).

A remaining question is whether a block can carry out several consumptions at once.
The answer depends on its routing, which we do not lay out.
We therefore use two rules that bracket it.
\begin{itemize}
\item \emph{Serial consumption} (the default): one consumption runs on a block at a time, whatever its support. This is the accounting of the data blocks of Ref.~\cite{litinski2019game}, where a Pauli-product measurement may use the routing space of the whole block.
\item \emph{Parallel consumption}: any number of consumptions run at the same time provided their logical supports on the data block are pairwise disjoint. Each still lasts $\delta$, consumptions that share a data qubit are still serial, and the dependencies of the circuit are still respected. No routing area is added for the simultaneous measurements.
\end{itemize}
Serial consumption is pessimistic for a block that can route several measurements at once; parallel consumption is optimistic, since it charges nothing for the extra routing.
The rule matters most for \Mfac{8,3,2}: its many single-$CCZ$ outputs often act on disjoint qubits, whereas a tailored output already packs $CCZ$ gates that could run together into one consumption.
The fixed-configuration tables use serial consumption, and rows marked fast$^\parallel$ use parallel consumption.
The worked example below has no data block (as in \cref{tab:standalone_factory_costs}), but still pays $\delta$ per consumption in the fast block, $\delta=\apxdeltar$ rounds.
The tables use serial consumption, and rows marked fast$^\parallel$ use parallel consumption.

\input{includes/generated/appendix_regimes.tex}

\subsection{An end-to-end example}\label{section:appendix_model_worked}

The target is the pair of CCZ gates on disjoint supports (the \Mfac{14,6,2} target) or on supports sharing one qubit (the \Mfac{12,5,2} target) of \cref{tab:standalone_factory_costs}, at $d=\apxd$ with $\apxlanes$ lanes.
Three strategies make them, each with one factory:
\begin{itemize}
\item \textbf{\Mfac{8,3,2}}: one \Mfac{8,3,2} factory, run twice;
\item \textbf{\Mfac{12,5,2}} / \textbf{\Mfac{14,6,2}}: one tailored factory, run once, whose output is consumed once.
\end{itemize}

Each schedule begins with a race among the lanes.
Nothing is prepared before time zero.
Idle lanes start whenever raw states are still missing and the raw buffer has room.
The buffer holds two ETD batches per factory, so at most $2n$ lanes per factory are in flight: \apxseightactive{} for \Mfac{8,3,2}, \apxttwelveactive{} for \Mfac{12,5,2} and \apxtfourteenactive{} for \Mfac{14,6,2}.
When the last needed state arrives, every attempt still running is aborted; its rounds so far are charged.
Let $t_{(j)}$ be the time of the $j$-th arrival among $L$ lanes started together (an order statistic).
Its mean follows exactly from $G(t)$, the probability that one lane has delivered its kept state by round $t$ (computed from the attempt-length distribution):
\begin{align}
\mathbb E\,t_{(j)}=\sum_{t\ge0}\,\sum_{i<j}\binom Li G(t)^i\left(1-G(t)\right)^{L-i}.
\end{align}
The fastest of many lanes arrive well before the mean $\bar t_T$.
On average the 12th of 24 arrives after $\apxttwelvetnexact$ rounds, and the 8th of 16 after $\apxseighttnexact$.

From these arrival times the latency of each strategy follows.
Write $\tau$ in rounds and $\delta=\apxdeltar$.
\begin{itemize}
\item \textbf{\Mfac{12,5,2}, \Mfac{14,6,2}}: the factory starts at $t_{(n)}$ and its output is consumed once:
\begin{align}
T=t_{(n)}+\tau+\delta,
\end{align}
so the latency is $\apxttwelvetnexact+\apxttwelvetaur+\apxdeltar=\apxttwelvelatexact$ rounds for \Mfac{12,5,2} and $\apxtfourteentnexact+\apxtfourteentaur+\apxdeltar=\apxtfourteenlatexact$ rounds for \Mfac{14,6,2}.
\item \textbf{\Mfac{8,3,2}}: batch one is full at $t_{(8)}=\apxseighttn$.
The eight lanes that delivered restart and fill batch two long before the first run ends at $\apxseightendone$.
The second run therefore starts when the first ends; the first output is consumed while the second run is in progress:
\begin{align}
T=t_{(8)}+2\tau+\delta=\apxseightlat\ \text{rounds}.
\end{align}
\end{itemize}
\Mfac{8,3,2} is limited by its single factory, and \Mfac{12,5,2} and \Mfac{14,6,2} by their longer ETD runs ($\tau=\apxttwelvetau$ and $\apxtfourteentau$ steps against $\apxseighttau$).
With one factory each, \Mfac{12,5,2} and \Mfac{14,6,2} finish before \Mfac{8,3,2} because one run replaces two.
The tailored resources also need one consumption instead of two, which matters when many consumptions queue on one data block (\cref{section:appendix_model_capacity}).
\cref{fig:appendix_timeline} draws the three schedules.

Turning from time to space, the volume adds four occupied areas:
\begin{align}
V=q_{\rm tile}\Big(\underbrace{F\tau\cdot\text{runs}}_{\text{factories}}+\underbrace{k\cdot t_{\rm stored}}_{\text{storage}}+\underbrace{\textstyle\sum t_{\rm wait}}_{\text{raw buffer}}\Big)+q_{\rm lane}\sum t_{\rm lane}.
\end{align}
The terms are the factory runs, the outputs held in storage until their consumption ends, and the raw states waiting for a batch (one tile each).
The last term is the lane-rounds charged in the race.
For \Mfac{8,3,2} these are $\apxseightfacTR+\apxseightstoreTR+\apxseightrawTR=\apxseightTR$ tile-rounds and $\apxseightlaneR$ lane-rounds.
This gives $V=\apxseighttileV+\apxseightlaneV$ qubit-rounds $=\apxseightvol\times10^6$.
The raw-buffer term is large for \Mfac{8,3,2} because batch two waits for the only factory.
For \Mfac{12,5,2} it is $\apxttwelvefacTR+\apxttwelvestoreTR+\apxttwelverawTR=\apxttwelveTR$ tile-rounds and $\apxttwelvelaneR$ lane-rounds, $V=\apxttwelvevol\times10^6$.

The footprint, by contrast, counts every installed qubit, busy or not: the factories, their storage, their raw buffers of two ETD batches ($2n$ tiles per factory) and all lanes: $(F+k+2n)\,q_{\rm tile}\cdot N_{\rm fac}+\apxlanes\,q_{\rm lane}$.
For \Mfac{8,3,2} this is $\apxseightfacfp+\apxseightbuffp+\apxseightlanefp=\apxseightfp$ qubits.

Finally, the total error per circuit is the sum of three terms:
\begin{itemize}
\item $E_1$: cultivated states consumed directly, without distillation (the $\sqrt T$ rotations of \cref{section:fourth_level}); zero here;
\item $E_2$: the $\epsilon_{\rm ETD}$ of every consumed output;
\item $E_3$: the topological error, i.e.\ the total tile-rounds times $p_L(d)$.
\end{itemize}
For \Mfac{8,3,2}, $E_2=2\epsilon_{\rm ETD}=\apxseightEtwo$ and $E_3=\apxseightTR\times\apxpLd=\apxseightEthree$.
For \Mfac{12,5,2}, $E_2=\apxttwelveEtwo$ and $E_3=\apxttwelveEthree$.
ETD rejections add $\tau(1/q-1)\approx\apxseightretry$ rounds per run on average and are negligible here.

As a check, \cref{tab:appendix_worked} evaluates these formulas on sampled lane races and compares them with the event-driven scheduler used for every table of the main text.
Latency agrees within $\apxmaxreldiff\%$ and volume within $\apxmaxrelvol\%$; footprints and $E_2$ agree exactly.

\input{includes/generated/appendix_worked.tex}
\input{includes/generated/appendix_timeline.tex}

\subsection{Supply and capacity}\label{section:appendix_model_capacity}

The worked example fixes the number of lanes; varying it shows how the supply limits the schedule.
With few lanes, the factories wait for $T$ states and every strategy is slow.
With one lane, \Mfac{8,3,2} needs $\apxseightlanesone$ rounds and \Mfac{14,6,2} needs $\apxtfourteenlanesone$.
Once the lanes can fill the raw buffer, the schedule is limited by the factories.
\Mfac{8,3,2} settles at $\apxseightlanessixtyfour$ rounds and \Mfac{14,6,2} at $\apxtfourteenlanessixtyfour$ with 64 lanes; at 8 lanes they are at $\apxseightlaneseight$ and $\apxtfourteenlaneseight$ (\cref{fig:cultivation_capacity}).
Because a lane's kept time is random, the relevant quantity is the order statistic $t_{(n)}$, not $n\bar t_T/L$.
Deterministic lanes would make every batch arrive at the same moment and hide this effect.

The worked example also fixes the distance at $d=\apxd$, where $E_3$ dominates the error of every strategy, so a strategy with more tile-rounds is less accurate as well as larger; all comparisons in this paper are at this distance.

%% file: includes/generated/appendix_factories.tex
\begin{table*}[t]\centering
\caption{ETD factory parameters. A run lasts $\tau=n-m_x$ logical steps on $F=1.5(k+m_x)+4$ tiles, and the output is stored on $k$ tiles. $A_w$ counts the weight-$w$ $Z$-error patterns on the $n$ inputs that are undetected and act non-trivially. $\epsilon_{\rm ETD}$ is the exact output error per accepted output, summed over $w\le4$ at $p_T=1.45\times10^{-6}$, and $1-q$ is the rejection probability.}\label{tab:appendix_factories}
\widetable{\begin{tabular}{ll|rrr|rrr|rr|rr}\toprule
 & & \multicolumn{3}{c|}{Code} & \multicolumn{3}{c|}{Run} & \multicolumn{2}{c|}{Undetected errors} & \multicolumn{2}{c}{Output} \\
Name & Code & $n$ & $k$ & $m_x$ & $\tau$ & $F$ & Store & $A_2$ & $A_4$ & $\epsilon_{\rm ETD}$ & $1-q$ \\\midrule
\Mfac{8,3,2} & $[[8,3,2]]$ & 8 & 3 & 1 & 7 & 10 & 3 & 28 & 56 & $5.9\times10^{-11}$ & $1.2\times10^{-5}$ \\
\Mfac{12,5,2} & $[[12,5,2]]$ & 12 & 5 & 1 & 11 & 13 & 5 & 66 & 480 & $1.4\times10^{-10}$ & $1.7\times10^{-5}$ \\
\Mfac{14,6,2} & $[[14,6,2]]$ & 14 & 6 & 1 & 13 & 14.5 & 6 & 91 & 987 & $1.9\times10^{-10}$ & $2.0\times10^{-5}$ \\
\Mfac{16,7,2} & $[[16,7,2]]$ & 16 & 7 & 1 & 15 & 16 & 7 & 120 & 1792 & $2.5\times10^{-10}$ & $2.3\times10^{-5}$ \\
\Mfac{20,9,2} & $[[20,9,2]]$ & 20 & 9 & 1 & 19 & 19 & 9 & 190 & 4800 & $4.0\times10^{-10}$ & $2.9\times10^{-5}$ \\
\Mfac{24,11,2} & $[[24,11,2]]$ & 24 & 11 & 1 & 23 & 22 & 11 & 276 & 10560 & $5.8\times10^{-10}$ & $3.5\times10^{-5}$ \\
\bottomrule\end{tabular}}\end{table*}

%% file: includes/generated/appendix_regimes.tex
\begin{table}[t]\centering
\caption{The data blocks of Ref.~\cite{litinski2019game} as used by the model. Consuming a resource is one Pauli-product measurement lasting $\delta$ logical steps ($\delta\cdot d$ rounds at $d=21$). The area is in tiles for $N$ logical data qubits: $1.5N+3$ for the compact, $2N+4$ for the intermediate and $2N+\sqrt{8N}+1$ for the fast block. The columns give it for $N=6$ (the worked example), $N=24$ (the $GF(2^8)$ multiplier) and $N=132$ (the 2-lane carry-lookahead adder).}\label{tab:appendix_regimes}
\begin{tabular}{lrrrrr}\toprule
Block & $\delta$ (steps) & $\delta$ (rounds) & $N=6$ & $N=24$ & $N=132$ \\\midrule
compact & 9 & 189 & 12.0 & 39.0 & 201.0 \\
intermediate & 5 & 105 & 16.0 & 52.0 & 268.0 \\
fast & 1 & 21 & 19.9 & 62.9 & 297.5 \\
\bottomrule\end{tabular}\end{table}

%% file: includes/generated/appendix_worked.tex
\begin{table*}[t]\centering
\caption{Worked example in which two $CCZ$ gates are produced and consumed with no data block, at $d=21$ with 32 cultivation lanes and fast consumption ($\delta=21$ rounds). ``Closed form'' evaluates the formulas of \cref{section:appendix_model_worked} on 20,000 samples of the cultivation lanes, and ``scheduler'' is the event-driven model of the main text (2,000 seeds). $n\times$runs is the number of $T$ inputs per ETD run times the number of runs, ``Lanes'' is the number of lanes in flight (capped by the two-batch raw buffer), $t_{(n)}$ the exact mean arrival of the $n$-th $T$ state, $\tau$ the ETD duration in rounds, $E_2$ the distillation error and $E_3$ the topological error per circuit.}\label{tab:appendix_worked}
\widetable{\begin{tabular}{llrrrrrrrrrrr}\toprule
 & & & & & & \multicolumn{2}{c}{Rounds} & \multicolumn{2}{c}{Qubit-rounds ($10^6$)} & & \multicolumn{2}{c}{$E_3$} \\
\cmidrule(lr){7-8}\cmidrule(lr){9-10}\cmidrule(lr){12-13}
Strategy & Target & $n\times$runs & Lanes & $t_{(n)}$ & $\tau$ & Closed form & Scheduler & Closed form & Scheduler & $E_2$ & Closed form & Scheduler \\\midrule
\Mfac{8,3,2} & Disjoint & 8$\times$2 & 16 & 45.4 & 147 & 360 & 361 & 4.68 & 4.68 & $1.2\times10^{-10}$ & $4.2\times10^{-9}$ & $4.2\times10^{-9}$ \\
\Mfac{12,5,2} & Shared & 12$\times$1 & 24 & 45.9 & 231 & 298 & 298 & 3.72 & 3.72 & $1.4\times10^{-10}$ & $3.3\times10^{-9}$ & $3.3\times10^{-9}$ \\
\Mfac{14,6,2} & Disjoint & 14$\times$1 & 28 & 46.0 & 273 & 340 & 340 & 4.74 & 4.75 & $1.9\times10^{-10}$ & $4.3\times10^{-9}$ & $4.3\times10^{-9}$ \\
\bottomrule\end{tabular}}\end{table*}

%% file: includes/generated/appendix_timeline.tex
\begin{figure}[t]\centering
\begin{tikzpicture}[x=0.0190cm,y=0.62cm,font=\scriptsize]
\node[anchor=east] at (-7.9,2) {\Mfac{8,3,2}};
\fill[gray!35] (0.0,1.78) rectangle (45.4,2.22);
\fill[blue!30] (45.4,1.78) rectangle (192.4,2.22);
\node at (118.9,2.00) {ETD 1};
\fill[blue!45] (192.4,1.78) rectangle (339.4,2.22);
\node at (265.9,2.00) {ETD 2};
\fill[red!60] (192.4,2.14) rectangle (213.4,2.58);
\fill[red!60] (339.4,2.14) rectangle (360.4,2.58);
\draw[thick] (360.4,1.70) -- (360.4,2.50);
\node[anchor=east] at (-7.9,1) {\Mfac{12,5,2}};
\fill[gray!35] (0.0,0.78) rectangle (45.9,1.22);
\fill[blue!30] (45.9,0.78) rectangle (276.9,1.22);
\node at (161.4,1.00) {ETD};
\fill[red!60] (276.9,1.14) rectangle (297.9,1.58);
\draw[thick] (297.9,0.70) -- (297.9,1.50);
\node[anchor=east] at (-7.9,0) {\Mfac{14,6,2}};
\fill[gray!35] (0.0,-0.22) rectangle (45.9,0.22);
\fill[blue!30] (45.9,-0.22) rectangle (318.9,0.22);
\node at (182.4,0.00) {ETD};
\fill[red!60] (318.9,0.14) rectangle (339.9,0.58);
\draw[thick] (339.9,-0.30) -- (339.9,0.50);
\draw[->] (0,-0.6) -- (420,-0.6) node[right] {rounds};
\draw (0,-0.55) -- (0,-0.65) node[below] {0};
\draw (100,-0.55) -- (100,-0.65) node[below] {100};
\draw (200,-0.55) -- (200,-0.65) node[below] {200};
\draw (300,-0.55) -- (300,-0.65) node[below] {300};
\draw (400,-0.55) -- (400,-0.65) node[below] {400};
\end{tikzpicture}
\caption{Mean event times of the worked example (\cref{tab:appendix_worked}). Grey bars show the cultivation lanes filling one ETD batch and blue bars the ETD runs. Red marks, drawn above the bars, show the consumption of a $CCZ$ into the target, one Pauli-product measurement of $\delta$ rounds each. The vertical line is the mean latency from the closed form.}\label{fig:appendix_timeline}
\end{figure}
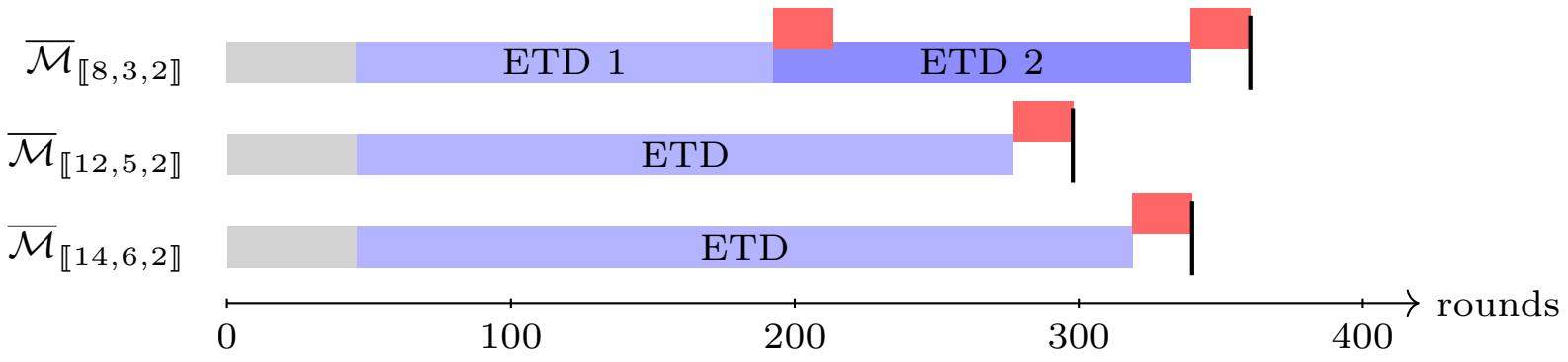

%% file: sections/H_extended_benchmarks.tex
\section{Extended Benchmark Results}\label{section:extended_tables}

This appendix collects the tables behind \cref{section:tailored_algorithmic_primitives} that the main text only summarizes.
All settings are those of \cref{section:comparison_framework}, and the colors follow the convention stated there.

\subsection{Adders in all data blocks}

\cref{table:cla_single_cla,table:cla_single_ripple} give the single-lane adders of \cref{section:single_lane_adders}, and \cref{table:batched_addition_cla,table:batched_addition_ripple} the shared-addend adders of \cref{section:carry_lookahead_addition}.
Each table lists, for every strategy, the compact, intermediate and fast data blocks at $d=21$ and the fast block under parallel consumption (rows fast$^{\parallel}$).
The fast-block rows are those of \cref{table:cla_single,table:batched_addition}.
The tables also separate the error into its distillation and topological parts and give the 95th percentile of the latency and the installed footprint.
\cref{fig:adder_ripple_shared,fig:adder_cla_shared} draw the width-4 instances of the two shared-addend adders.

\input{includes/generated/adder_cla.tex}
\input{includes/generated/adder_shared.tex}
\input{includes/adders_shared_figure}

\subsection{Dependence on cultivation capacity}

\Cref{fig:cultivation_capacity} shows how the latency depends on the number of factories and cultivation lanes in the fast block at $d=21$.
With one or two factories the schedule is limited by the factories from $8$ lanes on: the latency barely depends on the number of lanes and is ordered by the ETD time per $CCZ$, $\tau/R=7$, $6.5$, $5.5$, $5$, $4.75$ and $4.6$ steps for \Mfac{8,3,2}, \Mfac{14,6,2}, \Mfac{12,5,2} and the $3$-, $4$- and $5$-stars.
With many factories and few lanes it is limited by the $T$ supply and ordered approximately by the raw $T$ states per $CCZ$, $n/R$: $8$ for \Mfac{8,3,2}, $7$ for \Mfac{14,6,2}, $6$ for \Mfac{12,5,2}, and $5.3$ and $5$ for \Mfac{16,7,2} and \Mfac{20,9,2}.
With many factories and lanes it is limited by the number of consumptions: with eight \Mfac{8,3,2} factories and at least $32$ lanes, the serial data block carries $64$ consumptions of $\delta=21$ rounds, which the packed resources avoid by needing fewer consumptions.

\input{includes/generated/capacity_sweep.tex}

%% file: includes/generated/adder_cla.tex
\begin{table*}\centering
\caption{Single-lane 4-bit carry-lookahead adder with carry-in (16 AND gates, $N=42$ logical qubits) in all data blocks at $d=21$. Settings, columns and colors are those of \cref{section:comparison_framework}.}
\label{table:cla_single_cla}
\widetable{\begin{tabular}{llrr|rr|rr|rrr}\toprule
Strategy & Block & $d$ & Cons. & Rounds & 95\% & Qubit-rounds & Qubits & $\epsilon_{\rm ETD}$ & $\epsilon_{\rm topo}$ & $\epsilon_{\rm total}$ \\\midrule
\Mfac{8,3,2} & compact & 21 & 16 & 3,216 & 3,243 & 256,377,261 & 112,142 & $9.4\times10^{-10}$ & $2.8\times10^{-7}$ & $2.8\times10^{-7}$ \\
 & intermediate & 21 & 16 & 2,502 & 2,529 & 250,166,032 & 131,546 & $9.4\times10^{-10}$ & $2.8\times10^{-7}$ & $2.8\times10^{-7}$ \\
 & fast & 21 & 16 & 2,418 & 2,445 & 272,779,765 & 145,067 & $9.4\times10^{-10}$ & $3.0\times10^{-7}$ & $3.0\times10^{-7}$ \\
 & fast$^\parallel$ & 21 & 16 & 2,418 & 2,445 & 272,779,765 & 145,067 & $9.4\times10^{-10}$ & $3.0\times10^{-7}$ & $3.0\times10^{-7}$ \\
\midrule
\Mfac{12,5,2} & compact & 21 & 10 & \textcolor{darkred}{2,319} & 2,325 & \textcolor{darkblue}{201,183,745} & 123,608 & $1.1\times10^{-9}$ & $2.2\times10^{-7}$ & $2.2\times10^{-7}$ \\
 & intermediate & 21 & 10 & \textcolor{darkred}{2,109} & 2,115 & \textcolor{darkblue}{223,760,687} & 143,012 & $1.1\times10^{-9}$ & $2.5\times10^{-7}$ & $2.5\times10^{-7}$ \\
 & fast & 21 & 10 & \textcolor{darkred}{2,025} & 2,031 & \textcolor{darkblue}{241,503,444} & 156,533 & $1.1\times10^{-9}$ & $2.7\times10^{-7}$ & $2.7\times10^{-7}$ \\
 & fast$^\parallel$ & 21 & 10 & \textcolor{darkred}{2,025} & 2,031 & \textcolor{darkblue}{241,503,444} & 156,533 & $1.1\times10^{-9}$ & $2.7\times10^{-7}$ & $2.7\times10^{-7}$ \\
\midrule
\Mfac{14,6,2} & compact & 21 & 8 & \textcolor{darkred}{2,418} & 2,445 & \textcolor{darkblue}{221,716,044} & 129,341 & $1.5\times10^{-9}$ & $2.4\times10^{-7}$ & $2.5\times10^{-7}$ \\
 & intermediate & 21 & 8 & \textcolor{darkred}{2,334} & 2,361 & 258,567,680 & 148,745 & $1.5\times10^{-9}$ & $2.9\times10^{-7}$ & $2.9\times10^{-7}$ \\
 & fast & 21 & 8 & \textcolor{darkred}{2,250} & 2,277 & 278,920,783 & 162,266 & $1.5\times10^{-9}$ & $3.1\times10^{-7}$ & $3.1\times10^{-7}$ \\
 & fast$^\parallel$ & 21 & 8 & \textcolor{darkred}{2,250} & 2,277 & 278,920,783 & 162,266 & $1.5\times10^{-9}$ & $3.1\times10^{-7}$ & $3.1\times10^{-7}$ \\
\midrule
\Mfac{12,5,2}+\Mfac{14,6,2} (mixed) & compact & 21 & 8 & \textcolor{darkred}{2,014} & 2,025 & \textcolor{darkblue}{206,422,650} & 166,385 & $1.4\times10^{-9}$ & $2.3\times10^{-7}$ & $2.3\times10^{-7}$ \\
 & intermediate & 21 & 8 & \textcolor{darkred}{1,636} & 1,647 & \textcolor{darkblue}{205,056,480} & 185,789 & $1.4\times10^{-9}$ & $2.3\times10^{-7}$ & $2.3\times10^{-7}$ \\
 & fast & 21 & 8 & \textcolor{darkred}{1,468} & 1,479 & \textcolor{darkblue}{206,612,387} & 199,310 & $1.4\times10^{-9}$ & $2.3\times10^{-7}$ & $2.3\times10^{-7}$ \\
 & fast$^\parallel$ & 21 & 8 & \textcolor{darkred}{1,468} & 1,479 & \textcolor{darkblue}{206,612,387} & 199,310 & $1.4\times10^{-9}$ & $2.3\times10^{-7}$ & $2.3\times10^{-7}$ \\
\bottomrule\end{tabular}}\end{table*}

\begin{table*}\centering
\caption{Single-lane 4-bit ripple-carry~\cite{gidney2018halving} adder with carry-in (4 AND gates, $N=30$ logical qubits) in all data blocks at $d=21$. Settings, columns and colors are those of \cref{section:comparison_framework}.}
\label{table:cla_single_ripple}
\widetable{\begin{tabular}{llrr|rr|rr|rrr}\toprule
Strategy & Block & $d$ & Cons. & Rounds & 95\% & Qubit-rounds & Qubits & $\epsilon_{\rm ETD}$ & $\epsilon_{\rm topo}$ & $\epsilon_{\rm total}$ \\\midrule
\Mfac{8,3,2} & compact & 21 & 4 & 947 & 975 & 54,363,309 & 96,266 & $2.4\times10^{-10}$ & $6.0\times10^{-8}$ & $6.0\times10^{-8}$ \\
 & intermediate & 21 & 4 & 737 & 765 & 53,798,321 & 110,378 & $2.4\times10^{-10}$ & $5.9\times10^{-8}$ & $5.9\times10^{-8}$ \\
 & fast & 21 & 4 & 653 & 681 & 55,365,397 & 121,396 & $2.4\times10^{-10}$ & $6.1\times10^{-8}$ & $6.1\times10^{-8}$ \\
 & fast$^\parallel$ & 21 & 4 & 653 & 681 & 55,365,397 & 121,396 & $2.4\times10^{-10}$ & $6.1\times10^{-8}$ & $6.1\times10^{-8}$ \\
\midrule
\Mfac{12,5,2} & compact & 21 & 4 & \textcolor{darkred}{933} & 939 & 54,875,872 & 107,732 & $2.4\times10^{-10}$ & $6.0\times10^{-8}$ & $6.0\times10^{-8}$ \\
 & intermediate & 21 & 4 & \textcolor{darkred}{723} & 729 & 54,073,816 & 121,844 & $2.4\times10^{-10}$ & $5.9\times10^{-8}$ & $5.9\times10^{-8}$ \\
 & fast & 21 & 4 & \textcolor{darkred}{639} & 645 & 55,481,243 & 132,862 & $2.4\times10^{-10}$ & $6.1\times10^{-8}$ & $6.1\times10^{-8}$ \\
 & fast$^\parallel$ & 21 & 4 & \textcolor{darkred}{639} & 645 & 55,481,243 & 132,862 & $2.4\times10^{-10}$ & $6.1\times10^{-8}$ & $6.1\times10^{-8}$ \\
\midrule
\Mfac{14,6,2} & compact & 21 & 4 & \textcolor{darkred}{929} & 939 & 54,961,956 & 113,465 & $2.4\times10^{-10}$ & $6.0\times10^{-8}$ & $6.0\times10^{-8}$ \\
 & intermediate & 21 & 4 & \textcolor{darkred}{719} & 729 & 54,101,477 & 127,577 & $2.4\times10^{-10}$ & $5.9\times10^{-8}$ & $5.9\times10^{-8}$ \\
 & fast & 21 & 4 & \textcolor{darkred}{635} & 645 & 55,463,290 & 138,595 & $2.4\times10^{-10}$ & $6.1\times10^{-8}$ & $6.1\times10^{-8}$ \\
 & fast$^\parallel$ & 21 & 4 & \textcolor{darkred}{635} & 645 & 55,463,290 & 138,595 & $2.4\times10^{-10}$ & $6.1\times10^{-8}$ & $6.1\times10^{-8}$ \\
\midrule
\Mfac{12,5,2}+\Mfac{14,6,2} (mixed) & compact & 21 & 4 & \textcolor{darkred}{926} & 939 & 55,345,223 & 150,509 & $2.4\times10^{-10}$ & $6.0\times10^{-8}$ & $6.1\times10^{-8}$ \\
 & intermediate & 21 & 4 & \textcolor{darkred}{716} & 729 & 54,442,831 & 164,621 & $2.4\times10^{-10}$ & $5.9\times10^{-8}$ & $6.0\times10^{-8}$ \\
 & fast & 21 & 4 & \textcolor{darkred}{632} & 645 & 55,771,921 & 175,639 & $2.4\times10^{-10}$ & $6.1\times10^{-8}$ & $6.1\times10^{-8}$ \\
 & fast$^\parallel$ & 21 & 4 & \textcolor{darkred}{632} & 645 & 55,771,921 & 175,639 & $2.4\times10^{-10}$ & $6.1\times10^{-8}$ & $6.1\times10^{-8}$ \\
\bottomrule\end{tabular}}\end{table*}

%% file: includes/generated/adder_shared.tex
\begin{table*}\centering
\caption{2-lane shared-addend Sklansky carry-lookahead adder of width 8 (50 AND gates, $N=132$ logical qubits) in all data blocks at $d=21$. Settings, columns and colors are those of \cref{section:comparison_framework}.}
\label{table:batched_addition_cla}
\widetable{\begin{tabular}{llrr|rr|rr|rrr}\toprule
Strategy & Block & $d$ & Cons. & Rounds & 95\% & Qubit-rounds & Qubits & $\epsilon_{\rm ETD}$ & $\epsilon_{\rm topo}$ & $\epsilon_{\rm total}$ \\\midrule
\Mfac{8,3,2} & compact & 21 & 50 & 9,642 & 9,669 & 1,934,094,077 & 231,212 & $2.9\times10^{-9}$ & $2.2\times10^{-6}$ & $2.2\times10^{-6}$ \\
 & intermediate & 21 & 50 & 7,500 & 7,527 & 1,952,858,006 & 290,306 & $2.9\times10^{-9}$ & $2.2\times10^{-6}$ & $2.2\times10^{-6}$ \\
 & fast & 21 & 50 & 7,416 & 7,443 & 2,114,811,602 & 316,322 & $2.9\times10^{-9}$ & $2.4\times10^{-6}$ & $2.4\times10^{-6}$ \\
 & fast$^\parallel$ & 21 & 50 & 7,416 & 7,443 & 2,114,811,602 & 316,322 & $2.9\times10^{-9}$ & $2.4\times10^{-6}$ & $2.4\times10^{-6}$ \\
\midrule
\Mfac{12,5,2} & compact & 21 & 26 & \textcolor{darkred}{6,158} & 6,183 & \textcolor{darkblue}{1,298,295,939} & 242,678 & $3.4\times10^{-9}$ & $1.5\times10^{-6}$ & $1.5\times10^{-6}$ \\
 & intermediate & 21 & 26 & \textcolor{darkred}{5,990} & 6,015 & \textcolor{darkblue}{1,611,169,365} & 301,772 & $3.4\times10^{-9}$ & $1.8\times10^{-6}$ & $1.8\times10^{-6}$ \\
 & fast & 21 & 26 & \textcolor{darkred}{5,906} & 5,931 & \textcolor{darkblue}{1,735,635,326} & 327,788 & $3.4\times10^{-9}$ & $1.9\times10^{-6}$ & $2.0\times10^{-6}$ \\
 & fast$^\parallel$ & 21 & 26 & \textcolor{darkred}{5,906} & 5,931 & \textcolor{darkblue}{1,735,635,326} & 327,788 & $3.4\times10^{-9}$ & $1.9\times10^{-6}$ & $2.0\times10^{-6}$ \\
\midrule
\Mfac{14,6,2} & compact & 21 & 25 & \textcolor{darkred}{7,058} & 7,068 & \textcolor{darkblue}{1,523,250,353} & 248,411 & $4.8\times10^{-9}$ & $1.7\times10^{-6}$ & $1.7\times10^{-6}$ \\
 & intermediate & 21 & 25 & \textcolor{darkred}{6,974} & 6,984 & \textcolor{darkblue}{1,909,388,886} & 307,505 & $4.8\times10^{-9}$ & $2.1\times10^{-6}$ & $2.1\times10^{-6}$ \\
 & fast & 21 & 25 & \textcolor{darkred}{6,890} & 6,900 & \textcolor{darkblue}{2,057,677,263} & 333,521 & $4.8\times10^{-9}$ & $2.3\times10^{-6}$ & $2.3\times10^{-6}$ \\
 & fast$^\parallel$ & 21 & 25 & \textcolor{darkred}{6,890} & 6,900 & \textcolor{darkblue}{2,057,677,263} & 333,521 & $4.8\times10^{-9}$ & $2.3\times10^{-6}$ & $2.3\times10^{-6}$ \\
\midrule
\Mfac{12,5,2}+\Mfac{14,6,2} (mixed) & compact & 21 & 25 & \textcolor{darkred}{6,570} & 6,582 & \textcolor{darkblue}{1,556,986,100} & 285,455 & $4.6\times10^{-9}$ & $1.7\times10^{-6}$ & $1.7\times10^{-6}$ \\
 & intermediate & 21 & 25 & \textcolor{darkred}{6,171} & 6,183 & \textcolor{darkblue}{1,822,643,470} & 344,549 & $4.6\times10^{-9}$ & $2.0\times10^{-6}$ & $2.0\times10^{-6}$ \\
 & fast & 21 & 25 & \textcolor{darkred}{6,087} & 6,099 & \textcolor{darkblue}{1,949,893,658} & 370,565 & $4.6\times10^{-9}$ & $2.2\times10^{-6}$ & $2.2\times10^{-6}$ \\
 & fast$^\parallel$ & 21 & 25 & \textcolor{darkred}{6,087} & 6,099 & \textcolor{darkblue}{1,949,893,658} & 370,565 & $4.6\times10^{-9}$ & $2.2\times10^{-6}$ & $2.2\times10^{-6}$ \\
\midrule
\Mfac{16,7,2} & compact & 21 & 24 & \textcolor{darkred}{6,020} & 6,039 & \textcolor{darkblue}{1,307,443,048} & 254,144 & $3.6\times10^{-9}$ & $1.5\times10^{-6}$ & $1.5\times10^{-6}$ \\
 & intermediate & 21 & 24 & \textcolor{darkred}{5,852} & 5,871 & \textcolor{darkblue}{1,611,671,767} & 313,238 & $3.6\times10^{-9}$ & $1.8\times10^{-6}$ & $1.8\times10^{-6}$ \\
 & fast & 21 & 24 & \textcolor{darkred}{5,768} & 5,787 & \textcolor{darkblue}{1,732,675,848} & 339,254 & $3.6\times10^{-9}$ & $1.9\times10^{-6}$ & $1.9\times10^{-6}$ \\
 & fast$^\parallel$ & 21 & 24 & \textcolor{darkred}{5,768} & 5,787 & \textcolor{darkblue}{1,732,675,848} & 339,254 & $3.6\times10^{-9}$ & $1.9\times10^{-6}$ & $1.9\times10^{-6}$ \\
\midrule
\Mfac{20,9,2} & compact & 21 & 24 & \textcolor{darkred}{6,019} & 6,039 & \textcolor{darkblue}{1,341,354,704} & 265,610 & $3.7\times10^{-9}$ & $1.5\times10^{-6}$ & $1.5\times10^{-6}$ \\
 & intermediate & 21 & 24 & \textcolor{darkred}{5,851} & 5,871 & \textcolor{darkblue}{1,644,948,171} & 324,704 & $3.7\times10^{-9}$ & $1.8\times10^{-6}$ & $1.8\times10^{-6}$ \\
 & fast & 21 & 24 & \textcolor{darkred}{5,767} & 5,787 & \textcolor{darkblue}{1,765,933,521} & 350,720 & $3.7\times10^{-9}$ & $2.0\times10^{-6}$ & $2.0\times10^{-6}$ \\
 & fast$^\parallel$ & 21 & 24 & \textcolor{darkred}{5,767} & 5,787 & \textcolor{darkblue}{1,765,933,521} & 350,720 & $3.7\times10^{-9}$ & $2.0\times10^{-6}$ & $2.0\times10^{-6}$ \\
\midrule
\Mfac{24,11,2} & compact & 21 & 24 & \textcolor{darkred}{6,019} & 6,039 & \textcolor{darkblue}{1,375,002,920} & 277,076 & $3.7\times10^{-9}$ & $1.5\times10^{-6}$ & $1.5\times10^{-6}$ \\
 & intermediate & 21 & 24 & \textcolor{darkred}{5,851} & 5,871 & \textcolor{darkblue}{1,678,003,683} & 336,170 & $3.7\times10^{-9}$ & $1.9\times10^{-6}$ & $1.9\times10^{-6}$ \\
 & fast & 21 & 24 & \textcolor{darkred}{5,767} & 5,787 & \textcolor{darkblue}{1,798,989,033} & 362,186 & $3.7\times10^{-9}$ & $2.0\times10^{-6}$ & $2.0\times10^{-6}$ \\
 & fast$^\parallel$ & 21 & 24 & \textcolor{darkred}{5,767} & 5,787 & \textcolor{darkblue}{1,798,989,033} & 362,186 & $3.7\times10^{-9}$ & $2.0\times10^{-6}$ & $2.0\times10^{-6}$ \\
\bottomrule\end{tabular}}\end{table*}

\begin{table*}\centering
\caption{2-lane shared-addend ripple-carry~\cite{gidney2018halving} adder of width 8 (16 AND gates, $N=100$ logical qubits) in all data blocks at $d=21$. Settings, columns and colors are those of \cref{section:comparison_framework}.}
\label{table:batched_addition_ripple}
\widetable{\begin{tabular}{llrr|rr|rr|rrr}\toprule
Strategy & Block & $d$ & Cons. & Rounds & 95\% & Qubit-rounds & Qubits & $\epsilon_{\rm ETD}$ & $\epsilon_{\rm topo}$ & $\epsilon_{\rm total}$ \\\midrule
\Mfac{8,3,2} & compact & 21 & 16 & 3,216 & 3,243 & 503,126,180 & 188,876 & $9.4\times10^{-10}$ & $5.6\times10^{-7}$ & $5.6\times10^{-7}$ \\
 & intermediate & 21 & 16 & 2,502 & 2,529 & 506,113,824 & 233,858 & $9.4\times10^{-10}$ & $5.7\times10^{-7}$ & $5.7\times10^{-7}$ \\
 & fast & 21 & 16 & 2,418 & 2,445 & 541,358,778 & 256,159 & $9.4\times10^{-10}$ & $6.1\times10^{-7}$ & $6.1\times10^{-7}$ \\
 & fast$^\parallel$ & 21 & 16 & 2,418 & 2,445 & 541,358,778 & 256,159 & $9.4\times10^{-10}$ & $6.1\times10^{-7}$ & $6.1\times10^{-7}$ \\
\midrule
\Mfac{12,5,2} & compact & 21 & 15 & \textcolor{darkred}{3,112} & 3,138 & 504,124,132 & 200,342 & $9.6\times10^{-10}$ & $5.6\times10^{-7}$ & $5.6\times10^{-7}$ \\
 & intermediate & 21 & 15 & \textcolor{darkred}{2,440} & 2,466 & 507,529,653 & 245,324 & $9.6\times10^{-10}$ & $5.7\times10^{-7}$ & $5.7\times10^{-7}$ \\
 & fast & 21 & 15 & \textcolor{darkred}{2,356} & 2,382 & 541,479,877 & 267,625 & $9.6\times10^{-10}$ & $6.1\times10^{-7}$ & $6.1\times10^{-7}$ \\
 & fast$^\parallel$ & 21 & 15 & \textcolor{darkred}{2,356} & 2,382 & 541,479,877 & 267,625 & $9.6\times10^{-10}$ & $6.1\times10^{-7}$ & $6.1\times10^{-7}$ \\
\midrule
\Mfac{14,6,2} & compact & 21 & 9 & \textcolor{darkred}{2,462} & 2,472 & \textcolor{darkblue}{413,415,927} & 206,075 & $1.5\times10^{-9}$ & $4.6\times10^{-7}$ & $4.6\times10^{-7}$ \\
 & intermediate & 21 & 9 & \textcolor{darkred}{2,336} & 2,346 & \textcolor{darkblue}{496,774,400} & 251,057 & $1.5\times10^{-9}$ & $5.6\times10^{-7}$ & $5.6\times10^{-7}$ \\
 & fast & 21 & 9 & \textcolor{darkred}{2,252} & 2,262 & \textcolor{darkblue}{528,319,663} & 273,358 & $1.5\times10^{-9}$ & $5.9\times10^{-7}$ & $5.9\times10^{-7}$ \\
 & fast$^\parallel$ & 21 & 9 & \textcolor{darkred}{2,252} & 2,262 & \textcolor{darkblue}{528,319,663} & 273,358 & $1.5\times10^{-9}$ & $5.9\times10^{-7}$ & $5.9\times10^{-7}$ \\
\midrule
\Mfac{12,5,2}+\Mfac{14,6,2} (mixed) & compact & 21 & 8 & \textcolor{darkred}{2,281} & 2,283 & \textcolor{darkblue}{408,495,857} & 243,119 & $1.5\times10^{-9}$ & $4.6\times10^{-7}$ & $4.6\times10^{-7}$ \\
 & intermediate & 21 & 8 & \textcolor{darkred}{2,113} & 2,115 & \textcolor{darkblue}{473,077,361} & 288,101 & $1.5\times10^{-9}$ & $5.3\times10^{-7}$ & $5.3\times10^{-7}$ \\
 & fast & 21 & 8 & \textcolor{darkred}{1,993} & 2,004 & \textcolor{darkblue}{490,600,017} & 310,402 & $1.5\times10^{-9}$ & $5.5\times10^{-7}$ & $5.5\times10^{-7}$ \\
 & fast$^\parallel$ & 21 & 8 & \textcolor{darkred}{1,993} & 2,004 & \textcolor{darkblue}{490,600,017} & 310,402 & $1.5\times10^{-9}$ & $5.5\times10^{-7}$ & $5.5\times10^{-7}$ \\
\midrule
\Mfac{16,7,2} & compact & 21 & 15 & \textcolor{darkred}{3,101} & 3,120 & 517,595,341 & 211,808 & $9.6\times10^{-10}$ & $5.8\times10^{-7}$ & $5.8\times10^{-7}$ \\
 & intermediate & 21 & 15 & \textcolor{darkred}{2,429} & 2,448 & 517,535,344 & 256,790 & $9.6\times10^{-10}$ & $5.8\times10^{-7}$ & $5.8\times10^{-7}$ \\
 & fast & 21 & 15 & \textcolor{darkred}{2,345} & 2,364 & 551,236,691 & 279,091 & $9.6\times10^{-10}$ & $6.2\times10^{-7}$ & $6.2\times10^{-7}$ \\
 & fast$^\parallel$ & 21 & 15 & \textcolor{darkred}{2,345} & 2,364 & 551,236,691 & 279,091 & $9.6\times10^{-10}$ & $6.2\times10^{-7}$ & $6.2\times10^{-7}$ \\
\midrule
\Mfac{20,9,2} & compact & 21 & 15 & \textcolor{darkred}{3,100} & 3,102 & 531,094,175 & 223,274 & $9.6\times10^{-10}$ & $5.9\times10^{-7}$ & $6.0\times10^{-7}$ \\
 & intermediate & 21 & 15 & \textcolor{darkred}{2,428} & 2,430 & 528,318,429 & 268,256 & $9.6\times10^{-10}$ & $5.9\times10^{-7}$ & $5.9\times10^{-7}$ \\
 & fast & 21 & 15 & \textcolor{darkred}{2,344} & 2,346 & 561,995,691 & 290,557 & $9.6\times10^{-10}$ & $6.3\times10^{-7}$ & $6.3\times10^{-7}$ \\
 & fast$^\parallel$ & 21 & 15 & \textcolor{darkred}{2,344} & 2,346 & 561,995,691 & 290,557 & $9.6\times10^{-10}$ & $6.3\times10^{-7}$ & $6.3\times10^{-7}$ \\
\midrule
\Mfac{24,11,2} & compact & 21 & 15 & \textcolor{darkred}{3,100} & 3,102 & 543,233,944 & 234,740 & $9.6\times10^{-10}$ & $6.1\times10^{-7}$ & $6.1\times10^{-7}$ \\
 & intermediate & 21 & 15 & \textcolor{darkred}{2,428} & 2,430 & 538,087,381 & 279,722 & $9.6\times10^{-10}$ & $6.0\times10^{-7}$ & $6.0\times10^{-7}$ \\
 & fast & 21 & 15 & \textcolor{darkred}{2,344} & 2,346 & 571,764,644 & 302,023 & $9.6\times10^{-10}$ & $6.4\times10^{-7}$ & $6.4\times10^{-7}$ \\
 & fast$^\parallel$ & 21 & 15 & \textcolor{darkred}{2,344} & 2,346 & 571,764,644 & 302,023 & $9.6\times10^{-10}$ & $6.4\times10^{-7}$ & $6.4\times10^{-7}$ \\
\bottomrule\end{tabular}}\end{table*}

%% file: includes/adders_shared_figure.tex
\begin{figure*}[tp]
\centering
\includegraphics{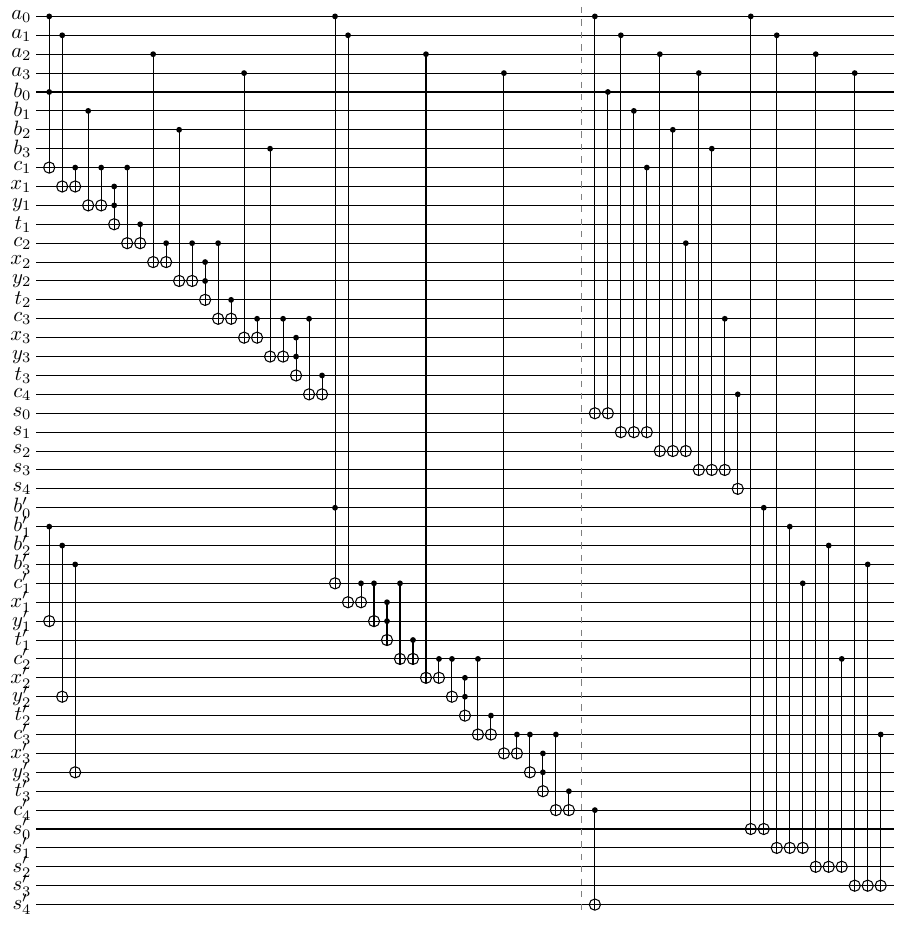}
\caption{Width-4 instance of the shared-addend ripple-carry adder of \cref{section:carry_lookahead_addition} ($8$ AND gates, $N=48$; the benchmark uses width $8$, with $16$ AND gates and $N=100$), drawn as in \cref{fig:adders}.
The two lanes share the addend $a$ and have no carry-in, so $c_1=a_0b_0$; labels of the second lane are primed, e.g.\ $b'_i=b^{(2)}_i$.}
\label{fig:adder_ripple_shared}
\end{figure*}

\begin{figure*}[tp]
\centering
\includegraphics{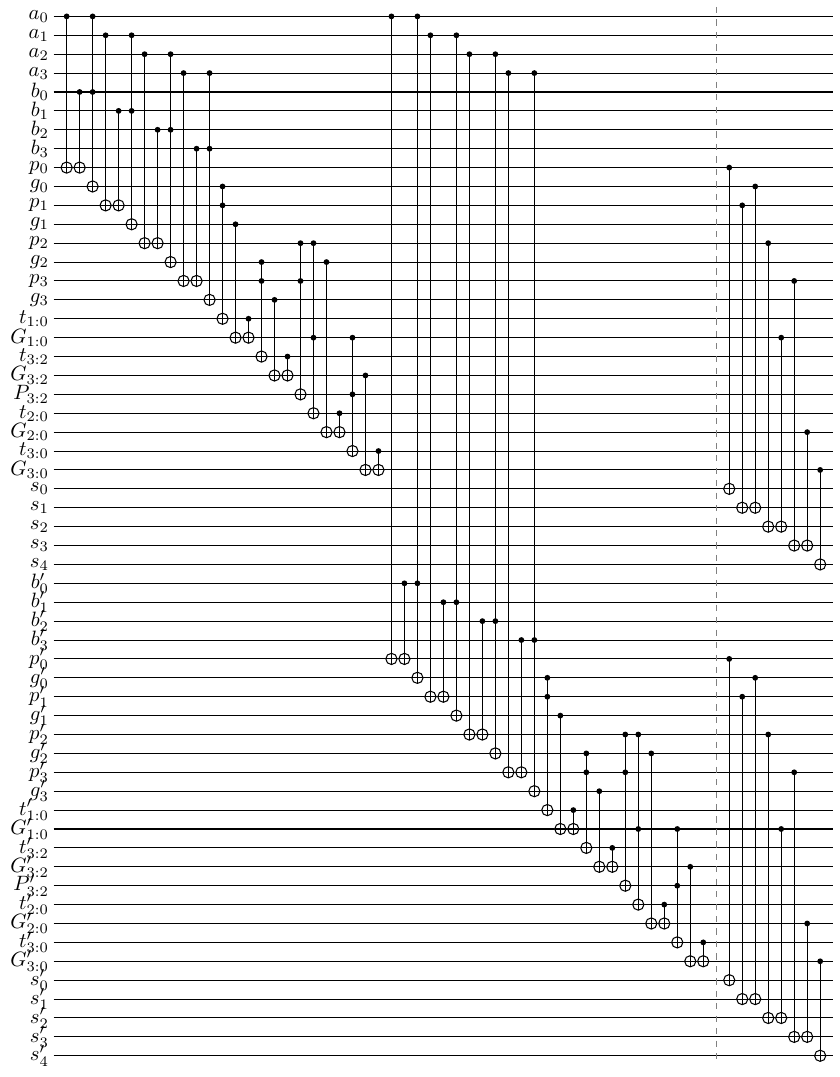}
\caption{Width-4 instance of the shared-addend Sklansky carry-lookahead adder of \cref{section:carry_lookahead_addition} ($18$ AND gates, $N=56$; the benchmark uses width $8$, with $50$ AND gates and $N=132$), drawn as in \cref{fig:adder_ripple_shared}.}
\label{fig:adder_cla_shared}
\end{figure*}

%% file: includes/generated/capacity_sweep.tex
\begin{figure*}[t]\centering
\begin{tikzpicture}\begin{axis}[width=0.48\textwidth,height=5.5cm,xmode=log,log basis x=2,xlabel={$T$-cultivation lanes},ylabel={Latency (rounds)},title={Standalone, disjoint target},legend pos=north east,legend style={font=\scriptsize}]
\addplot coordinates {(1,1156.2) (2,665.0) (4,447.8) (8,384.3) (16,359.1) (32,359.1) (64,359.1)};
\addplot coordinates {(1,1160.7) (2,737.7) (4,528.9) (8,415.5) (16,359.7) (32,340.8) (64,340.8)};
\legend{{\Mfac{8,3,2}},{\Mfac{14,6,2}}}\end{axis}\end{tikzpicture}\hfill
\begin{tikzpicture}\begin{axis}[width=0.48\textwidth,height=5.5cm,xmode=log,log basis x=2,xlabel={$T$-cultivation lanes},ylabel={Latency (rounds)},title={$GF(2^8)$ multiplier},legend pos=north east,legend style={font=\scriptsize},legend columns=2]
\addplot coordinates {(8,9492.9) (16,9469.5) (32,9469.5) (64,9469.5) (128,9469.5)};
\addplot coordinates {(8,4865.2) (16,4791.9) (32,4768.2) (64,4768.2) (128,4768.2)};
\addplot coordinates {(8,7519.2) (16,7467.0) (32,7458.9) (64,7458.9) (128,7458.9)};
\addplot coordinates {(8,8874.9) (16,8822.7) (32,8801.1) (64,8801.1) (128,8801.1)};
\addplot coordinates {(8,6569.7) (16,6490.5) (32,6454.5) (64,6450.0) (128,6450.0)};
\addplot coordinates {(8,4269.3) (16,2224.5) (32,1514.5) (64,1509.0) (128,1509.0)};
\addplot coordinates {(8,3306.6) (16,2065.5) (32,1976.5) (64,1950.0) (128,1950.0)};
\addplot coordinates {(8,2957.1) (16,1936.5) (32,1783.7) (64,1714.8) (128,1698.0)};
\legend{{\Mfac{8,3,2}$\times$1},{\Mfac{8,3,2}$\times$2},{\Mfac{12,5,2}$\times$1},{\Mfac{14,6,2}$\times$1},{\Mfac{20,9,2}$\times$1},{\Mfac{8,3,2}$\times$8},{\Mfac{12,5,2}$\times$4},{\Mfac{20,9,2}$\times$4}}\end{axis}\end{tikzpicture}
\caption{Latency against the number of cultivation lanes in the fast block at $d=21$ (mean over random seeds), for the standalone disjoint target (left) and the $GF(2^8)$ multiplier with one to eight factories of each family (right). Each factory runs the smallest ETD protocol that covers the requested resource, so a $CCZ$ that is not packed into a larger resource uses $\code{8,3,2}$ on the same factory.}\label{fig:cultivation_capacity}\end{figure*}
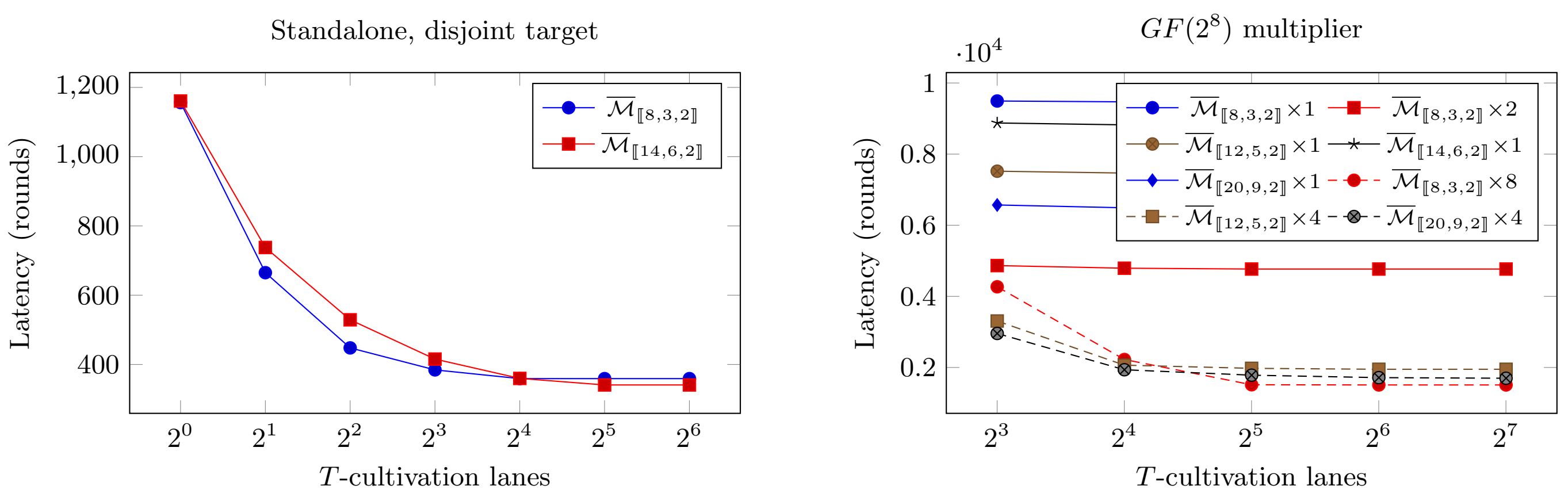

%% file: sections/I_fourth_level.tex
\section{Fourth-Level Resources}
\label{section:fourth_level}

The ETD factory on the $\code{30,8,2}$ code (\cref{fig:3082_etd_factory}), \Mfac{30,8,2}, is one of the two fourth-level resources in this paper; the other, the glued $CCCZ$ star \Mfac{40,10,2}, is costed in \cref{section:cccz_star_costing}.
Up to free operations its reduced logical action is two $CCCZ$ gates on disjoint qubits.
We compare the cost of applying these two $CCCZ$ gates with \Mfac{30,8,2} against the cost of applying them with $CCZ$ states only.
At the cultivation settings we consider, every costed \Mfac{30,8,2} configuration loses in latency, volume and footprint, and none reaches a target error of $10^{-7}$.
A set of optimistic parameters (\cref{section:fourth_level_optimistic}) identifies what would have to change: with a parallel-injection factory and an accurate $\sqrt T$ source, \Mfac{30,8,2} becomes a latency-for-volume trade, $5$--$24\%$ faster at $26$--$54\%$ more installed volume in slow data blocks, and slower in the fast block.

\subsection{What \texorpdfstring{\Mfac{30,8,2}}{the [[30,8,2]] factory} implements}

The transversal layer of \cref{section:further_examples_CSStp} consists of $30$ $\sqrt T$-type rotations ($15$ $\sqrt T$ and $15$ $\sqrt T^{\dagger}$, times Clifford phases; \cref{fig:3082_gates}); every layer that realizes the drawn logical phase polynomial modulo Cliffords needs a $\sqrt T$-type input on all $30$ legs.
The logical action of \Mfac{30,8,2} is equivalent under free operations (logical $CX$ gates and an $X$ frame) to
\begin{align}
CCCZ(0,1,2,3)\cdot CCCZ(4,5,6,7)~,
\label{eq:r30_minimal}
\end{align}
a Tier-1 resource.
From the logical phase polynomial of \cref{fig:3082_gates}, $11$ logical $CX$ gates and an $X$ frame on four qubits reach this form exactly, with no Clifford terms (\cref{fig:3082_gates}(b), up to relabeling); we checked the identity on all $2^{8}$ basis states.
The benchmark is therefore the state, or gate, $CCCZ\otimes CCCZ$ on eight qubits, and the baseline is the cheapest way to apply two disjoint $CCCZ$ gates with $CCZ$ states (\cref{section:fourth_level_baseline}).
\Mfac{30,8,2} cannot provide two $CCCZ$ gates that share a qubit; a star of $CCCZ$ gates would need a different code, such as the glued $CCCZ$ stars of \cref{section:cccz_star}, which we cost in \cref{section:cccz_star_costing}.

\subsection{Two \texorpdfstring{$CCCZ$}{CCCZ} gates from \texorpdfstring{$CCZ$}{CCZ} states}\label{section:fourth_level_baseline}

A $CCCZ$ costs three $CCZ$ gates with unitary uncomputation of an ancilla (compute the AND of two inputs, apply $CCZ$ to it and the other two, uncompute), or two with measurement-based uncomputation plus one reaction time~\cite{gidney2018halving}.
Two disjoint $CCCZ$ gates therefore need six or four $CCZ$ states; these are the two $CCZ$-only baselines below.
\cref{fig:r30_ccz_baseline} draws both, next to the \Mfac{30,8,2} construction.
Both are drawn for $CCCZ(0,1,2,3)$ with its ancilla $u$; $CCCZ(4,5,6,7)$ is identical, uses its own ancilla and runs in parallel, and every Toffoli and $CCZ$ gate consumes one $CCZ$ state from one of two \Mfac{8,3,2} factories.
In the unitary construction (\cref{fig:r30_ccz_baseline}(b)) a second Toffoli uncomputes the ancilla, so the two $CCCZ$ gates take six $CCZ$ states in three serial rounds.
In the measurement-based construction (\cref{fig:r30_ccz_baseline}(c)) the ancilla is measured in the $X$ basis and the phase $(-1)^{m\,x_0x_1}$ of outcome $m$ is removed by a classically controlled $CZ$, so the two gates take four $CCZ$ states in two serial rounds plus one reaction time.
With \Mfac{30,8,2} (\cref{fig:r30_ccz_baseline}(d)) one factory output is consumed on all eight qubits at once.
Each outcome of the teleportation leaves a byproduct $B$ on each group of four qubits, which is a Clifford for $1$ of the $16$ outcomes and one $CCZ$ gate up to Cliffords for the other $15$; the corrections are $CCZ$ states from a second factory (\Mfac{8,3,2}, or \Mfac{14,6,2}, which supplies both at once).
We checked the constructions (b) and (c) on the state vector, and the byproduct statistics of (d) on all $2^{8}$ outcomes.

\input{includes/r30_ccz_baseline}

\subsection{Costing \texorpdfstring{\Mfac{30,8,2}}{the [[30,8,2]] factory}}

We cost \Mfac{30,8,2} in the model of \cref{section:comparison_framework} with three additions.
\begin{itemize}
\item \emph{$\sqrt T$ supply.}
The $\sqrt T$ states come from the cultivation scheme of Ref.~\cite{chen2026efficient}, with $p_{\sqrt T}=1.5\times10^{-5}$.
A $\sqrt T$ lane has the footprint of a $T$ lane and needs $0.879$ times its time per kept state, the ratio of the two protocols' costs in Ref.~\cite{chen2026efficient}.
The output patch has distance $21$, as for $T$ states, and is grown to a larger working distance in two rounds.
\Mfac{30,8,2} and the $T$-based factories share one pool of $32$ lanes, split in the best of the ratios $\sqrt T{:}T\in\{16{:}16,24{:}8,28{:}4\}$.
As for $T$ states, we model a faulty $\sqrt T$ state as the ideal state followed by a $Z$ error with probability $p_{\sqrt T}$.
For $T$ states this form can be enforced by Clifford twirling; for $\sqrt T$ states, which are not eigenstates of a Clifford, it is an assumption, and other error components would add to $E_{1}$ and $E_{2}$.
\item \emph{$\sqrt T$ injection byproducts.}
Injecting a $P_{\pm\pi/16}$ rotation by teleportation applies its inverse with probability $1/2$, which must be fixed by a $P_{\pi/8}$ rotation, i.e.\ a $T$ state.
Of the $30$ rotations, $29$ are applied during the ETD run; the fix-ups commute with the rest of the run and are applied at its end, after one reaction time.
The \Mfac{30,8,2} run therefore takes $\tau=29+B$ logical steps with $B\sim\mathrm{Bin}(29,1/2)$, and each leg has input error $p_{\sqrt T}+p_{T}/2$.
\item \emph{Teleportation byproducts.}
Consuming a fourth-level resource leaves a third-level byproduct.
For \Mfac{30,8,2} it needs zero, one or two $CCZ$ corrections, with probabilities $1/256$, $30/256$ and $225/256$ (mean $1.875$).
The corrections are made either speculatively, alongside \Mfac{30,8,2}, and discarded if unused, or reactively, once the byproduct is known; they are consumed one reaction time ($10$ rounds) after the \Mfac{30,8,2} consumption.
A fourth error term $E_{4}$ collects the distillation and topological errors of the correction hardware.
\end{itemize}
The CCZ-only constructions keep one ancilla per $CCCZ$ for the whole schedule (data block $N=10$ against $N=8$ for \Mfac{30,8,2}).
Every construction installs two factories: two \Mfac{8,3,2} factories for the CCZ-only constructions, and \Mfac{30,8,2} plus one correction factory (\Mfac{8,3,2} or \Mfac{14,6,2}).

\subsection{Results}

\cref{tab:boolean_r30} compares the constructions at $d=21$ in the three data blocks.
In the fast block the best \Mfac{30,8,2} configuration (speculative corrections, with a \Mfac{14,6,2} correction factory) needs $1{,}049$ rounds and $5.4\times10^{7}$ occupied qubit-rounds.
The best CCZ-only construction (measurement-based, two \Mfac{8,3,2} factories) needs $372$ rounds and $1.9\times10^{7}$ qubit-rounds, and the unitary construction needs $509$ rounds.
\Mfac{30,8,2} is slower in every block: its ETD run alone takes $29$ steps plus about $14.5$ fix-up steps and a reaction time, roughly $920$ rounds, before its output exists, against $7$ steps for \Mfac{8,3,2}.
More cultivation does not help: with $256$ lanes the \Mfac{30,8,2} latency drops only from $1{,}049$ to $1{,}019$ rounds.

At a target error of $10^{-7}$, \Mfac{30,8,2} fails outright: its error floor as $d\to\infty$ is $1.1\times10^{-7}$, set by $435\,p_{\rm leg}^{2}$ per output plus the expected corrections.

\input{includes/generated/boolean_r30.tex}

\subsection{Optimistic parameters}\label{section:fourth_level_optimistic}

Three features of \Mfac{30,8,2} set its cost: the serial ETD run of $\tau=29$ steps, the $\sqrt T$ injection fix-ups (about $14.5$ further steps after a reaction time), and the $\sqrt T$ input error, which sets the floor.
The cultivation throughput is not a bottleneck.
\cref{tab:boolean_r30_optimistic} relaxes these features one at a time and together, in all three data blocks.
For the factory we assume a hypothetical parallel-injection layout that applies several of the $29$ rotations per logical step ($\tau=\lceil29/\text{ports}\rceil$, with the fix-up block parallelized in the same way) at a larger area.
No such layout has been designed; the number of ports and the area factor are scenario parameters.
For fairness we also give the same layout to the \Mfac{8,3,2} factories of the CCZ-only construction ($\tau=\lceil7/4\rceil=2$) and, in the last row, to the \Mfac{14,6,2} correction factory.

A better $\sqrt T$ source alone is not enough.
With $p_{\sqrt T}=p_{T}$ and no fix-ups the floor falls to $1.1\times10^{-9}$ and \Mfac{30,8,2} reaches the $10^{-7}$ target at $d=21$, but the serial layout keeps it slower than CCZ-only compilation (compact block: $1{,}067$ against $937$ rounds) with $1.9$ to $3.1$ times the installed volume.
The factory layout is the main lever.
With four ports and twice the area, and with the $\sqrt T$ fix-ups still included, the \Mfac{30,8,2} latency at $d=21$ drops from $1{,}381$ to $747$ rounds in the compact block and from $1{,}045$ to $427$ in the fast block; the floor, however, stays at $1.1\times10^{-7}$.

With all three relaxations (four ports, twice the area, $p_{\sqrt T}=p_{T}$, no fix-ups), \Mfac{30,8,2} meets the $10^{-7}$ target at $d=21$ in every block.
Against the measurement-based CCZ-only construction with the costed \Mfac{8,3,2} layout it is then faster in every block, but always at a larger installed volume: $638$ against $937$ rounds in the compact block ($-32\%$ latency, $+25\%$ volume), $496$ against $601$ in the intermediate block ($-17\%$, $+47\%$), and $360$ against $371$ in the fast block ($-3\%$, $+69\%$).
When the \Mfac{8,3,2} factories get the same parallel layout, and so does the \Mfac{14,6,2} correction factory, \Mfac{30,8,2} keeps a latency advantage only where the data block is slow: $-24\%$ latency at $+26\%$ volume in the compact block, $-5\%$ latency at $+54\%$ volume in the intermediate block, and $1.9\times$ the latency and $3.0\times$ the volume in the fast block, where the CCZ-only construction reaches the target at $d=21$ in $162$ rounds.
With only two ports ($\tau=15$, area $\times1.5$), \Mfac{30,8,2} at $d=21$ is $17\%$ faster than the measurement-based CCZ-only construction in the compact block, about level ($+3\%$) in the intermediate block and slower in the fast block.

The mechanism relies on the number of consumptions on the critical path.
The measurement-based CCZ-only construction consumes four $CCZ$ states in two serial rounds of two, each costing a data-block step $\delta$, plus one reaction time.
\Mfac{30,8,2} consumes one \Mfac{30,8,2} output and, in $255/256$ of the cases, one \Mfac{14,6,2} correction output.
In a slow data block ($\delta=9$ or $5$ steps) the saved consumptions outweigh the longer factory run, provided the run is short; in the fast block ($\delta=1$) the factory run dominates and the smaller CCZ factory wins.
\Mfac{30,8,2} is therefore a latency-for-volume trade in compact, consumption-bound layouts, the same kind of trade as the third-level resources but with a volume premium, if a parallel-injection \Mfac{30,8,2} factory with $\tau\lesssim10$ steps at about twice the serial area can be built and if the $\sqrt T$ source is accurate enough to lower the floor well below the target.
Neither condition is met at the cultivation settings of this paper, so we report the trade as conditional; a fourth-level resource whose $CCCZ$ factors share qubits is a separate route.

\input{includes/generated/boolean_r30_optimistic.tex}
\subsection{Three \texorpdfstring{$CCCZ$}{CCCZ} gates sharing a control: \texorpdfstring{\Mfac{40,10,2}}{the [[40,10,2]] factory}}\label{section:cccz_star_costing}

For $R=3$, the glued $CCCZ$ stars of \cref{section:cccz_star} constitute a $\code{40,10,2}$ code whose transversal $\sqrt T$-type layer implements three $CCCZ$ gates that share one control qubit.
Unlike \Mfac{30,8,2}, its output is therefore an irreducible fourth-level star rather than two disjoint factors.
We cost its ETD factory, \Mfac{40,10,2}, with the same additions as \Mfac{30,8,2}.
It has $m_{x}=1$, so a run takes $\tau=39+B$ logical steps with $B\sim\mathrm{Bin}(39,1/2)$ fix-up steps, and every weight-two input error is undetected and harmful, $A_{2}=\binom{40}{2}=780$.
Consuming its output leaves zero, one, two or three $CCZ$ corrections, with probabilities $1/1024$, $21/1024$, $147/1024$ and $855/1024$ (mean $2.81$); in each case the number of corrections equals the lower bound set by the stabilizer nullity of the byproduct.
The $CCZ$-only baselines apply each $CCCZ$ through a temporary AND of the control and one target: nine $CCZ$ states with unitary uncomputation, or six with measurement-based uncomputation plus one reaction time.
They keep one ancilla per $CCCZ$, so their data block holds $N=13$ qubits against $N=10$ for \Mfac{40,10,2}.
As before, every construction installs two factories and shares a pool of $32$ lanes.

\cref{tab:boolean_star40} gives the results.
\Mfac{40,10,2} loses in every regime: in the fast block at $d=21$ it needs $1{,}649$ rounds and $1.0\times10^{8}$ qubit-rounds with speculative corrections, against $517$ rounds and $3.2\times10^{7}$ for the measurement-based $CCZ$-only construction, and even in the compact block, where each saved consumption is worth nine logical steps, it needs $2{,}076$ rounds against $1{,}314$.
The ETD run of \Mfac{40,10,2}, $39$ steps plus about $19.5$ fix-up steps, outweighs the consumptions it saves.
At a target error of $10^{-7}$ its floor, $780\,p_{\rm leg}^{2}$ per output plus the corrections, is $1.9\times10^{-7}$, so no configuration reaches the target, whereas both $CCZ$-only constructions reach it at $d=21$.
Sharing a control qubit does not rescue a fourth-level resource whose factory is serial.

\input{includes/generated/boolean_star40.tex}

%% file: includes/r30_ccz_baseline.tex
\begin{figure*}[t]
\centering
\begin{minipage}[b]{0.12\textwidth}\centering (a)\\[4pt]
\begin{quantikz}[row sep={0.45cm,between origins},column sep=0.35cm]
\lstick{$0$} & \ctrl{1} & \qw \\
\lstick{$1$} & \ctrl{1} & \qw \\
\lstick{$2$} & \ctrl{1} & \qw \\
\lstick{$3$} & \control{} & \qw \\
\lstick{$4$} & \ctrl{1} & \qw \\
\lstick{$5$} & \ctrl{1} & \qw \\
\lstick{$6$} & \ctrl{1} & \qw \\
\lstick{$7$} & \control{} & \qw
\end{quantikz}
\end{minipage}\hfill
\begin{minipage}[b]{0.25\textwidth}\centering (b)\\[4pt]
\begin{quantikz}[row sep={0.5cm,between origins},column sep=0.35cm]
\lstick{$0$} & \ctrl{1} & \qw & \ctrl{1} & \qw \\
\lstick{$1$} & \ctrl{3} & \qw & \ctrl{3} & \qw \\
\lstick{$2$} & \qw & \ctrl{1} & \qw & \qw \\
\lstick{$3$} & \qw & \ctrl{1} & \qw & \qw \\
\lstick{$\ket0_u$} & \targ{} & \control{} & \targ{} & \rstick{$\ket0$}\qw
\end{quantikz}
\end{minipage}\hfill
\begin{minipage}[b]{0.27\textwidth}\centering (c)\\[4pt]
\begin{quantikz}[row sep={0.5cm,between origins},column sep=0.35cm]
\lstick{$0$} & \ctrl{1} & \qw & \qw & \ctrl{1} & \qw \\
\lstick{$1$} & \ctrl{3} & \qw & \qw & \control{}\vcw{3} & \qw \\
\lstick{$2$} & \qw & \ctrl{1} & \qw & \qw & \qw \\
\lstick{$3$} & \qw & \ctrl{1} & \qw & \qw & \qw \\
\lstick{$\ket0_u$} & \targ{} & \control{} & \meter{$X$} & \cw & 
\end{quantikz}
\end{minipage}\hfill
\begin{minipage}[b]{0.3\textwidth}\centering (d)\\[4pt]
\begin{quantikz}[row sep={0.45cm,between origins},column sep=0.3cm]
\lstick{$0$} & \gate[wires=8]{\begin{array}{c}\text{consume}\\ \Mfac{30,8,2}\\ \text{output}\end{array}} & \qw & \gate[wires=4]{B_{0123}} & \qw \\
\lstick{$1$} & & \qw & & \qw \\
\lstick{$2$} & & \qw & & \qw \\
\lstick{$3$} & & \qw & & \qw \\
\lstick{$4$} & & \qw & \gate[wires=4]{B_{4567}} & \qw \\
\lstick{$5$} & & \qw & & \qw \\
\lstick{$6$} & & \qw & & \qw \\
\lstick{$7$} & & \qw & & \qw
\end{quantikz}
\end{minipage}
\caption{The \Mfac{30,8,2} comparison of \cref{section:fourth_level}.
(a)~The benchmark, two $CCCZ$ gates on disjoint qubits.
(b), (c)~The $CCZ$-only constructions with unitary and measurement-based uncomputation of the ancilla $u$, drawn for $CCCZ(0,1,2,3)$.
(d)~The \Mfac{30,8,2} construction with its teleportation byproducts $B$.}
\label{fig:r30_ccz_baseline}
\end{figure*}
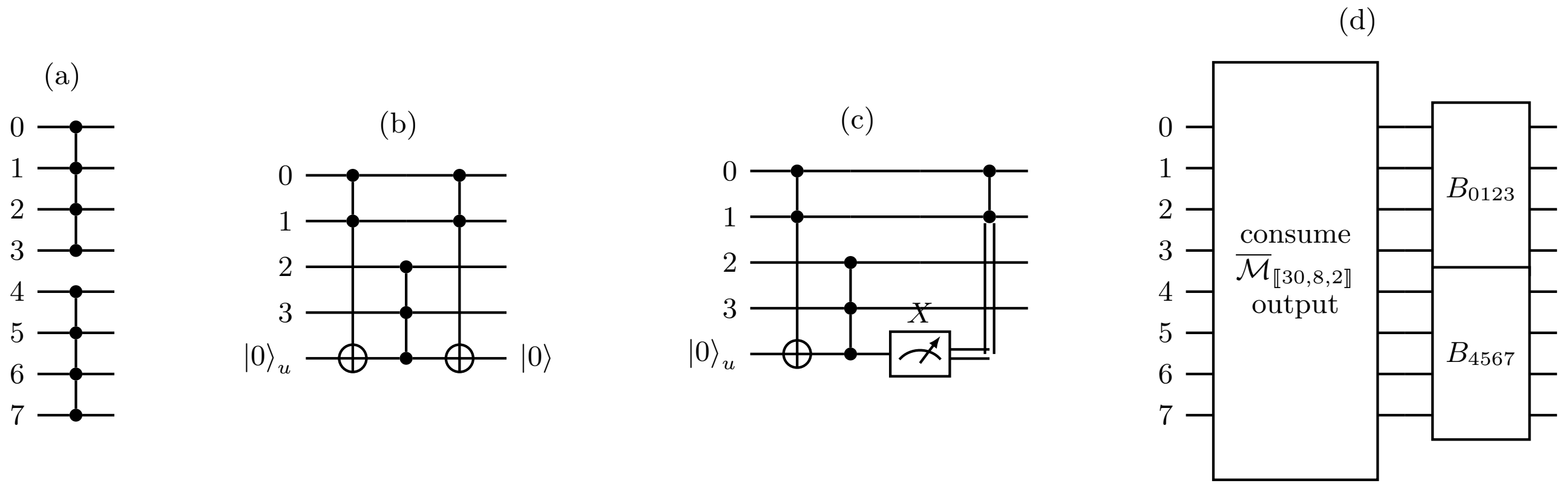

%% file: includes/generated/boolean_r30.tex
\begin{table*}[t]\centering
\caption{Two disjoint $CCCZ$ gates, the reduced form of \Mfac{30,8,2}, applied with \Mfac{30,8,2} or with $CCZ$ states only, at $d=21$ in the model of \cref{section:fourth_level}: every row installs two factories and shares a pool of 32 lanes, and column $\sqrt T$ gives the $\sqrt T$ lanes of the best split. $E_2$ is the error of the distilled outputs of the main factory, $E_3$ the topological error outside the correction hardware, and $E_4$ the error of the correction outputs and their factories. Latency is in rounds and volume in qubit-rounds. A latency in \textcolor{darkred}{dark red} or a volume in \textcolor{darkblue}{dark blue} is smaller than the best CCZ-only value in the same column.}\label{tab:boolean_r30}
\widetable{\begin{tabular}{llrrrrrrrrrrr}\toprule
 & & & \multicolumn{2}{c}{Compact} & \multicolumn{2}{c}{Intermediate} & \multicolumn{2}{c}{Fast} & \multicolumn{4}{c}{Fast} \\
\cmidrule(lr){4-5}\cmidrule(lr){6-7}\cmidrule(lr){8-9}\cmidrule(lr){10-13}
Construction & Hardware & $\sqrt T$ & Rounds & Volume & Rounds & Volume & Rounds & Volume & Footprint & $E_2$ & $E_3$ & $E_4$ \\\midrule
\begin{tabular}[t]{@{}l@{}}CCZ-only,\\unitary (6 CCZ)\end{tabular} & \ensuremath{2{\times}}\Mfac{8,3,2} & -- & 1,305 & $4.7\times10^{7}$ & 801 & $3.7\times10^{7}$ & 509 & $2.9\times10^{7}$ & 105,919 & $3.5\times10^{-10}$ & $3.0\times10^{-8}$ & 0 \\
\begin{tabular}[t]{@{}l@{}}CCZ-only,\\measurement (4 CCZ)\end{tabular} & \ensuremath{2{\times}}\Mfac{8,3,2} & -- & 937 & $2.9\times10^{7}$ & 601 & $2.4\times10^{7}$ & 372 & $1.9\times10^{7}$ & 105,919 & $2.4\times10^{-10}$ & $1.9\times10^{-8}$ & 0 \\
\midrule
\begin{tabular}[t]{@{}l@{}}\Mfac{30,8,2},\\speculative corrections\end{tabular} & \begin{tabular}[t]{@{}l@{}}\Mfac{30,8,2}+\\S8\end{tabular} & 28 & 1,545 & $5.7\times10^{7}$ & 1,340 & $5.9\times10^{7}$ & 1,172 & $5.9\times10^{7}$ & 176,969 & $1.1\times10^{-7}$ & $5.8\times10^{-8}$ & $5.4\times10^{-9}$ \\
\begin{tabular}[t]{@{}l@{}}\Mfac{30,8,2},\\speculative corrections\end{tabular} & \begin{tabular}[t]{@{}l@{}}\Mfac{30,8,2}+\\T14\end{tabular} & 28 & 1,384 & $5.2\times10^{7}$ & 1,217 & $5.3\times10^{7}$ & 1,049 & $5.4\times10^{7}$ & 194,168 & $1.1\times10^{-7}$ & $5.0\times10^{-8}$ & $7.4\times10^{-9}$ \\
\begin{tabular}[t]{@{}l@{}}\Mfac{30,8,2},\\reactive corrections\end{tabular} & \begin{tabular}[t]{@{}l@{}}\Mfac{30,8,2}+\\S8\end{tabular} & 24 & 1,776 & $5.5\times10^{7}$ & 1,572 & $5.8\times10^{7}$ & 1,404 & $6.0\times10^{7}$ & 176,969 & $1.1\times10^{-7}$ & $6.1\times10^{-8}$ & $3.0\times10^{-9}$ \\
\begin{tabular}[t]{@{}l@{}}\Mfac{30,8,2},\\reactive corrections\end{tabular} & \begin{tabular}[t]{@{}l@{}}\Mfac{30,8,2}+\\T14\end{tabular} & 16 & 1,741 & $5.5\times10^{7}$ & 1,573 & $5.9\times10^{7}$ & 1,406 & $6.1\times10^{7}$ & 194,168 & $1.1\times10^{-7}$ & $6.1\times10^{-8}$ & $3.9\times10^{-9}$ \\
\midrule
\begin{tabular}[t]{@{}l@{}}\Mfac{30,8,2},\\speculative,\\reaction $d$ rounds\end{tabular} & \begin{tabular}[t]{@{}l@{}}\Mfac{30,8,2}+\\S8\end{tabular} & 28 & 1,558 & $5.8\times10^{7}$ & 1,353 & $5.9\times10^{7}$ & 1,186 & $6.0\times10^{7}$ & 176,969 & $1.1\times10^{-7}$ & $5.8\times10^{-8}$ & $5.4\times10^{-9}$ \\
\begin{tabular}[t]{@{}l@{}}\Mfac{30,8,2},\\speculative,\\no $\sqrt T$ fix-up\end{tabular} & \begin{tabular}[t]{@{}l@{}}\Mfac{30,8,2}+\\S8\end{tabular} & 28 & 1,233 & $4.1\times10^{7}$ & 1,029 & $4.1\times10^{7}$ & 861 & $4.0\times10^{7}$ & 151,391 & $9.8\times10^{-8}$ & $3.8\times10^{-8}$ & $4.5\times10^{-9}$ \\
\midrule
\begin{tabular}[t]{@{}l@{}}\Mfac{30,8,2},\\speculative,\\$p_{\sqrt T}=10^{-5}$\end{tabular} & \begin{tabular}[t]{@{}l@{}}\Mfac{30,8,2}+\\S8\end{tabular} & 28 & 1,544 & $5.7\times10^{7}$ & 1,339 & $5.9\times10^{7}$ & 1,172 & $5.9\times10^{7}$ & 176,969 & $5.0\times10^{-8}$ & $5.8\times10^{-8}$ & $5.4\times10^{-9}$ \\
\begin{tabular}[t]{@{}l@{}}\Mfac{30,8,2},\\speculative,\\$p_{\sqrt T}=10^{-5}$\end{tabular} & \begin{tabular}[t]{@{}l@{}}\Mfac{30,8,2}+\\T14\end{tabular} & 28 & 1,384 & $5.2\times10^{7}$ & 1,217 & $5.4\times10^{7}$ & 1,049 & $5.4\times10^{7}$ & 194,168 & $5.0\times10^{-8}$ & $5.0\times10^{-8}$ & $7.4\times10^{-9}$ \\
\begin{tabular}[t]{@{}l@{}}\Mfac{30,8,2},\\speculative,\\$p_{\sqrt T}=5\times10^{-6}$\end{tabular} & \begin{tabular}[t]{@{}l@{}}\Mfac{30,8,2}+\\S8\end{tabular} & 28 & 1,555 & $5.8\times10^{7}$ & 1,350 & $5.9\times10^{7}$ & 1,182 & $6.0\times10^{7}$ & 176,969 & $1.4\times10^{-8}$ & $5.8\times10^{-8}$ & $5.4\times10^{-9}$ \\
\begin{tabular}[t]{@{}l@{}}\Mfac{30,8,2},\\speculative,\\$p_{\sqrt T}=5\times10^{-6}$\end{tabular} & \begin{tabular}[t]{@{}l@{}}\Mfac{30,8,2}+\\T14\end{tabular} & 28 & 1,384 & $5.2\times10^{7}$ & 1,216 & $5.3\times10^{7}$ & 1,048 & $5.4\times10^{7}$ & 194,168 & $1.4\times10^{-8}$ & $5.0\times10^{-8}$ & $7.4\times10^{-9}$ \\
\bottomrule\end{tabular}}\end{table*}

%% file: includes/generated/boolean_r30_optimistic.tex
\begin{table*}[t]\centering
\caption{\Mfac{30,8,2} under the optimistic parameters of \cref{section:fourth_level_optimistic} at $d=21$, with the workload, lane pool and model of \cref{tab:boolean_r30}. Starting from the best costed configurations, the rows set $p_{\sqrt T}=p_T=1.45\times10^{-6}$ at unchanged lane cost, drop the $P_{\pi/8}$ fix-ups, and use the hypothetical parallel-injection layout ($\tau=\lceil (n-m_x)/\text{ports}\rceil$); the second row gives this layout to the CCZ factories. Each configuration uses the $\sqrt T{:}T$ split of smallest installed volume, and the last block runs it at the smallest odd $d\ge21$ that meets the target error. Latency in rounds, volumes in qubit-rounds. A latency in \textcolor{darkred}{dark red} or a volume in \textcolor{darkblue}{dark blue} is smaller than the best CCZ-only value (first two rows, or their counterparts in the last block) in the same column.}\label{tab:boolean_r30_optimistic}
\widetable{\begin{tabular}{lrrrrrrrrr}\toprule
 & \multicolumn{3}{c}{Compact} & \multicolumn{3}{c}{Intermediate} & \multicolumn{3}{c}{Fast} \\
\cmidrule(lr){2-4}\cmidrule(lr){5-7}\cmidrule(lr){8-10}
Configuration & Rounds & Occupied & Installed & Rounds & Occupied & Installed & Rounds & Occupied & Installed \\\midrule
CCZ-only, measurement (4 CCZ), \ensuremath{2{\times}}\Mfac{8,3,2} & 936 & $2.9\times10^{7}$ & $8.9\times10^{7}$ & 600 & $2.4\times10^{7}$ & $6.0\times10^{7}$ & 370 & $1.9\times10^{7}$ & $3.9\times10^{7}$ \\
\quad parallel-injection \Mfac{8,3,2} (4 ports, area $\times2$) & 831 & $2.4\times10^{7}$ & $9.4\times10^{7}$ & 495 & $1.9\times10^{7}$ & $5.9\times10^{7}$ & 162 & $9.8\times10^{6}$ & $2.0\times10^{7}$ \\
\midrule
\Mfac{30,8,2} + \Mfac{14,6,2}, speculative (as \cref{tab:boolean_r30}) & 1,381 & $5.2\times10^{7}$ & $2.6\times10^{8}$ & 1,213 & $5.3\times10^{7}$ & $2.3\times10^{8}$ & 1,045 & $5.4\times10^{7}$ & $2.0\times10^{8}$ \\
\quad $p_{\sqrt T}=p_T$, no $\sqrt T$ fix-ups & 1,068 & $3.6\times10^{7}$ & $1.7\times10^{8}$ & 901 & $3.6\times10^{7}$ & $1.5\times10^{8}$ & 733 & $3.5\times10^{7}$ & $1.2\times10^{8}$ \\
\quad parallel-injection \Mfac{30,8,2} (4 ports, area $\times2$) & \textcolor{darkred}{747} & $3.1\times10^{7}$ & $1.5\times10^{8}$ & 579 & $3.0\times10^{7}$ & $1.2\times10^{8}$ & 427 & $2.8\times10^{7}$ & $9.0\times10^{7}$ \\
\quad 2 ports, area $\times1.5$, $p_{\sqrt T}=p_T$, no fix-ups & \textcolor{darkred}{776} & $2.8\times10^{7}$ & $1.3\times10^{8}$ & 616 & $2.8\times10^{7}$ & $1.1\times10^{8}$ & 450 & $2.5\times10^{7}$ & $7.9\times10^{7}$ \\
\quad 4 ports, area $\times2$, $p_{\sqrt T}=p_T$, no fix-ups & \textcolor{darkred}{638} & $2.4\times10^{7}$ & $1.1\times10^{8}$ & 500 & $2.3\times10^{7}$ & $9.0\times10^{7}$ & 361 & $2.1\times10^{7}$ & $6.6\times10^{7}$ \\
\quad same, \Mfac{14,6,2} also 4 ports, area $\times2$ & \textcolor{darkred}{626} & \textcolor{darkblue}{$2.3\times10^{7}$} & $1.2\times10^{8}$ & \textcolor{darkred}{465} & $2.2\times10^{7}$ & $8.9\times10^{7}$ & 300 & $1.9\times10^{7}$ & $5.9\times10^{7}$ \\
\midrule
\multicolumn{10}{l}{\emph{Target error $1.0\times10^{-7}$: distance, rounds, installed volume}} \\
CCZ-only, measurement (4 CCZ), \ensuremath{2{\times}}\Mfac{8,3,2} & $d=21$ & 937 & $8.9\times10^{7}$ & $d=21$ & 601 & $6.0\times10^{7}$ & $d=21$ & 371 & $3.9\times10^{7}$ \\
\quad parallel-injection \Mfac{8,3,2} (4 ports, area $\times2$) & $d=21$ & 831 & $9.4\times10^{7}$ & $d=21$ & 495 & $5.9\times10^{7}$ & $d=21$ & 162 & $2.0\times10^{7}$ \\
\Mfac{30,8,2} + \Mfac{14,6,2}, speculative (as \cref{tab:boolean_r30}) & -- & \multicolumn{2}{c}{floor $1.1\times10^{-7}$} & -- & \multicolumn{2}{c}{floor $1.1\times10^{-7}$} & -- & \multicolumn{2}{c}{floor $1.1\times10^{-7}$} \\
\quad $p_{\sqrt T}=p_T$, no $\sqrt T$ fix-ups & $d=21$ & 1,067 & $1.7\times10^{8}$ & $d=21$ & 900 & $1.5\times10^{8}$ & $d=21$ & 732 & $1.2\times10^{8}$ \\
\quad 4 ports, area $\times2$, $p_{\sqrt T}=p_T$, no fix-ups & $d=21$ & \textcolor{darkred}{638} & $1.1\times10^{8}$ & $d=21$ & 496 & $8.9\times10^{7}$ & $d=21$ & 360 & $6.6\times10^{7}$ \\
\quad same, \Mfac{14,6,2} also 4 ports, area $\times2$ & $d=21$ & \textcolor{darkred}{630} & $1.2\times10^{8}$ & $d=21$ & \textcolor{darkred}{468} & $9.0\times10^{7}$ & $d=21$ & 303 & $6.0\times10^{7}$ \\
\bottomrule\end{tabular}}\end{table*}

%% file: includes/generated/boolean_star40.tex
\begin{table*}[t]\centering
\caption{Three $CCCZ$ gates sharing one control qubit, applied with the glued $CCCZ$ star \Mfac{40,10,2} or with $CCZ$ states only, in the model of \cref{tab:boolean_r30} (\cref{section:cccz_star_costing}), at $d=21$ in the three data blocks. Latency is in rounds and volume in qubit-rounds. A latency in \textcolor{darkred}{dark red} or a volume in \textcolor{darkblue}{dark blue} is smaller than the best CCZ-only value in the same column.}\label{tab:boolean_star40}
\widetable{\begin{tabular}{l|rr|rr|rr}\toprule
 & \multicolumn{2}{c|}{Compact} & \multicolumn{2}{c|}{Intermediate} & \multicolumn{2}{c}{Fast} \\
Construction & Rounds & Volume & Rounds & Volume & Rounds & Volume \\\midrule
CCZ-only, unitary (9 CCZ) & 1,870 & $8.7\times10^{7}$ & 1,114 & $6.3\times10^{7}$ & 778 & $5.1\times10^{7}$ \\
CCZ-only, measurement (6 CCZ) & 1,314 & $5.3\times10^{7}$ & 810 & $4.1\times10^{7}$ & 517 & $3.2\times10^{7}$ \\
\midrule
\Mfac{40,10,2}, speculative corrections & 2,076 & $9.8\times10^{7}$ & 1,817 & $1.0\times10^{8}$ & 1,649 & $1.0\times10^{8}$ \\
\Mfac{40,10,2}, reactive corrections & 2,297 & $8.7\times10^{7}$ & 2,053 & $9.3\times10^{7}$ & 1,885 & $9.7\times10^{7}$ \\
\Mfac{40,10,2}, speculative, $p_{\sqrt T}=5\times10^{-6}$ & 2,079 & $9.8\times10^{7}$ & 1,820 & $1.0\times10^{8}$ & 1,652 & $1.0\times10^{8}$ \\
\bottomrule\end{tabular}}\end{table*}

%% file: main.bib
@article{ray2024implementing,
  title = {Implementing fault-tolerant non-Clifford gates using the [[8,3,2]] color code},
  author = {Honciuc Menendez, Daniel and Ray, Annie and Vasmer, Michael},
  journal = {Phys. Rev. A},
  volume = {109},
  issue = {6},
  pages = {062438},
  numpages = {13},
  year = {2024},
  month = {Jun},
  publisher = {American Physical Society},
  doi = {10.1103/PhysRevA.109.062438},
  url = {https://link.aps.org/doi/10.1103/PhysRevA.109.062438}
}

@article{hou2025stabilizer,
    author = {Hou, Zong-Yue and Cao, ChunJun and Yang, Zhi-Cheng},
    title = {Stabilizer Entanglement Enhances Magic Injection},
    journal = {npj Quantum Information},
    volume = {12},
    pages = {113},
    year = {2026},
    doi = {10.1038/s41534-026-01265-4},
    eprint = {2503.20873},
    archivePrefix = {arXiv}
}

@article{cao2022quantum,
  title={Quantum lego: Building quantum error correction codes from tensor networks},
  author={Cao, ChunJun and Lackey, Brad},
  journal={PRX Quantum},
  volume={3},
  number={2},
  pages={020332},
  year={2022},
  publisher={APS}
}

@article{rengaswamy2020optimality,
  title={On optimality of CSS codes for transversal T},
  author={Rengaswamy, Narayanan and Calderbank, Robert and Newman, Michael and Pfister, Henry D},
  journal={IEEE Journal on Selected Areas in Information Theory},
  volume={1},
  number={2},
  pages={499--514},
  year={2020},
  publisher={IEEE}
}

@article{bravyi2012magic,
  title={Magic-state distillation with low overhead},
  author={Bravyi, Sergey and Haah, Jeongwan},
  journal={Physical Review A—Atomic, Molecular, and Optical Physics},
  volume={86},
  number={5},
  pages={052329},
  year={2012},
  publisher={APS}
}

@article{jain2025transversal,
  title={Transversal Clifford and T-gate codes of short length and high distance},
  author={Jain, Shubham P and Albert, Victor V},
  journal={IEEE Journal on Selected Areas in Information Theory},
  year={2025},
  doi={10.1109/JSAIT.2025.3570832},
  eprint={2408.12752},
  archivePrefix={arXiv},
  publisher={IEEE}
}

@article{bravyi2015doubled,
  title={Doubled color codes},
  author={Bravyi, Sergey and Cross, Andrew},
  journal={arXiv preprint arXiv:1509.03239},
  year={2015}
}

@book{gottesman1997stabilizer,
  title={Stabilizer codes and quantum error correction},
  author={Gottesman, Daniel},
  year={1997},
  publisher={California Institute of Technology}
}

@article{vanhoof2020space,
  title={Space-efficient quantum multiplication of polynomials for binary finite fields with sub-quadratic Toffoli gate count},
  author={Van Hoof, Iggy},
  journal={arXiv preprint arXiv:1910.02849},
  year={2019}
}

@article{pato2025hyper,
  title={Hyper-optimized Quantum Lego Contraction Schedules},
  author={Pato, Balint and Vanlerberghe, June and Brown, Kenneth R},
  journal={arXiv preprint arXiv:2510.08210},
  year={2025}
}

@article{cao2025growing,
  title={Growing sparse quantum codes from a seed},
  author={Cao, ChunJun and Lackey, Brad},
  journal={arXiv preprint arXiv:2507.13496},
  year={2025}
}

@article{cao2024quantum,
  title={Quantum Lego expansion pack: Enumerators from tensor networks},
  author={Cao, ChunJun and Gullans, Michael J and Lackey, Brad and Wang, Zitao},
  journal={PRX Quantum},
  volume={5},
  number={3},
  pages={030313},
  year={2024},
  publisher={APS}
}

@article{webster2023transversal,
  title={Transversal diagonal logical operators for stabiliser codes},
  author={Webster, Mark A and Quintavalle, Armanda O and Bartlett, Stephen D},
  journal={New Journal of Physics},
  volume={25},
  number={10},
  pages={103018},
  year={2023},
  publisher={IOP Publishing}
}

@article{vasmer2022morphing,
  title={Morphing quantum codes},
  author={Vasmer, Michael and Kubica, Aleksander},
  journal={PRX Quantum},
  volume={3},
  number={3},
  pages={030319},
  year={2022},
  publisher={APS}
}

@article{hangleiter2025fault,
  title={Fault-Tolerant Compiling of Classically Hard Instantaneous Quantum Polynomial Circuits on Hypercubes},
  author={Hangleiter, Dominik and Kalinowski, Marcin and Bluvstein, Dolev and Cain, Madelyn and Maskara, Nishad and Gao, Xun and Kubica, Aleksander and Lukin, Mikhail D and Gullans, Michael J},
  journal={PRX Quantum},
  volume={6},
  number={2},
  pages={020338},
  year={2025},
  publisher={APS}
}

@article{wills2025constant,
  title={Constant-overhead magic state distillation},
  author={Wills, Adam and Hsieh, Min-Hsiu and Yamasaki, Hayata},
  journal={Nature Physics},
  volume={21},
  number={11},
  pages={1842--1846},
  year={2025},
  publisher={Nature Publishing Group UK London}
}

@article{vuillot2022quantum,
  title={Quantum pin codes},
  author={Vuillot, Christophe and Breuckmann, Nikolas P},
  journal={IEEE Transactions on Information Theory},
  volume={68},
  number={9},
  pages={5955--5974},
  year={2022},
  publisher={IEEE}
}

@article{hastings2018distillation,
  title={Distillation with sublogarithmic overhead},
  author={Hastings, Matthew B and Haah, Jeongwan},
  journal={Physical review letters},
  volume={120},
  number={5},
  pages={050504},
  year={2018},
  publisher={APS}
}

@article{krishna2018towards,
  title={Towards low overhead magic state distillation},
  author={Krishna, Anirudh and Tillich, Jean-Pierre},
  journal={arXiv preprint arXiv:1811.08461},
  year={2018}
}

@article{cui2017diagonal,
  title={Diagonal gates in the {Clifford} hierarchy},
  author={Cui, Shawn X. and Gottesman, Daniel and Krishna, Anirudh},
  journal={Physical Review A},
  volume={95},
  number={1},
  pages={012329},
  year={2017},
  doi={10.1103/PhysRevA.95.012329}
}

@article{campbell2017unified,
  title={Unified framework for magic state distillation and multiqubit gate synthesis with reduced resource cost},
  author={Campbell, Earl T and Howard, Mark},
  journal={Physical Review A},
  volume={95},
  number={2},
  pages={022316},
  year={2017},
  publisher={APS}
}

@article{campbell2017unifying,
  title={Unifying gate synthesis and magic state distillation},
  author={Campbell, Earl T and Howard, Mark},
  journal={Physical review letters},
  volume={118},
  number={6},
  pages={060501},
  year={2017},
  publisher={APS}
}

@inproceedings{mqt,
  title        = {The {{MQT}} Handbook: {{A}} Summary of Design Automation Tools and Software for Quantum Computing},
  shorttitle   = {{The MQT Handbook}},
  author       = {Wille, Robert and Berent, Lucas and Forster, Tobias and Kunasaikaran, Jagatheesan and Mato, Kevin and Peham, Tom and Quetschlich, Nils and Rovara, Damian and Sander, Aaron and Schmid, Ludwig and Schoenberger, Daniel and Stade, Yannick and Burgholzer, Lukas},
  year         = 2024,
  booktitle    = {IEEE International Conference on Quantum Software (QSW)},
  doi          = {10.1109/QSW62656.2024.00013},
  eprint       = {2405.17543},
  eprinttype   = {arxiv},
  addendum     = {A live version of this document is available at \url{https://mqt.readthedocs.io}}
}

@article{haah2018codes,
  title={Codes and protocols for distilling $ t $, controlled-$ s $, and toffoli gates},
  author={Haah, Jeongwan and Hastings, Matthew B},
  journal={Quantum},
  volume={2},
  pages={71},
  year={2018},
  publisher={Verein zur F{\"o}rderung des Open Access Publizierens in den Quantenwissenschaften}
}

@article{Shen2025quantumlegoxp,
  doi = {10.22331/q-2025-10-06-1876},
  url = {https://doi.org/10.22331/q-2025-10-06-1876},
  title = {Quantum {L}ego and {XP} {S}tabilizer {C}odes},
  author = {Shen, Ruohan and Wang, Yixu and Cao, ChunJun},
  journal = {{Quantum}},
  issn = {2521-327X},
  publisher = {{Verein zur F{\"{o}}rderung des Open Access Publizierens in den Quantenwissenschaften}},
  volume = {9},
  pages = {1876},
  month = oct,
  year = {2025}
}

@article{gidney2024magic,
  title={Magic state cultivation: growing T states as cheap as CNOT gates},
  author={Gidney, Craig and Shutty, Noah and Jones, Cody},
  journal={arXiv preprint arXiv:2409.17595},
  year={2024}
}

@article{litinski2019magic,
  title={Magic state distillation: Not as costly as you think},
  author={Litinski, Daniel},
  journal={Quantum},
  volume={3},
  pages={205},
  year={2019},
  publisher={Verein zur F{\"o}rderung des Open Access Publizierens in den Quantenwissenschaften}
}

@article{amy2014polynomial,
  title={Polynomial-time {T}-depth optimization of {Clifford+T} circuits via matroid partitioning},
  author={Amy, Matthew and Maslov, Dmitri and Mosca, Michele},
  journal={IEEE Transactions on Computer-Aided Design of Integrated Circuits and Systems},
  volume={33},
  number={10},
  pages={1476--1489},
  year={2014},
  doi={10.1109/TCAD.2014.2341953},
  publisher={IEEE}
}

@article{amy2019t,
  title={T-count optimization and Reed--Muller codes},
  author={Amy, Matthew and Mosca, Michele},
  journal={IEEE Transactions on Information Theory},
  volume={65},
  number={8},
  pages={4771--4784},
  year={2019},
  publisher={IEEE}
}

@article{gottesman1999quantum,
  title={Quantum teleportation is a universal computational primitive},
  author={Gottesman, Daniel and Chuang, Isaac L},
  journal={arXiv preprint quant-ph/9908010},
  year={1999}
}

@article{grassl2009generalized,
  title={Generalized concatenated quantum codes},
  author={Grassl, Markus and Shor, Peter and Smith, Graeme and Smolin, John and Zeng, Bei},
  journal={Physical Review A—Atomic, Molecular, and Optical Physics},
  volume={79},
  number={5},
  pages={050306},
  year={2009},
  publisher={APS}
}

@article{litinski2019game,
  title={A game of surface codes: Large-scale quantum computing with lattice surgery},
  author={Litinski, Daniel},
  journal={Quantum},
  volume={3},
  pages={128},
  year={2019},
  publisher={Verein zur F{\"o}rderung des Open Access Publizierens in den Quantenwissenschaften}
}

@article{cao2023quantum,
  title={Quantum weight enumerators and tensor networks},
  author={Cao, ChunJun and Lackey, Brad},
  journal={IEEE Transactions on Information Theory},
  volume={70},
  number={5},
  pages={3512--3528},
  year={2023},
  publisher={IEEE}
}

@article{beverland2020lower,
  title={Lower bounds on the non-{C}lifford resources for quantum computations},
  author={Beverland, Michael and Campbell, Earl and Howard, Mark and Kliuchnikov, Vadym},
  journal={Quantum Science and Technology},
  volume={5},
  number={3},
  pages={035009},
  year={2020},
  publisher={IOP Publishing}
}

@article{leone2024stabilizer,
  title={Stabilizer entropies are monotones for magic-state resource theory},
  author={Leone, Lorenzo and Bittel, Lennart},
  journal={Physical Review A},
  volume={110},
  number={4},
  pages={L040403},
  year={2024},
  publisher={APS}
}

@article{leone2022stabilizer,
  title={Stabilizer r{\'e}nyi entropy},
  author={Leone, Lorenzo and Oliviero, Salvatore FE and Hamma, Alioscia},
  journal={Physical Review Letters},
  volume={128},
  number={5},
  pages={050402},
  year={2022},
  publisher={APS}
}

@article{bravyi2005universal,
  title={Universal quantum computation with ideal Clifford gates and noisy ancillas},
  author={Bravyi, Sergey and Kitaev, Alexei},
  journal={Physical Review A—Atomic, Molecular, and Optical Physics},
  volume={71},
  number={2},
  pages={022316},
  year={2005},
  publisher={APS}
}

@article{cao2026quantum,
  title={Quantum Lego Power-up: Designing Transversal Gates with Tensor Networks},
  author={Cao, ChunJun and Lackey, Brad},
  journal={arXiv preprint arXiv:2603.03542},
  eprint={2603.03542},
  archivePrefix={arXiv},
  year={2026}
}

@article{cowtan2020phase,
  title={Phase Gadget Synthesis for Shallow Circuits},
  author={Cowtan, Alexander and Dilkes, Silas and Duncan, Ross and Simmons, Will and Sivarajah, Seyon},
  journal={Electronic Proceedings in Theoretical Computer Science},
  volume={318},
  pages={213--228},
  year={2020},
  doi={10.4204/EPTCS.318.13},
  eprint={1906.01734},
  archivePrefix={arXiv}
}

@article{kubica2015unfolding,
  title={Unfolding the color code},
  author={Kubica, Aleksander and Yoshida, Beni and Pastawski, Fernando},
  journal={New Journal of Physics},
  volume={17},
  number={8},
  pages={083026},
  year={2015},
  publisher={IOP Publishing}
}

@misc{perlin2023qldpc,
  author = {Perlin, Michael A.},
  title = {{qLDPC}},
  year = {2023},
  publisher = {GitHub},
  journal = {GitHub repository},
  howpublished = {\url{https://github.com/qLDPCOrg/qLDPC}},
}

@misc{pato_2025_planqtn,
  author       = {Pato, Balint and
                  Vanlerberghe, June and
                  Cao, ChunJun and
                  Lackey, Brad and
                  Brown, Kenneth},
  title        = {PlanqTN, a Python library and interactive web app
                   implementing the quantum LEGO framework
                  },
  month        = aug,
  year         = 2025,
  publisher    = {Zenodo},
  doi          = {10.5281/zenodo.16761072},
  url          = {https://doi.org/10.5281/zenodo.16761072},
}

@article{wang2025comprehensive,
  title={A comprehensive study of quantum arithmetic circuits},
  author={Wang, Siyi and Li, Xiufan and Lee, Wei Jie Bryan and Deb, Suman and Lim, Eugene and Chattopadhyay, Anupam},
  journal={Philosophical Transactions of the Royal Society A: Mathematical, Physical and Engineering Sciences},
  volume={383},
  number={2288},
  year={2025},
  publisher={The Royal Society}
}

@article{shor1999polynomial,
  title={Polynomial-time algorithms for prime factorization and discrete logarithms on a quantum computer},
  author={Shor, Peter W},
  journal={SIAM review},
  volume={41},
  number={2},
  pages={303--332},
  year={1999},
  publisher={SIAM}
}

@article{amy2013meet,
  title={A meet-in-the-middle algorithm for fast synthesis of depth-optimal quantum circuits},
  author={Amy, Matthew and Maslov, Dmitri and Mosca, Michele and Roetteler, Martin},
  journal={IEEE Transactions on Computer-Aided Design of Integrated Circuits and Systems},
  volume={32},
  number={6},
  pages={818--830},
  year={2013},
  publisher={IEEE}
}

@book{nielsen2010quantum,
  title={Quantum computation and quantum information},
  author={Nielsen, Michael A and Chuang, Isaac L},
  year={2010},
  publisher={Cambridge university press}
}

@article{campbell2017roads,
  title={Roads towards fault-tolerant universal quantum computation},
  author={Campbell, Earl T and Terhal, Barbara M and Vuillot, Christophe},
  journal={Nature},
  volume={549},
  number={7671},
  pages={172--179},
  year={2017},
  publisher={Nature Publishing Group UK London}
}

@article{eastin2009restrictions,
  title={Restrictions on transversal encoded quantum gate sets},
  author={Eastin, Bryan and Knill, Emanuel},
  journal={Physical review letters},
  volume={102},
  number={11},
  pages={110502},
  year={2009},
  publisher={APS}
}

@article{jochym2014using,
  title={Using concatenated quantum codes for universal fault-tolerant quantum gates},
  author={Jochym-O’Connor, Tomas and Laflamme, Raymond},
  journal={Physical review letters},
  volume={112},
  number={1},
  pages={010505},
  year={2014},
  publisher={APS}
}

@article{ogorman2017quantum,
  title={Quantum computation with realistic magic-state factories},
  author={O'Gorman, Joe and Campbell, Earl T},
  journal={Physical Review A},
  volume={95},
  number={3},
  pages={032338},
  year={2017},
  publisher={APS}
}

@article{steinberg2025far,
  title={Far from perfect: Quantum error correction with (hyperinvariant) evenbly codes},
  author={Steinberg, Matthew and Fan, Junyu and Harris, Robert J and Elkouss, David and Feld, Sebastian and Jahn, Alexander},
  journal={Quantum},
  volume={9},
  pages={1826},
  year={2025},
  publisher={Verein zur F{\"o}rderung des Open Access Publizierens in den Quantenwissenschaften}
}

@article{steinberg2025universal,
  title={Universal fault-tolerant logic with heterogeneous holographic codes},
  author={Steinberg, Matthew and Fan, Junyu and Eisert, Jens and Feld, Sebastian and Jahn, Alexander and Cao, Chunjun},
  journal={arXiv preprint arXiv:2504.10386},
  year={2025}
}

@article{fan2024lego_hqec,
  title={LEGO\_HQEC: Automating the Analysis, Construction, and Decoding of Holographic Quantum Codes},
  author={Fan, Junyu and Steinberg, Matthew and Jahn, Alexander and Cao, Chunjun and Sarkar, Aritra and Feld, Sebastian},
  journal={arXiv preprint arXiv:2410.22861},
  year={2024}
}

@article{rajchel2025absolutely,
  title={Absolutely maximally entangled pure states of multipartite quantum systems},
  author={Rajchel-Mieldzio{\'c}, Grzegorz and Bistro{\'n}, Rafa{\l} and Rico, Albert and Lakshminarayan, Arul and {\.Z}yczkowski, Karol},
  journal={arXiv preprint arXiv:2508.04777},
  year={2025}
}

@article{goyeneche2014genuinely,
  title={Genuinely multipartite entangled states and orthogonal arrays},
  author={Goyeneche, Dardo and {\.Z}yczkowski, Karol},
  journal={Physical review A},
  volume={90},
  number={2},
  pages={022316},
  year={2014},
  publisher={APS}
}

@article{steane1996multiple,
  title={Multiple-particle interference and quantum error correction},
  author={Steane, Andrew},
  journal={Proceedings of the Royal Society of London. Series A: Mathematical, Physical and Engineering Sciences},
  volume={452},
  number={1954},
  pages={2551--2577},
  year={1996},
  publisher={The Royal Society London}
}

@article{ni2015non,
  title={A non-commuting stabilizer formalism},
  author={Ni, Xiaotong and Buerschaper, Oliver and Van den Nest, Maarten},
  journal={Journal of Mathematical Physics},
  volume={56},
  number={5},
  year={2015},
  publisher={AIP Publishing}
}

@article{webster2022xp,
  title={The XP stabiliser formalism: a generalisation of the pauli stabiliser formalism with arbitrary phases},
  author={Webster, Mark A and Brown, Benjamin J and Bartlett, Stephen D},
  journal={Quantum},
  volume={6},
  pages={815},
  year={2022},
  publisher={Verein zur F{\"o}rderung des Open Access Publizierens in den Quantenwissenschaften}
}

@article{cuccaro2004new,
  title={A new quantum ripple-carry addition circuit},
  author={Cuccaro, Steven A and Draper, Thomas G and Kutin, Samuel A and Moulton, David Petrie},
  journal={arXiv preprint quant-ph/0410184},
  year={2004}
}

@article{farrelly2021tensor,
  title={Tensor-network codes},
  author={Farrelly, Terry and Harris, Robert J and McMahon, Nathan A and Stace, Thomas M},
  journal={Physical Review Letters},
  volume={127},
  number={4},
  pages={040507},
  year={2021},
  publisher={APS}
}

@article{farrelly2022local,
  title={Local tensor-network codes},
  author={Farrelly, Terry and Tuckett, David K and Stace, Thomas M},
  journal={New Journal of Physics},
  volume={24},
  number={4},
  pages={043015},
  year={2022},
  publisher={IOP Publishing}
}

@article{cain2026shor,
  title={Shor's algorithm is possible with as few as 10,000 reconfigurable atomic qubits},
  author={Cain, Madelyn and Xu, Qian and King, Robbie and Picard, Lewis RB and Levine, Harry and Endres, Manuel and Preskill, John and Huang, Hsin-Yuan and Bluvstein, Dolev},
  journal={arXiv preprint arXiv:2603.28627},
  eprint={2603.28627},
  archivePrefix={arXiv},
  year={2026}
}

@article{huang2025hybrid,
  title={Hybrid lattice surgery: Non-Clifford gates via non-Abelian surface codes},
  author={Huang, Sheng-Jie and Warman, Alison and Schafer-Nameki, Sakura and Chen, Yanzhu},
  journal={arXiv preprint arXiv:2510.20890},
  year={2025}
}

@article{rains2002quantum,
  title={Quantum weight enumerators},
  author={Rains, Eric M},
  journal={IEEE Transactions on Information Theory},
  volume={44},
  number={4},
  pages={1388--1394},
  year={2002},
  publisher={IEEE}
}

@article{horsman2012surface,
  title = {Surface code quantum computing by lattice surgery},
  author = {Horsman, Clare and Fowler, Austin G. and Devitt, Simon and Van Meter, Rodney},
  journal = {New Journal of Physics},
  volume = {14},
  number = {12},
  pages = {123011},
  year = {2012}
}

@inproceedings{cheung2008design,
  title={On the design and optimization of a quantum polynomial-time attack on elliptic curve cryptography},
  author={Cheung, Donny and Maslov, Dmitri and Mathew, Jimson and Pradhan, Dhiraj K},
  booktitle={Workshop on Quantum Computation, Communication, and Cryptography},
  pages={96--104},
  year={2008},
  organization={Springer}
}

@article{gossett1998quantum,
  title={Quantum carry-save arithmetic},
  author={Gossett, Phil},
  journal={arXiv preprint quant-ph/9808061},
  year={1998}
}

@article{draper2004logarithmic,
  title={A logarithmic-depth quantum carry-lookahead adder},
  author={Draper, Thomas G and Kutin, Samuel A and Rains, Eric M and Svore, Krysta M},
  journal={arXiv preprint quant-ph/0406142},
  year={2004}
}

@article{sklansky1960conditional,
  title={Conditional-sum addition logic},
  author={Sklansky, Jack},
  journal={IRE Transactions on Electronic computers},
  number={2},
  pages={226--231},
  year={1960},
  publisher={IEEE}
}

@article{breuckmann2026cups,
  title={Cups and Gates I: Cohomology Invariants and Logical Quantum Operations},
  author={Breuckmann, Nikolas P and Davydova, Margarita and Eberhardt, Jens N and Tantivasadakarn, Nathanan},
  journal={Communications in Mathematical Physics},
  volume={407},
  number={5},
  pages={86},
  year={2026},
  publisher={Springer},
  doi={10.1007/s00220-026-05570-z},
  eprint={2410.16250},
  archivePrefix={arXiv}
}

@article{breuckmann2021quantum,
  title={Quantum low-density parity-check codes},
  author={Breuckmann, Nikolas P and Eberhardt, Jens Niklas},
  journal={PRX quantum},
  volume={2},
  number={4},
  pages={040101},
  year={2021},
  publisher={APS}
}

@article{orus2014practical,
  title={A practical introduction to tensor networks: Matrix product states and projected entangled pair states},
  author={Or{\'u}s, Rom{\'a}n},
  journal={Annals of physics},
  volume={349},
  pages={117--158},
  year={2014},
  publisher={Elsevier}
}

@misc{openai_chatgpt_2026,
  author       = {{OpenAI}},
  title        = {{ChatGPT, GPT-5.5}},
  year         = {2026},
  howpublished = {\url{https://chat.openai.com/}},
  note         = {Large language model.}
}

@misc{anthropic_claude_2026,
  author       = {{Anthropic}},
  title        = {{Claude Opus 5.5}},
  year         = {2026},
  howpublished = {\url{https://claude.ai/}},
  note         = {Large language model.}
}

@article{cross2024improved,
  title={Improved QLDPC surgery: Logical measurements and bridging codes},
  author={Cross, Andrew W and He, Zhiyang and Rall, Patrick J and Yoder, Theodore J},
  journal={arXiv preprint arXiv:2407.18393},
  year={2024}
}

@article{he2025extractors,
  title={Extractors: QLDPC architectures for efficient Pauli-based computation},
  author={He, Zhiyang and Cowtan, Alexander and Williamson, Dominic J and Yoder, Theodore J},
  journal={arXiv preprint arXiv:2503.10390},
  year={2025}
}

@inproceedings{grover1996fast,
  title={A fast quantum mechanical algorithm for database search},
  author={Grover, Lov K},
  booktitle={Proceedings of the twenty-eighth annual ACM symposium on Theory of computing},
  pages={212--219},
  year={1996}
}

@article{anqiwills2026,
    title={Magic State Distillation via Codes over Binary Extension Fields},
    author={Gong, Anqi and Pattison, Christopher A. and Rall, Patrick and Wills, Adam},
    journal={arXiv preprint arXiv:2608.09727},
    eprint={2608.09727},
    archivePrefix={arXiv},
    year={2026}
}

@article{chen2026efficient,
  title={Efficient Magic State Cultivation for $\sqrt{T}$ Gates},
  author={Chen, I and Fonseca, Matheus da Silva and Sornborger, Andrew},
  journal={arXiv preprint arXiv:2606.10430},
  eprint={2606.10430},
  archivePrefix={arXiv},
  year={2026}
}

@article{kobayashi2026clifford,
  title={Clifford hierarchy stabilizer codes: transversal non-Clifford gates and magic states},
  author={Kobayashi, Ryohei and Zhu, Guanyu and Hsin, Po-Shen},
  journal={Physical Review Letters},
  volume={136},
  number={25},
  pages={250802},
  year={2026},
  publisher={APS}
}

@article{maslov2009depth,
  title={An {$O(m^2)$}-depth quantum algorithm for the elliptic curve discrete logarithm problem over {$GF(2^m)$}},
  author={Maslov, Dmitri and Mathew, Jimson and Cheung, Donny and Pradhan, Dhiraj K.},
  journal={Quantum Information and Computation},
  volume={9},
  number={7--8},
  pages={610--621},
  year={2009}
}

@article{gidney2018halving,
  title={Halving the cost of quantum addition},
  author={Gidney, Craig},
  journal={Quantum},
  volume={2},
  pages={74},
  year={2018}
}

@article{jones2013multilevel,
  title={Multilevel distillation of magic states for quantum computing},
  author={Jones, Cody},
  journal={Physical Review A},
  volume={87},
  number={4},
  pages={042305},
  year={2013}
}

@article{gidney2019efficient,
  title={Efficient magic state factories with a catalyzed $|CCZ\rangle\to2|T\rangle$ transformation},
  author={Gidney, Craig and Fowler, Austin G.},
  journal={Quantum},
  volume={3},
  pages={135},
  year={2019},
  doi={10.22331/q-2019-04-30-135}
}

@phdthesis{steinberg2026towards,
  author  = {Steinberg, Matthew A.},
  title   = {Towards Fault-Tolerant Implementation of Holographic Quantum Error Correction},
  school  = {Delft University of Technology},
  address = {Delft, The Netherlands},
  year    = {2026},
  month   = nov,
  type    = {{PhD} dissertation},
  doi     = {10.4233/uuid:57831325-397c-4af6-87be-f3c4b466b61a}
}

@inproceedings{kissinger2020Pyzx,
  author = {Kissinger, Aleks and van de Wetering, John},
  title = {{PyZX: Large Scale Automated Diagrammatic Reasoning}},
  year = {2020},
  booktitle = {{\rm Proceedings 16th International Conference on} Quantum Physics and Logic, {\rm Chapman University, Orange, CA, USA., 10-14 June 2019}},
  editor = {Coecke, Bob and Leifer, Matthew},
  series = {Electronic Proceedings in Theoretical Computer Science},
  volume = {318},
  pages = {229-241},
  publisher = {Open Publishing Association},
  doi = {10.4204/EPTCS.318.14}
}

@article{berezutskii2025tensor,
  title={Tensor networks for quantum computing},
  author={Berezutskii, Aleksandr and Liu, Minzhao and Acharya, Atithi and Ellerbrock, Roman and Gray, Johnnie and Haghshenas, Reza and He, Zichang and Khan, Abid and Kuzmin, Viacheslav and Lyakh, Dmitry and others},
  journal={Nature Reviews Physics},
  volume={7},
  number={10},
  pages={581--593},
  year={2025},
  publisher={Nature Publishing Group UK London}
}

@article{zhu2023non,
  title={Non-Clifford and parallelizable fault-tolerant logical gates on constant and almost-constant rate homological quantum LDPC codes via higher symmetries},
  author={Zhu, Guanyu and Sikander, Shehryar and Portnoy, Elia and Cross, Andrew W and Brown, Benjamin J},
  journal={arXiv preprint arXiv:2310.16982},
  year={2023}
}

@article{schotte2022quantum,
  title={Quantum error correction thresholds for the universal Fibonacci Turaev-Viro code},
  author={Schotte, Alexis and Zhu, Guanyu and Burgelman, Lander and Verstraete, Frank},
  journal={Physical Review X},
  volume={12},
  number={2},
  pages={021012},
  year={2022},
  publisher={APS}
}

@article{williamson2026low,
  title={Low-overhead fault-tolerant quantum computation by gauging logical operators},
  author={Williamson, Dominic J and Yoder, Theodore J},
  journal={Nature Physics},
  volume={22},
  pages={598--603},
  year={2026},
  doi={10.1038/s41567-026-03220-8},
  eprint={2410.02213},
  archivePrefix={arXiv},
  publisher={Nature Publishing Group UK London}
}

@article{swaroop2024universal,
  title={Universal adapters between quantum LDPC codes},
  author={Swaroop, Esha and Jochym-O'Connor, Tomas and Yoder, Theodore J},
  journal={arXiv preprint arXiv:2410.03628},
  year={2024}
}

@article{cowtan2026parallel,
  title={Parallel logical measurements via quantum code surgery},
  author={Cowtan, Alexander and He, Zhiyang and Williamson, Dominic J and Yoder, Theodore J},
  journal={PRX Quantum},
  volume={7},
  number={2},
  pages={020325},
  year={2026},
  publisher={APS}
}

@article{peham2026synthesis,
  title={Synthesis and Optimization of Encoding Circuits for Fault-Tolerant Quantum Computation},
  author={Peham, Tom and Steinberg, Matthew and Wille, Robert and Heu{\ss}en, Sascha},
  journal={arXiv preprint arXiv:2605.15266},
  year={2026}
}

@article{perlin2026quantum,
  title={Quantum codes from classical annealing},
  author={Perlin, Michael A. and Steinberg, Matthew and Criger, Ben},
  journal={arXiv preprint arXiv:2607.19574},
  year={2026}
}
